\documentclass[amsmath,amssymb,nofootinbib,prd]{revtex4}
\pdfoutput=1

\usepackage{graphicx}
\usepackage{epsfig}
\usepackage{rotate}
\usepackage{amsmath}
\usepackage{amssymb}
\usepackage{amsfonts}
\usepackage{enumerate}
\usepackage{afterpage}
\usepackage{color}

\newcommand{\ud}{\mathrm{d}}
\newcommand{\p}{\partial}
\newcommand{\cH}{\mathcal{H}}
\newcommand{\Perp}{\mathcal{P}}
\newcommand{\Tr}{\rm Tr}

\newcommand{\Det}{\rm det}

\newcommand{\Em}{\mathcal{E}_m}

\begin{document}

\title{Observed galaxy number counts on the lightcone up to second order: II. Derivation}

\author{Daniele Bertacca$^{a}$, Roy Maartens$^{a,b}$, Chris Clarkson$^{c}$\\~}

\affiliation{
$^a$Physics Department, University of the Western
Cape, Cape Town 7535, South Africa\\
$^b$Institute of Cosmology \& Gravitation, University of
Portsmouth,
Portsmouth PO1 3FX, UK \\
$^c$Centre for Astrophysics, Cosmology \& Gravitation, and, Department of Mathematics \& Applied Mathematics, University of Cape Town, Cape Town 7701, South Africa}

\begin{abstract}

We present a detailed derivation of the observed galaxy number over-density on cosmological scales up to second order in perturbation theory. 
We include all relativistic effects that arise from observing on the past lightcone.  
The derivation is in a general gauge, and applies to all dark energy models (including interacting dark energy) and to metric theories of modified gravity. 
The result will be important for accurate cosmological parameter estimation, including non-Gaussianity, since all projection effects need to be taken into account. 
It also offers the potential for new probes of General Relativity, dark energy and modified gravity. 
This paper accompanies Paper I which presents the key results for the concordance model in Poisson gauge. 

\end{abstract}

\date{\today}

\maketitle



\section{Introduction}

The Newtonian prediction for the galaxy number over-density is accurate only on small scales. On cosmological scales, relativistic effects alter the observed number over-density through projection onto our past lightcone. This gives the well-known corrections from redshift space distortions and gravitational lensing convergence, but there are further Doppler, Sachs-Wolfe, integrated SW and time-delay type terms. The full relativistic effects have been calculated to first order in perturbation theory by~\cite{Yoo:2009au, Yoo:2010ni, Bonvin:2011bg, Challinor:2011bk, Jeong:2011as}. 
The nature, importance and possible detectability of the relativistic corrections to the Newtonian approximation at first order have been considered by~\cite{Bruni:2011ta}--\cite{Raccanelli:2013gja}.

We derive  the formula for the observed galaxy number over-density up to second order on cosmological scales. The result in Poisson gauge and for the concordance model is presented in the companion Paper I \cite{Bertacca:2014dra}. Paper I also considers the implications of the result for cosmological observables.  Here we  give the detailed derivation both in a general gauge and in Poisson gauge, for a flat Roberston-Walker background which allows for general dark energy models, including those where dark energy interacts non-gravitationally with cold dark matter. These interacting models have momentum exchange between dark energy and cold dark matter, which can lead to a velocity bias between galaxies and cold dark matter. Our results allow for velocity bias.


Our results also apply to metric theories of modified gravity as an alternative to dark energy, since we do not impose any field equations (in particular, we do not assume that the gauge invariant metric perturbations $\Phi$ and $\Psi$ are equal). 
The main result contains all relativistic effects up to second order that arise from observing on the past light cone, including all redshift and lensing distortions -- due to convergence, shear and rotation -- and all contributions from velocities, SW, ISW and time-delay terms (for example, see \cite{Pyne:1995bs, Mollerach:1997up}). This is related to the second-order perturbations of the cosmological distance-redshift relation \cite{Umeh:2012pn}--\cite{Clarkson:2014pda}, and to the weak lensing shear up to second order \cite{Bernardeau:2009bm,Bernardeau:2011tc}.

The second-order effects that we derive, especially those involving integrals along the line of sight, may make a non-negligible contribution to the observed number counts. This could be important for removing bias  on parameter estimation in precision cosmology with galaxy surveys. Recently, it has been shown that second-order effects on the distance-redshift relation induce a shift in the background that may have a significant effect on estimates of the Hubble constant and matter density parameter \cite{Ben-Dayan:2014swa,Clarkson:2014pda}.

We follow the ``cosmic rulers" approach of  \cite{Jeong:2011as,Schmidt:2012ne}, generalizing it from first to second order.
Consequently,  we use only the observed redshift $z$ in our analysis. In particular, all background quantities are evaluated at the observed, not background, redshift. Thus we do not need to identify the perturbations of redshift (these are derived in full detail up to second order by  \cite{Umeh:2012pn,Umeh:2014ana}). We neglect magnification bias, leaving this for future work  \cite{Bertacca:2014hwa}.

The paper is organised as follows:
in Section~\ref{Sec:CosmicRulers} we briefly review and generalize the cosmic rulers from first to second order,
and apply it to obtain the second-order perturbations of galaxy number counts in redshift space.
We perturb a flat Robertson-Walker universe in a general gauge in Section~\ref{Sec:GeneralGauge} and specialize to Poisson Gauge in Section~\ref{Sec:PoissonGauge}.
In Section~\ref{Sec:GalaxyBias} we describe how to relate the fluctuations of galaxy number density to the underlying matter density fluctuation $\delta_m$.
Finally, Section~\ref{Sec:Conclusions} is devoted to conclusions. 

The derivation of the second-order solutions is careful to include all steps needed for an independent verification of the results. It is therefore lengthy and technical.  
\begin{itemize}
\item For readers who want to understand the derivation of the general formula, see Section~\ref{Sec:CosmicRulers} and Section~\ref{Sec:GeneralGauge}. The main results are Eqs. \eqref{maindel} and \eqref{maindel2}.
\item For readers who are happy  to skip the proof and look at the main result for galaxy number count fluctuations at second order in the Poisson gauge -- see Paper I for the concordance model, or Section \ref{Sec:PoissonGauge} for general dark energy and modified gravity models. The main results are Eqs. (\ref{Poiss-Deltag-2}) and Eq. (\ref{Poiss-Deltag-3}).
 
\end{itemize}
Conventions: units $c=G = 1$;  signature is $(-, +, +, +)$; Greek indices run over $0, 1, 2, 3$, and Latin over $1, 2, 3$.\\

{\em Note added in version 4:} In the previous version 3 (and in the published paper), we mistakenly omitted some terms which arise from the integration of a first-order quantity taking into account perturbations of the direction of the null geodesic (so-called post-Born terms). This error has been corrected here and in the companion paper 1405.4403v5. Our results are now in agreement, in the appropriate limit, with those of \cite{Nielsen:2016ldx}.

\section{Cosmic rulers and the observed overdensity}
\label{Sec:CosmicRulers}

The cosmic rulers formalism of \cite{Jeong:2011as,Schmidt:2012ne} provides a map between redshift-space and real-space, without introducing a metric. Here we generalize it to second order.  We denote quantities in the redshift frame with a bar. 

\begin{figure}[!htbp]
\centering
\includegraphics[width=9 cm]{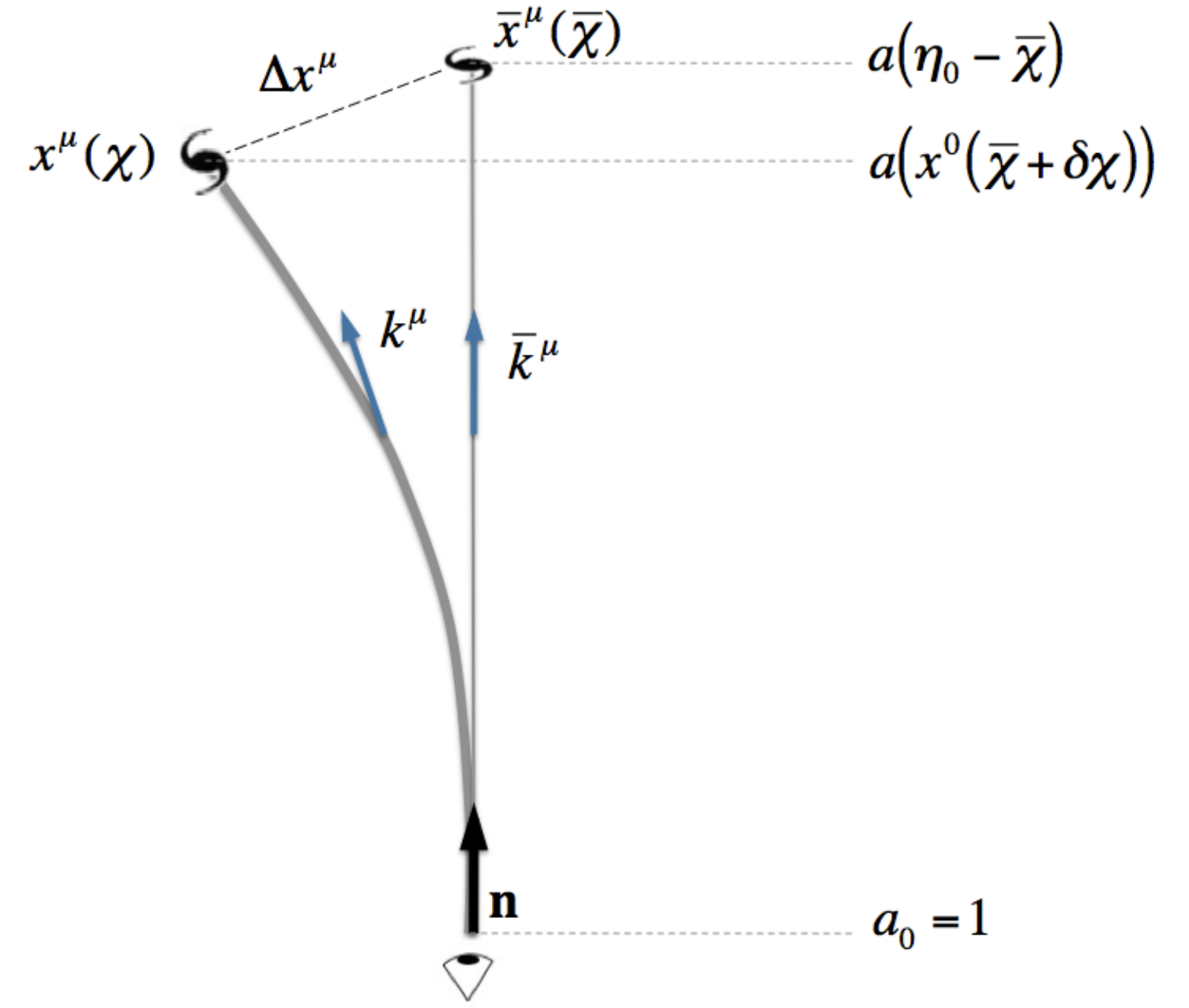}
\caption{
The real-space and redshift-space views (adapted from \cite{Schmidt:2012ne}). 
\label{fig:1}}
\end{figure}

Redshift-space is the ``cosmic laboratory'' where we probe the observations. We perform perturbations in real-space and not in redshift-space. In redshift-space we use coordinates which effectively flatten our past lightcone so that
the photon geodesic from an observed galaxy has the following conformal space-time coordinates (see Fig. \ref{fig:1}):
\begin{equation}
\bar{x}^\mu=(\bar \eta,\; \bar {\bf x})=(\eta_0-\bar \chi, \; \bar \chi \, {\bf n}).
\end{equation}
Here $\bar \chi(z)$ is the comoving distance to the observed redshift in redshift-space, calculated in the background , and ${\bf n}$ is the observed direction to the  galaxy, 
$n^i=\bar x^i/\bar \chi=\delta^{ij} (\p \bar \chi/\p \bar x^j)$. 
Using $\bar \chi$ as an  affine parameter, the total derivative along the past light cone is
\begin{equation}
\frac{\ud }{\ud \bar \chi} = - \frac{\p }{\p \bar \eta} + n^i \frac{\p}{\p \bar x^i} \;.
\end{equation}

To map from redshift-space to real-space (the ``physical frame"), we introduce coordinates $x^\mu=x^\mu(\chi)$, where $\chi$ is the physical comoving distance of the source  (see Fig. \ref{fig:1}). 
Then, to second order
 \begin{eqnarray}
\label{xph}
x^\mu (\chi) = \bar{x}^\mu (\chi)+ \delta x^\mu (\chi) \quad {\rm where} \quad \delta x^\mu (\chi)= \delta x^{\mu (1)}  (\chi)+\frac{1}{2}\delta x^{\mu (2)}  (\chi).
\end{eqnarray}
We map the real-space frame to the redshift frame perturbatively via
\begin{eqnarray}
\label{xph2}
x^\mu (\chi) = \bar{x}^\mu (\bar \chi)+ \Delta x^\mu (\bar \chi) \quad {\rm where} \quad \Delta x^\mu (\bar \chi)= \Delta x^{\mu (1)} (\bar \chi)+\frac{1}{2}\Delta x^{\mu (2)} (\bar \chi).
\end{eqnarray}
The physical comoving distance to the source is a perturbation about the value in the redshift frame:
\begin{eqnarray}
\label{chi}
\chi = \bar \chi+ \delta \chi  \quad {\rm where} \quad \delta \chi= \delta \chi^{(1)} +\frac{1}{2}\delta \chi ^{(2)}\,.
\end{eqnarray}
Then we have
 \begin{eqnarray}
\label{xph3}
x^\mu ( \chi) &=& \bar{x}^\mu (\chi)+ \delta x^{\mu (1)}  (\chi)+\frac{1}{2}\delta x^{\mu (2)}  (\chi) \nonumber \\
                            &=& \bar{x}^\mu (\bar \chi)+ \frac{\ud  \bar{x}^\mu }{\ud \bar \chi} \delta \chi^{(1)}+
                                     \delta x^{\mu (1)} (\bar \chi) +\frac{1}{2}\frac{\ud  \bar{x}^\mu }{\ud \bar \chi} \delta \chi^{(2)}(\bar \chi)   +
                                     \frac{1}{2}\frac{\ud^2  \bar{x}^\mu }{\ud \bar \chi^2} \left(\delta \chi^{(1)}\right)^2  
                            + \frac{\ud \delta x^{\mu (1)}}{\ud \bar \chi} \delta \chi^{(1)} +  \frac{1}{2} \delta x^{\mu (2)} (\bar \chi)\;,
\end{eqnarray}
which implies
\begin{eqnarray}
\label{Deltax1}
\Delta x^{\mu (1)} (\bar \chi) &=&  \frac{\ud  \bar{x}^\mu }{\ud \bar \chi} \delta \chi^{(1)}+ \delta x^{\mu (1)} (\bar \chi) \\
\label{Deltax2}
\Delta x^{\mu (2)} (\bar \chi)&=& \frac{\ud  \bar{x}^\mu }{\ud \bar \chi} \delta \chi^{(2)}   + \frac{\ud^2  \bar{x}^\mu }{\ud \bar \chi^2} \left(\delta \chi^{(1)}\right)^2  + 2 \frac{\ud \delta x^{\mu (1)}}{\ud \bar \chi} \delta \chi^{(1)} +  \delta x^{\mu (2)} (\bar \chi)\;.
\end{eqnarray}

The photon 4-momentum is
\begin{equation}
\label{pmu}
p^\mu=\frac{\nu (a)}{a} k^\mu
\end{equation}
where $a$ is the scale factor, $\nu \propto 1/a $ is the frequency in a homogeneous and isotropic space-time, and $k^\mu$ is a null geodesic vector. In the redshift frame
\begin{equation}
\label{kmu-0}
 \bar{k}^\mu=\frac{\ud  \bar{x}^\mu }{\ud \bar \chi}=\left(-1, \; {\bf n} \right)\;,
\end{equation}
while the physical $k^\mu$ evaluated at $\bar \chi$ is 
\begin{equation}
\label{kmu}
 k^\mu(\bar \chi) = \frac{\ud  x^\mu }{\ud \bar \chi}(\bar \chi)= \frac{\ud }{\ud \bar \chi}  \left(\bar{x}^\mu + \delta x^\mu\right)(\bar \chi) = \left(-1+\delta \nu^{(1)}+\frac{1}{2}\delta \nu^{(2)},\; n^i+\delta n^{i (1)}+\frac{1}{2}\delta n^{i (2)} \right)(\bar \chi).
\end{equation}
For $\mu=0$ we have 
\begin{eqnarray}
\label{Deltax0-1}
\Delta x^{0 (1)} (\bar \chi) &=& - \delta \chi^{(1)}+ \delta x^{0 (1)}  \\
\label{Deltax0-2}
\Delta x^{0 (2)} (\bar \chi)&=& - \delta \chi^{(2)}   +2 \, \delta \nu^{(1)} \delta \chi^{(1)} +  \delta x^{0 (2)} \;,
\end{eqnarray}
where $\ud \delta x^{0 (n)}/ \ud \bar \chi =  \delta \nu^{(n)}$, and for $\mu=i$, 
\begin{eqnarray}
\label{Deltaxi-1}
\Delta x^{i (1)} (\bar \chi) &=& n^i  \delta \chi^{(1)} + \delta x^{i (1)}  \\
\label{Deltaxi-2}
\Delta x^{i (2)} (\bar \chi)&=& n^i \delta \chi^{(2)}  +2\,\delta n^{i (1)} \delta \chi^{(1)} + \delta x^{i (2)} \;.
\end{eqnarray}
where $\ud \delta x^{i (n)}/ \ud \bar \chi =  \delta n^{i (n)}$.
From Eq. (\ref{kmu}), we obtain explicitly 
\begin{eqnarray}
\label{deltax0-n}
\delta x^{0 (n)}(\bar \chi)= \int_0^{\bar \chi} \ud \tilde{\chi} \; \delta \nu^{(n)} (\tilde \chi)\;,
\\
\label{deltaxi-n}
\delta x^{i (n)}(\bar \chi)= \int_0^{\bar \chi} \ud \tilde{\chi} \; \delta n^{i(n)} (\tilde \chi)\;,
\end{eqnarray}
where we have imposed the boundary conditions at the observer: $\delta x^{0 (n)}_o=0$ and $\delta x^{i (n)}_o=0$.


\subsection{The scale factor}

In real-space the scale factor is 
\begin{equation}
a = a(x^0(\chi))=a(\bar x^0+ \Delta x^0)=\bar a \left[ 1 + \cH   \Delta x^{0(1)} +\frac{1}{2} \cH  \Delta x^{0(2)}
+ \frac{1}{2} (\cH' + \cH^2) \left( \Delta x^{0(1)} \right)^2\right]\;,
\end{equation}
where $\bar a = a (\bar x^0)$, prime is $\p/\p \bar x^0 = \p/\p \bar \eta$ and $\cH=\bar a'/\bar a$.
Defining
\begin{equation}
\label{a}
\frac{a}{\bar a} = 1 + \Delta \ln a^{(1)} + \frac{1}{2} \Delta \ln a^{(2)}\;, 
\end{equation}
we find
\begin{eqnarray}
\label{DeltaLna-1}
\Delta \ln a^{(1)} &=&  \cH \,  \Delta x^{0(1)} =  \cH  \left(  - \delta \chi^{(1)}+ \delta x^{0 (1)} \right)
\\
\label{DeltaLna-2}
 \Delta \ln a^{(2)} &=&  (\cH' + \cH^2) \left(\Delta x^{0 (1)} \right)^2 + \cH  \Delta x^{0(2)} 
                                                 = \frac{(\cH' + \cH^2)}{\cH^2} \left( \Delta \ln a^{(1)} \right)^2 + \cH  \Delta x^{0(2)} \nonumber \\
                                                 &=& (\cH' + \cH^2) \left(- \delta \chi^{(1)}+ \delta x^{0 (1)} \right)^2 -\cH \delta \chi^{(2)}   +2 \, \cH \delta \nu^{(1)} \delta \chi^{(1)} + \cH \delta x^{0 (2)} \;.
\end{eqnarray}

\subsection{Four-vectors and tetrads}

The galaxy four-velocity can be given as
\begin{eqnarray}
u^\mu = \frac{\ud x^\mu}{\ud s}=\frac{\ud x^{\hat{\alpha}}}{\ud s}\Lambda^\mu_{\hat{\alpha}}=u^{\hat{\alpha}}\Lambda^\mu_{\hat{\alpha}}\;,
\end{eqnarray}
where $s$ is proper time and $\Lambda^\mu_{\hat \alpha}$ is an orthonormal tetrad. If we choose $u^\mu$ as the timelike basis vector, then 
\begin{equation}
\label{E0mu}
u_\mu= \Lambda_{\hat{0} \mu}=a \, E_{\hat{0} \mu} \quad \quad {\rm and}  \quad \quad u^\mu=\Lambda^\mu_{\hat 0}= a^{-1}E_{\hat{0}}^\mu\;,
\end{equation}
where $E^\mu_{\hat{\alpha}}$ is the tetrad in the comoving frame.
In the background 
\begin{equation}
 E_{\hat{0}\mu}^{(0)}=(-1, {\bf 0})\;,
\end{equation}
and perturbing, we obtain
\begin{eqnarray}
\label{E^mu_hata}
E_{\hat{0}\mu}\left(x^\nu(\chi)\right)=E_{\hat{0}\mu}\left( \bar x^\nu(\bar \chi)+\Delta x^\nu\right)= E_{\hat{0}\mu}^{(0)}(\bar \chi)+E_{\hat{0}\mu}^{(1)}(\bar \chi)+\left(\frac{\p E_{\hat{0}\mu}}{\p \bar x^\nu}\right)^{(1)}\,\Delta x^{\nu (1)}+\frac{1}{2}E_{\hat{0}\mu}^{(2)}(\bar \chi)\;.
\end{eqnarray}

From $k^\mu$, the map from redshift to real-space is given by
\begin{eqnarray}
\label{pertkmu}
k^{\mu}\left(\chi\right)= \frac{\ud x^{\mu}(\chi)}{\ud \chi}=k^{\mu}\left(\bar \chi+\delta \chi \right)&=& k^{\mu(0)}(\bar \chi)+k^{\mu (1)}(\bar \chi)+\left(\frac{\ud k^{\mu}}{\ud \bar \chi}\right)^{(1)} \, \delta \chi^{(1)}+\frac{1}{2}k^{\mu (2)}(\bar \chi)  \;,
\end{eqnarray}
where $k^{\mu(0)}(\bar \chi)=\bar k^\mu$.

\subsection{The observed redshift}

The observed redshift is given by 
\begin{eqnarray}
(1+z)\big|_{\chi} = \frac{(u_\mu p^\mu)\big|_{\chi} }{(u_\mu p^\mu)|_o}=\frac{\nu(\chi)}{\nu_o}\frac{(E_{\hat{0}\mu} k^\mu)\big|_{\chi} }{(E_{\hat{0}\mu} k^\mu)|_o}=\frac{a_o}{a(\chi)}\frac{(E_{\hat{0}\mu} k^\mu)\big|_{\chi} }{(E_{\hat{0}\mu} k^\mu)|_o}\;,
\end{eqnarray}
where we used $\nu \propto 1/a$. Quantities evaluated at the observer have a subscript o, while other quantities are assumed to be evaluated at the emitter (we suppress a subscript e for convenience). Choosing\footnote{Another possibility is the generalization $a_o \neq \bar a_o = a(\bar x^0)=1$. Then
$a_o=a(\eta)|_o=1 + \Delta \ln a^{(1)}_o + \Delta \ln a^{(2)}_o/2$,
and we should  add $-\Delta \ln a^{(1)}_o$  to the right side of Eq.\ (\ref{Ek-1}) and $-\Delta \ln a^{(2)}_o/2$  to the right side of Eq.\ (\ref{Ek-2}). In  order to obtain $\Delta \ln a^{(1)}_o$ and $\Delta \ln a^{(2)}_o/2$  we can follow the prescription used in  \cite{Jeong:2013psa} (which computes  $a_o$ only at first order). In our case, i.e. when we assume  $a_o = 1$, by construction  we automatically have  $\Delta \ln a^{(1)}_o=\Delta \ln a^{(2)}_o/2=0$.
} 
$a_o = 1$ and given\footnote{From Eq.\ (\ref{pmu}), for $\bar \chi=0$ we have
$p_{\hat{0} o} =( \Lambda_{\hat{0} \mu}p^\mu)|_o=\nu_o\;,~
p_{\hat{a}  o} = (\Lambda_{\hat{a} \mu}p^\mu)|_o=n_{\hat{a}} \nu_o$.
} 
\begin{eqnarray}
\label{Ek_o}
(E_{\hat{0}\mu} k^\mu)|_o=1 \;,
\end{eqnarray}
then
\begin{eqnarray}
1+z=\frac{E_{\hat{0}\mu} k^\mu}{a}\;.
\end{eqnarray}

From Eq.\ (\ref{a}), $\bar a$ is the scale factor in redshift-space. Then $\bar a = 1/(1+z)$. From Eqs.\ (\ref{a}), (\ref{E^mu_hata}) and (\ref{pertkmu}), we get 
\begin{eqnarray}
\label{Ek-total}
1=\frac{1+ (E_{\hat{0}\mu} k^\mu)^{(1)}+\frac{1}{2}(E_{\hat{0}\mu} k^\mu)^{(2)}}{ 1 + \Delta \ln a^{(1)} + \frac{1}{2} \Delta \ln a^{(2)}}\;,
\end{eqnarray}
where
\begin{eqnarray}
(E_{\hat{0}\mu} k^\mu)^{(0)} = 1\;.
\end{eqnarray}
Then from Eq.\ (\ref{Ek-total}) we can find $\Delta \ln a^{(1)}$ and $\Delta \ln a^{(2)}$, such that\footnote{In Eq.\ (\ref{Ek-2}) we have used 
$k^{\nu (0)} [ {\p E_{\hat{0}\mu}}/{\p \bar x^\nu}]^{(1)} = ({\ud\bar{x}^{\nu}}/{\ud \bar \chi}) [ {\p E_{\hat{0}\mu}}/{\p \bar x^\nu}]^{(1)} =   [  {\ud E_{\hat{0}\mu}}/{\ud \bar \chi} ]^{(1)}$.
}
\begin{eqnarray}
\label{Ek-1}
\Delta \ln a^{(1)} &=& (E_{\hat{0}\mu} k^\mu)^{(1)} = E_{\hat{0}\mu}^{(1)} k^{\mu (0)}+E_{\hat{0}\mu}^{(0)} k^{\mu (1)}=- E_{\hat{0}0}^{(1)} + n^i E_{\hat{0}i}^{(1)} - \delta \nu^{(1)}\;,
\\
 \Delta \ln a^{(2)} &=& (E_{\hat{0}\mu} k^\mu)^{(2)} =
2 E_{\hat{0}\mu}^{(1)} k^{\mu (1)}+E_{\hat{0}\mu}^{(2)} k^{\mu (0)}+E_{\hat{0}\mu}^{(0)} k^{\mu (2)}
+ 2 \delta \chi^{(1)} E_{\hat{0}\mu}^{(0)}\left(\frac{\ud k^{\mu}}{\ud \bar \chi}\right)^{(1)}+ 2 \delta \chi^{(1)} k^{\mu (0)}  \left(\frac{\ud E_{\hat{0}\mu}}{\ud \bar \chi}\right)^{(1)} \nonumber \\
&+& 2 k^{\mu (0)}  \left( \frac{\p E_{\hat{0}\mu}}{\p \bar x^\nu}\right)^{(1)}  \delta x^{\nu (1)} 
\nonumber \\
&= & 2 E_{\hat{0}\mu}^{(1)} k^{\mu (1)}+E_{\hat{0}\mu}^{(2)} k^{\mu (0)}
+ E_{\hat{0}\mu}^{(0)} k^{\mu (2)}+2 k^{\mu (0)} \left( \frac{\p E_{\hat{0}\mu}}{\p \bar x^\nu}\right)^{(1)}   \delta x^{\nu (1)} +2 \delta \chi^{(1)} \frac{\ud}{\ud \bar \chi} \Delta \ln a^{(1)}  \nonumber \\
&=& 2 E_{\hat{0}0}^{(1)} \delta \nu^{ (1)}+ 2  E_{\hat{0}i}^{(1)} \delta n^{i (1)}-  \delta \nu^{ (2)}- E_{\hat{0}0}^{(2)} +  n^i E_{\hat{0}i}^{(2)} +2 \left[- \left( \frac{\p E_{\hat{0}0}}{\p \bar x^\nu}\right)^{(1)} + n^i \left( \frac{\p E_{\hat{0}i}}{\p \bar x^\nu}\right)^{(1)}  \right]  \delta x^{\nu (1)}
 \nonumber \\ \label{Ek-2}
&+& 2  \delta \chi^{(1)} \frac{\ud}{\ud \bar \chi} \Delta \ln a^{(1)} \;.
\end{eqnarray}

Using Eqs.\ (\ref{DeltaLna-1}) and (\ref{Ek-1}) at first order, and Eqs.\ (\ref{DeltaLna-2}) and (\ref{Ek-2})  at second order we obtain 
\begin{eqnarray}
\label{chi_1}
\delta \chi^{(1)} & = &  \delta x^{0 (1)} - \frac{\Delta \ln a^{(1)}}{\cH}= \delta x^{0 (1)}- \Delta x^{0 (1)} \;,\\ 
\label{chi_2}
\delta \chi^{(2)} & = &-\frac{1}{\cH} \Delta \ln a^{(2)}+ \frac{(\cH' + \cH^2)}{\cH^3} \left( \Delta \ln a^{(1)} \right)^2 - \frac{2}{\cH}\delta \nu^{(1)} \Delta \ln a^{(1)} +2 \delta \nu^{(1)} \delta x^{0 (1)}+  \delta x^{0 (2)}\;.
\end{eqnarray}
Given $\Delta \ln a^{(1)}$ from Eq.\ (\ref{Ek-1}), and $\Delta \ln a^{(2)}$ from Eq.\ (\ref{Ek-2}), it is useful to rewrite Eqs.\ (\ref{DeltaLna-1}) and (\ref{DeltaLna-2}) as
\begin{eqnarray}
\label{Dx0_1}
\Delta x^{0(1)} & = & \frac{\Delta \ln a^{(1)}}{ \cH }\;, \\
\label{Dx0_2}
\Delta x^{0(2)}& = &\frac{1}{\cH} \Delta \ln a^{(2)}-  \frac{(\cH' + \cH^2)}{\cH^3} \left( \Delta \ln a^{(1)} \right)^2  \;.
\end{eqnarray}

\subsection{Number density}

The physical number density of galaxies $n_g$ as a function of physical comoving coordinates $x^\mu$ is defined
by the  observed number of galaxies contained within a volume $\mathcal{\bar V}$:
 \begin{equation}
\mathcal{N}=
\int_{ \mathcal{\bar V}}  \sqrt{-g(x^\alpha)}\: n_g(x^\alpha)\: \varepsilon_{\mu\nu\rho\sigma}
u^\mu(x^\alpha) \frac{\partial x^\nu}{\partial \bar x^1} \frac{\partial x^\rho}{\partial \bar x^2} \frac{\partial x^\sigma}{\partial \bar x^3}
 \ud^3\bar {\bf x}\;,
\end{equation}
where 
$\varepsilon_{\mu\nu\rho\sigma}$ is the Levi-Civita tensor, and 
 $u^\mu$ is the four velocity vector as a function of comoving location.
In the redshift frame 
\begin{equation}
\mathcal{N} =
\int_{ \mathcal{\bar V}} {\bar a^3(\bar x^0) \,  n_g\left(\bar x^0,\bar {\bf x}\right)} \, \ud^3\bar {\bf x} \;,
\end{equation}
so that
\begin{equation}
\label{eqng}
 {a(\bar x^0)^3 \, n_g (\bar x^0, \bar{\bf x})} = \sqrt{-g(x^\alpha)}\: n_g(x^\alpha)\: \varepsilon_{\mu\nu\rho\sigma}
u^\mu(x^\alpha) \frac{\partial x^\nu}{\partial \bar x^1} \frac{\partial x^\rho}{\partial \bar x^2}\frac{\partial x^\sigma}{\partial \bar x^3} \;.
\end{equation}

Using the cosmic rulers defined in the previous section we expand all quantities on the left side of Eq. (\ref{eqng})  in the observed coordinates, i.e. in the redshift frame.
Using Eq.\ (\ref{E0mu}), 
\begin{equation}
\label{eqng2}
 \sqrt{-\hat g(x^\alpha)}\: a^3(x^0) n_g(x^\alpha)\; 
 \varepsilon_{\mu\nu\rho\sigma} \, E_{\hat 0}^\mu(x^\alpha) \frac{\partial x^\nu}{\partial \bar x^1} \frac{\partial x^\rho}{\partial \bar x^2}\frac{\partial x^\sigma}{\partial \bar x^3} \;,
\end{equation}
where $\hat g$ is the determinant of the comoving metric $\hat g_{\mu \nu}=g_{\mu \nu}/a^2$. We separate Eq. (\ref{eqng2}) into three parts:
\begin{equation}
\label{dV}
{\rm (1)} ~ \sqrt{-\hat g(x^\alpha)}\;,\quad \quad {\rm (2)}~ \varepsilon_{\mu\nu\rho\sigma} \, E_{\hat 0}^\mu(x^\alpha) \frac{\partial x^\nu}{\partial \bar x^1} \frac{\partial x^\rho}{\partial \bar x^2}\frac{\partial x^\sigma}{\partial \bar x^3} \quad \quad {\rm and \quad (3)}~ a^3(x^0) n_g(x^\alpha)\;.\nonumber
\end{equation}
Then\footnote{For a rank two tensor $\mathbb{M}$,
\begin{eqnarray*}
\mathnormal{M} &=& \Det{(\mathbb{M})}=\mathnormal{M}^{(0)}+\mathnormal{M}^{(1)}+\mathnormal{M}^{(2)}/2\;, \quad {\rm where}  \quad  \mathnormal{M}^{(0)}=\Det{(\mathbb{M}^{(0)})}\;, \quad \quad \mathnormal{M}^{(1)} = \mathnormal{M}^{(0)} \Tr[\mathbb{M}^{(0)-1}\mathbb{M}^{(1)}] \;, \nonumber\\
\mathnormal{M}^{(2)} &=&  \mathnormal{M}^{(0)}\{ [{\mathnormal{M}^{(1)}}/{\mathnormal{M}^{(0)}}]^{2}  -   \Tr [(\mathbb{M}^{(0)-1}\mathbb{M}^{(1)})(\mathbb{M}^{(0)-1}\mathbb{M}^{(1)})]+  \Tr[\mathbb{M}^{(0)-1}\mathbb{M}^{(2)}] \} \;.
\end{eqnarray*}
}:
\begin{itemize}
\item[(1)] Splitting $ \sqrt{-\hat g(x^\alpha)}$ as 
\begin{eqnarray}
\sqrt{-\hat g(x^\alpha)}=\sqrt{-\hat g (x^\alpha)}^{\,(0)}+\delta\sqrt{-\hat g (x^\alpha)}^{\,(1)}+\frac{1}{2}\delta\sqrt{-\hat  g (x^\alpha)}^{\,(2)}
\end{eqnarray}
we have
\begin{eqnarray}
\delta\sqrt{-\hat g (x^\alpha)}^{\,(1)}&=&\frac{1}{2} \sqrt{-\hat g (x^\alpha)}^{\,(0)}\hat g_\mu^{\mu (1)}(x^\alpha) \;,\\
\delta \sqrt{-\hat g (x^\alpha)}^{\,(2)}&=& \frac{1}{2} \sqrt{-\hat g (x^\alpha)}^{\,(0)}\left(\frac{1}{2} \hat g_\mu^{\mu (1)} \hat g_\nu^{\nu (1)}+ \hat g_\mu^{\mu (2)}-\hat g_\mu^{\nu (1)}\hat g_\nu^{\mu (1)}\right)(x^\alpha)\;.
\end{eqnarray}
Here $\hat g_\nu^{\mu (n)} = \hat g^{\mu\sigma (0)} \hat g_{\sigma \nu}^{(n)}$ and $ \hat g_{\mu\nu}= \hat g_{\mu\nu}^{(0)} + \hat g_{\mu\nu}^{(1)}+ \hat g_{\mu\nu}^{(2)}/2$.

Mapping all these terms from real- to redshift-space, we find
\begin{eqnarray}
\sqrt{-\hat g (x^\alpha)}^{\,(0)}&=&1\;,\\
\delta\sqrt{-\hat g (x^\alpha)}^{\,(1)}&=&\delta\sqrt{-\hat g (\bar x^\alpha)}^{\,(1)} + \left(\frac{\p}{\p x^\nu}\delta\sqrt{-\hat g (\bar x^\alpha)}\right)^{(1)} \Delta x^{\nu (1)} \nonumber\\ &=&\frac{1}{2} \hat g_\mu^{\mu (1)}(\bar x^\alpha)+\frac{1}{2}  \left(\frac{\p  \hat g_\mu^{\mu}}{\p \bar x^\nu} \right)^{(1)} (\bar x^\alpha) \; \Delta x^{\nu (1)}\;, \\
\delta \sqrt{-\hat g (x^\alpha)}^{\,(2)}&=& \delta \sqrt{-\hat g (\bar x^\alpha)}^{\,(2)}\;.
\end{eqnarray}
Rewriting
\begin{eqnarray}
\sqrt{-\hat g(x^\alpha)}=1+\Delta\sqrt{-\hat g (\bar x^\alpha)}^{\,(1)}+\frac{1}{2}\Delta\sqrt{-\hat  g (\bar x^\alpha)}^{\,(2)}\;,
\end{eqnarray}
we find
\begin{eqnarray}
\label{deltaSqrtg-1}
\Delta\sqrt{-\hat g (\bar x^\alpha)}^{\,(1)}&=&\frac{1}{2} \hat g_\mu^{\mu (1)}(\bar x^\alpha) \;,\\
\label{deltaSqrtg-2}
\Delta \sqrt{-\hat g (\bar x^\alpha)}^{\,(2)}&=& 
\frac{1}{4} \hat g_\mu^{\mu (1)} (\bar x^\alpha) \; \hat g_\nu^{\nu (1)} (\bar x^\alpha)+\frac{1}{2} \hat g_\mu^{\mu (2)} (\bar x^\alpha)-\frac{1}{2}\hat g_\mu^{\nu (1)} (\bar x^\alpha) \; \hat g_\nu^{\mu (1)}  (\bar x^\alpha) + \left(\frac{\p  \hat g_\mu^{\mu}}{\p \bar x^\nu} \right)^{(1)} (\bar x^\alpha) \;  \Delta x^{\nu (1)}. 
\end{eqnarray}
\item[(2)] We write Eq. (\ref{dV}) as
\begin{eqnarray}
\label{dV2}
\varepsilon_{\mu\nu\rho\sigma} \, E_{\hat 0}^\mu(x^\alpha) \frac{\p x^\nu}{\p \bar x^1} \frac{\p x^\rho}{\p \bar x^2}\frac{\p x^\sigma}{\p \bar x^3}=E_{\hat 0}^{0} (x^\alpha) \; \left|\frac{\p {\bf x}}{\p \bar{\bf x}}\right| + E_{\hat 0}^{i}(x^\alpha) \; \Sigma_i \;,
\end{eqnarray}
where
\begin{eqnarray} 
\label{Sigma_i}
 \left|\frac{\p {\bf x}}{\p \bar{\bf x}}\right|  = \Det\left(\frac{\p {\it x}^{\it i}}{\p  {\it  \bar{x}}^{\it j}}\right)\quad \quad {\rm and}\quad \quad \Sigma_{\it i} = \epsilon_{\it ijk} \left(-\frac{\partial  {\it  x}^0}{\partial  {\it  \bar x}^1} \frac{\partial  {\it  x}^{\it j}}{\partial  {\it  \bar x}^2}\frac{\partial  {\it  x}^{\it k}}{\partial  {\it  \bar x}^3} + \frac{\partial  {\it  x}^{\it j}}{\partial  {\it  \bar x}^1} \frac{\partial  {\it  x}^0}{\partial  {\it  \bar x}^2}\frac{\partial  {\it  x}^{\it k}}{\partial  {\it  \bar x}^3} -\frac{\partial  {\it  x}^{\it j}}{\partial  {\it  \bar x}^1} \frac{\partial  {\it  x}^{\it k}}{\partial  {\it  \bar x}^2}\frac{\partial  {\it  x}^0}{\partial  {\it  \bar x}^3}\right) \;.
\end{eqnarray}
The first term on the right of  Eq.\ (\ref{dV2}) is
\begin{eqnarray}
E_{\hat 0}^{0} (x^\alpha) \; \left|\frac{\p {\bf x}}{\p \bar{\bf x}}\right|  = \left[E_{\hat 0}^{0(0)} (x^\alpha)+E_{\hat 0}^{0(1)} (x^\alpha)+ \frac{1}{2}E_{\hat 0}^{0(2)} (x^\alpha)\right]\left[ \left|\frac{\p {\bf x}}{\p \bar{\bf x}}\right|^{(0)}+ \left|\frac{\p {\bf x}}{\p \bar{\bf x}}\right|^{(1)} + \frac{1}{2} \left|\frac{\p {\bf x}}{\p \bar{\bf x}}\right|^{(2)} \right]\;,
\end{eqnarray}
where 
\begin{eqnarray}
E_{\hat 0}^{0(0)}&=&1\; \quad \quad E_{\hat 0}^{0(1)} (x^\alpha)=E_{\hat 0}^{0(1)} (\bar x^\alpha) + \left(\frac{\p E_{\hat 0}^0 }{\p \bar x^\mu}\right)^{(1)} (\bar x^\alpha) \; \Delta x^{\mu (1)}\;, \\
 \left|\frac{\p {\bf x}}{\p \bar{\bf x}}\right|^{(0)} &=& 1\;,  \quad \quad  \left|\frac{\p {\bf x}}{\p \bar{\bf x}}\right|^{(1)} =  \left(\frac{\p \Delta x^i }{\p \bar x^i}\right)^{(1)}\;, \\
\left|\frac{\p {\bf x}}{\p \bar{\bf x}}\right|^{(2)}  &=& \left(\frac{\p \Delta x^i }{\p  \bar x^i}\right)^{(1)}\left(\frac{\p \Delta x^j }{\p  \bar x^j}\right)^{(1)}- \left(\frac{\p \Delta x^i }{\p  \bar x^j}\right)^{(1)}\left(\frac{\p \Delta x^j }{\p  \bar x^i}\right)^{(1)}+ \left(\frac{\p \Delta x^i }{\p  \bar x^i}\right)^{(2)}\;.
\end{eqnarray}
The second term on the right of Eq.\ (\ref{dV2})  is
\begin{eqnarray}
 E_{\hat 0}^{i}(x^\alpha)\; \Sigma_i = \left[E_{\hat 0}^{i(0)} (x^\alpha)+E_{\hat 0}^{i(1)} (x^\alpha)+ \frac{1}{2}E_{\hat 0}^{i(2)} (x^\alpha)\right]\left[\Sigma_i^{(0)}+\Sigma_i^{(1)} + \frac{1}{2} \Sigma_i^{(2)} \right]\;,
\end{eqnarray}
where 
\begin{eqnarray}
E_{\hat 0}^{i(0)}&=&0\;  \quad \quad  E_{\hat 0}^{i(1)} (x^\alpha)=E_{\hat 0}^{i(1)} (\bar x^\alpha) + \left(\frac{\p E_{\hat 0}^i }{\p  \bar x^\mu}\right)^{(1)} (\bar x^\alpha)  \; \Delta x^{\mu (1)}\;,\\
\Sigma_i^{(0)} &=& n_i\;,  \quad \quad  \Sigma_i^{(1)} =  - \left(\frac{\p \Delta x^0 }{\p \bar x^i}\right)^{(1)} + \epsilon_{\it ijr}  \epsilon^{\it pqr} n_p \left(\frac{\p \Delta x^j }{\p \bar x^q}\right)^{(1)}\;.
\end{eqnarray}
Here we do not compute $\Sigma_i^{(2)}$ because $E_{\hat 0}^{i(0)}=0$.

\item[(3)] From Eq. (\ref{a}), 
\begin{equation}
\label{a3}
a^3 = \bar a^3 \left[1 + 3 \; \Delta \ln a^{(1)} + 3 \left(\Delta \ln a^{(1)}\right)^2 + \frac{3}{2} \Delta \ln a^{(2)}\right]\;.
\end{equation}
Writing $n_g(x^\alpha)= n_g^{(0)}(x^0) +  n_g^{(1)}(x^\alpha) +n_g^{(2)}(x^\alpha)/2 $, we have
\begin{eqnarray}
n_g^{(0)}(x^0)  = n_g^{(0)}(\bar x^0 + \Delta x^0) =  \bar n_g (\bar x^0) +  \frac{\p \bar n_g }{\p \bar x^0} \, \Delta x^{0(1)} + \frac{1}{2}  \frac{\p^2 \bar n_g }{{\p \bar x^0}^2}  \left(\Delta x^{0(1)} \right)^2+   \frac{1}{2} \frac{\p \bar n_g }{\p \bar x^0}  \, \Delta x^{0(2)}\;,
\end{eqnarray}
where $n_g^{(0)}(\bar x^0)=\bar n_g (\bar x^0)$ and
\begin{eqnarray}
n_g^{(1)}(x^\alpha)  = n_g^{(1)}(\bar x^\alpha + \Delta x^{\alpha(1)}) =  n_g^{(1)}(\bar x^\alpha) +  \left(\frac{\p  n_g }{\p \bar x^\alpha}\right)^{(1)}  \Delta x^{\alpha(1)}\;,  \quad \quad  \frac{1}{2} n_g^{(2)}(x^\alpha)=\frac{1}{2} n_g^{(2)}(\bar x^\alpha) \;.
\end{eqnarray}
Defining $\delta_g^{(n)}= n_g^{(n)}(\bar x^\alpha)/\bar n_g  (\bar x^0)$ and considering Eqs.\ (\ref{Dx0_1}) and (\ref{Dx0_2}),  we have
\begin{eqnarray}
n_g(x^\alpha) & = &  \bar n_g \left\{1 + \frac{\ud \ln \bar n_g }{\ud  \ln \bar a} \, \Delta \ln a^{(1)} + \delta_g^{(1)}  + \left(\frac{\p \delta_g }{\p \bar x^\mu}\right)^{(1)}  \Delta x^{\mu(1)} + \frac{\ud \ln \bar n_g }{\ud  \ln \bar a}  \, \Delta \ln a^{(1)} \, \delta_g^{(1)}  \right. \nonumber \\
                          & + &     \left. \frac{1}{2} \left[- \frac{\ud \ln \bar n_g }{\ud  \ln \bar a} + \left(\frac{\ud \ln \bar n_g }{\ud  \ln \bar a} \right)^2+\frac{\ud^2 \ln \bar n_g }{{\ud  \ln \bar a}^2}\right] \left( \Delta \ln a^{(1)}\right)^2+     \frac{1}{2}  \frac{\ud \ln \bar n_g }{\ud  \ln \bar a} \, \Delta \ln a^{(2)}
                         +   \frac{1}{2}  \delta_g^{(2)}  \right\} \;.
\end{eqnarray}
\end{itemize}

At this point, it is useful to define the parallel and perpendicular projection operators to the observed line-of-sight direction (see also \cite{Jeong:2011as,Schmidt:2012ne}). For any  spatial vectors and tensors:
\begin{eqnarray}
\label{Projection}
A_{\parallel} = n^{i} n^{j} A_{ij } \;, \quad
B_{\perp}^i =  \Perp^{ij} B_j = B^i -n^i B_{\parallel}\;,  \quad 
A_{\perp} =  \Perp^{ij}  A_{ij }\;, \quad \Perp^i_{j}= \delta^{i}_j-n^in_j\:.
\end{eqnarray}
The directional derivatives are defined as
\begin{eqnarray}
\label{Projection2}
\p_\parallel = n^i  {\p \over \p \bar x^i}\;,  \quad \quad \p^2_\parallel  = \p_\parallel \p_\parallel,\quad \quad  \p_{\perp i} =  \Perp^j_i \p_j= {\p \over \p \bar x^i} -  n_i \p_\parallel\;,
\end{eqnarray}
and we have 
\begin{eqnarray}
\label{Projection3}
{\p n^j \over \p \bar x^i} &=& \frac{1}{\bar \chi}\Perp_i^j\;,   \quad \quad  \quad \quad n^i B_{\perp i} = 0\;, \quad \quad  \quad  \quad n^i \p_{\perp i}=0\;, \nonumber \\
\frac{\ud}{\ud \bar \chi} \p_{\perp}^i &=&  \p_{\perp}^i  \frac{\ud}{\ud \bar \chi}-\frac{1}{\bar \chi} \p_{\perp}^i\;, \quad \quad  \quad  \p_{\perp i} \left(n^k \Perp^i_j\right)=\frac{1}{\bar \chi}\left(\delta^k_j-3n^kn_j\right)\;,\nonumber \\
 {\p B^i \over   \p \bar x^j} &=& n^i n_j \p_{\parallel} B_{\parallel} + n^i \p_{\perp j} B_{\parallel} +\p_{\perp j} B_{\perp}^i+ n_j \p_{\parallel} B_{\perp}^i +\frac{1}{\bar \chi} \Perp^i_j B_{\parallel}\;,\nonumber \\
{\rm and}& &\quad \quad  \nabla^2_\perp = \p_{\perp i}\p_\perp^i =\delta^{ij} {\p \over   \p \bar x^i} {\p  \over   \p \bar x^j}- \partial_\parallel^2 - \frac{2}{\bar \chi} \p_\parallel\;.
\end{eqnarray}
We also have
\begin{eqnarray} 
\label{Projection4}
\Tr [\bar \p {\it B}] &=&  {\p B^i \over   \p \bar x^i} = \p_{\perp i} B_{\perp}^i+  \p_{\parallel} B_{\parallel} + \frac{2}{\bar \chi} B_{\parallel}  \\
\Tr [(\bar \p {\it B})^2] &=& \left({\p B^i \over   \p \bar x^j}\right)  \left( {\p B^j \over   \p \bar x^i} \right)  = (\p_{\parallel} B_{\parallel})(\p_{\parallel} B_{\parallel})+ (\p_{\perp j} B_{\perp}^i)(\p_{\perp i} B_{\perp}^j) - \frac{2}{\bar \chi} B_{\perp i}  \p_{\parallel} B_{\perp}^i+ \frac{2}{\bar \chi} B_{\parallel} \p_{\perp i} B_{\perp}^i \nonumber \\
&+& 2 (\p_{\parallel} B_{\perp}^i) (\p_{\perp i} B_{\parallel}) + \frac{2}{\bar \chi^2} B_{\parallel} B_{\parallel}\;.
\end{eqnarray}
Using Eq.\ (\ref{Projection}), we find $\Delta x^{i(1)}=n^i \Delta x_{\|}^{(1)}+ \Perp^i_j  \Delta x^{j(1)} =n^i \Delta x_{\|}^{(1)}+\Delta x_\perp^{i(1)} $ where
\begin{eqnarray} 
\label{Dx_||-1}
\Delta x_{\parallel}^{(1)} &=&  \delta \chi^{(1)}+ \delta x_{\parallel}^{(1)}=  \delta x^{0 (1)} - \frac{\Delta \ln a^{(1)}}{\cH} + \delta x_{\parallel}^{(1)}=  \delta x^{0 (1)} + \delta x_{\parallel}^{(1)}-\Delta x^{0 (1)}  \;, \\
\label{Dx_perp-1}
\Delta x_{\perp}^{i(1)} &=& \delta x_{\perp}^{i(1)}\;.
\end{eqnarray}
For $\Delta x^{i(2)}=n^i \Delta x_{\|}^{(2)}+ \Perp^i_j  \Delta x^{j(2)} =n^i \Delta x_{\|}^{(2)}+\Delta x_\perp^{i(2)} $  we have 
\begin{eqnarray} 
\label{Dx_||-2}
\Delta x_{\parallel}^{(2)}&=& \delta \chi^{(2)} + 2 \delta n_{\parallel}^{(1)} \delta \chi^{(1)} +  \delta x_{\parallel}^{(2)}= -\frac{1}{\cH} \Delta \ln a^{(2)}+ \frac{(\cH' + \cH^2)}{\cH^3} \left( \Delta \ln a^{(1)} \right)^2 - \frac{2}{\cH}\delta \nu^{(1)} \Delta \ln a^{(1)} \nonumber \\
&&+2 \delta \nu^{(1)} \delta x^{0 (1)}+   \delta x^{0 (2)}  +2 \delta n_{\parallel}^{(1)}   \delta x^{0 (1)} -  \frac{2}{\cH} \delta n_{\parallel}^{(1)} \Delta \ln a^{(1)}  +   \delta x_{\parallel}^{(2)} \nonumber \\
&=& \delta x^{0 (2)} +  \delta x_{\parallel}^{(2)} -  \Delta x^{0 (2)}+ 2 \left(\delta \nu^{(1)}+ \delta n_{\parallel}^{(1)} \right) \delta\chi^{(1)}\;, \\
\label{Dx_perp-2}
\Delta x_{\perp}^{i(2)}&=&2 \delta n_{\perp}^{i (1)} \delta \chi^{(1)}  + \delta x_{\perp}^{i(2)} = 2 \delta n_{\perp}^{i(1)}   \delta x^{0 (1)} -  \frac{2}{\cH} \delta n_{\perp}^{i(1)} \Delta \ln a^{(1)} +  \delta x_{\perp}^{i(2)}  \;.
\end{eqnarray}

We define the  parallel and perpendicular parts of the tetrad in the comoving frame\footnote{Note that in general $E_{\hat \alpha}^{i}$ is not a  3-space tensor in the index $i$, so that  $E_{\hat 0}^\| \neq E_{\hat 0 \|}$ and $E_{\hat 0}^{\perp i} \neq \delta^{ij} E_{\hat 0 \perp j}$.
}:
\begin{eqnarray}
E_{\hat \alpha}^i &=&n^i  E_{\hat \alpha}^\| +E_{\hat \alpha}^{\perp i}\;,  \quad \quad {\rm where}  \quad \quad E_{\hat \alpha}^\| = n_i E_{\hat \alpha}^i  \quad \quad {\rm and}  \quad \quad E_{\hat \alpha}^{\perp i} = \Perp^i_j E_{\hat \alpha}^j \;,\\
E_{\hat \alpha i} &=& n_i  E_{\hat \alpha \|} +E_{\hat \alpha \perp i}\;,  \quad \quad {\rm where}  \quad \quad E_{\hat \alpha \|} = n^i E_{\hat \alpha i}  \quad \quad {\rm and}  \quad \quad E_{\hat \alpha \perp i} = \Perp^j_i E_{\hat \alpha j} \;.
\end{eqnarray}

Taking into account Eqs.\ (\ref{Projection}), (\ref{Projection2}),  (\ref{Projection3}) and (\ref{Projection4})  and observing that  $\epsilon_{\it ijr}  \epsilon^{\it pqr}=(\delta^p_i \delta^q_j - \delta^p_j \delta^q_i )$, we  obtain the observed fractional number over-density
\begin{eqnarray}
\label{Delta_g}
\Delta_g=\frac{ n_g (\bar x^0, \bar{\bf x}) - \bar n_g(\bar x^0)}{ \bar n_g(\bar x^0)} = \Delta_g^{(1)} + \frac{1}{2}\Delta_g^{(2)}\;,
\end{eqnarray}
where
\begin{eqnarray}
\label{Deltag-1}
\Delta_g^{(1)} &=&  \delta_g^{(1)} + \frac{1}{2} \hat g_\mu^{\mu (1)} + b_e \, \Delta \ln a^{(1)}  +   \p_{\parallel} \Delta x_{\parallel}^{(1)}  + \frac{2}{\bar \chi} \Delta x_{\parallel}^{(1)}  - 2 \kappa^{(1)} +  E_{\hat 0}^{0(1)} +  E_{\hat 0}^{ \parallel(1)} \;,   \\
\label{Deltag-2}
 \Delta_g^{(2)}&=&  \delta_g^{(2)} + \frac{1}{2} \hat g_\mu^{\mu (2)} +   b_e \, \Delta \ln a^{(2)} +  \p_{\parallel} \Delta x_{\parallel}^{(2)}  + \frac{2}{\bar \chi} \Delta x_{\parallel}^{(2)} - 2\kappa^{(2)}  +   E_{\hat 0}^{0 (2)}   +  E_{\hat 0 }^{\| (2)}  \nonumber \\
 &+& \frac{1}{4} \hat g_\mu^{\mu (1)} \; \hat g_\nu^{\nu (1)}  -\frac{1}{2}\hat g_\mu^{\nu (1)}  \; \hat g_\nu^{\mu (1)}  +    \frac{1}{\cH} \hat g_\mu^{\mu (1)}{'} \Delta \ln a^{(1)} + \left(\p_{\parallel}\hat g_\mu^{\mu (1)}\right) \Delta x_{\parallel}^{(1)} +   \left(\p_{\perp i}\hat g_\mu^{\mu (1)}\right) \Delta x_{\perp}^{i (1)} \nonumber \\
& +&   \left(- b_e + b_e^2 +  \frac{\ud \ln b_e }{\ud  \ln \bar a} \right) \left( \Delta \ln a^{(1)}\right)^2 +  2 b_e \, \Delta \ln a^{(1)}  \delta_g^{(1)} +  \frac{2}{\cH}  \delta_g^{(1)}{'} \Delta \ln a^{(1)}  + 2 \p_{\parallel}\delta_g^{(1)} \Delta x_{\parallel}^{(1)} \nonumber \\
& +&  2 \p_{\perp}^{i} \delta_g^{(1)}  \Delta x_{\perp i}^{(1)}  +4   \left(\kappa^{(1)}\right)^2 + \frac{2}{\bar \chi^2} \left( \Delta x_{\parallel}^{(1)} \right)^2 -4 \kappa^{(1)}  \p_{\parallel} \Delta x_{\parallel}^{(1)} -  \frac{4}{\bar \chi} \kappa^{(1)}  \Delta x_{\parallel}^{(1)} +   \frac{4}{\bar \chi}  \Delta x_{\parallel}^{(1)} \p_{\parallel} \Delta x_{\parallel}^{(1)}\nonumber \\
 & -&    \left( \p_{\perp j}   \Delta x_{\perp}^{i (1)}  \right)  \left( \p_{\perp i}   \Delta x_{\perp}^{j (1)}  \right) + \frac{2}{\bar \chi}  \Delta x_{\perp i}^{(1)} \p_{\parallel}   \Delta x_{\perp}^{i (1)} - 2\left( \p_{\parallel}   \Delta x_{\perp}^{i (1)}  \right)  \left( \p_{\perp i}   \Delta x_{\parallel}^{(1)}  \right)+  \frac{2}{\cH} E_{\hat 0}^{0 (1)}{'} \Delta \ln a^{(1)}  \nonumber \\
 &+&   2\p_{\parallel} E_{\hat 0}^{0 (1)} \Delta x_{\parallel}^{(1)} + 2 \p_{\perp i}E_{\hat 0}^{0 (1)}\Delta x_{\perp}^{i (1)} + 2 E_{\hat 0}^{0 (1)}  \p_{\parallel} \Delta x_{\parallel}^{(1)} - 4 E_{\hat 0}^{0 (1)}  \kappa^{(1)} +  \frac{4}{\bar \chi} E_{\hat 0}^{0 (1)}   \Delta x_{\parallel}^{(1)}+  \frac{2}{\cH} E_{\hat 0}^{\| (1)}{'} \Delta \ln a^{(1)}  \nonumber \\
 & +&   2\p_{\parallel} E_{\hat 0 }^{\| (1)} \Delta x_{\parallel}^{(1)} +  2 \p_{\perp i}E_{\hat 0}^{\|(1)}\Delta x_{\perp}^{i (1)} 
  -2E_{\hat 0}^{\|(1)} \p_{\parallel}  \Delta x^{0 (1)} -2E_{\hat 0 }^{\perp i (1)} \p_{\perp i}  \Delta x^{0 (1)}-4  E_{\hat 0 }^{\| (1)}  \kappa^{(1)} +  \frac{4}{\bar \chi} E_{\hat 0}^{\| (1)}  \Delta x_{\parallel}^{(1)}  \nonumber \\
 & -& 2 E_{\hat 0}^{\perp i (1)}   \p_{\perp i} \Delta x_{\parallel}^{(1)}
   + b_e \, \hat g_\mu^{\mu (1)} \Delta \ln a^{(1)} +  \hat g_\mu^{\mu (1)}  \delta_g^{(1)} +2 \left( \frac{1}{2}  \hat g_\mu^{\mu (1)}+ b_e  \Delta \ln a^{(1)} + \delta_g^{(1)} \right) \nonumber \\
  &\times& \left( \p_{\parallel}\Delta x_{\parallel}^{(1)} - 2   \kappa^{(1)} + \frac{2}{\bar \chi} \Delta x_{\parallel}^{(1)} +E_{\hat 0}^{0 (1)} +  E_{\hat 0}^{\| (1)}   \right)\nonumber \\
 &=&  \delta_g^{(2)} + \frac{1}{2} \hat g_\mu^{\mu (2)} +   b_e \, \Delta \ln a^{(2)} +  \p_{\parallel} \Delta x_{\parallel}^{(2)}  + \frac{2}{\bar \chi} \Delta x_{\parallel}^{(2)} - 2\kappa^{(2)}  +  E_{\hat 0}^{0 (2)}   + E_{\hat 0 }^{\| (2)} +\left(\Delta_g^{(1)} \right)^2  \nonumber \\
 &-& \frac{1}{2}\hat g_\mu^{\nu (1)}  \; \hat g_\nu^{\mu (1)} +    \frac{1}{\cH} \hat g_\mu^{\mu (1)}{'} \Delta \ln a^{(1)} +   \left(\p_{\parallel}\hat g_\mu^{\mu (1)}\right) \Delta x_{\parallel}^{(1)} + \left(\p_{\perp i}\hat g_\mu^{\mu (1)}\right) \Delta x_{\perp}^{i (1)} +  \frac{2}{\cH}  \delta_g^{(1)}{'} \Delta \ln a^{(1)}   \nonumber \\
 &+&  2 \p_{\parallel}\delta_g^{(1)} \Delta x_{\parallel}^{(1)} +2 \p_{\perp}^{i} \delta_g^{(1)}  \Delta x_{\perp i}^{(1)}  + \left(- b_e +  \frac{\ud \ln b_e }{\ud  \ln \bar a} \right) \left( \Delta \ln a^{(1)}\right)^2 - \frac{2}{\bar \chi^2} \left( \Delta x_{\parallel}^{(1)} \right)^2 + \frac{4}{\bar \chi} \kappa^{(1)}  \Delta x_{\parallel}^{(1)} \nonumber \\
  & -&   \left( \p_{\perp j}   \Delta x_{\perp}^{i (1)}  \right)  \left( \p_{\perp i}   \Delta x_{\perp}^{j (1)}  \right) + \frac{2}{\bar \chi}  \Delta x_{\perp i}^{(1)} \p_{\parallel}   \Delta x_{\perp}^{i (1)} - 2\left( \p_{\parallel}   \Delta x_{\perp}^{i (1)}  \right)  \left( \p_{\perp i}   \Delta x_{\parallel}^{(1)}  \right) +  \frac{2}{\cH} E_{\hat 0}^{0 (1)}{'} \Delta \ln a^{(1)}  \nonumber \\
   &+& 2 \p_{\parallel} E_{\hat 0}^{0 (1)} \Delta x_{\parallel}^{(1)}+  \frac{2}{\cH} E_{\hat 0}^{\| (1)}{'} \Delta \ln a^{(1)} + 2 \p_{\parallel} E_{\hat 0 }^{\| (1)} \Delta x_{\parallel}^{(1)} +  2 \p_{\perp i}\left( E_{\hat 0}^{0(1)}+ E_{\hat 0}^{\|(1)} \right)\Delta x_{\perp}^{i (1)}-   2 \left(\delta n_\|^{ (1)} +  \delta\nu^{(1)} \right)  E_{\hat 0}^{\| (1)}\nonumber \\
 &-&  2 E_{\hat 0 }^{\perp i (1)} \p_{\perp i} \left(  \Delta x^{0(1)}+ \Delta x_{\parallel}^{(1)} \right) - \left(\delta_g^{(1)}\right)^2 - \left(\p_{\parallel} \Delta x_{\parallel}^{(1)}\right)^2 - \left(E_{\hat 0}^{0(1)}+ E_{\hat 0}^{\|(1)} \right)^2\;.   
 \end{eqnarray}
Here
\begin{eqnarray}
b_e = \frac{\ud \ln \left(\bar a^3 \bar n_g\right)}{\ud \ln \bar a}
\end{eqnarray}
 is the evolution bias  term related to the comoving number density~\cite{Jeong:2011as}.
 Note that 
 \begin{eqnarray} 
 \label{partialparallep}
\p_{\parallel}  \Delta x^{\mu (n)} (\bar \chi, {\bf n})=   \p_{\bar \chi}  \Delta x^{\mu (n)} \;,
\end{eqnarray}
where $\p_{\bar \chi} $ is applied to all terms that are functions of $\bar x^0=\bar \eta(\bar \chi)$ and/ or $\bar x^i=\bar x^i (\bar \chi)$.

\subsection{Weak lensing terms }

In Eq. \eqref{Deltag-2}, we introduced the coordinate weak lensing convergence terms at order $n$:
 \begin{eqnarray}
 \label{kappa-n}
\kappa^{(n)}=- \frac{1}{2}  \p_{\perp i} \Delta x_{\perp}^{i (n)} .
\end{eqnarray}
Then  the second-order transverse part of the volume distortion, which appears in Eq.\ (\ref{Deltag-2}), is  
\begin{equation}
\label{volume-effect1}
- \kappa^{(2)}+ 2 \left(\kappa^{(1)}\right)^2-   \frac{1}{2}\left( \p_{\perp j}   \Delta x_{\perp}^{i (1)}  \right)  \left( \p_{\perp i}   \Delta x_{\perp}^{j (1)}  \right) \;.
\end{equation}
The coordinate weak lensing shear $\gamma$ and rotation $\vartheta$ do not contribute to the observed number counts at first order but quadratic products do contribute at second order, via the third term above.
They are defined by splitting
$ \p_{\perp i}   \Delta x_{\perp j}^{(1)} $ 
into its trace, tracefree and antisymmetric parts: 
\begin{equation}
\label{volumedistortion1}
   \p_{\perp i}   \Delta x_{\perp j}^{(1)}  =-\gamma_{ij}^{(1)}-\Perp_{ij}\kappa^{(1)}-\vartheta_{ij}^{(1)}\;,
\end{equation}
where
\begin{equation}
\label{shear}
\gamma_{ij}^{(1)}=- \p_{\perp (i}   \Delta x_{\perp j)}^{(1)}-\Perp_{ij}\kappa^{(1)},~~
\vartheta_{ij}^{(1)}=-\p_{\perp [i}   \Delta x_{\perp j]}^{(1)}.
\end{equation}
 Then    
\begin{equation}
\label{volume-effect}
\left( \p_{\perp j}   \Delta x_{\perp}^{i (1)}  \right)  \left( \p_{\perp i}   \Delta x_{\perp}^{j (1)}  \right) = 2 \left(\kappa^{(1)}\right)^2 +2 \big|\gamma^{(1)}\big|^2 - \vartheta_{ij}^{(1)}\vartheta^{ij(1)} \;,
\end{equation}
where $2|\gamma^{(1)}|^2=\gamma_{ij}^{(1)}\gamma^{ij (1)} $. 
Explicit expressions for $\gamma_{ij}^{(1)}$  and $\vartheta_{ij}^{(1)}\vartheta^{ij(1)}$ are given in Eqs. \eqref{gamg} and \eqref{varg}  in a general gauge.


\section{Perturbed flat Robertson-Walker background in a general gauge}
\label{Sec:GeneralGauge}

The results obtained in the previous section have not yet used the specific form of the  metric.
Here we assume a spatially flat RW background, perturbed in a general gauge to second order:
 \begin{eqnarray} 
 \label{metric}
 \ud s^2 = a(\eta)^2\left[-\left(1 + 2A^{(1)}+A^{(2)}\right)\ud\eta^2-2\left(B_{i}^{(1)}+\frac{1}{2}B_{i}^{(2)}\right)\ud\eta\ud x^i+\left(\delta_{ij} +h_{ij}^{(1)}+\frac{1}{2}h_{ij}^{(2)}\right)\ud x^i\ud x^j\right] \;,
\end{eqnarray}
where $B^{i(n)}= \p_i B^{(n)} + \hat B_i^{(n)}$ and $ \hat B_i^{(n)}$ is a solenoidal vector, i.e. $\p^i\hat B_i^{(n)}=0$. The 3-tensor is
$h^{(n)}_{ij} = 2 D^{(n)} \delta_{ij} + F_{ij}^{(n)}$, where $F^{(n)}_{ij}= (\p_i\p_j- \delta_{ij}\nabla^2/3)F^{(n)}+\p_i \hat F^{(n)}_j+ \p_j \hat F^{(n)}_i+\hat h^{(n)}_{ij}$. Here $D^{(n)}$ and $F^{(n)}$ are scalars and $\hat F^{(n)}_i$ is a solenoidal vector field,   $\p^i \hat h^{(n)}_{ij}=\hat h_i^{i(n)}=0$.

Up to second order, the geodesic equation for $k^\mu$ is \cite{Pyne:1995bs}
\begin{eqnarray} 
\label{Eqphoton}
\frac{\ud k^\mu}{ \ud \bar \chi}+\left(\hat \Gamma^\mu_{\alpha \beta}+ \delta x^{\sigma} \frac{\partial \hat \Gamma^\mu_{\alpha \beta}}{\partial \bar x^{\sigma}}\right)k^\alpha k^\beta=0 \nonumber
\end{eqnarray}

where $\hat \Gamma^\mu_{\alpha \beta}$ are the Christoffel symbols defined using the comoving metric $\hat g_{\mu \nu}$.
At  zeroth order, we obtain Eq.\ (\ref{kmu-0}).
At first order, Eq.\ (\ref{Eqphoton}) yields 
\begin{eqnarray} 
\label{dnu-1}
\frac{\ud}{\ud\bar \chi} \left(\delta \nu^{(1)} - 2 A^{(1)} +  B_\|^{(1)}\right) &=& A^{(1)}{'} - B_\|^{(1)}{'} - \frac{1}{2} h_{\|}^{(1)}{'} \;,\\
\label{de-1}
\frac{\ud}{\ud\bar \chi} \left( \delta n^{i(1)} + B^{i(1)} + h_j^{i(1)} n^j \right) &=& 
- \p^i A^{(1)} + \p^i B_\|^{(1)}- \frac{1}{\bar \chi}B_{\perp}^{i(1)} + \frac{1}{2} \p^i h_\|^{(1)} - \frac{1}{\bar \chi} \Perp^{ij} h_{jk}^{(1)} n^k \;, 
\end{eqnarray}
in agreement with \cite{Schmidt:2012ne}.
At second order, we find
\begin{eqnarray} 
\label{dnu-2}
&&\frac{\ud}{\ud\bar \chi} \left[\delta \nu^{(2)} - 2 A^{(2)} +  B_\|^{(2)}+4A^{(1)} \delta \nu^{(1)}+2 B_i^{(1)} \delta n^{i(1)}  \right] =A^{(2)}{'} - B_\|^{(2)}{'} - \frac{1}{2} h_{\|}^{(2)}{'} + 2\delta n^{i(1)}  \bigg[\p_i \left(2  A^{(1)}- B_{\|}^{(1)}\right) \nonumber\\
&& -\left( B_i^{(1)}{'} +n^j h_{ij}^{(1)}{'} \right)+ \frac{1}{\bar \chi} B_{\perp i}^{(1)}  \bigg]   + 2 \left[\frac{\ud}{\ud \bar \chi}\left(2 {A^{(1)}}' -{B_\|^{(1)}}'\right)+  \left(A^{(1)}{''} - B^{(1)}_{\| }{''} - \frac{1}{2}h^{(1)}_{\| }{''} \right)    \right] \left(\delta x^{0(1)} + \delta x_{\|}^{(1)}\right)\nonumber \\
&& +2  \frac{\ud}{\ud \bar \chi}  \left[\frac{\ud}{\ud \bar \chi}\left(2 A^{(1)} - B^{(1)}_{\| }\right)+\left(A^{(1)}{'} - B^{(1)}_{\| }{'} - \frac{1}{2}h^{(1)}_{\| }{'} \right) \right] \delta x_{\|}^{(1)}+ 2 \Bigg\{\p_{\perp i}\bigg[\frac{\ud}{\ud \bar \chi}\left(2 A^{(1)} - B^{(1)}_{\| }\right)\nonumber \\
&&+\left(A^{(1)}{'} - B^{(1)}_{\| }{'} - \frac{1}{2}h^{(1)}_{\| }{'} \right) \bigg] + \frac{1}{\bar \chi} \left[ \frac{\ud B_{\perp i}^{(1)}}{\ud \bar \chi} - \p_{\perp i}\left(2 A^{(1)} - B^{(1)}_{\| }\right)+  B_{\perp i}^{(1)}{'} +  n^j h_{ j k}^{(1)}{'} \Perp_i^k - \frac{1}{\bar \chi} B_{\perp i}^{(1)} \right] \Bigg\} \delta x_{\perp}^{i (1)} \;,  
 \end{eqnarray} 
 \begin{eqnarray} 
\label{de-2}
&&\frac{\ud}{\ud\bar \chi} \left[ \delta n^{i(2)} + B^{i(2)} + h_j^{i(2)} n^j - 2 \delta \nu^{(1)}  B^{i(1)} + 2 \delta n^{j(1)}  h_j^{i(1)} \right] = 
- \p^i A^{(2)} + \p^i B_\|^{(2)}- \frac{1}{\bar \chi}B_{\perp}^{i(2)} + \frac{1}{2} \p^i h_\|^{(2)} - \frac{1}{\bar \chi} \Perp^{ij} h_{jk}^{(2)} n^k  \nonumber \\
&&+ 2 \delta \nu^{(1)}\left[2 \p^i A^{(1)} -B^{i(1)}{'}- \p^i B_{\|}^{(1)} +\frac{1}{\bar \chi} B_{\perp}^{i(1)} - n^j h_{j}^{i(1)}{'} \right] + 2 \delta n^{j(1)} \left[ \p^i B_j^{(1)} -\p_j B^{i(1)}+ n^k \p^i h_{jk}^{(1)} -  n^k  \p_j  h_k^{i(1)}  \right]\nonumber \\
 && -2   {\left[\p^i\left( A^{(1)} - B^{(1)}_{\| } - \frac{1}{2} h^{(1)}_{\| } \right)  + \frac{\ud}{\ud \bar \chi}\left( B^{i(1)}+  n^j h_{ j }^{i(1)} \right)  + \frac{1}{\bar \chi} \left(B_{\perp}^{i(1)} +   n^j h_{ j k}^{(1)} \Perp_i^k \right)\right]}'  \left(\delta x^{0(1)} + \delta x_{\|}^{(1)}\right)  \nonumber\\
  \nonumber\\
 &&-2 \frac{\ud}{\ud \bar \chi} \left[\p^i\left( A^{(1)} - B^{(1)}_{\| } - \frac{1}{2} h^{(1)}_{\| } \right)  + \frac{\ud}{\ud \bar \chi}\left( B^{i(1)}+  n^j h_{ j}^{i(1)} \right)  + \frac{1}{\bar \chi} \left(B_{\perp}^{i(1)} +   n^j h_{ j k}^{(1)}\Perp_i^k \right)\right] \delta x_{\|}^{(1)}
 \nonumber \\
 && -2 \Bigg\{ \p_{\perp l} \left[\p^i\left( A^{(1)} - B^{(1)}_{\| } - \frac{1}{2} h^{(1)}_{\| } \right)  + \frac{\ud}{\ud \bar \chi}\left( B^{i(1)}+  n^j h_{ j }^{i(1)} \right)  + \frac{1}{\bar \chi} \left(B_{\perp}^{i(1)} +   n^j h_{ j k}^{(1)}\Perp^{ik} \right)\right] \nonumber \\
 &&  + \frac{1}{\bar \chi} \Perp_l^j \left( \p^{i} B_{j}^{(1)} - \p_{j} B^{i(1)}  - \frac{\ud}{\ud \bar \chi}   h_{j}^{i(1)}  - n^k  \p_j h_{k}^{i(1)} +  n^k \p^i h_{jk}^{(1)}\right) \Bigg\}  \delta x_{\perp}^{l (1)} \;.
\end{eqnarray}

To solve Eqs.\  (\ref{dnu-1}),  (\ref{de-1}),  (\ref{dnu-2})  and (\ref{de-2}) we require the values of $\delta \nu^{(1)}$, $\delta \nu^{(2)}$, $\delta n^{i (1)}$ and  $\delta n^{i (2)}$ today. In this case we need all the components of the tetrads $\Lambda^{\hat \alpha}_\mu$ and $E^{\hat \alpha}_\mu$ (see Appendix \ref{A}, Eq.\ (\ref{LambdaE-1-2})) which are defined through the following  relations
\begin{eqnarray}
\label{LambdaE}
g^{\mu \nu}\Lambda^{\hat \alpha}_\mu \Lambda^{\hat \beta}_\nu &=& \eta^{\hat  \alpha \hat \beta}\;,  \quad  \quad  \eta_{\hat  \alpha \hat \beta} \Lambda^{\hat \alpha}_\mu \Lambda^{\hat \beta}_\nu = g_{\mu \nu}\;,  \quad  \quad g^{\mu \nu} \Lambda^{\hat \beta}_\nu = \Lambda^{\hat \beta \mu}\;,    \quad  \quad \eta_{\hat  \alpha \hat \beta}  \Lambda^{\hat \beta}_\nu  =  \Lambda_{\hat \beta \nu} \;,\nonumber\\
\hat g^{\mu \nu} E^{\hat \alpha}_\mu E^{\hat \beta}_\nu& = &\eta^{\hat  \alpha \hat \beta}\;, \quad  \quad  \eta_{\hat  \alpha \hat \beta} E^{\hat \alpha}_\mu E^{\hat \beta}_\nu = \hat g_{\mu \nu}\;,  \quad  \quad \hat g^{\mu \nu} E^{\hat \beta}_\nu = E^{\hat \beta \mu}\;,    \quad  \quad \eta_{\hat  \alpha \hat \beta}  E^{\hat \beta}_\nu  =  E_{\hat \beta \nu}  \;,
\end{eqnarray}
where $ \eta_{\hat  \alpha \hat \beta} $ the comoving Minkowski metric.
Using Eq. (\ref{Ek_o}) we have, at first order,
\begin{eqnarray}
\label{dnude-1o}
\delta\nu^{(1)}_o&=&A^{(1)}_o+v^{(1)}_{\| o}-B^{(1)}_{\| o}\;, \\
\delta n^{\hat a (1)}_o&=&-v^{\hat a (1)}_o-\frac{1}{2} n^i h^{\hat a (1)}_{i \, o} \;,
\end{eqnarray}
and, at second order,
\begin{eqnarray}
\label{dnude-2o}
  \delta\nu^{(2)}_o&=&  A^{(2)}_o+ v^{(2)}_{\| o}- B^{(2)}_{\| o}- 3 \left(A^{(1)}_o\right)^2 -2 v^{(1)}_{\| o} A^{(1)}_o +  4 B^{(1)}_{\| o} A^{(1)}_o -  v_{k \, o}^{(1)} v_o^{k (1)}+n^i h^{(1)}_{i k \, o}  v_o^{k (1)}\nonumber \\
&+& n^i h^{(1)}_{i k \, o}  B_o^{k (1)}  +2 B_{k \, o}^{(1)} v_o^{k (1)}  \;, \\
 \delta n^{\hat a (2)}_o&=&- v^{\hat a (2)}_o-\frac{1}{2} n^i h^{\hat a (2)}_{i \, o} +  v_o^{\hat a (1)}  v_{\| \, o}^{(1)}- v_o^{\hat a (1)}  B_{\| \, o}^{(1)} + B_o^{\hat a (1)}  v_{\| \, o}^{(1)} - B_o^{\hat a (1)}  B_{\| \, o}^{(1)} +  \frac{3}{4}   n^i h^{k(1)}_{i \, o}  h_{k \, o}^{\hat a (1)} \;.
\end{eqnarray}
From Eqs.\ (\ref{dnu-1}), (\ref{de-1}) and the constraint from Eq.\ (\ref{dnude-1o}), we obtain at first order
\begin{eqnarray}
\label{dnude-1}
\delta\nu^{(1)}&=&- \left (A^{(1)}_o-v^{(1)}_{\| \, o}\right)+ 2 A^{(1)} - B^{(1)}_{\| } + \int_0^{\bar \chi} \ud \tilde \chi \left(A^{(1)}{'} - B^{(1)}_{\| }{'} - \frac{1}{2}h^{(1)}_{\| }{'} \right) \nonumber \\
&=&- \left (A^{(1)}_o-v^{(1)}_{\| \, o}\right)+ 2 A^{(1)} - B^{(1)}_{\| } - 2I^{(1)} \;, \\
\delta n^{i (1)}&=&+ B^{i (1)}_{o}-v^{i (1)}_o+\frac{1}{2} n^j h^{i (1)}_{j \, o} - B^{i (1)}-n^j h^{i (1)}_{j} \nonumber \\
&&- \int_0^{\bar \chi} \ud \tilde \chi  \left[\tilde \p^i \left( A^{(1)} -   B^{(1)}_{\| } -\frac{1}{2}  h^{(1)}_{\| }  \right)+ \frac{1}{\tilde \chi}\left( B^{i (1)}_{\perp } + \Perp^{ij}h^{(1)}_{j k}  n^k\right) \right]\nonumber \\
&=& n^i \delta n_\|^{ (1)}+\delta n_\perp^{i (1)}\;,
\end{eqnarray}
where
\begin{eqnarray}
\label{dnue-||perp-1}
\delta n_\|^{ (1)}&=&A^{(1)}_o-v^{(1)}_{\| \, o}-A^{(1)}- \frac{1}{2}  h^{(1)}_{\| }+2I^{(1)}\;,   \\
\delta n_\perp^{i (1)}&=&  B^{i (1)}_{\perp \, o }-v^{i (1)}_{\perp \, o }+ \frac{1}{2} n^k h_{k\,o}^{j(1)} \Perp^i_j-\left( B^{i (1)}_{\perp }+ n^k h_{k}^{j(1)} \Perp^i_j\right) + 2S_{\perp}^{i(1)}  \;.
\end{eqnarray}
Here  we have defined
\begin{eqnarray}
\label{iota}
I^{(n)}& =& -\frac{1}{2} \int_0^{\bar \chi} \ud \tilde \chi \left(A^{(n)}{'} - B^{(n)}_{\| }{'} - \frac{1}{2}h^{(n)}_{\| }{'} \right)\;,
\\
S^{i(n)} &=& -\frac{1}{2} \int_0^{\bar \chi} \ud \tilde \chi \left[ \tilde\p^i \left( A^{(n)} - B^{(n)}_{\| } - \frac{1}{2}h^{(n)}_{\| } \right) + \frac{1}{\tilde \chi} \left(B^{i (n)}+  n^k h_{k}^{i(n)} \right)\right]\;, 
\label{varsigma}
\end{eqnarray}
where $I^{(n)}$ is the ISW term at order $n$, $\tilde \p_i= \p /\p \tilde x^i$ and 
\begin{eqnarray}
S_{\perp}^{i(n)} &=& -\frac{1}{2} \int_0^{\bar \chi} \ud \tilde \chi \left[ \tilde\p^i_\perp \left( A^{(n)} - B^{(n)}_{\| } - \frac{1}{2}h^{(n)}_{\| } \right) + \frac{1}{\tilde \chi} \left(B^{i (n)}_{\perp }+  n^k h_{kj}^{(n)} \Perp^{ij}  \right)\right]\;,\\
S_{\|}^{(n)} &=& \frac{1}{2} \left( A^{(n)}_o - B^{(n)}_{\| \, o} - \frac{1}{2}h^{(n)}_{\| \, o} \right)- \frac{1}{2} \left( A^{(n)} - B^{(n)}_{\| } - \frac{1}{2}h^{(n)}_{\| } \right)+ I^{(n)}- \frac{1}{2} \int_0^{\bar \chi} \ud \tilde \chi \frac{1}{\tilde \chi} \left( B^{(n)}_{\| }+h^{(n)}_{\| } \right)\;. 
\end{eqnarray}
Note the following useful relation 
\begin{eqnarray}
\label{du+de}
\delta n_\|^{ (1)} +  \delta\nu^{(1)}&=&A^{(1)}-B^{(1)}_{\| }- \frac{1}{2} h^{(1)}_{\| }\;.
\end{eqnarray}

At second order we find
\begin{eqnarray}
\label{dnu2-2}
 \delta\nu^{(2)}&=&- A^{(2)}_o+ v^{(2)}_{\| \, o}+\left(A^{(1)}_o\right)^2 - 2A^{(1)}_o B^{(1)}_{\| \, o}+ \left(B^{(1)}_{\| \, o}\right)^2+ 6A^{(1)}_o v^{(1)}_{\| \, o}-2B^{(1)}_{\| \, o}v^{(1)}_{\| \, o}- v^{(1)}_{k\, o} v^{k (1)}_o  \nonumber \\
&&+n^i h_{ik\,o}^{(1)} v^{k (1)}_o +4 \left (A^{(1)}_o-v^{(1)}_{\| \, o}\right) \left ( 2 A^{(1)} - B^{(1)}_{\| } -2 I^{(1)}\right) - 2 \left(B_{\perp \, o}^{i(1)} -v_{\perp \, o}^{i(1)}+\frac{1}{2}n^k h_{k\,o}^{j(1)} \Perp^i_j  \right)B_{\perp i}^{(1)}
\nonumber \\
&&+ 2\left(B_{\perp \, o}^{i(1)} -v_{\perp \, o}^{i(1)}+\frac{1}{2}n^k h_{k\,o}^{j(1)} \Perp^i_j  \right)  \int_0^{\bar \chi} \ud \tilde \chi \left[\tilde\p_{\perp i}\left( 2 A^{(1)} - B^{(1)}_{\| }\right)- \left(B_{\perp i}^{(1)}{'}
+ n^j h_{ j k}^{(1)}{'} \Perp_i^k  \right) + \frac{1}{\tilde \chi} B_{\perp i}^{(1)}\right] \nonumber \\
&&+\left(2A^{(2)}- B^{(2)}_{\| }\right)-3 \left( 2 A^{(1)} - B^{(1)}_{\| }\right)^2-2 B^{(1)}_{\| }\left(A^{(1)} - B^{(1)}_{\| } - \frac{1}{2}h^{(1)}_{\| } \right)+ 2 B_{\perp i}^{(1)}\left(B_{\perp}^{i(1)} + \Perp^i_j h_{k}^{j(1)} n^k\right)  \nonumber \\
&&+8\left( 2 A^{(1)} - B^{(1)}_{\| }\right) I^{(1)}-4B_{\perp i}^{(1)} S_{\perp}^{i(1)} - 2 I^{(2)} 
 + 2 \int_0^{\bar \chi} \ud \tilde \chi \left\{ \left(A^{(1)} - B^{(1)}_{\| } - \frac{1}{2}h^{(1)}_{\| } \right)  \frac{\ud}{\ud \tilde \chi}\left( 2 A^{(1)} - B^{(1)}_{\| }\right)\right. \nonumber \\
&& + \left.\left(A^{(1)}{'} - B^{(1)}_{\| }{'} - \frac{1}{2}h^{(1)}_{\| }{'} \right) \left[-  \left(B^{(1)}_{\|} + h^{(1)}_{\|} \right)+ 4 I^{(1)} \right]  +\left[- \left(B_{\perp}^{i(1)}+ n^j h_{ j k}^{(1)} \Perp^{ik}  \right) +2 S_{\perp }^{i(1)}  \right]   \right.  \nonumber \\
&&  \left.  \times   \left[\tilde\p_{\perp i}\left( 2 A^{(1)} - B^{(1)}_{\| } \right)- \left(B_{\perp i}^{(1)}{'} + n^j h_{ j k}^{(1)}{'} \Perp_i^k  \right) + \frac{1}{\tilde \chi} B_{\perp i}^{(1)}\right]  \right\} +  \delta\nu^{(2)}_{\rm post-Born}\;,
\end{eqnarray}
where
   \begin{eqnarray}
  &&\delta\nu^{(2)}_{\rm post-Born} = -2 \left(2 A^{(1)}_o -  B^{(1)}_{\| o}\right) \left(3 v_{\| \, o}^{(1)} + 2 h^{(1)}_{\| \, o} - \frac{1}{2} h^{j (1)}_{j \, o} \right) -2 \left(A^{(1)}_o-v^{(1)}_{\| \, o}\right)\left( 2 A^{(1)} - B^{(1)}_{\| }   -2   I^{(1)} \right) \nonumber\\
        &&    - 4 \left(- B^{(1)}_{\| o} + v_{\| \, o}^{(1)} - \frac34 h^{(1)}_{\| \, o} + \frac14 h^{j (1)}_{j \, o} \right) \int_0^{\bar \chi}   \frac{\ud \tilde{\chi}}{\tilde \chi} \left(2 A^{(1)} - B^{(1)}_{\| }\right)   -2 \left( B^{i (1)}_{\perp \, o }-v^{i (1)}_{\perp \, o }+ \frac{1}{2} n^k h_{k\,o}^{j(1)} \Perp^i_j \right)    \nonumber\\
          && \times \int_0^{\bar \chi}  \ud \tilde{\chi}  \left[ \tilde \p_{\perp i} \left(2 A^{(1)} - B^{(1)}_{\| }\right) + \frac{1}{\tilde\chi} B_{\perp i}^{(1)} \right]  +2 \left(2 A^{(1)} - B^{(1)}_{\| }\right) \left( A^{(1)}+ \frac{1}{2}  h^{(1)}_{\| }-2I^{(1)} -2\kappa^{(1)} \right)    \nonumber\\
    && + 2\left[\frac{\ud}{\ud \bar \chi}\left(2 A^{(1)} - B^{(1)}_{\| }\right)+\left(A^{(1)}{'} - B^{(1)}_{\| }{'} - \frac{1}{2}h^{(1)}_{\| }{'} \right) \right] \delta x_{\|}^{(1)}  +2 \left(2 A^{(1)}{'} - B^{(1)}_{\| }{'}\right)\left(\delta x^{0(1)} + \delta x_{\|}^{(1)}\right)  \nonumber\\
    &&+2  \left[\p_{\perp i}\left(2 A^{(1)} - B^{(1)}_{\| }\right)+\frac{1}{\bar \chi} B_{\perp i}^{(1)} \right] \delta x_{\perp}^{i (1)}   +2 \int_0^{\bar \chi}  \ud \tilde{\chi} \Bigg\{   \left(A^{(1)}{''} - B^{(1)}_{\| }{''} - \frac{1}{2}h^{(1)}_{\| }{''} \right)   \left(\delta x^{0(1)} + \delta x_{\|}^{(1)}\right)  \nonumber\\
      &&  - \left(2 A^{(1)}{'} - B^{(1)}_{\| }{'}\right) \left(A^{(1)}-B^{(1)}_{\| }- \frac{1}{2} h^{(1)}_{\| }\right)   +  \left(A^{(1)}{'} - B^{(1)}_{\| }{'} - \frac{1}{2}h^{(1)}_{\| }{'} \right)   \left(A^{(1)} + \frac{1}{2}  h^{(1)}_{\| }-2I^{(1)}\right) \nonumber \\
 &&    - \left(2 A^{(1)} - B^{(1)}_{\| }\right)  \left[\frac{\ud}{\ud \tilde \chi}\left(A^{(1)} + \frac{1}{2}  h^{(1)}_{\| }\right)+\left(A^{(1)}{'} - B^{(1)}_{\| }{'} - \frac{1}{2}h^{(1)}_{\| }{'} \right) \right] + \left( B^{i (1)}_{\perp }+ n^k h_{k}^{j(1)} \Perp^i_j- 2S_{\perp}^{i(1)} \right)  \nonumber \\
 && \times \left[ \tilde \p_{\perp i} \left(2 A^{(1)} - B^{(1)}_{\| }\right) + \frac{1}{\tilde\chi} B_{\perp i}^{(1)} \right]  +\bigg[ \tilde \p_{\perp j} \left(B^{j (1)} + n^k h_{k}^{j(1)} \right)    \left(2 A^{(1)} - B^{(1)}_{\| }\right)   - \frac{2}{\tilde \chi} \left( B^{(1)}_{\| } + h^{(1)}_{\|}\right)    \left(2 A^{(1)} - B^{(1)}_{\| }\right)    \nonumber \\
 &&-  2    \left(2 A^{(1)} - B^{(1)}_{\| }\right)   \tilde \p_{\perp j}S_{\perp}^{j(1)}  \bigg]  + 2 \bigg[ \frac{\ud}{\ud \tilde \chi}\left(2 A^{(1)} - B^{(1)}_{\| }\right)   -  \frac{1}{\tilde\chi} \left(2 A^{(1)} - B^{(1)}_{\| }\right)  \bigg] \kappa^{(1)} + \bigg[  \tilde \p_{\perp i}   \left(A^{(1)}{'} - B^{(1)}_{\| }{'} - \frac{1}{2}h^{(1)}_{\| }{'} \right)     \nonumber \\
 &&+  \frac{1}{\tilde\chi}  \left( B_{\perp i}^{(1)}{'} +  n^j h_{ j k}^{(1)}{'} \Perp_i^k \right) \bigg]   \delta x_{\perp}^{i (1)} \Bigg\} \;.  \nonumber
 \end{eqnarray}

Splitting $\delta n^{i(2)}= n^i \delta n_\|^{ (2)}+\delta n_\perp^{i (2)}$, we obtain
\begin{eqnarray}
\label{de_||-2}  
 \delta n_\|^{(2)}&=&+A^{(2)}_o- v^{(2)}_{\| \, o} +  \left(v^{(1)}_{\| \, o}\right)^2-\frac{1}{4} \left(h^{(1)}_{\| \, o}\right)^2- \frac{1}{4} n^i h_{ij\,o}^{(1)}\,  \Perp^j_k \, h^{k (1)}_{p \, o} n^p-4A^{(1)}_o v^{(1)}_{\| \, o}-2 v^{(1)}_{\| \, o} h^{(1)}_{\| \, o}-2 n^i h_{ik\,o}^{(1)} \Perp^k_j v^{j (1)}_o \nonumber \\
&& - 2\left (A^{(1)}_o-v^{(1)}_{\| \, o}\right)\left ( 2 A^{(1)} + h^{(1)}_{\| } \right) -  2 \left(B_{\perp \, o}^{i(1)} -v_{\perp \, o}^{i(1)}+\frac{1}{2}n^k h_{k\,o}^{j(1)} \Perp^i_j  \right)h_{i}^{p(1)} n_p 
+ 8  \left (A^{(1)}_o-v^{(1)}_{\| \, o}\right) I^{(1)}
\nonumber \\
&&+ 2\left(B_{\perp \, o}^{i(1)} -v_{\perp \, o}^{i(1)}+\frac{1}{2}n^k h_{k\,o}^{j(1)} \Perp^i_j  \right) \int_0^{\bar \chi} \ud \tilde \chi \left[ \p_\| \left(B_{\perp i}^{(1)}
+ n^j h_{ j k}^{(1)}\Perp_i^k \right)- \p_{\perp i}\left(B^{(1)}_{\| }+h^{(1)}_{\| } \right) \right. \nonumber \\
&& + \left.  \frac{1}{\tilde \chi} \left(B_{\perp i}^{(1)}+  2 n^j h_{kj}^{(1)} \Perp_i^k  \right)\right] -A^{(2)}-\frac{1}{2}h_\|^{(2)}- 2 h^{(1)}_{\| }  \left(A^{(1)}- B^{(1)}_{\| }- \frac{1}{2}h^{(1)}_{\| }\right) +\left( 2 A^{(1)} - B^{(1)}_{\| } \right)^2 \nonumber \\
&&+2\left( 2 A^{(1)} - B^{(1)}_{\| } \right)\left(B^{(1)}_{\| } + h^{(1)}_{\| } \right)+2 \left(B_{\perp}^{i(1)}+ n^k h_{ k}^{j(1)} \Perp^{i}_j  \right) h_{i}^{p(1)} n_p  -4 \left ( 2 A^{(1)} + h^{(1)}_{\| } \right) I^{(1)}-4n_j h_ i^{j(1)} S_{\perp}^{i(1)}\nonumber \\
 &&+  2 I^{(2)} + 2 \int_0^{\bar \chi} \ud \tilde \chi \left\{ \left( 2 A^{(1)} - B^{(1)}_{\| } - 4 I^{(1)}\right) \left(A^{(1)}{'} - B^{(1)}_{\| }{'} - \frac{1}{2}h^{(1)}_{\| }{'} \right)
 + \left[- \left(B_{\perp}^{i(1)}+ n^j h_{ j k}^{(1)} \Perp^{ik}  \right) +2 S_{\perp }^{i(1)}  \right]    \right.\nonumber \\
&&  \left.  \times \left[ \tilde \p_\| \left(B_{\perp i}^{(1)} + n^j h_{ j k}^{(1)}\Perp_i^k \right) - \tilde \p_{\perp i}\left(B^{(1)}_{\| }+h^{(1)}_{\| } \right)  + \frac{1}{\tilde \chi} \left(B_{\perp i}^{(1)}+  2 n^j h_j^{k(1)} \Perp_{ik}  \right)\right]  \right\} +  \delta n^{(2)}_{\| \rm post-Born} \;, 
\end{eqnarray}
where
    \begin{eqnarray}
  && \delta n^{(2)}_{\| \rm post-Born} = 2 \left( A^{(1)}_o + \frac{1}{2} h^{(1)}_{\| o}\right) \left(3 v_{\| \, o}^{(1)} + 2 h^{(1)}_{\| \, o}- \frac{1}{2} h^{j (1)}_{j \, o}  \right)
-  \left(A^{(1)}_o + \frac{1}{2}  h^{(1)}_{\| o }\right)^2  +2  \left(A^{(1)}_o-v^{(1)}_{\| \, o} \right) \bigg(A^{(1)} + \frac{1}{2}  h^{(1)}_{\| } \nonumber\\
 && -2  I^{(1)} \bigg) +  4   \left(- B^{(1)}_{\| o} + v_{\| \, o}^{(1)} - \frac34 h^{(1)}_{\| \, o} + \frac14 h^{j (1)}_{j \, o} \right)  \int_0^{\bar \chi}   \frac{\ud \tilde{\chi} }{\tilde \chi} \left( A^{(1)} + \frac{1}{2}  h^{(1)}_{\| }\right)    + 2 \left(  B^{i (1)}_{\perp \, o }-v^{i (1)}_{\perp \, o }+ \frac{1}{2} n^k h_{k\,o}^{j(1)} \Perp^i_j  \right) \nonumber\\
&& \times \int_0^{\bar \chi}  \ud \tilde{\chi} \left[ \tilde \p_{\perp i} \left(A^{(1)} + \frac{1}{2}  h^{(1)}_{\| } \right) -  \frac{1}{\tilde \chi} n^j h_{ j}^{k(1)} \Perp_{ik} \right]   
- 2\bigg[\frac{\ud}{\ud \bar \chi}\left(A^{(1)} + \frac{1}{2}  h^{(1)}_{\| }\right) +\left(A^{(1)}{'} - B^{(1)}_{\| }{'} - \frac{1}{2}h^{(1)}_{\| }{'} \right) \bigg] \delta x_{\|}^{(1)}\nonumber\\
&&  -2 \left(A^{(1)}{'} + \frac{1}{2}  h^{(1)}_{\| }{'} \right)\left(\delta x^{0(1)} + \delta x_{\|}^{(1)}\right)- \left(A^{(1)} + \frac{1}{2}  h^{(1)}_{\| }\right) \left(A^{(1)}+ \frac{1}{2}  h^{(1)}_{\| }-4I^{(1)}\right) + 4  \left(A^{(1)} + \frac{1}{2}  h^{(1)}_{\| }\right)  \kappa^{(1)} \nonumber\\
    && -2  \left[\p_{\perp i}\left(A^{(1)} + \frac{1}{2}  h^{(1)}_{\| } \right) - \frac{1}{\bar \chi} n^j h_{ j}^{k(1)} \Perp_{ik} \right] \delta x_{\perp}^{i (1)}  +2 \int_0^{\bar \chi}  \ud \tilde{\chi} \Bigg\{ - \left(A^{(1)}{''} - B^{(1)}_{\| }{''} - \frac{1}{2}h^{(1)}_{\| }{''} \right)   \left(\delta x^{0(1)} + \delta x_{\|}^{(1)}\right)  \nonumber \\
 && + \left(A^{(1)}{'} + \frac{1}{2}  h^{(1)}_{\| }{'}\right) \left(A^{(1)}-B^{(1)}_{\| }- \frac{1}{2} h^{(1)}_{\| }\right)  +2  \left(A^{(1)}{'} - B^{(1)}_{\| }{'} - \frac{1}{2}h^{(1)}_{\| }{'} \right) I^{(1)}      -\left( B^{i (1)}_{\perp }+ n^k h_{k}^{j(1)} \Perp^i_j  - 2S_{\perp}^{i(1)} \right) \nonumber \\
 && \times \left[ \tilde \p_{\perp i} \left(A^{(1)} + \frac{1}{2}  h^{(1)}_{\| } \right) -  \frac{1}{\tilde \chi} n^j h_{ j}^{k(1)} \Perp_{ik} \right]   -  \left( A^{(1)} + \frac{1}{2}  h^{(1)}_{\| }\right)  \bigg[   \tilde \p_{\perp j} \left(B^{j (1)} + n^k h_{k}^{j(1)} \right)   - \frac{2}{\tilde \chi} \left( B^{(1)}_{\| } + h^{(1)}_{\|}\right) \nonumber \\
 && -  2   \tilde \p_{\perp j}S_{\perp}^{j(1)}  \bigg]  + 2 \bigg[ - \frac{\ud}{\ud \tilde \chi}\left(A^{(1)} + \frac{1}{2}  h^{(1)}_{\| } \right)  +  \frac{1}{\tilde\chi} \left(A^{(1)} + \frac{1}{2}  h^{(1)}_{\| } \right)  \bigg] \kappa^{(1)}  - \bigg[  \tilde \p_{\perp i}  
 \left(A^{(1)}{'} - B^{(1)}_{\| }{'} - \frac{1}{2}h^{(1)}_{\| }{'} \right)    \nonumber \\
 &&  +  \frac{1}{\tilde\chi}  \left( B_{\perp i}^{(1)}{'} +  n^j h_{ j k}^{(1)}{'} \Perp_i^k \right) \bigg]   \delta x_{\perp}^{i (1)} \Bigg\} \;,
 \nonumber
 \end{eqnarray}

and\footnote{From Eq.\ (\ref{Projection3}), 
\begin{eqnarray*}
 \Perp_k^{i} \Perp^{jl}[ \p^k B_l^{(1)} -\p_l B^{k(1)}+ n^m \p^k h_{lm}^{(1)} -  n^m  \p_l  h_m^{k(1)}  ]= 2[  \p_{\perp}^{[i} B_{\perp}^{j](1)}  + \p_{\perp}^{[i}     ( \Perp^{j]}_m h^{m(1)}_q n^q )  -  {\tilde \chi}^{-1} (n^{[i} B_{\perp}^{j](1)}+ n^{[i}  \Perp^{j]}_m h^{m(1)}_q n^q )   ].
\end{eqnarray*}
}
\begin{eqnarray}
\label{de_perp-2}  
 \delta n_\perp^{i(2)}&=&+ B^{i(2)}_{\perp \, o}-  v^{i(2)}_{\perp \, o} + \frac{1}{2} n^j h_{ j k\, o}^{(2)} \Perp^{ki}+   v^{(1)}_{\| \, o}v^{i(1)}_{\perp \, o} - 3 v^{(1)}_{\| \, o}B^{i(1)}_{\perp \, o} +  B^{(1)}_{\| \, o}B^{i(1)}_{\perp \, o}-  B^{(1)}_{\| \, o}v^{i(1)}_{\perp \, o} + 2 A^{(1)}_o n^j h_{ j k\, o}^{(1)} \Perp^{ki} \nonumber \\
&&-4 v^{(1)}_{\| \, o} n^j h_{ j k\, o}^{(1)} \Perp^{ki}  -2 v^{j (1)}_{\perp o}\Perp^{l}_j  \, h^{k (1)}_{l \, o} \Perp_{k}^{i(1)}-\frac{1}{4} h^{(1)}_{\| \, o} n^j h_{ j k\, o}^{(1)} \Perp^{ki}  -\frac{1}{4} n^j h^{k (1)}_{j \, o}  \Perp^{l}_k  \, h^{p (1)}_{l \, o} \Perp_{p}^{i} \nonumber \\
&& - 4\left (A^{(1)}_o-v^{(1)}_{\| \, o}\right) \left(B_{\perp}^{i(1)}+ n^k h_{ k}^{j(1)} \Perp^{i}_j  - 2 S_{\perp}^{i(1)} \right) + 2 \left( B^{j (1)}_{\perp \, o }-v^{j (1)}_{\perp \, o }+ \frac{1}{2} n^k h_{k\,o}^{p(1)} \Perp^j_p \right) \bigg\{  - \Perp^{l}_j  \, h^{k (1)}_{l} \Perp_{k}^{i} \nonumber \\
&&+ \left. 2  \int_0^{\bar \chi} \ud \tilde \chi \left[  \tilde\p_{\perp}^{[i} B_{\perp}^{j](1)}  +\tilde \p_{\perp}^{[i}     \left( \Perp^{j]}_m h^{m(1)}_q n^q \right)   - \frac{1}{\tilde \chi} \left(n^{[i} B_{\perp}^{i](1)} + n^{[i}  \Perp^{j]}_m h^{m(1)}_q n^q \right)   \right] \right\}  - B^{i(2)}_{\perp} -   n^j h_{ j k}^{(2)} \Perp^{ki} \nonumber \\
&& + 4 A^{(1)} B^{i(1)}_{\perp} -2  B^{(1)}_{\|}B^{i(1)}_{\perp}  +2 A^{(1)} n^j h_{ j k}^{(1)} \Perp^{ki} +  h_\|^{(1)} n^j h_{ j k}^{(1)} \Perp^{ki} +2 \left( B^{j (1)}_{\perp }+ n^k h_{k}^{p(1)} \Perp^j_p\right)  \Perp^{l}_j  \, h^{k (1)}_{l} \Perp_{k}^{i} \nonumber \\
&&- 8 \left(B_{\perp}^{i(1)}+ n^j h_{ j k}^{(1)} \Perp^{ik}  \right) I^{(1)}- 4\,  \Perp^{l}_j  \, h^{k (1)}_{l} \Perp_{k}^{i} S_{\perp}^{j(1)}+2 S_{\perp }^{i(2)}
 + 2 \int_0^{\bar \chi} \ud \tilde \chi \left\{ -    \left(A^{(1)}{'} - B^{(1)}_{\| }{'} - \frac{1}{2}h^{(1)}_{\| }{'} \right)    \right. \nonumber \\
&& \left.   \times \left(B_{\perp}^{i(1)}+ n^j h_{ j k}^{(1)} \Perp^{ik}  \right)  + \left(2 A^{(1)} -B_\|^{(1)}\right)  \frac{\ud}{\ud \tilde \chi} \left(B_{\perp}^{i(1)}+ n^j h_{ j k}^{(1)} \Perp^{ik}  \right) +  2 \left(2 A^{(1)} -B_\|^{(1)} - 2 I^{(1)}\right)  \right. \nonumber \\
&&  \left. \times  \left[ \tilde\p^i_\perp \left( A^{(1)} - B^{(1)}_{\| } - \frac{1}{2}h^{(1)}_{\| } \right) + \frac{1}{\tilde \chi} \left(B^{i (1)}_{\perp }+  n^k h_{kj}^{(1)} \Perp^{ij}  \right) \right] +2 \left[ - \left(B_{\perp j}^{(1)}               + n^p h_{pk}^{(1)} \Perp_j^{k}  \right)+ 2 \delta_{jp}  S_{\perp}^{p(1)} \right]  \right. \nonumber \\
&& \left.  \times \left[ \tilde \p_{\perp}^{[i} B_{\perp}^{j](1)}  + \tilde \p_{\perp}^{[i}     \left( \Perp^{j]}_m h^{m(1)}_q n^q \right)   -  \frac{1}{\tilde \chi} \left(n^{[i} B_{\perp}^{j](1)}+ n^{[i}  \Perp^{j]}_m h^{m(1)}_q n^q \right)   \right]+\left( A^{(1)} - B^{(1)}_{\| } - \frac{1}{2}h^{(1)}_{\| } \right) \right.  \nonumber\\ 
&& \left.  \times \bigg[ \tilde\p^i_\perp \left(B^{(1)}_{\| }+ h^{(1)}_{\| } \right) - \p_\| \left(B_{\perp}^{i (1)}+n^p h_{pq}^{(1)} \Perp^{iq} \right) - \frac{1}{\tilde \chi} \left(B^{i (1)}_{\perp }+  2 n^p h_{pq}^{(1)} \Perp^{iq}  \right) \bigg]\right\} +  \delta n^{(2)}_{\perp \rm post-Born}\;,
\end{eqnarray}

where

\begin{eqnarray}
 &&  \delta n^{(2)}_{\perp \rm post-Born} = + 2 \left(   B^{i (1)}_{\perp \, o } +  n^k h_{k\,o}^{j(1)} \Perp^i_j \right) \left(3 v_{\| \, o}^{(1)} + 2 h^{(1)}_{\| \, o} - \frac{1}{2} h^{j (1)}_{j \, o} \right)   -2 \left( A^{(1)}_o  - B^{(1)}_{\| o} - \frac{1}{2} h^{(1)}_{\| o}\right) \left(   v^{i (1)}_{\perp \, o } + \frac{1}{2} n^k h_{k\,o}^{j(1)} \Perp^i_j \right)\nonumber\\
 && + 2  \left(A^{(1)}_o-v^{(1)}_{\| \, o} \right) \left(B^{i (1)}_{\perp }+ n^k h_{k}^{j(1)} \Perp^i_j  - 2 S_{\perp}^{i(1)}  \right) + 4  \left(- B^{(1)}_{\| o} + v_{\| \, o}^{(1)} - \frac34 h^{(1)}_{\| \, o} + \frac14 h^{m (1)}_{m \, o} \right) \nonumber \\ 
 && \times  \int_0^{\bar \chi}   \frac{\ud \tilde{\chi} }{\tilde \chi} \left( B^{i (1)}_{\perp }+ n^k h_{k}^{j(1)} \Perp^i_j \right)   + 2   \left(  B^{j (1)}_{\perp \, o }-v^{j (1)}_{\perp \, o }+ \frac{1}{2} n^k h_{k\,o}^{p(1)} \Perp^j_p \right) \int_0^{\bar \chi}  \ud \tilde{\chi}\bigg[  \Perp^i_m \p_{\perp j} \left(B^{m (1)} + n^k h_{k}^{m(1)} \right)   \nonumber\\
 &&  -  \frac{1}{\tilde\chi} \Perp^i_m   h_{j}^{m(1)} + \frac{1}{\tilde \chi} \Perp^i_j \left(A^{(1)} - B^{(1)}_{\| } - \frac{1}{2}h^{(1)}_{\| } \right) \bigg]   - 2 \left(B^{i (1)}_{\perp }+ n^k h_{k}^{j(1)} \Perp^i_j \right) \left( A^{(1)} + \frac{1}{2}  h^{(1)}_{\| } - 2I^{(1)}\right)  \nonumber \\
&& -2\bigg[\frac{\ud}{\ud \bar \chi}\left(B^{i (1)}_{\perp }+ n^k h_{k}^{j(1)} \Perp^i_j \right)+ \p_{\perp i}  \left(A^{(1)} - B^{(1)}_{\| } - \frac{1}{2}h^{(1)}_{\| } \right)  +  \frac{1}{\bar \chi} \left(B^{i (1)}_{\perp }+ n^k h_{k}^{j(1)} \Perp^i_j \right) \bigg] \delta x_{\|}^{(1)} \nonumber\\
&& -2  \left( B_{\perp i}^{(1)}{'} +  n^j h_{ j k}^{(1)}{'} \Perp_i^k \right) \left(\delta x^{0(1)} + \delta x_{\|}^{(1)}\right)  + 4 \left(B^{i (1)}_{\perp }+ n^k h_{k}^{j(1)} \Perp^i_j \right)  \kappa^{(1)}  - 2  \bigg[  \Perp^i_m \p_{\perp j} \left(B^{m (1)} + n^k h_{k}^{m(1)} \right)   \nonumber\\
&& -  \frac{1}{\bar \chi} \Perp^i_m   h_{j}^{m(1)} + \frac{1}{\bar \chi} \Perp^i_j \left(A^{(1)} - B^{(1)}_{\| } - \frac{1}{2}h^{(1)}_{\| } \right) \bigg]  \delta x_{\perp}^{j (1)}  +2 \int_0^{\bar \chi}  \ud \tilde{\chi} \Bigg\{ - \bigg[ \tilde \p_{\perp }^i    \left(A^{(1)}{'} - B^{(1)}_{\| }{'} - \frac{1}{2}h^{(1)}_{\| }{'} \right)  \nonumber\\
&&  +  \frac{1}{\tilde\chi^2} \left(B^{i (1)}_{\perp }+ n^k h_{k}^{j(1)} \Perp^i_j \right)    +  \frac{1}{\tilde \chi}  \left( B_{\perp }^{i(1)}{'} +  n^j h_{ j }^{k(1)}{'} \Perp_k^i \right) \bigg]  \left(\delta x^{0(1)} + \delta x_{\|}^{(1)}\right) + \left(A^{(1)}-B^{(1)}_{\| }- \frac{1}{2} h^{(1)}_{\| }\right) 
\nonumber\\
&& \times \left( B_{\perp }^{i(1)}{'} +  n^j h_{ j }^{k(1)}{'} \Perp_k^i \right)  - \left[  \tilde \p_{\perp }^i    \left(A^{(1)} - B^{(1)}_{\| } - \frac{1}{2}h^{(1)}_{\| }  \right)   +  \frac{1}{\tilde\chi}  \left(B^{i (1)}_{\perp }+ n^k h_{k}^{j(1)} \Perp^i_j \right)  \right]  \left(A^{(1)} + \frac{1}{2}  h^{(1)}_{\| } -2 I^{(1)}\right) \nonumber\\
&&  +  \left(B^{i (1)}_{\perp }+ n^k h_{k}^{j(1)} \Perp^i_j \right)  \left[\frac{\ud}{\ud \tilde \chi}\left(A^{(1)} + \frac{1}{2}  h^{(1)}_{\| }\right)+\left(A^{(1)}{'} - B^{(1)}_{\| }{'} - \frac{1}{2}h^{(1)}_{\| }{'} \right) \right]  \nonumber\\
&& -   \left[  \Perp^i_m  \tilde\p_{\perp j} \left(B^{m (1)} + n^k h_{k}^{m(1)} \right) -  \frac{1}{\tilde\chi} \Perp^i_m   h_{j}^{m(1)} + \frac{1}{\tilde \chi} \Perp^i_j \left(A^{(1)} - B^{(1)}_{\| } - \frac{1}{2}h^{(1)}_{\| } \right) \right]  \left[  \left( B^{j (1)}_{\perp }+ n^k h_{k}^{m(1)} \Perp^j_m\right) - 2S_{\perp}^{j(1)} \right]
   \nonumber \\
 &&   +  \left( B^{i (1)}_{\perp }+ n^k h_{k}^{j(1)} \Perp^i_j \right)   \left[-  \tilde \p_{\perp m} \left(B^{m (1)} + n^l h_{l}^{m(1)} \right) + \frac{2}{\tilde \chi} \left( B^{(1)}_{\| } + h^{(1)}_{\|}\right) + 2  \tilde \p_{\perp m}S_{\perp}^{m(1)}  \right]  \nonumber\\
&&+ 2 \bigg[ - \frac{\ud}{\ud \tilde \chi}\left(B^{i (1)}_{\perp }+ n^k h_{k}^{j(1)} \Perp^i_j \right)    +  \frac{1}{\tilde\chi} \left(B^{i (1)}_{\perp }+ n^k h_{k}^{j(1)} \Perp^i_j \right)  \bigg] \kappa^{(1)}  \nonumber \\
 &&  -  \bigg[     \Perp^{im} \tilde \p_{\perp j}   \tilde \p_{\perp m}    \left(A^{(1)} - B^{(1)}_{\| } - \frac{1}{2}h^{(1)}_{\| } \right)   + \frac{1}{\tilde \chi} \Perp^i_j  \left(A^{(1)}{'} - B^{(1)}_{\| }{'} - \frac{1}{2}h^{(1)}_{\| }{'} \right)  +    \frac{1}{\tilde \chi^2} \Perp^i_j  \left(A^{(1)} - B^{(1)}_{\| } - \frac{1}{2}h^{(1)}_{\| } \right)  \nonumber \\
 && +\frac{1}{\tilde\chi}  \Perp^i_m \tilde \p_{\perp j} \left(B^{m (1)}_{\perp }+ n^k   h_{k}^{l(1)} \Perp^{m}_l \right)  + \frac{1}{\tilde\chi}   \Perp^k_j  \p^i_{\perp} B^{ (1)}_k +  \frac{1}{\tilde\chi}   \p^i_{\perp} \left(h_{jl}^{(1)} n^l \right) - \frac{1}{\tilde\chi^2}\Perp^i_l h_{j}^{l(1)} \bigg] \delta x_{\perp}^{j (1)}     \Bigg\}   \;.  \nonumber
 \end{eqnarray}

Combining Eqs.\  (\ref{dnu-2}) and (\ref{de_||-2})  we  obtain the  useful relation

\begin{eqnarray} 
\label{dnu+de_||-2}
 \delta\nu^{(2)} +  \delta n_\|^{(2)}&=&+\left(A^{(1)}_o\right)^2 - 2A^{(1)}_o B^{(1)}_{\| \, o}+ \left(B^{(1)}_{\| \, o}\right)^2 + 2A^{(1)}_o v^{(1)}_{\| \, o} -2B^{(1)}_{\| \, o}v^{(1)}_{\| \, o}-  v^{(1)}_{\| \, o} h^{(1)}_{\| \, o} -\frac{1}{4} \left(h^{(1)}_{\| \, o}\right)^2   -   v^{(1)}_{\perp k \, o} v^{k (1)}_{\perp \, o}\nonumber \\
&& +  n^i h_{ik\,o}^{(1)} \Perp^k_j v^{j (1)}_o - n^i h_{ik\,o}^{(1)} \Perp^k_j B^{j (1)}_o - \frac{1}{4} n^i h_{ij\,o}^{(1)}\,  \Perp^j_k \, h^{k (1)}_{p \, o} n^p   -  B_{\perp \, o}^{i(1)}B_{\perp i \, o}^{(1)}+ 2 v_{\perp \, o}^{i(1)}B_{\perp i \, o}^{(1)}\nonumber \\
&&+4 \left (A^{(1)}_o-v^{(1)}_{\| \, o}\right) \left (  A^{(1)} - B^{(1)}_{\| } - \frac{1}{2} h^{(1)}_{\| } \right)  -8 \left(B_{\perp \, o}^{i(1)} -v_{\perp \, o}^{i(1)}+\frac{1}{2}n^k h_{k\,o}^{j(1)} \Perp^i_j  \right) \delta_{il} S_{\perp }^{l(1)} \nonumber \\
&&+ A^{(2)}- B^{(2)}_{\| } -\frac{1}{2}h_\|^{(2)}-2\left( 2 A^{(1)} - B^{(1)}_{\| } \right)^2+  \left(B^{(1)}_{\| } + h^{(1)}_{\| } \right) \left( 2 A^{(1)} + h^{(1)}_{\| } \right)\nonumber \\
&& + \left(B_{\perp}^{i(1)} +  n^k h_{k}^{j(1)}  \Perp^i_j \right) \left(B_{\perp i}^{(1)}+ n^p h_{p m}^{(1)} \Perp^m_i  \right) +8 \left(A^{(1)} - B^{(1)}_{\| }- \frac{1}{2}h^{(1)}_{\| }\right)  I^{(1)} -8 S_{\perp }^{i(1)}S_{\perp }^{j(1)} \delta_{ij}     \nonumber \\
&&+ 2 \int_0^{\bar \chi} \ud \tilde \chi \left\{ \left(A^{(1)} - B^{(1)}_{\| } - \frac{1}{2}h^{(1)}_{\| } \right) \left[2\left(A^{(1)}{'} - B^{(1)}_{\| }{'} - \frac{1}{2}h^{(1)}_{\| }{'} \right)+  \frac{\ud}{\ud \tilde \chi}\left( 2 A^{(1)} - B^{(1)}_{\| }\right) \right]  \right. \nonumber \\
&& - \left(B_{\perp}^{i(1)} + n^k h_{k}^{j(1)} \Perp^i_j  \right)\left[ \tilde \p_{\perp i}\left (  A^{(1)} - B^{(1)}_{\| } - \frac{1}{2} h^{(1)}_{\| } \right)  + \frac{1}{\tilde \chi} \left(B_{\perp i}^{(1)}+   n^m h_{mp}^{(1)} \Perp^p_i  \right)\right]\bigg\}\nonumber \\
&& +\left(\delta\nu^{(2)} +  \delta n_\|^{(2)}\right)_{ \rm post-Born}\;,
\end{eqnarray}
where
 \begin{eqnarray}
&&\left(\delta\nu^{(2)} +  \delta n_\|^{(2)}\right)_{ \rm post-Born} = -2  \left( A^{(1)}_o -  B^{(1)}_{\| o} - \frac{1}{2} h^{(1)}_{\| o}\right) \left(3 v_{\| \, o}^{(1)} + 2 h^{(1)}_{\| \, o} - \frac{1}{2} h^{j (1)}_{j \, o} \right) \nonumber \\
&& -2 \left(A^{(1)}_o-v^{(1)}_{\| \, o} \right)  \left(A^{(1)}   -  B^{(1)}_{\| } - \frac{1}{2}  h^{(1)}_{\| }\right) + 4\left( B^{j (1)}_{\perp \, o }-v^{j (1)}_{\perp \, o }+ \frac{1}{2} n^l h_{l\,o}^{p(1)} \Perp^j_p   \right)  S_{\perp}^{j(1)}    \nonumber \\
 &&   - 4 \left(- B^{(1)}_{\| o} + v_{\| \, o}^{(1)} - \frac34 h^{(1)}_{\| \, o} + \frac14 h^{m (1)}_{m \, o} \right) \int_0^{\bar \chi}   \frac{\ud \tilde{\chi}}{\tilde \chi} \left( A^{(1)} - B^{(1)}_{\| } - \frac{1}{2} h^{(1)}_{\|} \right)     \nonumber \\
 && + 2 \frac{\ud}{\ud \bar \chi}\left(A^{(1)}  - B^{(1)}_{\| } - \frac{1}{2}  h^{(1)}_{\| }\right) \delta x_{\|}^{(1)}  + 2 \left(A^{(1)}{'} - B^{(1)}_{\| }{'}  - \frac{1}{2}  h^{(1)}_{\| }{'} \right)\left(\delta x^{0(1)} + \delta x_{\|}^{(1)}\right)  \nonumber \\
 &&  + 2 \left(A^{(1)}   -  B^{(1)}_{\| } - \frac{1}{2}  h^{(1)}_{\| }\right) \left(A^{(1)} + \frac{1}{2}  h^{(1)}_{\| }- 2 I^{(1)}\right) - 4 \left(A^{(1)}   -  B^{(1)}_{\| } - \frac{1}{2}  h^{(1)}_{\| }\right)\kappa^{(1)} \nonumber \\
 && +2  \left[\p_{\perp i}\left(A^{(1)} - B^{(1)}_{\| } - \frac{1}{2}  h^{(1)}_{\| } \right) + \frac{1}{\bar \chi} \left(B^{(1)}_{\perp i}+  n^j h_{ j}^{k(1)} \Perp_{ik}  \right) \right] \delta x_{\perp}^{i (1)}  \nonumber \\
 && +2 \int_0^{\bar \chi}  \ud \tilde{\chi} \Bigg\{   - \left( A^{(1)} - B^{(1)}_{\| } - \frac{1}{2} h^{(1)}_{\|} \right)  \left[\frac{\ud}{\ud \tilde \chi}\left(A^{(1)} + \frac{1}{2}  h^{(1)}_{\| }\right) + 2 \left(A^{(1)}{'} - B^{(1)}_{\| }{'} - \frac{1}{2}h^{(1)}_{\| }{'} \right) \right]   \nonumber \\
 &&-  \left[ \tilde \p_{\perp j}\left( A^{(1)} - B^{(1)}_{\| } - \frac{1}{2} h^{(1)}_{\|} \right) +  \frac{1}{\tilde \chi} \left(B^{(1)}_{\perp j}+  n^m h_{ m}^{k(1)} \Perp_{jk}  \right) \right] \left[   - \left( B^{j (1)}_{\perp }+ n^l h_{l}^{p(1)} \Perp^j_p\right) + 2S_{\perp}^{j(1)} \right]   \nonumber \\
 &&  + \left( A^{(1)} - B^{(1)}_{\| } - \frac{1}{2} h^{(1)}_{\|} \right)  \left[   \tilde \p_{\perp m} \left(B^{m (1)} + n^l h_{l}^{m(1)} \right) - \frac{2}{\tilde \chi} \left( B^{(1)}_{\| } + h^{(1)}_{\|}\right) - 2  \tilde \p_{\perp m}S_{\perp}^{m(1)}  \right]  \nonumber \\
 && + 2 \bigg[ \frac{\ud}{\ud \tilde \chi}\left( A^{(1)} - B^{(1)}_{\| } - \frac{1}{2} h^{(1)}_{\|} \right)    -  \frac{1}{\tilde\chi} \left( A^{(1)} - B^{(1)}_{\| } - \frac{1}{2} h^{(1)}_{\|} \right)  \bigg] \kappa^{(1)}  \Bigg\} \;.   \nonumber
\end{eqnarray}

From Eqs. (\ref{deltax0-n}), (\ref{deltaxi-n}) and (\ref{du+de}), we find to first order that
\begin{eqnarray}
\label{dx0-1}
\delta x^{0(1)}&=& -\bar \chi \left (A^{(1)}_o-v^{(1)}_{\| \, o}\right)+ \int_0^{\bar \chi} \ud \tilde \chi \left[ 2 A^{(1)} - B^{(1)}_{\| } + \left(\bar \chi-\tilde \chi\right) \left(A^{(1)}{'} - B^{(1)}_{\| }{'} - \frac{1}{2}h^{(1)}_{\| }{'} \right) \right] \; \\
\label{dx||-1}
\delta x_{\|}^{(1)}&=& \bar \chi \left(A^{(1)}_o-v^{(1)}_{\| \, o}\right)-\int_0^{\bar \chi} \ud \tilde \chi \left[ \left(A^{(1)}+ \frac{1}{2}  h^{(1)}_{\| }\right) +  \left(\bar \chi-\tilde \chi\right) \left(A^{(1)}{'} - B^{(1)}_{\| }{'} - \frac{1}{2}h^{(1)}_{\| }{'} \right) \right] \;,   \\
\label{dxperp-1}
\delta x_{\perp}^{i (1)}&=& \bar \chi \left(B^{i (1)}_{\perp \, o }-v^{i (1)}_{\perp \, o }+ \frac{1}{2} n^k h_{k\,o}^{j(1)} \Perp^i_j\right)-\int_0^{\bar \chi} \ud \tilde \chi \left\{ \left( B^{i (1)}_{\perp }+ n^k h_{k}^{j(1)} \Perp^i_j\right) \right. \nonumber \\
&&+\left .\left(\bar \chi-\tilde \chi\right)\left[\tilde \p^i_\perp \left( A^{(1)} - B^{(1)}_{\| } - \frac{1}{2}h^{(1)}_{\| } \right) + \frac{1}{\tilde \chi} \left(B^{i (1)}_{\perp }+  n^k h_{kj}^{(1)} \Perp^{ij}  \right)\right] \right\}\;,\\
\label{s-1}
\delta x^{0(1)} + \delta x_{\|}^{(1)} &=&   \int_0^{\bar \chi} \ud \tilde \chi \left(A^{(1)} - B^{(1)}_{\| }- \frac{1}{2}h^{(1)}_{\| }\right) = -  T^{(1)}\;,
\end{eqnarray}
where we defined 
\begin{eqnarray}
\label{s-n}
T^{(n)}= - \int_0^{\bar \chi} \ud \tilde \chi \left(A^{(n)} - B^{(n)}_{\| }- \frac{1}{2}h^{(n)}_{\| }\right) 
\end{eqnarray}
which is an integrated radial displacement corresponding to a time delay term at order $n$, which generalizes the usual (Shapiro) time delay  term $T^{(1)}$ \cite{Challinor:2011bk}.

At second order,
\begin{eqnarray}
\label{dx0-2}
 \delta x^{0(2)}&=& \bar \chi\left[-A^{(2)}_o+ v^{(2)}_{\| \, o}+\left(A^{(1)}_o\right)^2 -2 A^{(1)}_o B^{(1)}_{\| \, o}+ \left(B^{(1)}_{\| \, o}\right)^2+ 6A^{(1)}_o v^{(1)}_{\| \, o}-2 B^{(1)}_{\| \, o}v^{(1)}_{\| \, o}- v^{(1)}_{k\, o} v^{k (1)}_o \right. \nonumber \\
&& \left. +n^i h_{ik\,o}^{(1)} v^{k (1)}_o\right]+4  \left (A^{(1)}_o-v^{(1)}_{\| \, o}\right)  \int_0^{\bar \chi} \ud \tilde \chi \left[\left ( 2 A^{(1)} - B^{(1)}_{\| } \right) + \left(\bar \chi-\tilde \chi\right) \left(A^{(1)}{'} - B^{(1)}_{\| }{'} - \frac{1}{2}h^{(1)}_{\| }{'} \right) \right]\nonumber \\
&&+  2\left(B_{\perp \, o}^{i(1)} -v_{\perp \, o}^{i(1)}+\frac{1}{2}n^k h_{k\,o}^{j(1)} \Perp^i_j  \right) \int_0^{\bar \chi} \ud \tilde \chi \left\{- B_{\perp i}^{(1)}+    \left(\bar \chi-\tilde \chi\right) \left[ \tilde \p_{\perp i}\left( 2 A^{(1)} - B^{(1)}_{\| }\right)- \left(B_{\perp i}^{(1)}{'}+ n^j h_{ j k}^{(1)}{'} \Perp_i^k  \right)\right. \right.  \nonumber \\
&& \left. \left.  + \frac{1}{\tilde \chi} B_{\perp i}^{(1)}\right]\right\} + \int_0^{\bar \chi} \ud \tilde \chi \left[ 2A^{(2)}- B^{(2)}_{\| }-12\left(A^{(1)}\right)^2+10A^{(1)} B^{(1)}_{\|} -  \left(B^{(1)}_{\|}\right)^2+ B^{(1)}_{\|} h^{(1)}_{\| }  \right. \nonumber \\
&& +2 B_{\perp i}^{(1)}\left(B_{\perp}^{i(1)} + \Perp^i_j h_{k}^{j(1)} n^k\right) + 8 \left( 2 A^{(1)} - B^{(1)}_{\| }\right) I^{(1)}-4B_{\perp i}^{(1)} S_{\perp}^{i(1)} \bigg]  \nonumber \\
&& + \int_0^{\bar \chi} \ud \tilde \chi\left(\bar \chi-\tilde \chi\right) \left\{A^{(2)}{'} - B^{(2)}_{\| }{'} - \frac{1}{2}h^{(2)}_{\| }{'} + 2 \left(A^{(1)} - B^{(1)}_{\| } - \frac{1}{2}h^{(1)}_{\| } \right)  \frac{\ud}{\ud \tilde \chi}\left( 2 A^{(1)} - B^{(1)}_{\| }\right)\right. \nonumber \\
&& \left.  + 2 \left(A^{(1)}{'} - B^{(1)}_{\| }{'} - \frac{1}{2}h^{(1)}_{\| }{'} \right) \left[-  \left(B^{(1)}_{\|} + h^{(1)}_{\|} \right)+ 4 I^{(1)} \right]   +  2 \left[- \left(B_{\perp}^{i(1)}+ n^j h_{ j k}^{(1)} \Perp^{ik}  \right) +2 S_{\perp }^{i(1)}  \right]   \right.  \nonumber \\
&& \left.  \times  \left[\tilde \p_{\perp i}\left( 2 A^{(1)} - B^{(1)}_{\| } \right)- \left(B_{\perp i}^{(1)}{'} + n^j h_{ j k}^{(1)}{'} \Perp_i^k  \right) + \frac{1}{\tilde \chi} B_{\perp i}^{(1)}\right]  \right\} + \delta x^{0(2)}_{\rm post-Born}\;,
\end{eqnarray}
where
    \begin{eqnarray}
    && \delta x^{0(2)}_{\rm post-Born} = -2\bar \chi \left(2 A^{(1)}_o -  B^{(1)}_{\| o}\right) \left(3 v_{\| \, o}^{(1)} + 2 h^{(1)}_{\| \, o} - \frac{1}{2} h^{j (1)}_{j \, o} \right) + 4  \left(A^{(1)}_o-v^{(1)}_{\| \, o}\right) \bigg[ \frac{\bar \chi}{2} \left(2 A^{(1)} - B^{(1)}_{\| }\right)  \nonumber \\
 &&  -   \int_0^{\bar \chi} \ud \tilde \chi  \left(2 A^{(1)} - B^{(1)}_{\| }- I^{(1)} \right)    \bigg]   
  - 4 \left(- B^{(1)}_{\| o} + v_{\| \, o}^{(1)} - \frac34 h^{(1)}_{\| \, o} + \frac14 h^{m (1)}_{m \, o} \right) \int_0^{\bar \chi}  \ud \tilde{\chi} ~ \frac{ (\bar \chi - \tilde \chi)}{\tilde \chi} \left(2 A^{(1)} - B^{(1)}_{\| }\right)    \nonumber\\
      && -2 \left(B^{i (1)}_{\perp \, o }-v^{i (1)}_{\perp \, o }+ \frac{1}{2} n^k h_{k\,o}^{j(1)} \Perp^i_j \right) \int_0^{\bar \chi}  \ud \tilde{\chi} ~ (\bar \chi - \tilde \chi)  \left[ \tilde \p_{\perp i} \left(2 A^{(1)} - B^{(1)}_{\| }\right) + \frac{1}{\tilde\chi} B_{\perp i}^{(1)} \right] \nonumber\\
 && - 2\left(2 A^{(1)} - B^{(1)}_{\| }\right) \int_0^{\bar \chi} \ud \tilde \chi  \left(A^{(1)}+ \frac{1}{2}  h^{(1)}_{\| } - 2I^{(1)} \right)    
+2\int_0^{\bar \chi} \ud \tilde{\chi} \Bigg\{  \left(A^{(1)}{'} - B^{(1)}_{\| }{'} - \frac{1}{2}h^{(1)}_{\| }{'} \right) \delta x_{\|}^{(1)}  \nonumber \\
 &&- \left(2 A^{(1)}{'} - B^{(1)}_{\| }{'}\right)  T^{(1)}  + 2\left(2 A^{(1)} - B^{(1)}_{\| }\right)    \left(A^{(1)} + \frac{1}{2}  h^{(1)}_{\| }- 2I^{(1)} - \kappa^{(1)}\right) \nonumber \\
    &&+  \left[ \tilde \p_{\perp i}\left(2 A^{(1)} - B^{(1)}_{\| }\right)+\frac{1}{\tilde \chi} B_{\perp i}^{(1)} \right] \delta x_{\perp}^{ i (1)} \Bigg\}  +2 \int_0^{\bar \chi}  \ud \tilde{\chi} ~ (\bar \chi - \tilde \chi) \Bigg\{  - \left(A^{(1)}{''} - B^{(1)}_{\| }{''} - \frac{1}{2}h^{(1)}_{\| }{''} \right)   T^{(1)} \nonumber\\
    &&     - \left(2 A^{(1)}{'} - B^{(1)}_{\| }{'}\right) \left(A^{(1)}-B^{(1)}_{\| }- \frac{1}{2} h^{(1)}_{\| } \right)  - \left(A^{(1)}{'} - B^{(1)}_{\| }{'} - \frac{1}{2}h^{(1)}_{\| }{'} \right)  \left( -A^{(1)}- \frac{1}{2}  h^{(1)}_{\| }+2I^{(1)}\right)   \nonumber \\
 && - \left(2 A^{(1)} - B^{(1)}_{\| }\right)   \bigg[\frac{\ud}{\ud \tilde \chi}\left(A^{(1)} + \frac{1}{2}  h^{(1)}_{\| }\right)  +\left(A^{(1)}{'} - B^{(1)}_{\| }{'} - \frac{1}{2}h^{(1)}_{\| }{'} \right) \bigg]     \nonumber \\
 &&+\left[ \tilde \p_{\perp i} \left(2 A^{(1)} - B^{(1)}_{\| }\right) + \frac{1}{\tilde\chi} B_{\perp i}^{(1)} \right]\left( B^{i (1)}_{\perp }+ n^k h_{k}^{j(1)} \Perp^i_j - 2S_{\perp}^{i(1)}\right) \nonumber \\
 && -  \left(2 A^{(1)} - B^{(1)}_{\| }\right) \left[  -  \tilde \p_{\perp m} \left(B^{m (1)} + n^l h_{l}^{m(1)} \right) + \frac{2}{\tilde \chi} \left( B^{(1)}_{\| } + h^{(1)}_{\|}\right) + 2  \tilde \p_{\perp m}S_{\perp}^{m(1)}  \right]  \nonumber \\
 && + 2 \bigg[ \frac{\ud}{\ud \tilde \chi}\left(2 A^{(1)} - B^{(1)}_{\| }\right)    -  \frac{1}{\tilde\chi} \left(2 A^{(1)} - B^{(1)}_{\| }\right)  \bigg] \kappa^{(1)} + \bigg[  \tilde \p_{\perp i}  
 \left(A^{(1)}{'} - B^{(1)}_{\| }{'} - \frac{1}{2}h^{(1)}_{\| }{'} \right)     +  \frac{1}{\tilde\chi}  \left( B_{\perp i}^{(1)}{'} +  n^j h_{ j k}^{(1)}{'} \Perp_i^k \right) \bigg]   \delta x_{\perp}^{i (1)} \Bigg\}  \;,  \nonumber
 \end{eqnarray}

\begin{eqnarray}
\label{dx_||-2}  
\delta x_\|^{(2)}&=&\bar \chi \left[ A^{(2)}_o-  v^{(2)}_{\| \, o} + \left(v^{(1)}_{\| \, o}\right)^2-\frac{1}{4} \left(h^{(1)}_{\| \, o}\right)^2- \frac{1}{4} n^i h_{ij\,o}^{(1)}\,  \Perp^j_k \, h^{k (1)}_{p \, o} n^p-4A^{(1)}_o v^{(1)}_{\| \, o}- 2 v^{(1)}_{\| \, o} h^{(1)}_{\| \, o}-2 n^i h_{ik\,o}^{(1)} \Perp^k_j v^{j (1)}_o \right] \nonumber \\
&& - 2 \left (A^{(1)}_o-v^{(1)}_{\| \, o}\right)  \int_0^{\bar \chi} \ud \tilde \chi \left[\left ( 2 A^{(1)} + h^{(1)}_{\| } \right)+2  \left(\bar \chi-\tilde \chi\right) \left(A^{(1)}{'} - B^{(1)}_{\| }{'} - \frac{1}{2}h^{(1)}_{\| }{'} \right) \right]\nonumber \\
&& +2  \left(B_{\perp \, o}^{i(1)} -v_{\perp \, o}^{i(1)}+\frac{1}{2}n^k h_{k\,o}^{j(1)} \Perp^i_j  \right)  \int_0^{\bar \chi} \ud \tilde \chi \bigg\{ - h_{i}^{p(1)} n_p + \left(\bar \chi-\tilde \chi\right)
 \bigg[ \tilde \p_\| \left(B_{\perp i}^{(1)} + n^j h_{ j k}^{(1)}\Perp_i^k \right)- \tilde \p_{\perp i}\left(B^{(1)}_{\| }+h^{(1)}_{\| } \right) \nonumber \\
 && + \left. \left.  \frac{1}{\tilde \chi} \left(B_{\perp i}^{(1)}+  2 n^j h_{j}^{k(1)} \Perp^i_k  \right)\right] \right\} + \int_0^{\bar \chi} \ud \tilde \chi \bigg\{- A^{(2)}-\frac{1}{2}h_\|^{(2)}- 2 h^{(1)}_{\| }  \left(A^{(1)}- B^{(1)}_{\| }- \frac{1}{2}h^{(1)}_{\| }\right)  \nonumber \\
 &&+ \left( 2 A^{(1)} - B^{(1)}_{\| } \right)^2+2 \left( 2 A^{(1)} - B^{(1)}_{\| } \right)\left(B^{(1)}_{\| } + h^{(1)}_{\| } \right)+2 \left(B_{\perp}^{i(1)}+ n^k h_{ k}^{j(1)} \Perp^{i}_j  \right) h_{i}^{p(1)} n_p  -4 \left ( 2 A^{(1)} + h^{(1)}_{\| } \right) I^{(1)} \nonumber \\
 &&-4 n_j h_ i^{j(1)} S_{\perp}^{i(1)}\bigg\}+ \int_0^{\bar \chi} \ud \tilde \chi  \left(\bar \chi-\tilde \chi\right)  \left\{- \left(A^{(2)}{'} - B^{(2)}_{\| }{'} - \frac{1}{2}h^{(2)}_{\| }{'} \right)+ 2 \left( 2 A^{(1)} - B^{(1)}_{\| } - 4 I^{(1)}\right)  \right.\nonumber \\
&& \times \left. \left(A^{(1)}{'} - B^{(1)}_{\| }{'}  - \frac{1}{2}h^{(1)}_{\| }{'} \right)    + 2 \left[- \left(B_{\perp}^{i(1)}+ n^j h_{ j k}^{(1)} \Perp^{ik}  \right) +2 S_{\perp }^{i(1)}  \right]  \left[ \tilde \p_\| \left(B_{\perp i}^{(1)} + n^j h_{ j k}^{(1)}\Perp_i^k \right) \right. \right. \nonumber \\
&& - \left. \left. \tilde \p_{\perp i}\left(B^{(1)}_{\| }+h^{(1)}_{\| } \right) + \frac{1}{\tilde \chi} \left(B_{\perp i}^{(1)}+  2 n^j h_{j}^{k(1)} \Perp^i_k  \right)\right]  \right\}+  \delta x^{(2)}_{\| \rm post-Born} \;,
\end{eqnarray}
where
\begin{eqnarray}
 &&  \delta x^{(2)}_{\| \rm post-Born} =+ 2\bar \chi \left( A^{(1)}_o + \frac{1}{2} h^{(1)}_{\| o}\right) \left(3 v_{\| \, o}^{(1)} + 2 h^{(1)}_{\| \, o} - \frac{1}{2} h^{j (1)}_{j \, o} \right)    - \bar \chi \left(A^{(1)}_o + \frac{1}{2}  h^{(1)}_{\| o}\right)^2  + 4\left(A^{(1)}_o-v^{(1)}_{\| \, o} \right) \bigg[-\frac{\bar \chi}{2}\left(A^{(1)} + \frac{1}{2}  h^{(1)}_{\| }\right) 
  \nonumber\\
    && +  \int_0^{\bar \chi} \ud \tilde{\chi}   \left(A^{(1)} + \frac{1}{2}  h^{(1)}_{\| } - I^{(1)} \right) \bigg]
 +4\left(- B^{(1)}_{\| o} + v_{\| \, o}^{(1)} - \frac34 h^{(1)}_{\| \, o} + \frac14 h^{m (1)}_{m \, o} \right)   \int_0^{\bar \chi}   \ud \tilde{\chi} ~ \frac{(\bar \chi - \tilde \chi)  }{\tilde \chi}\left( A^{(1)} + \frac{1}{2}  h^{(1)}_{\| }\right)    \nonumber\\
 &&  +2 \left(B^{i (1)}_{\perp \, o }-v^{i (1)}_{\perp \, o }+ \frac{1}{2} n^k h_{k\,o}^{j(1)} \Perp^i_j  \right) \int_0^{\bar \chi}   \ud \tilde{\chi} ~ (\bar \chi - \tilde \chi) \left[ \tilde \p_{\perp i} \left(A^{(1)} + \frac{1}{2}  h^{(1)}_{\| } \right) -  \frac{1}{\bar \chi} n^j h_{ j}^{k(1)} \Perp_{ik} \right]  \nonumber\\
    &&   +2\left(A^{(1)} + \frac{1}{2}  h^{(1)}_{\| }\right) \int_0^{\bar \chi} \ud \tilde \chi \left(A^{(1)}+ \frac{1}{2}  h^{(1)}_{\| }  - 2I^{(1)} \right)  +2\int_0^{\bar \chi} \ud \tilde{\chi} \Bigg\{    -\left(A^{(1)}{'} - B^{(1)}_{\| }{'} - \frac{1}{2}h^{(1)}_{\| }{'} \right)  \delta x_{\|}^{(1)} \nonumber\\
    &&  - \left(A^{(1)} + \frac{1}{2}  h^{(1)}_{\| }\right) \left[ \frac{3}{2}\left(A^{(1)} + \frac{1}{2}  h^{(1)}_{\| }\right)  -4I^{(1)}\right] 
    + \left(A^{(1)}{'} + \frac{1}{2}  h^{(1)}_{\| }{'} \right)  T^{(1)}   \nonumber\\
         &&  +2 \left(A^{(1)} + \frac{1}{2}  h^{(1)}_{\| } \right)   \kappa^{(1)} - \bigg[  \tilde \p_{\perp j}  
 \left(A^{(1)}+\frac{1}{2}   h^{(1)}_{\| } \right)     -  \frac{1}{\tilde\chi}    n^j h_j^{k(1)} \Perp_{jk} \ \bigg]   \delta x_{\perp}^{j (1)} \Bigg\} \nonumber\\
&&  +2 \int_0^{\bar \chi}   \ud \tilde{\chi} ~ (\bar \chi - \tilde \chi)\Bigg\{ + \left(A^{(1)}{'} + \frac{1}{2}  h^{(1)}_{\| }{'}\right) \left(A^{(1)}-B^{(1)}_{\| }- \frac{1}{2} h^{(1)}_{\| } \right)   +  \left(A^{(1)}{''} - B^{(1)}_{\| }{''} - \frac{1}{2}h^{(1)}_{\| }{''} \right)  T^{(1)}  \nonumber \\
 &&  +2\left(A^{(1)}{'} - B^{(1)}_{\| }{'} - \frac{1}{2}h^{(1)}_{\| }{'} \right) I^{(1)}    -\left[ \tilde \p_{\perp i} \left(A^{(1)} + \frac{1}{2}  h^{(1)}_{\| } \right) -  \frac{1}{\tilde \chi} n^j h_{ j}^{k(1)} \Perp_{ik} \right] \left( B^{i (1)}_{\perp }+ n^k h_{k}^{j(1)} \Perp^i_j - 2S_{\perp}^{i(1)} \right) \nonumber\\
    && +  \left( A^{(1)} + \frac{1}{2}  h^{(1)}_{\| }\right)  \left[ -  \tilde \p_{\perp m} \left(B^{m (1)} + n^l h_{l}^{m(1)} \right) + \frac{2}{\tilde \chi} \left( B^{(1)}_{\| } + h^{(1)}_{\|}\right) + 2  \tilde \p_{\perp m}S_{\perp}^{m(1)}  \right]    \nonumber\\
    && + 2 \bigg[ - \frac{\ud}{\ud \tilde \chi}\left(A^{(1)} + \frac{1}{2}  h^{(1)}_{\| } \right)    +  \frac{1}{\tilde\chi} \left(A^{(1)} + \frac{1}{2}  h^{(1)}_{\| } \right)  \bigg] \kappa^{(1)}\nonumber\\
    &&  - \bigg[  \tilde \p_{\perp i}  
 \left(A^{(1)}{'} - B^{(1)}_{\| }{'} - \frac{1}{2}h^{(1)}_{\| }{'} \right)     +  \frac{1}{\tilde\chi}  \left( B_{\perp i}^{(1)}{'} +  n^j h_{ j k}^{(1)}{'} \Perp_i^k \right) \bigg]   \delta x_{\perp}^{i (1)} \Bigg\} \;,  \nonumber
 \end{eqnarray}

and

\begin{eqnarray}
\label{dx_perp-2}  
 \delta x_\perp^{i(2)}&=& \bar \chi \bigg[  B^{i(2)}_{\perp \, o}-  v^{i(2)}_{\perp \, o} + \frac{1}{2} n^j h_{ j k\, o}^{(2)} \Perp^{ki}+   v^{(1)}_{\| \, o}v^{i(1)}_{\perp \, o} - 3 v^{(1)}_{\| \, o}B^{i(1)}_{\perp \, o} +  B^{(1)}_{\| \, o}B^{i(1)}_{\perp \, o}-  B^{(1)}_{\| \, o}v^{i(1)}_{\perp \, o} + 2 A^{(1)}_o n^j h_{ j k\, o}^{(1)} \Perp^{ki} \nonumber \\
&&-4 v^{(1)}_{\| \, o} n^j h_{ j k\, o}^{(1)} \Perp^{ki}  -2 v^{j (1)}_{\perp o}\Perp^{l}_j  \, h^{k (1)}_{l \, o} \Perp_{k}^{i}-\frac{1}{4} h^{(1)}_{\| \, o} n^j h_{ j k\, o}^{(1)} \Perp^{ki}  -\frac{1}{4} n^j h^{k (1)}_{j \, o}  \Perp^{l}_k  \, h^{p (1)}_{l \, o} \Perp_{p}^{i} \bigg] \nonumber \\
&& - 4  \left (A^{(1)}_o-v^{(1)}_{\| \, o}\right) \int_0^{\bar \chi} \ud \tilde \chi \bigg\{\left(B_{\perp}^{i(1)}+ n^k h_{ k}^{j(1)} \Perp^{i}_j \right)+ \left(\bar \chi-\tilde \chi\right)\left[ \tilde \p^i_\perp \left( A^{(1)} - B^{(1)}_{\| } - \frac{1}{2}h^{(1)}_{\| } \right) \right. \nonumber \\
&& +\left. \frac{1}{\tilde \chi} \left(B^{i (1)}_{\perp }+  n^k h_{kj}^{(1)} \Perp^{ij}  \right)\right] \bigg\} + 2 \left( B^{j (1)}_{\perp \, o }-v^{j (1)}_{\perp \, o }+ \frac{1}{2} n^k h_{k\,o}^{l(1)} \Perp^j_l \right) \int_0^{\bar \chi} \ud \tilde \chi \bigg\{  - \Perp^{m}_j  \, h^{p (1)}_{m} \Perp_{p}^{i} \nonumber \\
&& +\left. 2\left(\bar \chi-\tilde \chi\right) \left[ \tilde \p_{\perp}^{[i} B_{\perp}^{j](1)}  + \tilde\p_{\perp}^{[i}     \left( \Perp^{j]}_m h^{m(1)}_q n^q \right)   - \frac{1}{\tilde \chi} \left(n^{[i} B_{\perp}^{j](1)} +n^{[i}  \Perp^{j]}_m h^{m(1)}_q n^q \right)   \right] \right\}  \nonumber \\
&& + \int_0^{\bar \chi} \ud \tilde \chi \bigg\{  -   B^{i(2)}_{\perp} -  n^j h_{ j k}^{(2)} \Perp^{ki} + 4 A^{(1)} B^{i(1)}_{\perp} - 2 B^{(1)}_{\|}B^{i(1)}_{\perp}  + 2 A^{(1)} n^j h_{ j k}^{(1)} \Perp^{ki}+ h_\|^{(1)} n^j h_{ j k}^{(1)} \Perp^{ki} \nonumber \\
&& +2  \left( B^{j (1)}_{\perp }+ n^k h_{k}^{p(1)} \Perp^j_p\right)  \Perp^{l}_j  \, h^{k (1)}_{l} \Perp_{k}^{i(1)}- 8 \left(B_{\perp}^{i(1)}+ n^j h_{ j k}^{(1)} \Perp^{ik}  \right) I^{(1)}- 4\,  \Perp^{l}_j  \, h^{k (1)}_{l} \Perp_{k}^{i} S_{\perp}^{j(1)} \bigg\} \nonumber \\
&& + \int_0^{\bar \chi} \ud \tilde \chi \left(\bar \chi-\tilde \chi\right) \left\{-\left[\tilde \p^i_\perp \left( A^{(2)} - B^{(2)}_{\| } - \frac{1}{2}h^{(2)}_{\| } \right) + \frac{1}{\tilde \chi} \left(B^{i (2)}_{\perp }+  n^k h_{kj}^{(2)} \Perp^{ij}  \right)\right]  -    2 \left(A^{(1)}{'} - B^{(1)}_{\| }{'} - \frac{1}{2}h^{(1)}_{\| }{'} \right)  \right. \nonumber \\
&& \left.   \times \left(B_{\perp}^{i(1)}+ n^j h_{ j k}^{(1)} \Perp^{ik}  \right)  + 2 \left(2 A^{(1)} -B_\|^{(1)}\right)  \frac{\ud}{\ud \tilde \chi} \left(B_{\perp}^{i(1)}+ n^j h_{ j k}^{(1)} \Perp^{ik}  \right) +  4 \left(2 A^{(1)} -B_\|^{(1)} - 2 I^{(1)}\right)  \right. \nonumber \\
&&  \left. \times  \left[\tilde\p^i_\perp \left( A^{(1)} - B^{(1)}_{\| } - \frac{1}{2}h^{(1)}_{\| } \right) + \frac{1}{\tilde \chi} \left(B^{i (1)}_{\perp }+  n^k h_{kj}^{(1)} \Perp^{ij}  \right) \right] +4 \left[ - \left(B_{\perp j}^{(1)}+ n^p h_{p k}^{(1)} \Perp_j^{k}  \right)+ 2 \delta_{jp}  S_{\perp}^{p(1)} \right]  \right. \nonumber \\
&& \left.  \times \left[ \tilde \p_{\perp}^{[i} B_{\perp}^{j](1)}  + \tilde \p_{\perp}^{[i}     \left( \Perp^{j]}_m h^{m(1)}_q n^q \right)   - \frac{1}{\tilde \chi} \left(n^{[i} B_{\perp}^{j](1)}+ n^{[i}  \Perp^{j]}_m h^{m(1)}_q n^q \right)   \right]+2\left( A^{(1)} - B^{(1)}_{\| } - \frac{1}{2}h^{(1)}_{\| } \right) \right.  \nonumber \\
&& \left.  \times \bigg[ \tilde\p^i_\perp \left(B^{(1)}_{\| }+ h^{(1)}_{\| } \right) - \p_\| \left(B_{\perp}^{i (1)}+n^p h_{pq}^{(1)} \Perp^{iq} \right) - \frac{1}{\tilde \chi} \left(B^{i (1)}_{\perp }+  2 n^p h_{pq}^{(1)} \Perp^{iq}  \right) \bigg]\right\} +  \delta x_{\perp  \rm post-Born}^{i(2)}\;,
\end{eqnarray}
where
 \begin{eqnarray}
 &&  \delta x_{\perp  \rm post-Born}^{i(2)}= 2\bar \chi \left(   B^{i (1)}_{\perp \, o } +  n^k h_{k\,o}^{j(1)} \Perp^i_j \right) \left(3 v_{\| \, o}^{(1)} + 2 h^{(1)}_{\| \, o}    - \frac{1}{2} h^{j (1)}_{j \, o} \right)    -2 \bar \chi \left( A^{(1)}_o  - B^{(1)}_{\| o} - \frac{1}{2} h^{(1)}_{\| o}\right) \left(   v^{i (1)}_{\perp \, o } + \frac{1}{2} n^k h_{k\,o}^{j(1)} \Perp^i_j \right)  \nonumber \\
&&  +4 \left(A^{(1)}_o-v^{(1)}_{\| \, o}  \right) \bigg[- \frac{\bar \chi}{2} \ \left(B^{i (1)}_{\perp }+ n^k h_{k}^{j(1)} \Perp^i_j \right)    +  \int_0^{\bar \chi} \ud \tilde{\chi}  \left(B^{i (1)}_{\perp }+ n^k h_{k}^{j(1)} \Perp^i_j  - S_{\perp}^{i(1)}  \right) \bigg]   \nonumber\\
&& +4\left(- B^{(1)}_{\| o} + v_{\| \, o}^{(1)} - \frac34 h^{(1)}_{\| \, o} + \frac14 h^{m (1)}_{m \, o} \right) \int_0^{\bar \chi} \ud \tilde{\chi} ~ \frac{ (\bar \chi - \tilde \chi) }{\tilde \chi}    \left( B^{i (1)}_{\perp }+ n^k h_{k}^{j(1)} \Perp^i_j \right)    + 2 \left(  B^{j (1)}_{\perp \, o }-v^{j (1)}_{\perp \, o }+ \frac{1}{2} n^l h_{l\,o}^{p(1)} \Perp^j_p  \right)   \nonumber\\
&& \times   \int_0^{\bar \chi} \ud \tilde{\chi} ~ (\bar \chi - \tilde \chi)     \left[  \Perp^i_m  \tilde \p_{\perp j} \left(B^{m (1)} + n^k h_{k}^{m(1)} \right) -  \frac{1}{\tilde\chi} \Perp^i_m   h_{j}^{m(1)} + \frac{1}{\tilde \chi} \Perp^i_j \left(A^{(1)} - B^{(1)}_{\| } - \frac{1}{2}h^{(1)}_{\| } \right) \right]   \nonumber\\
&& \nonumber\\
&&  + 2\left(B^{i (1)}_{\perp }+ n^k h_{k}^{j(1)} \Perp^i_j \right)   \int_0^{\bar \chi} \ud \tilde \chi  \left(A^{(1)}+ \frac{1}{2}  h^{(1)}_{\| } - 2I^{(1)}  \right) +2 \int_0^{\bar \chi} \ud \tilde{\chi} \Bigg\{   - \bigg[ \tilde \p^i_{\perp}  \left(A^{(1)} - B^{(1)}_{\| } - \frac{1}{2}h^{(1)}_{\| } \right)   \nonumber\\
&&    +  \frac{1}{\tilde \chi} \left(B^{i (1)}_{\perp }+ n^k h_{k}^{j(1)} \Perp^i_j \right) \bigg] \delta x_{\|}^{(1)} +  \left( B_{\perp i}^{(1)}{'} +  n^j h_{ j k}^{(1)}{'} \Perp_i^k \right) T^{(1)}   - 2 \left(B^{i (1)}_{\perp }+ n^k h_{k}^{j(1)} \Perp^i_j \right) \bigg( A^{(1)} + \frac{1}{2}  h^{(1)}_{\| } - 2I^{(1)}  \nonumber\\
&&      -  \kappa^{(1)} \bigg)  -    \left[  \Perp^i_m  \tilde \p_{\perp j} \left(B^{m (1)} + n^k h_{k}^{m(1)} \right) -  \frac{1}{\tilde\chi} \Perp^i_m   h_{j}^{m(1)} + \frac{1}{\tilde \chi} \Perp^i_j \left(A^{(1)} - B^{(1)}_{\| } - \frac{1}{2}h^{(1)}_{\| } \right) \right] \delta x_{\perp}^{j (1)}  \Bigg\}   \nonumber\\
&& +2 \int_0^{\bar \chi} \ud \tilde{\chi} ~ (\bar \chi - \tilde \chi)\Bigg\{ - \bigg[ \tilde \p_{\perp }^i    \left(A^{(1)}{'} - B^{(1)}_{\| }{'} - \frac{1}{2}h^{(1)}_{\| }{'} \right)   -  \frac{1}{\tilde\chi^2} \left(B^{i (1)}_{\perp }+ n^k h_{k}^{j(1)} \Perp^i_j \right)     +  \frac{1}{\tilde\chi}  \left( B_{\perp }^{i(1)}{'} +  n^j h_{ j }^{k(1)}{'} \Perp_k^i \right) \bigg] T^{(1)}  \nonumber\\
&&       + \left( B_{\perp }^{i(1)}{'} +  n^j h_{ j }^{k(1)}{'} \Perp_k^i \right)\left( A^{(1)}-B^{(1)}_{\| }- \frac{1}{2} h^{(1)}_{\| } \right)   - \left[  \tilde \p_{\perp }^i    \left(A^{(1)} - B^{(1)}_{\| } - \frac{1}{2}h^{(1)}_{\| }  \right)   +  \frac{1}{\tilde\chi}  \left(B^{i (1)}_{\perp }+ n^k h_{k}^{j(1)} \Perp^i_j \right)  \right]   \nonumber\\
&& \times  \left(A^{(1)} +  \frac{1}{2}  h^{(1)}_{\| }  - 2I^{(1)} \right) + \left(B^{i (1)}_{\perp }+ n^k h_{k}^{j(1)} \Perp^i_j \right)  \bigg[\frac{\ud}{\ud \tilde \chi}\left(A^{(1)} + \frac{1}{2}  h^{(1)}_{\| }\right)  +\left(A^{(1)}{'} - B^{(1)}_{\| }{'} - \frac{1}{2}h^{(1)}_{\| }{'} \right) \bigg]    
\nonumber\\
&& -   \left[  \Perp^i_m  \tilde \p_{\perp j} \left(B^{m (1)} + n^k h_{k}^{m(1)} \right) -  \frac{1}{\tilde\chi} \Perp^i_m   h_{j}^{m(1)} + \frac{1}{\tilde \chi} \Perp^i_j \left(A^{(1)} - B^{(1)}_{\| } - \frac{1}{2}h^{(1)}_{\| } \right) \right]  \left[  \left( B^{j (1)}_{\perp }+ n^l h_{l}^{p(1)} \Perp^j_p\right) - 2S_{\perp}^{j(1)} \right]    \nonumber\\
&&  +  \left( B^{i (1)}_{\perp }+ n^k h_{k}^{j(1)} \Perp^i_j \right)   \left[ -  \tilde \p_{\perp m} \left(B^{m (1)} + n^l h_{l}^{m(1)} \right) + \frac{2}{\tilde \chi} \left( B^{(1)}_{\| } + h^{(1)}_{\|}\right) + 2  \tilde \p_{\perp m}S_{\perp}^{m(1)}  \right] 
   \nonumber \\
 &&   + 2 \bigg[ - \frac{\ud}{\ud \tilde \chi}\left(B^{i (1)}_{\perp }+ n^k h_{k}^{j(1)} \Perp^i_j \right)    +  \frac{1}{\tilde\chi} \left(B^{i (1)}_{\perp }+ n^k h_{k}^{j(1)} \Perp^i_j \right)  \bigg] \kappa^{(1)}     -  \bigg[     \Perp^{im} \tilde \p_{\perp j}   \tilde \p_{\perp m}    \left(A^{(1)} - B^{(1)}_{\| } - \frac{1}{2}h^{(1)}_{\| } \right)   \nonumber\\
&&  + \frac{1}{\tilde \chi} \Perp^i_j  \left(A^{(1)}{'} - B^{(1)}_{\| }{'} - \frac{1}{2}h^{(1)}_{\| }{'} \right)  +    \frac{1}{\tilde \chi^2} \Perp^i_j  \left(A^{(1)} - B^{(1)}_{\| } - \frac{1}{2}h^{(1)}_{\| } \right)  +\frac{1}{\tilde\chi}  \Perp^i_m \tilde \p_{\perp j} \left(B^{m (1)}_{\perp }+ n^k   h_{k}^{l(1)} \Perp^{m}_l \right)  \nonumber \\
 && + \frac{1}{\tilde\chi}   \Perp^k_j  \p^i_{\perp} B^{ (1)}_k +  \frac{1}{\tilde\chi}   \p^i_{\perp} \left(h_{jl}^{(1)} n^l \right) - \frac{1}{\tilde\chi^2}\Perp^i_l h_{j}^{l(1)} \bigg] \delta x_{\perp}^{j (1)}  \Bigg\} \;. \nonumber
 \end{eqnarray}
 
Combining Eqs.\  (\ref{dx0-2}) and (\ref{dx_||-2})  [or integrating Eq.\ (\ref{dnu+de_||-2})], we have
\begin{eqnarray} 
\label{dx0+dx_||-2}
 \delta x^{0 (2)} + \delta x_\|^{(2)}&=&
 \bar \chi \bigg [\left(A^{(1)}_o\right)^2 - 2 A^{(1)}_o B^{(1)}_{\| \, o}+ \left(B^{(1)}_{\| \, o}\right)^2 +2  A^{(1)}_o v^{(1)}_{\| \, o} - 2 B^{(1)}_{\| \, o}v^{(1)}_{\| \, o}-   v^{(1)}_{\| \, o} h^{(1)}_{\| \, o} -\frac{1}{4} \left(h^{(1)}_{\| \, o}\right)^2   -  v^{(1)}_{\perp k \, o} v^{k (1)}_{\perp \, o}\nonumber \\
&& +  n^i h_{ik\,o}^{(1)} \Perp^k_j v^{j (1)}_o - n^i h_{ik\,o}^{(1)} \Perp^k_j B^{j (1)}_o - \frac{1}{4} n^i h_{ij\,o}^{(1)}\,  \Perp^j_k \, h^{k (1)}_{p \, o} n^p   - B_{\perp \, o}^{i(1)}B_{\perp i \, o}^{(1)}+2  v_{\perp \, o}^{i(1)}B_{\perp i \, o}^{(1)} \bigg]\nonumber \\
&&-4 \left (A^{(1)}_o-v^{(1)}_{\| \, o}\right)T^{(1)}  +4 \left(B_{\perp  i \, o}^{(1)} -v_{\perp i \, o}^{(1)}+\frac{1}{2}n^k h_{k\,o}^{j(1)} \Perp_{ij} \right)\nonumber \\
&&\times \int_0^{\bar \chi} \ud \tilde \chi\left(\bar \chi-\tilde\chi\right)\left[\tilde \p^i_\perp \left( A^{(1)} - B^{(1)}_{\| } - \frac{1}{2}h^{(1)}_{\| } \right) + \frac{1}{\tilde \chi} \left(B^{i (1)}_{\perp }+  n^k h_{kj}^{(1)} \Perp^{ij}  \right)\right]  \nonumber \\
&& - T^{(2)}+ 2 \int_0^{\bar \chi} \ud \tilde \chi \bigg[-\left( 2 A^{(1)} - B^{(1)}_{\| } \right)^2+ \frac{1}{2} \left(B^{(1)}_{\| } + h^{(1)}_{\| } \right) \left( 2 A^{(1)} + h^{(1)}_{\| } \right)\nonumber \\
&& + \frac{1}{2} \left(B_{\perp}^{i(1)} +  n^k h_{k}^{j(1)}  \Perp^i_j \right)  \left(B_{\perp i}^{(1)}+ n^p h_{p m}^{(1)} \Perp^m_i  \right)  +4 \left(A^{(1)} - B^{(1)}_{\| }- \frac{1}{2}h^{(1)}_{\| }\right)  I^{(1)}  -4 S_{\perp }^{i(1)}S_{\perp }^{j(1)} \delta_{ij}  \bigg]   \nonumber \\
&&+ 2 \int_0^{\bar \chi} \ud \tilde \chi\left(\bar \chi-\tilde\chi\right) \left\{ \left(A^{(1)} - B^{(1)}_{\| } - \frac{1}{2}h^{(1)}_{\| } \right) \left[2\left(A^{(1)}{'} - B^{(1)}_{\| }{'} - \frac{1}{2}h^{(1)}_{\| }{'} \right)+  \frac{\ud}{\ud \tilde \chi}\left( 2 A^{(1)} - B^{(1)}_{\| }\right) \right]  \right. \nonumber \\
&&  - \left(B_{\perp}^{i(1)} + n^k h_{k}^{j(1)} \Perp^i_j  \right)\left[ \tilde \p_{\perp i}\left (  A^{(1)} - B^{(1)}_{\| } - \frac{1}{2} h^{(1)}_{\| } \right)  + \frac{1}{\tilde \chi} \left(B_{\perp i}^{(1)}+   n^m h_{mp}^{(1)} \Perp^p_i  \right)\right]\bigg\}\nonumber \\
&&  + \left(\delta x^{0 (2)} + \delta x_\|^{(2)}\right)_{\rm post-Born}\;,
\end{eqnarray}
where
\begin{eqnarray}
 && \left(\delta x^{0 (2)} + \delta x_\|^{(2)}\right)_{\rm post-Born}= -2 \bar \chi \left( A^{(1)}_o -  B^{(1)}_{\| o} - \frac{1}{2} h^{(1)}_{\| o}\right) \left(3 v_{\| \, o}^{(1)} + 2 h^{(1)}_{\| \, o}  - \frac{1}{2} h^{j (1)}_{j \, o} \right) +  2 \left(A^{(1)}_o-v^{(1)}_{\| \, o}\right)   \nonumber \\
 &&\times \left[ \bar \chi  \left(A^{(1)}  - B^{(1)}_{\| } - \frac{1}{2}  h^{(1)}_{\| }\right) +2 T^{(1)} \right]  -4 \left(- B^{(1)}_{\| o} + v_{\| \, o}^{(1)} - \frac34 h^{(1)}_{\| \, o} + \frac14 h^{m (1)}_{m \, o} \right) \int_0^{\bar \chi}  \ud \tilde{\chi} ~  \frac{(\bar \chi - \tilde \chi)}{\tilde \chi}   \left( A^{(1)} - B^{(1)}_{\| } - \frac{1}{2} h^{(1)}_{\|} \right)    \nonumber \\
  && + 4 \left(  B^{i (1)}_{\perp \, o }-v^{i (1)}_{\perp \, o }+ \frac{1}{2} n^k h_{k\,o}^{j(1)} \Perp^i_j \right) \int_0^{\bar \chi}  \ud \tilde{\chi} S_{\perp i}^{(1)}  - 2  \left(A^{(1)}  - B^{(1)}_{\| } - \frac{1}{2}  h^{(1)}_{\| }\right) \int_0^{\bar \chi} \ud \tilde \chi \left[ \left(A^{(1)}+ \frac{1}{2}  h^{(1)}_{\| }\right) - 2I^{(1)} \right]   \nonumber \\
 &&   +2 \int_0^{\bar \chi}  \ud \tilde{\chi} \Bigg\{ -  \left(A^{(1)}{'} - B^{(1)}_{\| }{'}  - \frac{1}{2}  h^{(1)}_{\| }{'} \right)T^{(1)} +2 \left(A^{(1)}   -  B^{(1)}_{\| } - \frac{1}{2}  h^{(1)}_{\| }\right) \left(  A^{(1)} + \frac{1}{2}  h^{(1)}_{\| } -2I^{(1)} - \kappa^{(1)}  \right) \nonumber \\
 && +  \left[ \tilde \p_{\perp i}\left(A^{(1)} - B^{(1)}_{\| } - \frac{1}{2}  h^{(1)}_{\| } \right) + \frac{1}{\tilde \chi} \left(B^{(1)}_{\perp i}+  n^j h_{ j}^{k(1)} \Perp_{ik}  \right) \right] \delta x_{\perp}^{i (1)}  \Bigg\} \nonumber \\
 && +2 \int_0^{\bar \chi}  \ud \tilde{\chi} ~ (\bar \chi - \tilde \chi) \Bigg\{  - \left( A^{(1)} - B^{(1)}_{\| } - \frac{1}{2} h^{(1)}_{\|} \right)    \bigg[\frac{\ud}{\ud \tilde \chi}\left(A^{(1)} + \frac{1}{2}  h^{(1)}_{\| }\right)  + 2 \left(A^{(1)}{'} - B^{(1)}_{\| }{'} - \frac{1}{2}h^{(1)}_{\| }{'} \right) \bigg]     \nonumber \\
 && -  \left[ \tilde \p_{\perp j}\left( A^{(1)} - B^{(1)}_{\| } - \frac{1}{2} h^{(1)}_{\|} \right) +  \frac{1}{\tilde \chi} \left(B^{(1)}_{\perp j}+  n^m h_{ m}^{k(1)} \Perp_{jk}  \right) \right] \left[  -\left( B^{i (1)}_{\perp }+ n^k h_{k}^{j(1)} \Perp^i_j\right) + 2S_{\perp}^{i(1)} \right] \nonumber \\
 &&  - \left( A^{(1)} - B^{(1)}_{\| } - \frac{1}{2} h^{(1)}_{\|} \right)     \left[ -  \tilde \p_{\perp m} \left(B^{m (1)} + n^l h_{l}^{m(1)} \right) + \frac{2}{\tilde \chi} \left( B^{(1)}_{\| } + h^{(1)}_{\|}\right) + 2  \tilde \p_{\perp m}S_{\perp}^{m(1)}  \right]   \nonumber \\
 && + 2 \bigg[ \frac{\ud}{\ud \tilde \chi}\left( A^{(1)} - B^{(1)}_{\| } - \frac{1}{2} h^{(1)}_{\|} \right)    -  \frac{1}{\tilde\chi} \left( A^{(1)} - B^{(1)}_{\| } - \frac{1}{2} h^{(1)}_{\|} \right)  \bigg] \kappa^{(1)}  \nonumber \Bigg\}\;.
\end{eqnarray}

To obtain all the second order terms 
we require $\Delta \ln a^{(1)}$ (or $\Delta x^{0(1)}$),  $\delta \chi^{(1)}$,  $\Delta x^{0(1)}$, $ \Delta x^{(1)}_\|$,  $\Delta x_{\perp}^{i (1)}$ and $\Delta_g^{(1)}$. From Eqs.  (\ref{Ek-1}) and (\ref{chi_1}) we have
\begin{eqnarray}
\label{Deltalna-1}
\Delta \ln a^{(1)}&=&- E_{\hat{0}0}^{(1)} + E_{\hat{0}\|}^{(1)} - \delta \nu^{(1)}=\left (A^{(1)}_o-v^{(1)}_{\| \, o}\right) - A^{(1)}+ v_\|^{(1)}+ 2I^{(1)} \nonumber \\
&=&+\left (A^{(1)}_o-v^{(1)}_{\| \, o}\right) - A^{(1)}+ v_\|^{(1)}- \int_0^{\bar \chi} \ud \tilde \chi \left(A^{(1)}{'} - B^{(1)}_{\| }{'} - \frac{1}{2}h^{(1)}_{\| }{'} \right)\;, \\
\delta \chi^{(1)} &=& \delta x^{0(1)}- \Delta x^{0(1)}= \delta x^{0(1)}-\frac{1}{\cH} \Delta  \ln a^{(1)}=-\left(\bar \chi+\frac{1}{\cH}\right)\left(A^{(1)}_o-v^{(1)}_{\| \, o}\right)+  \frac{1}{\cH}\left(A^{(1)}- v_\|^{(1)}\right) \nonumber \\
&& + \int_0^{\bar \chi} \ud \tilde \chi \left[ 2 A^{(1)} - B^{(1)}_{\| }+\left(\bar \chi-\tilde \chi\right) \left(A^{(1)}{'} - B^{(1)}_{\| }{'} - \frac{1}{2}h^{(1)}_{\| }{'} \right) \right] -\frac{2}{\cH}I^{(1)} \;.
\end{eqnarray}
Then, from Eq. (\ref{Dx0_1})
\begin{eqnarray}
\label{Dx^0-1}
\Delta x^{0(1)}&=&\frac{1}{\cH}\left[\left (A^{(1)}_o-v^{(1)}_{\| \, o}\right) - A^{(1)}+ v_\|^{(1)}+ 2I^{(1)}\right] \nonumber \\
&=&  \frac{1}{\cH}\left[\left (A^{(1)}_o-v^{(1)}_{\| \, o}\right) - A^{(1)}+ v_\|^{(1)}- \int_0^{\bar \chi} \ud \tilde \chi \left(A^{(1)}{'} - B^{(1)}_{\| }{'} - \frac{1}{2}h^{(1)}_{\| }{'} \right) \right]\;, 
\end{eqnarray}
and from Eqs.\ (\ref{Dx_||-1}) and (\ref{s-1})
\begin{eqnarray}
\label{Dx||-1}
&&\Delta x^{(1)}_\| = - T^{(1)} - \Delta x^{0(1)}=- T^{(1)}- \frac{1}{\cH}\Delta \ln a^{(1)} =- T^{(1)} - \frac{1}{\cH}\left[\left (A^{(1)}_o-v^{(1)}_{\| \, o}\right) - A^{(1)}+ v_\|^{(1)}+ 2I^{(1)}\right]\nonumber \\
&&= \int_0^{\bar \chi} \ud \tilde \chi \left(A^{(1)} - B^{(1)}_{\| }- \frac{1}{2}h^{(1)}_{\| }\right)- \frac{1}{\cH}\left[\left (A^{(1)}_o-v^{(1)}_{\| \, o}\right) - A^{(1)}+ v_\|^{(1)}- \int_0^{\bar \chi} \ud \tilde \chi \left(A^{(1)}{'} - B^{(1)}_{\| }{'} - \frac{1}{2}h^{(1)}_{\| }{'} \right) \right]. 
 \end{eqnarray}
 Using Eq. (\ref{Dx_perp-1}), we have
 \begin{eqnarray}
 \Delta x_{\perp}^{i (1)}&=& \bar \chi \left(B^{i (1)}_{\perp \, o }-v^{i (1)}_{\perp \, o }+ \frac{1}{2} n^k h_{k\,o}^{j(1)} \Perp^i_j\right)-\int_0^{\bar \chi} \ud \tilde \chi \bigg\{ \left( B^{i (1)}_{\perp }+ n^k h_{k}^{j(1)} \Perp^i_j\right)    \nonumber \\
&& + \left(\bar \chi-\tilde \chi\right) \left[ \tilde \p^i_\perp \left( A^{(1)} - B^{(1)}_{\| } - \frac{1}{2}h^{(1)}_{\| } \right) + \frac{1}{\tilde \chi} \left(B^{i (1)}_{\perp }+  n^k h_{kj}^{(1)} \Perp^{ij}  \right)\right]\bigg\} \;.
\end{eqnarray}
In Eq.\ (\ref{Dx^0-1}) there is an ISW contribution and in Eq.\ (\ref{Dx||-1}) we have both time-delay and ISW contributions.

Now we can obtain $\Delta_g^{(1)} $. Using Eq.  (\ref{partialparallep}) for $\Delta x^{(1)}_\| $, we find
\begin{eqnarray}
\label{partial_||Dx||-1}
\p_\| \Delta x^{(1)}_\| &=& \p_{\bar \chi} \Delta x^{(1)}_\|= - \p_{\bar \chi} \left( T^{(1)} + \Delta x^{0(1)}\right)= \left(A^{(1)} - B^{(1)}_{\| }- \frac{1}{2}h^{(1)}_{\| }\right) - \frac{\cH'}{\cH^2}\Delta \ln a^{(1)}
- \frac{1}{\cH}\left(\frac{\ud \, \Delta \ln a}{\ud \bar \chi}\right)^{(1)} \nonumber \\
&=&\left(A^{(1)} - B^{(1)}_{\| }- \frac{1}{2}h^{(1)}_{\| }\right)+\frac{1}{\cH}\left[\frac{\ud \,}{\ud \bar \chi} \left( A^{(1)}-  v_\|^{(1)}\right)+ \left(A^{(1)}{'} - B^{(1)}_{\| }{'} - \frac{1}{2}h^{(1)}_{\| }{'} \right)\right]- \frac{\cH'}{\cH^2}\Delta \ln a^{(1)}. 
\end{eqnarray}
From Eqs.\ (\ref{Deltalna-1}), (\ref{Dx||-1}) and  (\ref{partial_||Dx||-1}), we find that Eq. (\ref{Deltag-1}) becomes
\begin{eqnarray}
\label{Deltag-1_2}
\Delta_g^{(1)}
 &=&  \delta_g^{(1)} + \left( b_e  - \frac{\cH'}{\cH^2} - \frac{2}{\bar \chi \cH}\right)\Delta \ln a^{(1)}  +\frac{1}{\cH}\left[\frac{\ud \,}{\ud \bar \chi} \left( A^{(1)}-  v_\|^{(1)}\right)+ \left(A^{(1)}{'} - B^{(1)}_{\| }{'} -  \frac{1}{2}h^{(1)}_{\| }{'} \right)\right] \nonumber \\
&& - \frac{2}{\bar \chi} T^{(1)}  - 2 \kappa^{(1)}   +   A^{(1)}+v_\|^{(1)} - B^{(1)}_{\| }- \frac{1}{2}h^{(1)}_{\| }+\frac{1}{2} h^{i(1)}_i ,
\end{eqnarray}
in agreement with  \cite{Yoo:2009au, Yoo:2010ni, Bonvin:2011bg, Challinor:2011bk,Jeong:2011as}.

Making explicit at first order the coordinate convergence lensing term  defined in Eq. (\ref{kappa-n}) (see also \cite{Schmidt:2012ne}), we find\footnote{To compute correctly  the lensing term, we need further properties of the parallel and orthogonal derivatives.
If
$\tilde x^j$ is not necessarily the same as $\bar x^i$ (i.e. $\tilde \chi$ can be different from $\bar \chi$),
then
${\p \bar x^i(\bar \chi)}/{\p \tilde x^j}= {\p \bar \chi}/{\p \tilde x^j} n^i+ \bar \chi {\p n^i}/{\p \tilde x^j}= ({\p \bar \chi}/{\p \tilde x^j}) n^i+ ({\bar \chi}/{\tilde \chi}) \Perp^i_j .$
(We used $\p n^i/\p \tilde x^j =\Perp^i_j/\tilde \chi$.) If $\bar \chi=\tilde \chi$, then  $\p \bar \chi / \p \tilde x^j=n_j$, and  if $\bar \chi \neq \tilde \chi$, we have $\p \bar \chi/ \p \tilde x^j=0$. Thus
${\p \bar x^i(\bar \chi)}/{\p \tilde x^j}=
\delta^i_j $ for $\bar \chi=\tilde \chi,$ and $=
({\bar \chi}/{\tilde \chi}) \Perp^i_j$ for $\bar \chi \neq \tilde \chi.$
Note that for $\bar \chi \neq \tilde \chi$,  the orthogonal part survives. Then
\begin{eqnarray*}
\tilde\p_{\perp j}F(\bar x^i)=\Perp^k_j ({\p \bar x^i(\bar \chi)}/{\p \tilde x^k}) {\p F}/{\p \bar x^i}= ({\bar \chi}/{\tilde \chi}) \Perp^i_j  {\p F}/{\p  \bar x^i}= ({\bar \chi}/{\tilde \chi}) \p_{\perp j} F \;.
\end{eqnarray*}
}
 \begin{eqnarray}
 \label{kappa-1}
 \kappa^{(1)}=- \frac{1}{2}  \p_{\perp i} \Delta x_{\perp}^{i (1)}= \kappa_1^{(1)}+\kappa_2^{(1)}+\kappa_3^{(1)}  
  \end{eqnarray}
 where, using $\p_{\perp i} \Perp^i_j=-2n_j/\bar \chi$,
 \begin{eqnarray}
 \label{kappa1-1}
 \kappa_1^{(1)} =  \frac{1}{2}   \p_{\perp i} \int_0^{\bar \chi} \ud \tilde \chi  \left(\bar \chi-\tilde \chi\right) \tilde \p^i_\perp \left( A^{(1)} - B^{(1)}_{\| } - \frac{1}{2}h^{(1)}_{\| } \right) =  \frac{1}{2}  \int_0^{\bar \chi} \ud \tilde \chi  \left(\bar \chi-\tilde \chi\right) \frac{\tilde \chi}{ \bar \chi}   \tilde \nabla^2_\perp \left( A^{(1)} - B^{(1)}_{\| } - \frac{1}{2}h^{(1)}_{\| } \right)\;, \\
 \label{kappa2-1}
 \kappa_2^{(1)} = \frac{1}{2} \p_{\perp i}  \int_0^{\bar \chi} \ud \tilde \chi \frac{\bar \chi}{\tilde \chi} \left(B^{i (1)}_{\perp }+  n^k h_{kj}^{(1)} \Perp^{ij}  \right)= \frac{1}{2}  \int_0^{\bar \chi} \ud \tilde \chi \bigg[ \tilde \p_\perp^i B_i^{ (1)} - \frac{2}{\tilde\chi}B_\|^{ (1)}+ \Perp^{ij} n^k  \tilde \p_i h_{jk}^{ (1)}+ \frac{1}{\tilde \chi} \left(h_i^{i (1)}-3h_\|^{ (1)} \right)\bigg] \;,\\
 \label{kappa3-1}
  \kappa_3^{(1)} = -\frac{\bar \chi}{2}\left[ \p_{\perp i}B^{i (1)}_{\perp \, o }- \p_{\perp i}v^{i (1)}_{\perp \, o }+ \frac{1}{2}h_{k\,o}^{j(1)}  \p_{\perp i} \left(n^k \Perp^i_j\right)\right]  = -\frac{1}{4}   \left(h_{i \, o}^{i (1)}-3h_{\| \, o}^{ (1)} \right)+ \left(B_{\| \, o}^{(1)} - v_{\| \, o}^{(1)}\right)\;.
 \end{eqnarray}

We can finally compute $\Delta \ln a^{(2)}$,  $\Delta x^{0(2)}$, $ \Delta x^{(2)}_\|$ and  $\Delta x_{\perp}^{i (2)}$.
From Eq. (\ref{Ek-2}) we find
\begin{eqnarray}
\label{Deltalna-2}
 \Delta\ln a^{(2)}&=& - \delta\nu^{(2)}+ E_{\hat{0}\|}^{(2)} - E_{\hat{0}0}^{(2)}-  2 E_{\hat{0}\|}^{(1)} \left(\frac{\ud T}{\ud \bar \chi}\right)^{(1)} -2 \p_\|\left( E_{\hat{0}\|}^{(1)} -  E_{\hat{0} 0}^{(1)}\right) T^{(1)} - \frac{2}{\cH} \left( E_{\hat{0}\|}^{(1)} -  E_{\hat{0} 0}^{(1)}\right) \left(\frac{\ud \, \Delta \ln a}{\ud \bar \chi}\right)^{(1)} \nonumber \\
&&+2 \left[ - \left( E_{\hat{0}\|}^{(1)} -  E_{\hat{0} 0}^{(1)}\right) + \frac{1}{\cH} \left(\frac{\ud \, \Delta \ln a}{\ud \bar \chi}\right)^{(1)} \right] \delta \nu^{(1)} -2 \delta x^{0 (1)} \left( \frac{\ud \delta \nu}{\ud \bar \chi}\right)^{(1)}  + 2 E_{\hat{0}\perp i}^{(1)} \delta n_\perp^{i (1)}  \nonumber \\
&&+2 \left[\p_{\perp i} \left( E_{\hat{0}\|}^{(1)} -  E_{\hat{0} 0}^{(1)}\right) - \frac{1}{\bar \chi} E_{\hat{0}\perp i}^{(1)}  \right] \delta x_\perp^{i (1)} \nonumber \\
&=&  A^{(2)}_o- v^{(2)}_{\| \, o}-\left(A^{(1)}_o\right)^2+2 A^{(1)}_o B^{(1)}_{\| \, o}- \left(B^{(1)}_{\| \, o}\right)^2-6A^{(1)}_o v^{(1)}_{\| \, o}+2 B^{(1)}_{\| \, o}v^{(1)}_{\| \, o} +  v^{(1)}_{k\, o} v^{k (1)}_o \nonumber \\
&&-n^i h_{ij\,o}^{(1)} v^{j (1)}_o +2 \left(A^{(1)}_o-v^{(1)}_{\| \, o}\right) \bigg\{-2\left( 2 A^{(1)} - B^{(1)}_{\| } \right) + \left(A^{(1)} +  v^{(1)}_{\|} -B^{(1)}_{\|}\right)  \nonumber \\
&&+\left(\bar \chi + \frac{1}{\cH}\right)\frac{\ud \,}{\ud \bar \chi}\left( 2 A^{(1)} - B^{(1)}_{\| } \right) - \frac{1}{\cH}\frac{\ud \,}{\ud \bar \chi} \left(A^{(1)} +  v^{(1)}_{\|} -B^{(1)}_{\|}\right) +\left(\bar \chi + \frac{1}{\cH}\right)\left(A^{(1)}{'} - B^{(1)}_{\| }{'} - \frac{1}{2}h^{(1)}_{\| }{'} \right)  \nonumber \\
&& + 4 I^{(1)} \bigg\} +  2\left(B_{\perp \, o}^{i(1)} -v_{\perp \, o}^{i(1)}+\frac{1}{2}n^k h_{k\,o}^{j(1)} \Perp^i_j  \right) \bigg\{ B_{\perp i}^{(1)}+ \bar \chi \, \p_{\perp i} \left(A^{(1)} +  v^{(1)}_{\|} -B^{(1)}_{\|}\right) \nonumber \\
&& -  \left. \int_0^{\bar \chi} \ud \tilde \chi \left[\tilde\p_{\perp i}\left( 2 A^{(1)} - B^{(1)}_{\| }\right)- \left(B_{\perp i}^{(1)}{'}+ n^j h_{ j k}^{(1)}{'} \Perp_i^k  \right) + \frac{1}{\tilde \chi} B_{\perp i}^{(1)}\right] \right\}-A^{(2)}
+   v^{(2)}_{\|} +  7 \left(A^{(1)}\right)^2   \nonumber \\
&& +  \left(B^{(1)}_{\|}\right)^2-4A^{(1)} B^{(1)}_{\|}+  \left(v^{(1)}_{\|}\right)^2+ v_{\perp i}^{(1)} v_{\perp}^{i(1)}+v^{(1)}_{\|} h^{(1)}_{\| } - 2 A^{(1)} v^{(1)}_{\| }   - 2 v_{\perp i}^{(1)} B_{\perp}^{i(1)}  \nonumber \\
&&-  \frac{2}{\cH}\left( A^{(1)} - v^{(1)}_{\| }\right) \left[\frac{\ud \,}{\ud \bar \chi} \left(A^{(1)} -  v^{(1)}_{\|} \right) + \left(A^{(1)}{'} - B^{(1)}_{\| }{'} - \frac{1}{2}h^{(1)}_{\| }{'} \right)\right]-4 \bigg[ 2\left( 2 A^{(1)} - B^{(1)}_{\| } \right)  \nonumber \\
&&- \left(A^{(1)} +  v^{(1)}_{\|} -B^{(1)}_{\|}\right) - \frac{1}{\cH}\frac{\ud \,}{\ud \bar \chi}\left( 2 A^{(1)} - B^{(1)}_{\| } \right)+  \frac{1}{\cH}\frac{\ud \,}{\ud \bar \chi} \left(A^{(1)} +  v^{(1)}_{\|} -B^{(1)}_{\|}\right)  \nonumber \\
&& - \frac{1}{\cH}\left(A^{(1)}{'} - B^{(1)}_{\| }{'} - \frac{1}{2}h^{(1)}_{\| }{'} \right) \bigg] I^{(1)}+ 4 v_{\perp i}^{(1)}S_{\perp}^{i(1)}- 2 \p_\| \left(A^{(1)} +  v^{(1)}_{\|} -B^{(1)}_{\|}\right) T^{(1)}
- 2 \bigg[ \frac{\ud \,}{\ud \bar \chi}\left( 2 A^{(1)} - B^{(1)}_{\| } \right)  \nonumber \\
&& + \left. \left(A^{(1)}{'} - B^{(1)}_{\| }{'} - \frac{1}{2}h^{(1)}_{\| }{'} \right)\right] \int_0^{\bar \chi} \ud \tilde \chi\left[ 2 A^{(1)} - B^{(1)}_{\| } + \left(\bar \chi-\tilde \chi\right) \left(A^{(1)}{'} - B^{(1)}_{\| }{'} - \frac{1}{2}h^{(1)}_{\| }{'} \right) \right]   \nonumber \\
&& - 2 \left[\p_{\perp i}\left(A^{(1)} +  v^{(1)}_{\|} -B^{(1)}_{\|}\right)  - \frac{1}{\bar \chi} \left( v^{(1)}_{\perp i}- B^{ (1)}_{\perp i }\right)\right]\int_0^{\bar \chi} \ud \tilde \chi \left\{ \left( B^{i (1)}_{\perp }+ n^k h_{k}^{j(1)} \Perp^i_j\right) \right. \nonumber \\
&&+ \left. \left(\bar \chi-\tilde \chi\right)\left[\tilde \p^i_\perp \left( A^{(1)} - B^{(1)}_{\| } - \frac{1}{2}h^{(1)}_{\| } \right) + \frac{1}{\tilde \chi} \left(B^{i (1)}_{\perp }+  n^k h_{kj}^{(1)} \Perp^{ij}  \right)\right] \right\} + 2 I^{(2)} \nonumber \\
&& +2 \int_0^{\bar \chi} \ud \tilde \chi  \left\{ \left( B^{(1)}_{\| } +  h^{(1)}_{\| } - 4 I^{(1)}\right) \left(A^{(1)}{'} - B^{(1)}_{\| }{'} - \frac{1}{2}h^{(1)}_{\| }{'} \right) - \left(A^{(1)} - B^{(1)}_{\| } - \frac{1}{2}h^{(1)}_{\| } \right)  \right. \nonumber \\
&& \times \frac{\ud}{\ud \tilde \chi}\left( 2 A^{(1)} - B^{(1)}_{\| }\right) +  \left[ \left(B_{\perp}^{i(1)}+ n^j h_{ j k}^{(1)} \Perp^{ik}  \right) - 2 S_{\perp }^{i(1)}  \right] \left[\tilde\p_{\perp i}\left( 2 A^{(1)} - B^{(1)}_{\| } \right) \right. \nonumber \\
 && -  \left. \left.  \left(B_{\perp i}^{(1)}{'} + n^j h_{ j k}^{(1)}{'} \Perp_i^k  \right) + \frac{1}{\tilde \chi} B_{\perp i}^{(1)}\right]  \right\}
 + \Delta\ln a^{(2)}_{\rm post-Born} \;,
 \end{eqnarray}
where
   \begin{eqnarray}
   &&
\Delta\ln a^{(2)}_{\rm post-Born}= + 2 \left(2 A^{(1)}_o -  B^{(1)}_{\| o}\right) \left(3 v_{\| \, o}^{(1)} + 2 h^{(1)}_{\| \, o} - \frac{1}{2} h^{j (1)}_{j \, o} \right) + 2\left(A^{(1)}_o-v^{(1)}_{\| \, o}\right)  \Bigg\{    2 A^{(1)} - B^{(1)}_{\| }   -2   I^{(1)}  \nonumber\\
   && -  \bar \chi   \left[\frac{\ud}{\ud \bar \chi}\left(2 A^{(1)} - B^{(1)}_{\| }\right)+\left(A^{(1)}{'} - B^{(1)}_{\| }{'} - \frac{1}{2}h^{(1)}_{\| }{'} \right) \right]  \Bigg\}   + 4 \left(- B^{(1)}_{\| o} + v_{\| \, o}^{(1)} - \frac34 h^{(1)}_{\| \, o} + \frac14 h^{j (1)}_{j \, o} \right)    \nonumber\\
         && \times  \int_0^{\bar \chi}   \frac{\ud \tilde{\chi}}{\tilde \chi} \left(2 A^{(1)} - B^{(1)}_{\| }\right)  + 2   \left(B^{i (1)}_{\perp \, o }-v^{i (1)}_{\perp \, o }+ \frac{1}{2} n^k h_{k\,o}^{j(1)} \Perp^i_j\right) \Bigg\{ -   \bar \chi \left[\p_{\perp i}\left(2 A^{(1)} - B^{(1)}_{\| }\right)+\frac{1}{\bar \chi} B_{\perp i}^{(1)} \right]  \nonumber\\
         &&  +  \int_0^{\bar \chi}  \ud \tilde{\chi}  \left[ \tilde \p_{\perp i} \left(2 A^{(1)} - B^{(1)}_{\| }\right) + \frac{1}{\tilde\chi} B_{\perp i}^{(1)} \right]  \Bigg\}  - 2 \left(2 A^{(1)} - B^{(1)}_{\| }\right) \left( A^{(1)}+ \frac{1}{2}  h^{(1)}_{\| }-2I^{(1)} -2\kappa^{(1)} \right)    \nonumber\\
  && + 2  \left[\frac{\ud}{\ud \bar \chi}\left(2 A^{(1)} - B^{(1)}_{\| }\right)+\left(A^{(1)}{'} - B^{(1)}_{\| }{'} - \frac{1}{2}h^{(1)}_{\| }{'} \right) \right] \int_0^{\bar \chi} \ud \tilde \chi     
  \left(A^{(1)}+ \frac{1}{2}  h^{(1)}_{\| }  -2I^{(1)} \right)  +  2 \left(2 A^{(1)}{'} - B^{(1)}_{\| }{'}\right)T^{(1)}  \nonumber\\
        &&   +2   \left[\p_{\perp i}\left(2 A^{(1)} - B^{(1)}_{\| }\right)+\frac{1}{\bar \chi} B_{\perp i}^{(1)} \right] \int_0^{\bar \chi} \ud \tilde \chi  \left( B^{i (1)}_{\perp }+ n^k h_{k}^{j(1)} \Perp^i_j  - 2S_{\perp}^{i(1)}  \right) \nonumber\\
    && +2 \int_0^{\bar \chi}  \ud \tilde{\chi} \Bigg\{  + \left(A^{(1)}{''} - B^{(1)}_{\| }{''} - \frac{1}{2}h^{(1)}_{\| }{''} \right)  T^{(1)} + \left(2 A^{(1)}{'} - B^{(1)}_{\| }{'}\right) \left(A^{(1)}-B^{(1)}_{\| }- \frac{1}{2} h^{(1)}_{\| }\right)   \nonumber\\
      &&    -  \left(A^{(1)}{'} - B^{(1)}_{\| }{'} - \frac{1}{2}h^{(1)}_{\| }{'} \right)   \left(A^{(1)} + \frac{1}{2}  h^{(1)}_{\| }-2I^{(1)}\right) + \left(2 A^{(1)} - B^{(1)}_{\| }\right)  \bigg[\frac{\ud}{\ud \tilde \chi}\left(A^{(1)} + \frac{1}{2}  h^{(1)}_{\| }\right) \nonumber \\
    &&     + \left(A^{(1)}{'} - B^{(1)}_{\| }{'} - \frac{1}{2}h^{(1)}_{\| }{'} \right) \bigg] - \left( B^{i (1)}_{\perp }+ n^k h_{k}^{j(1)} \Perp^i_j - 2S_{\perp}^{i(1)} \right)   \left[ \tilde \p_{\perp i} \left(2 A^{(1)} - B^{(1)}_{\| }\right) + \frac{1}{\tilde\chi} B_{\perp i}^{(1)} \right]    \nonumber \\
 && - \bigg[ \tilde \p_{\perp j} \left(B^{j (1)} + n^k h_{k}^{j(1)} \right)    \left(2 A^{(1)} - B^{(1)}_{\| }\right)  - \frac{2}{\tilde \chi} \left( B^{(1)}_{\| } + h^{(1)}_{\|}\right)    \left(2 A^{(1)} - B^{(1)}_{\| }\right)   -  2    \left(2 A^{(1)} - B^{(1)}_{\| }\right)   \tilde \p_{\perp j}S_{\perp}^{j(1)}  \bigg]   \nonumber \\
 && - 2 \bigg[ \frac{\ud}{\ud \tilde \chi}\left(2 A^{(1)} - B^{(1)}_{\| }\right)   -  \frac{1}{\tilde\chi} \left(2 A^{(1)} - B^{(1)}_{\| }\right)  \bigg] \kappa^{(1)} - \bigg[  \tilde \p_{\perp i}   \left(A^{(1)}{'} - B^{(1)}_{\| }{'} - \frac{1}{2}h^{(1)}_{\| }{'} \right)   \nonumber \\
 && +  \frac{1}{\tilde\chi}  \left( B_{\perp i}^{(1)}{'} +  n^j h_{ j k}^{(1)}{'} \Perp_i^k \right) \bigg]   \delta x_{\perp}^{i (1)} \Bigg\} \;.  \nonumber
 \end{eqnarray}

Using Eqs\ (\ref{Deltalna-1}) and (\ref{Deltalna-2}), Eq.\ (\ref{Dx0_2}) yields
\begin{eqnarray}
\label{Dx0_2-2}
\Delta x^{0(2)}&=& \frac{1}{\cH} \Delta \ln a^{(2)}-  \frac{(\cH' + \cH^2)}{\cH^3} \left( \Delta \ln a^{(1)} \right)^2 \nonumber \\
&=& + \frac{1}{\cH} A^{(2)}_o-\frac{1}{\cH} v^{(2)}_{\| \, o}-\left(\frac{\cH'}{\cH^3}+\frac{2}{\cH}\right)\left(A^{(1)}_o\right)^2-\frac{1}{\cH} \left(B^{(1)}_{\| \, o}\right)^2
+\frac{2}{\cH}A^{(1)}_o B^{(1)}_{\| \, o}+2\left( \frac{\cH'}{\cH^3}-\frac{2}{\cH}\right)A^{(1)}_o v^{(1)}_{\| \, o}\nonumber \\
&& +\frac{2}{\cH}B^{(1)}_{\| \, o}v^{(1)}_{\| \, o}- \frac{\cH'}{\cH^3}\left(v^{(1)}_{\| \, o}\right)^2+ \frac{1}{\cH}v^{(1)}_{\perp i \, o}v^{i (1)}_{\perp \, o}-\frac{1}{\cH}n^i h_{ij\,o}^{(1)} v^{j (1)}_o + 2 \left(A^{(1)}_o-v^{(1)}_{\| \, o}\right) \nonumber \\
&& \times  \bigg\{\left( \frac{\cH'}{\cH^3}- \frac{1}{\cH}\right)\left( 2 A^{(1)} - B^{(1)}_{\| } \right)  - \frac{\cH'}{\cH^3}\left(A^{(1)} +  v^{(1)}_{\|} -B^{(1)}_{\|}\right) +\left(\frac{\bar \chi}{\cH} + \frac{1}{\cH^2}\right) \frac{\ud \,}{\ud \bar \chi}\left( 2 A^{(1)} - B^{(1)}_{\| } \right) \nonumber \\
&&- \frac{1}{\cH^2} \frac{\ud \,}{\ud \bar \chi} \left(A^{(1)} +  v^{(1)}_{\|} -B^{(1)}_{\|}\right) +\left(\frac{\bar \chi}{\cH} + \frac{1}{\cH^2}\right) \left(A^{(1)}{'} - B^{(1)}_{\| }{'} - \frac{1}{2}h^{(1)}_{\| }{'} \right)  -  2 \left( \frac{\cH'}{\cH^3}-\frac{1}{\cH}\right) I^{(1)} \bigg\}  \nonumber \\ 
&&+ 2 \left(B_{\perp \, o}^{i(1)} -v_{\perp \, o}^{i(1)}+\frac{1}{2}n^k h_{k\,o}^{j(1)} \Perp^i_j  \right) \bigg\{\frac{1}{\cH} B_{\perp i}^{(1)}+ \frac{\bar \chi}{\cH}  \p_{\perp i} \left(A^{(1)} +  v^{(1)}_{\|} -B^{(1)}_{\|}\right) \nonumber \\  
&&  -  \frac{1}{\cH} \left. \int_0^{\bar \chi} \ud \tilde \chi \left[\tilde \p_{\perp i}\left( 2 A^{(1)} - B^{(1)}_{\| }\right)- \left(B_{\perp i}^{(1)}{'}+ n^j h_{ j k}^{(1)}{'} \Perp_i^k  \right) +\frac{1}{\tilde \chi} B_{\perp i}^{(1)}\right] \right\}  -\frac{1}{\cH} A^{(2)} + \frac{1}{\cH}  v^{(2)}_{\|}  \nonumber \\
&&+   \left(- \frac{\cH'}{\cH^3}+\frac{6}{\cH}\right) \left(A^{(1)}\right)^2    +  \frac{1}{\cH}  \left(B^{(1)}_{\|}\right)^2  -\frac{4}{\cH}A^{(1)} B^{(1)}_{\|}  -\frac{\cH'}{ \cH^3} \left(v^{(1)}_{\|}\right)^2+  \frac{1}{\cH} v_{\perp i}^{(1)} v_{\perp}^{i(1)}  \nonumber \\
&&+ \frac{1}{\cH}  v^{(1)}_{\|} h^{(1)}_{\| } + 2\frac{\cH'}{\cH^3}A^{(1)} v^{(1)}_{\| } -  \frac{2}{\cH} v_{\perp i}^{(1)} B_{\perp}^{i(1)}- \frac{2}{\cH^2} \left( A^{(1)} - v^{(1)}_{\| }\right)  \left[\frac{\ud \,}{\ud \bar \chi} \left(A^{(1)} -  v^{(1)}_{\|} \right) \right. \nonumber \\
&& \left.+ \left(A^{(1)}{'} - B^{(1)}_{\| }{'} - \frac{1}{2}h^{(1)}_{\| }{'} \right)\right] -4 \bigg[ - \left( \frac{\cH'}{\cH^3}- \frac{1}{\cH}\right)\left( 2 A^{(1)} - B^{(1)}_{\| } \right) +  \frac{\cH'}{\cH^3}\left(A^{(1)} +  v^{(1)}_{\|} -B^{(1)}_{\|}\right)  \nonumber \\
&& - \frac{1}{\cH^2}\frac{\ud \,}{\ud \bar \chi}\left( A^{(1)} - v^{(1)}_{\| } \right)  - \frac{1}{\cH^2}\left(A^{(1)}{'} - B^{(1)}_{\| }{'} - \frac{1}{2}h^{(1)}_{\| }{'} \right) + \left( \frac{\cH'}{\cH^3}+ \frac{1}{\cH}\right)I^{(1)} \bigg] I^{(1)} \nonumber \\
&& + \frac{4}{\cH} v_{\perp i}^{(1)}S_{\perp}^{i(1)} -  \frac{2}{\cH} \p_\| \left(A^{(1)} +  v^{(1)}_{\|} -B^{(1)}_{\|}\right) T^{(1)}- \frac{2}{\cH} \left[ \frac{\ud \,}{\ud \bar \chi}\left( 2 A^{(1)} - B^{(1)}_{\| } \right) +  \left(A^{(1)}{'} - B^{(1)}_{\| }{'} - \frac{1}{2}h^{(1)}_{\| }{'} \right)\right] \nonumber \\
&& \times   \int_0^{\bar \chi} \ud \tilde \chi\left[ 2 A^{(1)} - B^{(1)}_{\| } + \left(\bar \chi-\tilde \chi\right) \left(A^{(1)}{'} - B^{(1)}_{\| }{'} - \frac{1}{2}h^{(1)}_{\| }{'} \right) \right]  -  \frac{2}{\cH} \left[\p_{\perp i}\left(A^{(1)} +  v^{(1)}_{\|} -B^{(1)}_{\|}\right)  \right.\nonumber \\
&& \left. - \frac{1}{\bar \chi} \left( v^{(1)}_{\perp i}- B^{ (1)}_{\perp i }\right)\right] \int_0^{\bar \chi} \ud \tilde \chi \left\{ \left( B^{i (1)}_{\perp }+ n^k h_{k}^{j(1)} \Perp^i_j\right)+  \left(\bar \chi-\tilde \chi\right)\left[ \tilde \p^i_\perp \left( A^{(1)} - B^{(1)}_{\| } - \frac{1}{2}h^{(1)}_{\| } \right)  \right. \right. \nonumber \\
&&\left. \left. + \frac{1}{\tilde \chi} \left(B^{i (1)}_{\perp }+  n^k h_{kj}^{(1)} \Perp^{ij}  \right)\right] \right\} +  \frac{2}{\cH} I^{(2)}  + \frac{2}{\cH} \int_0^{\bar \chi} \ud \tilde \chi  \left\{ \left( B^{(1)}_{\| } +  h^{(1)}_{\| } - 4 I^{(1)}\right) \left(A^{(1)}{'} - B^{(1)}_{\| }{'} - \frac{1}{2}h^{(1)}_{\| }{'} \right) \right.\nonumber \\
&& \left. - \left(A^{(1)} - B^{(1)}_{\| } - \frac{1}{2}h^{(1)}_{\| } \right)  \frac{\ud}{\ud \tilde \chi}\left( 2 A^{(1)} - B^{(1)}_{\| }\right) + \left[ \left(B_{\perp}^{i(1)}+ n^j h_{ j k}^{(1)} \Perp^{ik}  \right) - 2 S_{\perp }^{i(1)}  \right]  \right. \nonumber \\
&& \left. \times \left[\tilde\p_{\perp i}\left( 2 A^{(1)} - B^{(1)}_{\| } \right)- \left(B_{\perp i}^{(1)}{'} + n^j h_{ j k}^{(1)}{'} \Perp_i^k  \right) +\frac{1}{\tilde \chi} B_{\perp i}^{(1)}\right]  \right\} +\Delta x^{0(2)}_{\rm post-Born} \;,
\end{eqnarray}
where
 \begin{eqnarray} 
 &&\Delta x^{0(2)}_{\rm post-Born}=
 + \frac{2} {\cH} \left(2 A^{(1)}_o -  B^{(1)}_{\| o}\right) \left(3 v_{\| \, o}^{(1)} + 2 h^{(1)}_{\| \, o} - \frac{1}{2} h^{j (1)}_{j \, o} \right) + \frac{2}{\cH}\left(A^{(1)}_o-v^{(1)}_{\| \, o}\right)  \Bigg\{    2 A^{(1)} - B^{(1)}_{\| }   -2   I^{(1)}  \nonumber\\
  && -  \bar \chi   \left[\frac{\ud}{\ud \bar \chi}\left(2 A^{(1)} - B^{(1)}_{\| }\right)+\left(A^{(1)}{'} - B^{(1)}_{\| }{'} - \frac{1}{2}h^{(1)}_{\| }{'} \right) \right]  \Bigg\}   + \frac{4}{\cH} \left(- B^{(1)}_{\| o} + v_{\| \, o}^{(1)} - \frac34 h^{(1)}_{\| \, o} + \frac14 h^{j (1)}_{j \, o} \right)    \nonumber\\
        && \times  \int_0^{\bar \chi}   \frac{\ud \tilde{\chi}}{\tilde \chi} \left(2 A^{(1)} - B^{(1)}_{\| }\right)  + \frac{2 }{\cH}   \left(B^{i (1)}_{\perp \, o }-v^{i (1)}_{\perp \, o }+ \frac{1}{2} n^k h_{k\,o}^{j(1)} \Perp^i_j\right) \Bigg\{ -   \bar \chi \left[\p_{\perp i}\left(2 A^{(1)} - B^{(1)}_{\| }\right)+\frac{1}{\bar \chi} B_{\perp i}^{(1)} \right]  \nonumber\\
         &&  +  \int_0^{\bar \chi}  \ud \tilde{\chi}  \left[ \tilde \p_{\perp i} \left(2 A^{(1)} - B^{(1)}_{\| }\right) + \frac{1}{\tilde\chi} B_{\perp i}^{(1)} \right]  \Bigg\}  - \frac{2}{\cH} \left(2 A^{(1)} - B^{(1)}_{\| }\right) \left( A^{(1)}+ \frac{1}{2}  h^{(1)}_{\| }-2I^{(1)} -2\kappa^{(1)} \right)    \nonumber\\
 && + \frac{2}{\cH}  \left[\frac{\ud}{\ud \bar \chi}\left(2 A^{(1)} - B^{(1)}_{\| }\right)+\left(A^{(1)}{'} - B^{(1)}_{\| }{'} - \frac{1}{2}h^{(1)}_{\| }{'} \right) \right] \int_0^{\bar \chi} \ud \tilde \chi     
  \left(A^{(1)}+ \frac{1}{2}  h^{(1)}_{\| }  -2I^{(1)} \right)  +  \frac{2}{\cH} \left(2 A^{(1)}{'} - B^{(1)}_{\| }{'}\right)T^{(1)}  \nonumber\\
        &&   + \frac{2} {\cH}    \left[\p_{\perp i}\left(2 A^{(1)} - B^{(1)}_{\| }\right)+\frac{1}{\bar \chi} B_{\perp i}^{(1)} \right] \int_0^{\bar \chi} \ud \tilde \chi  \left( B^{i (1)}_{\perp }+ n^k h_{k}^{j(1)} \Perp^i_j  - 2S_{\perp}^{i(1)}  \right) \nonumber\\
    && +\frac{2}{\cH}  \int_0^{\bar \chi}  \ud \tilde{\chi} \Bigg\{   \left(A^{(1)}{''} - B^{(1)}_{\| }{''} - \frac{1}{2}h^{(1)}_{\| }{''} \right)  T^{(1)} + \left(2 A^{(1)}{'} - B^{(1)}_{\| }{'}\right) \left(A^{(1)}-B^{(1)}_{\| }- \frac{1}{2} h^{(1)}_{\| }\right)   \nonumber\\
          &&    -  \left(A^{(1)}{'} - B^{(1)}_{\| }{'} - \frac{1}{2}h^{(1)}_{\| }{'} \right)   \left(A^{(1)} + \frac{1}{2}  h^{(1)}_{\| }-2I^{(1)}\right) + \left(2 A^{(1)} - B^{(1)}_{\| }\right)  \bigg[\frac{\ud}{\ud \bar \chi}\left(A^{(1)} + \frac{1}{2}  h^{(1)}_{\| }\right) \nonumber \\
    &&     - \left(A^{(1)}{'} - B^{(1)}_{\| }{'} - \frac{1}{2}h^{(1)}_{\| }{'} \right) \bigg] - \left( B^{i (1)}_{\perp }+ n^k h_{k}^{j(1)} \Perp^i_j - 2S_{\perp}^{i(1)} \right)   \left[ \tilde \p_{\perp i} \left(2 A^{(1)} - B^{(1)}_{\| }\right) + \frac{1}{\tilde\chi} B_{\perp i}^{(1)} \right]    \nonumber \\
 && - \bigg[ \tilde \p_{\perp j} \left(B^{j (1)} + n^k h_{k}^{j(1)} \right)    \left(2 A^{(1)} - B^{(1)}_{\| }\right)  - \frac{2}{\tilde \chi} \left( B^{(1)}_{\| } + h^{(1)}_{\|}\right)    \left(2 A^{(1)} - B^{(1)}_{\| }\right)   -  2    \left(2 A^{(1)} - B^{(1)}_{\| }\right)   \tilde \p_{\perp j}S_{\perp}^{j(1)}  \bigg]   \nonumber \\
 && - 2 \bigg[ \frac{\ud}{\ud \tilde \chi}\left(2 A^{(1)} - B^{(1)}_{\| }\right)   -  \frac{1}{\tilde\chi} \left(2 A^{(1)} - B^{(1)}_{\| }\right)  \bigg] \kappa^{(1)} - \bigg[  \tilde \p_{\perp i}   \left(A^{(1)}{'} - B^{(1)}_{\| }{'} - \frac{1}{2}h^{(1)}_{\| }{'} \right)   \nonumber \\
 && +  \frac{1}{\tilde\chi}  \left( B_{\perp i}^{(1)}{'} +  n^j h_{ j k}^{(1)}{'} \Perp_i^k \right) \bigg]   \delta x_{\perp}^{i (1)} \Bigg\} \;.  \nonumber
 \end{eqnarray}

From Eqs.\  (\ref{Dx_||-2}) and   (\ref{dx0+dx_||-2}) we deduce
\begin{eqnarray} 
\label{Dx_||-2_2}
\Delta x_{\parallel}^{(2)}&=&  \delta x^{0 (2)} +   \delta x_{\parallel}^{(2)} -\frac{1}{\cH} \Delta \ln a^{(2)}+ \left(\frac{\cH' }{\cH^3} +\frac{1}{\cH} \right)\left( \Delta \ln a^{(1)} \right)^2  - 2  \left(\frac{\ud T}{\ud \bar \chi}\right)^{(1)} \delta\chi^{(1)}  \nonumber \\
&=& \bar \chi \bigg [\left(A^{(1)}_o\right)^2 - 2A^{(1)}_o B^{(1)}_{\| \, o}+\left(B^{(1)}_{\| \, o}\right)^2 +2 A^{(1)}_o v^{(1)}_{\| \, o} -2 B^{(1)}_{\| \, o}v^{(1)}_{\| \, o}-  v^{(1)}_{\| \, o} h^{(1)}_{\| \, o} -\frac{1}{4} \left(h^{(1)}_{\| \, o}\right)^2   -   v^{(1)}_{\perp k \, o} v^{k (1)}_{\perp \, o}\nonumber \\
&& + n^i h_{ik\,o}^{(1)} \Perp^k_j v^{j (1)}_o - n^i h_{ik\,o}^{(1)} \Perp^k_j B^{j (1)}_o - \frac{1}{4} n^i h_{ij\,o}^{(1)}\,  \Perp^j_k \, h^{k (1)}_{p \, o} n^p   -  B_{\perp \, o}^{i(1)}B_{\perp i \, o}^{(1)}+2 v_{\perp \, o}^{i(1)}B_{\perp i \, o}^{(1)} \bigg]\nonumber \\
&&- 2 \left (A^{(1)}_o-v^{(1)}_{\| \, o}\right) \left[ \bar \chi \left (  A^{(1)} - B^{(1)}_{\| } - \frac{1}{2} h^{(1)}_{\| } \right) + 2T^{(1)}\right]  +4 \left(B_{\perp \, o}^{i(1)} -v_{\perp \, o}^{i(1)}+\frac{1}{2}n^k h_{k\,o}^{j(1)} \Perp^i_j  \right)\nonumber \\
&&\times \int_0^{\bar \chi} \ud \tilde \chi\left(\bar \chi-\tilde\chi\right)\left[ \tilde\p^i_\perp \left( A^{(1)} - B^{(1)}_{\| } - \frac{1}{2}h^{(1)}_{\| } \right) + \frac{1}{\tilde \chi} \left(B^{i (1)}_{\perp }+  n^k h_{kj}^{(1)} \Perp^{ij}  \right)\right]  \nonumber \\
&& + 2 \left (  A^{(1)} - B^{(1)}_{\| } - \frac{1}{2} h^{(1)}_{\| } \right)  \int_0^{\bar \chi} \ud \tilde \chi \left[ 2 A^{(1)} - B^{(1)}_{\| } + \left(\bar \chi-\tilde \chi\right) \left(A^{(1)}{'} - B^{(1)}_{\| }{'} - \frac{1}{2}h^{(1)}_{\| }{'} \right) \right]  \nonumber \\ 
&& -  T^{(2)}+2 \int_0^{\bar \chi} \ud \tilde \chi \bigg[ -\left( 2 A^{(1)} - B^{(1)}_{\| } \right)^2+ \frac{1}{2} \left(B^{(1)}_{\| } + h^{(1)}_{\| } \right) \left( 2 A^{(1)} + h^{(1)}_{\| } \right)\nonumber \\
&& + \frac{1}{2} \left(B_{\perp}^{i(1)} + \Perp^i_j h_{k}^{j(1)} n^k\right) \left(B_{i \perp}^{(1)}+ n^p h_{p m}^{(1)} \Perp^m_i  \right)   +4 \left(A^{(1)} - B^{(1)}_{\| }- \frac{1}{2}h^{(1)}_{\| }\right)  I^{(1)}-4 S_{\perp }^{i(1)}S_{\perp }^{j(1)} \delta_{ij}  \bigg]   \nonumber \\
&&+ 2 \int_0^{\bar \chi} \ud \tilde \chi\left(\bar \chi-\tilde\chi\right) \left\{ \left(A^{(1)} - B^{(1)}_{\| } - \frac{1}{2}h^{(1)}_{\| } \right) \left[2\left(A^{(1)}{'} - B^{(1)}_{\| }{'} - \frac{1}{2}h^{(1)}_{\| }{'} \right)+  \frac{\ud}{\ud \tilde \chi}\left( 2 A^{(1)} - B^{(1)}_{\| }\right) \right]  \right. \nonumber \\
&& - \left(B_{\perp}^{i(1)} + n^k h_{k}^{j(1)} \Perp^i_j  \right)\left[ \tilde \p_{\perp i}\left (  A^{(1)} - B^{(1)}_{\| } - \frac{1}{2} h^{(1)}_{\| } \right)  + \frac{1}{\tilde \chi} \left(B_{\perp i}^{(1)}+   n^m h_{mp}^{(1)} \Perp^p_i  \right)\right]\bigg\}\nonumber \\
&&- \frac{2}{\cH} \left (  A^{(1)} - B^{(1)}_{\| } - \frac{1}{2} h^{(1)}_{\| } \right) \Delta \ln a^{(1)}  -\frac{1}{ \cH} \Delta \ln a^{(2)}+ \left(\frac{\cH' }{\cH^3} +\frac{1}{\cH} \right)\left( \Delta \ln a^{(1)} \right)^2 \nonumber \\
&&+ \left(\delta x^{0 (2)} + \delta x_\|^{(2)}\right)_{\rm post-Born} \;,
\end{eqnarray}
and from Eqs.\  (\ref{Dx_perp-2}) and   (\ref{dx_perp-2}) we find
\begin{eqnarray} 
\label{Dx_perp-2_2}
\Delta x_{\perp}^{i(2)}&=& \delta x_{\perp}^{i(2)} + 2\delta n_{\perp}^{i(1)}   \delta x^{0 (1)} -  \frac{2}{\cH} \delta n_{\perp}^{i(1)} \Delta \ln a^{(1)} \nonumber \\
&=& \bar \chi \bigg[ B^{i(2)}_{\perp \, o}-  v^{i(2)}_{\perp \, o} + \frac{1}{2} n^j h_{ j k\, o}^{(2)} \Perp^{ki}  - 2 A^{(1)}_o B^{i (1)}_{\perp \, o } + 2 A^{(1)}_o v^{i (1)}_{\perp \, o } + A^{(1)}_on^k h_{k\,o}^{j(1)} \Perp^i_j -  v^{(1)}_{\| \, o}v^{i(1)}_{\perp \, o}  \nonumber\\
&& - v^{(1)}_{\| \, o}B^{i(1)}_{\perp \, o}+ B^{(1)}_{\| \, o}B^{i(1)}_{\perp \, o}-  B^{(1)}_{\| \, o}v^{i(1)}_{\perp \, o} - 3v^{(1)}_{\| \, o}n^k h_{k\,o}^{j(1)} \Perp^i_j  - 2 v^{j (1)}_{\perp o}\Perp^{l}_j  \, h^{k (1)}_{l \, o} \Perp_{k}^{i}-\frac{1}{4} h^{(1)}_{\| \, o} n^j h_{ j k\, o}^{(1)} \Perp^{ki}  \nonumber \\
&&  -\frac{1}{4} n^j h^{k (1)}_{j \, o}  \Perp^{l}_k  \, h^{p (1)}_{l \, o} \Perp_{p}^{i(1)} \bigg] + 2 \bar \chi \left(A^{(1)}_o-v^{(1)}_{\| \, o}\right) \left[\left( B^{i (1)}_{\perp }+ n^k h_{k}^{j(1)} \Perp^i_j\right)  
- 2S_{\perp}^{i(1)} \right] \nonumber \\
&& - 4  \left (A^{(1)}_o-v^{(1)}_{\| \, o}\right) \int_0^{\bar \chi} \ud \tilde \chi \bigg\{\left(B_{\perp}^{i(1)}+ n^k h_{ k}^{j(1)} \Perp^{i}_j \right)+ \left(\bar \chi-\tilde \chi\right)\left[\tilde \p^i_\perp \left( A^{(1)} - B^{(1)}_{\| } - \frac{1}{2}h^{(1)}_{\| } \right) \right. \nonumber \\
&& +\left. \frac{1}{\tilde \chi} \left(B^{i (1)}_{\perp }+  n^k h_{kj}^{(1)} \Perp^{ij}  \right)\right] \bigg\} + 2 \left( B^{ (1)}_{\perp j \, o }-v^{ (1)}_{\perp j \, o }+ \frac{1}{2} n^k h_{k\,o}^{l(1)} \Perp_{j l} \right) \int_0^{\bar \chi} \ud \tilde \chi \bigg\{  - \Perp^{jm}  \, h^{p (1)}_{m} \Perp_{p}^{i} \nonumber \\
&& +\left. 2\left(\bar \chi-\tilde \chi\right) \left[  \tilde \p_{\perp}^{[i} B_{\perp}^{j](1)}  + \tilde \p_{\perp}^{[i}     \left( \Perp^{j]}_m h^{m(1)}_q n^q \right)   - \frac{1}{\tilde \chi} \left(n^{[i} B_{\perp}^{j](1)} +n^{[i}  \Perp^{j]}_m h^{m(1)}_q n^q \right)   \right] \right\}  \nonumber \\
&&+ 2 \left(  B^{i (1)}_{\perp \, o }-v^{i (1)}_{\perp \, o }+ \frac{1}{2} n^k h_{k\,o}^{j(1)} \Perp^i_j\right) \bigg\{ \int_0^{\bar \chi} \ud \tilde \chi \left[ 2 A^{(1)} - B^{(1)}_{\| } + \left(\bar \chi-\tilde \chi\right) \left(A^{(1)}{'} - B^{(1)}_{\| }{'} - \frac{1}{2}h^{(1)}_{\| }{'} \right) \right] \nonumber \\
&&  -  \frac{1}{\cH}\Delta \ln a^{(1)} \bigg\} +  \frac{2}{\cH} \left( B^{i (1)}_{\perp }+ n^k h_{k}^{j(1)} \Perp^i_j- 2S_{\perp}^{i(1)}\right) \Delta \ln a^{(1)}  \nonumber \\
&&- 2\left[\left( B^{i (1)}_{\perp }+ n^k h_{k}^{j(1)} \Perp^i_j\right) - 2S_{\perp}^{i(1)} \right]\int_0^{\bar \chi} \ud \tilde \chi \left[ 2 A^{(1)} - B^{(1)}_{\| } + \left(\bar \chi-\tilde \chi\right) \left(A^{(1)}{'} - B^{(1)}_{\| }{'} - \frac{1}{2}h^{(1)}_{\| }{'} \right) \right] \nonumber \\
&& + \int_0^{\bar \chi} \ud \tilde \chi \bigg\{  -  B^{i(2)}_{\perp} - n^j h_{ j k}^{(2)} \Perp^{ki} + 4 A^{(1)} B^{i(1)}_{\perp} - 2 B^{(1)}_{\|}B^{i(1)}_{\perp}  + 2 A^{(1)} n^j h_{ j k}^{(1)} \Perp^{ki}+  h_\|^{(1)} n^j h_{ j k}^{(1)} \Perp^{ki} \nonumber \\
&& + 2\left( B^{j (1)}_{\perp }+ n^k h_{k}^{p(1)} \Perp^j_p\right)  \Perp^{l}_j  \, h^{k (1)}_{l} \Perp_{k}^{i(1)}- 8 \left(B_{\perp}^{i(1)}+ n^j h_{ j k}^{(1)} \Perp^{ik}  \right) I^{(1)}- 4\,  \Perp^{l}_j  \, h^{k (1)}_{l} \Perp_{k}^{i} S_{\perp}^{j(1)} \bigg\} \nonumber \\
&& + \int_0^{\bar \chi} \ud \tilde \chi \left(\bar \chi-\tilde \chi\right) \left\{-\left[ \tilde \p^i_\perp \left( A^{(2)} - B^{(2)}_{\| } - \frac{1}{2}h^{(2)}_{\| } \right) + \frac{1}{\tilde \chi} \left(B^{i (2)}_{\perp }+  n^k h_{kj}^{(2)} \Perp^{ij}  \right)\right]   \right. \nonumber \\
&& \left.   - 2 \left(A^{(1)}{'} - B^{(1)}_{\| }{'} - \frac{1}{2}h^{(1)}_{\| }{'} \right) \left(B_{\perp}^{i(1)}+ n^j h_{ j k}^{(1)} \Perp^{ik}  \right)  + 2 \left(2 A^{(1)} -B_\|^{(1)}\right)  \frac{\ud}{\ud \tilde \chi} \left(B_{\perp}^{i(1)}+ n^j h_{ j k}^{(1)} \Perp^{ik}  \right) \right. \nonumber \\
&&  \left.  +  4 \left(2 A^{(1)} -B_\|^{(1)} - 2 I^{(1)}\right) \left[\tilde \p^i_\perp \left( A^{(1)} - B^{(1)}_{\| } - \frac{1}{2}h^{(1)}_{\| } \right) + \frac{1}{\tilde \chi} \left(B^{i (1)}_{\perp }+  n^k h_{kj}^{(1)} \Perp^{ij}  \right) \right] \right. \nonumber \\
&& \left. -4 \left[  \left(B_{\perp j}^{(1)}+ n^p h_{p}^{k(1)} \Perp_{j k}  \right)- 2 \delta_{jp}  S_{\perp}^{p(1)} \right]   \left[ \tilde \p_{\perp}^{[i} B_{\perp}^{j](1)}  +\tilde  \p_{\perp}^{[i}     \left( \Perp^{j]}_m h^{m(1)}_q n^q \right)  - \frac{1}{\tilde \chi} \left(n^{[i} B_{\perp}^{j](1)}+ n^{[i}  \Perp^{j]}_m h^{m(1)}_q n^q \right)   \right]\right.  \nonumber \\
&& \left.+2\left( A^{(1)} - B^{(1)}_{\| } - \frac{1}{2}h^{(1)}_{\| } \right) \bigg[ \tilde\p^i_\perp \left(B^{(1)}_{\| }+ h^{(1)}_{\| } \right) - \p_\| \left(B_{\perp}^{i (1)}+n^p h_{pq}^{(1)} \Perp^{iq} \right)   - \frac{1}{\tilde \chi} \left(B^{i (1)}_{\perp }+  2 n^p h_{pq}^{(1)} \Perp^{iq}  \right) \bigg]\right\} \nonumber \\
&& +  \delta x_{\perp  \rm post-Born}^{i(2)} \;.
\end{eqnarray}

The next step is to compute $\p_\| \Delta x_{\parallel}^{(2)}$ and $\kappa^{(2)}$.
From Eq. (\ref{partialparallep}) we have $\p_\| \Delta x_{\parallel}^{(2)} (\eta, \bar {\bf x})=\p_{\bar \chi} \Delta x_{\parallel}^{(2)} (\bar \chi, {\bf n})$, so that
\begin{eqnarray} 
\label{Dx_||-2_2}
\p_\| \Delta x_{\parallel}^{(2)}&=&  \delta \nu^{(2)} + \delta n_{\parallel}^{(2)} - \frac{\ud }{\ud \bar \chi} \left( \frac{1}{\cH} \Delta \ln a^{(2)} \right)+  \frac{\ud }{\ud \bar \chi} \left[\left(\frac{\cH' }{\cH^3} +\frac{1}{\cH} \right)\left( \Delta \ln a^{(1)} \right)^2 \right] - 2 \frac{\ud }{\ud \bar \chi} \left[ \left(\frac{\ud T}{\ud \bar \chi}\right)^{(1)} \delta\chi^{(1)} \right] \nonumber \\
&=& -  \frac{\cH'}{\cH^2} \Delta \ln a^{(2)} +\left[-\frac{\cH'' }{\cH^3} +3\left( \frac{\cH' }{\cH^2} \right)^2 + \frac{\cH' }{\cH^2} \right] \left( \Delta \ln a^{(1)} \right)^2
+2 \left[\frac{\cH' }{\cH^2} \left(\frac{\ud T}{\ud \bar \chi}\right)^{(1)} +\frac{1}{\cH} \left(\frac{\ud^2 T}{{\ud \bar \chi}^2}\right)^{(1)} \right] \Delta \ln a^{(1)}  \nonumber \\
&&+ 2 \left(\frac{\cH' }{\cH^3} +\frac{1}{\cH} \right) \Delta \ln a^{(1)} \left(\frac{\ud \, \Delta \ln a}{\ud \bar \chi}\right)^{(1)} -  \frac{1}{ \cH}\left(\frac{\ud \, \Delta \ln a}{\ud \bar \chi}\right)^{(2)}  - 2 \left(\frac{\ud^2 T}{{\ud \bar \chi}^2}\right)^{(1)} \delta x^{0(1)} - 2  \left(\frac{\ud T}{\ud \bar \chi}\right)^{(1)}  \delta \nu^{(1)}  \nonumber \\
&& + \frac{2}{\cH}  \left(\frac{\ud T}{\ud \bar \chi}\right)^{(1)}   \left(\frac{\ud \, \Delta \ln a}{\ud \bar \chi}\right)^{(1)} + \delta \nu^{(2)} +   \delta n_{\parallel}^{(2)}\;.
\end{eqnarray}
 To obtain explicitly Eq.\ (\ref{Dx_||-2_2}) we have to determine 
\begin{eqnarray}
\left(\frac{\ud \, \Delta \ln a}{\ud \bar \chi}\right)^{(2)} &=& +2 \left(A^{(1)}_o-v^{(1)}_{\| \, o}\right) \bigg\{\left(\frac{\cH' }{\cH^2}-1\right) \left[\frac{\ud \,}{\ud \bar \chi}\left(  A^{(1)} - v^{(1)}_{\| } \right)+\left(A^{(1)}{'} - B^{(1)}_{\| }{'} - \frac{1}{2}h^{(1)}_{\| }{'} \right)\right] \nonumber \\
&& +   \frac{1}{\cH}  \frac{\ud \,}{\ud \bar \chi} \left[ \frac{\ud \,}{\ud \bar \chi}\left(  A^{(1)} - v^{(1)}_{\| } \right) + \left(A^{(1)}{'} - B^{(1)}_{\| }{'} - \frac{1}{2}h^{(1)}_{\| }{'} \right)\right]+\bar \chi \frac{\ud \,}{\ud \bar \chi}\left[ \frac{\ud \,}{\ud \bar \chi} \left( 2 A^{(1)} - B^{(1)}_{\| } \right) \right. \nonumber \\
&& \left. + \left(A^{(1)}{'} - B^{(1)}_{\| }{'} - \frac{1}{2}h^{(1)}_{\| }{'} \right)\right]\bigg\} + 2 \left(B_{\perp \, o}^{i(1)} -v_{\perp \, o}^{i(1)}+\frac{1}{2}n^k h_{k\,o}^{j(1)} \Perp^i_j  \right) \bigg[  \frac{\ud \,}{\ud \bar \chi}  B_{\perp i}^{(1)}   -  \p_{\perp i} \left(A^{(1)} -  v^{(1)}_{\|} \right)\nonumber \\
&&+\bar \chi  \frac{\ud \,}{\ud \bar \chi} \p_{\perp i} \left(A^{(1)} +  v^{(1)}_{\|} -B^{(1)}_{\|}\right)  +\left(B_{\perp i}^{(1)}{'}+ n^j h_{ j k}^{(1)}{'} \Perp_i^k  \right) - \frac{1}{\tilde \chi} B_{\perp i}^{(1)}\bigg] +  \frac{\ud \,}{\ud \bar \chi}\left(- A^{(2)}+   v^{(2)}_{\|} \right)\nonumber \\
&& - \left(A^{(2)}{'} - B^{(2)}_{\| }{'} - \frac{1}{2}h^{(2)}_{\| }{'} \right) +  \frac{\ud \,}{\ud \bar \chi}\ \bigg[ 7 \left(A^{(1)}\right)^2  + \left(B^{(1)}_{\|}\right)^2-4A^{(1)} B^{(1)}_{\|}- 2A^{(1)} v^{(1)}_{\| }  \nonumber \\
&& + \left(v^{(1)}_{\|}\right)^2+ v^{(1)}_{\|} h^{(1)}_{\| }  \bigg] - 2 \bigg[ \frac{\ud \,}{\ud \bar \chi}\left( 2 A^{(1)} - B^{(1)}_{\| } \right)   + \left. \left(A^{(1)}{'} - B^{(1)}_{\| }{'} - \frac{1}{2}h^{(1)}_{\| }{'} \right)\right] \left(2 A^{(1)} - B^{(1)}_{\| } \right) \nonumber \\
&& -2\bigg\{ \frac{\cH'}{\cH^2}\left( A^{(1)} - v^{(1)}_{\| }\right) +  \frac{1}{\cH}\left[\frac{\ud \,}{\ud \bar \chi}\left( A^{(1)} - v^{(1)}_{\| }\right) + \left(A^{(1)}{'} - B^{(1)}_{\| }{'} - \frac{1}{2}h^{(1)}_{\| }{'} \right) \right]\bigg\} \left[\frac{\ud \,}{\ud \bar \chi} \left(A^{(1)} -  v^{(1)}_{\|} \right) \right.  \nonumber \\
&& \left. + \left(A^{(1)}{'} - B^{(1)}_{\| }{'} - \frac{1}{2}h^{(1)}_{\| }{'} \right)\right] +2 \bigg[\left( 2 A^{(1)} +  h^{(1)}_{\| }  \right)+ \left(A^{(1)} -  v^{(1)}_{\|}\right) \bigg] \left(A^{(1)}{'} - B^{(1)}_{\| }{'} - \frac{1}{2}h^{(1)}_{\| }{'} \right)\nonumber \\
&& -  \frac{2}{\cH}\left( A^{(1)} - v^{(1)}_{\| }\right)\frac{\ud \,}{\ud \bar \chi}\left[\frac{\ud \,}{\ud \bar \chi} \left(A^{(1)} -  v^{(1)}_{\|} \right) + \left(A^{(1)}{'} - B^{(1)}_{\| }{'} - \frac{1}{2}h^{(1)}_{\| }{'} \right)\right]  +2 \left(A^{(1)} - B^{(1)}_{\| } - \frac{1}{2}h^{(1)}_{\| } \right)\nonumber \\
&&\times \left[-\frac{\ud \,}{\ud \bar \chi}\left( 2 A^{(1)} - B^{(1)}_{\| }\right)  + \p_\| \left(A^{(1)} +  v^{(1)}_{\|} -B^{(1)}_{\|}\right)\right]+  2\left(B_{\perp}^{i(1)}+ n^j h_{ j k}^{(1)} \Perp^{ik}  \right)\bigg[\p_{\perp i}\left(  A^{(1)} - v^{(1)}_{\| } \right)\nonumber \\
&&- \left(B_{\perp i}^{(1)}{'} + n^j h_{ j k}^{(1)}{'} \Perp_i^k  \right)  \bigg]  -2v_{\perp i}^{(1)} \p^i_\perp \left( A^{(1)} - B^{(1)}_{\| } - \frac{1}{2}h^{(1)}_{\| } \right) +  2 \frac{\ud \,}{\ud \bar \chi}  \left[ \frac{1}{2} v_{\perp i}^{(1)} v_{\perp}^{i(1)}  - v_{\perp i}^{(1)} B_{\perp}^{i(1)} \right] \nonumber \\
     &&  + 4 \bigg\{ \left(\frac{\cH' }{\cH^2} -1\right) \left[\frac{\ud \,}{\ud \bar \chi}\left(A^{(1)} - v^{(1)}_{\| } \right)   +\left(A^{(1)}{'} - B^{(1)}_{\| }{'} - \frac{1}{2}h^{(1)}_{\| }{'} \right)\right]+ \frac{1}{\cH} \frac{\ud \,}{\ud \bar \chi}\left[\frac{\ud \,}{\ud \bar \chi}\left(A^{(1)} - v^{(1)}_{\| } \right) \right.\nonumber \\    
 && \left. +\left(A^{(1)}{'} - B^{(1)}_{\| }{'} - \frac{1}{2}h^{(1)}_{\| }{'} \right) \right]\bigg\} I^{(1)}  - 2 \left[\frac{\ud \,}{\ud \bar \chi} \p_\| \left(A^{(1)} +  v^{(1)}_{\|} -B^{(1)}_{\|}\right) \right] T^{(1)}  - 2 \frac{\ud \,}{\ud \bar \chi} \bigg[ \frac{\ud \,}{\ud \bar \chi} \left( 2 A^{(1)} - B^{(1)}_{\| } \right) \nonumber \\
&& + \left(A^{(1)}{'} - B^{(1)}_{\| }{'} - \frac{1}{2}h^{(1)}_{\| }{'} \right)\bigg]  \int_0^{\bar \chi} \ud \tilde \chi\left[ 2 A^{(1)} - B^{(1)}_{\| } + \left(\bar \chi-\tilde \chi\right) \left(A^{(1)}{'} - B^{(1)}_{\| }{'} - \frac{1}{2}h^{(1)}_{\| }{'} \right) \right]   \nonumber \\
&&  -  4 \left[\p_{\perp i}\left(  A^{(1)} - v^{(1)}_{\| } \right)- \left(B_{\perp i}^{(1)}{'} + n^j h_{ j k}^{(1)}{'} \Perp_i^k  \right)  + \frac{1}{\tilde \chi} v_{\perp i}^{(1)}  -\frac{\ud \,}{\ud \bar \chi} v_{\perp i}^{(1)} \right]   S_{\perp }^{i(1)} \nonumber \\
&&- 2 \frac{\ud \,}{\ud \bar \chi} \bigg[\p_{\perp i}\left(A^{(1)} +  v^{(1)}_{\|} -B^{(1)}_{\|}\right)  - \frac{1}{\bar \chi} \left( v^{(1)}_{\perp i}- B^{ (1)}_{\perp i }\right)\bigg]\int_0^{\bar \chi} \ud \tilde \chi \left\{ \left( B^{i (1)}_{\perp }+ n^k h_{k}^{j(1)} \Perp^i_j\right) \right. \nonumber \\
&&+\left. \left(\bar \chi-\tilde \chi\right)\left[ \tilde \p^i_\perp \left( A^{(1)} - B^{(1)}_{\| } - \frac{1}{2}h^{(1)}_{\| } \right) + \frac{1}{\tilde \chi} \left(B^{i (1)}_{\perp }+  n^k h_{kj}^{(1)} \Perp^{ij}  \right)\right] \right\}+ \left(\frac{\ud \, \Delta \ln a}{\ud \bar \chi}\right)^{(2)}_{ \rm post-Born}\;,
 \end{eqnarray}

where
\begin{eqnarray} 
&&\left(\frac{\ud \, \Delta \ln a}{\ud \bar \chi}\right)^{(2)}_{ \rm post-Born}= -2 \bar \chi \left(A^{(1)}_o-v^{(1)}_{\| \, o}\right)  \frac{\ud}{\ud \bar \chi} \bigg[\frac{\ud}{\ud \bar \chi}\left( 2 A^{(1)} - B^{(1)}_{\| }\right)    + \left(A^{(1)}{'} - B^{(1)}_{\| }{'} - \frac{1}{2}h^{(1)}_{\| }{'} \right) \bigg] \nonumber \\
 && - 2  \bar \chi \left(B^{i (1)}_{\perp \, o }-v^{i (1)}_{\perp \, o }+ \frac{1}{2} n^k h_{k\,o}^{j(1)} \Perp^i_j\right)  \Bigg\{  \p_{\perp i}\frac{\ud}{\ud \bar \chi} \left(2 A^{(1)} - B^{(1)}_{\| }\right) + \p_{\perp i}   \left(A^{(1)}{'} - B^{(1)}_{\| }{'} - \frac{1}{2}h^{(1)}_{\| }{'} \right) \nonumber \\
 &&  -  \frac{1}{\bar \chi} \bigg[\p_{\perp i} \left(2 A^{(1)} - B^{(1)}_{\| }\right) + \left(  \frac{1}{\bar \chi} B_{\perp i}^{(1)} - \frac{\ud}{\ud \bar \chi} B_{\perp i}^{(1)} \right)    -    \left( B_{\perp i}^{(1)}{'} +  n^j h_{ j k}^{(1)}{'} \Perp_i^k \right) \bigg]  \Bigg\}     \nonumber \\
&& + 2 \frac{\ud}{\ud \bar \chi} \bigg[\frac{\ud}{\ud \bar \chi}\left( 2 A^{(1)} - B^{(1)}_{\| }\right)    + \left(A^{(1)}{'} - B^{(1)}_{\| }{'} - \frac{1}{2}h^{(1)}_{\| }{'} \right) \bigg] \int_0^{\bar \chi} \ud \tilde \chi \left(A^{(1)}+ \frac{1}{2}  h^{(1)}_{\| } -2I^{(1)} \right) \nonumber \\
 && + 2 \left[ \frac{\ud}{\ud \bar \chi}\left(2 A^{(1)}{'} - B^{(1)}_{\| }{'}\right)   + \left(A^{(1)}{''} - B^{(1)}_{\| }{''} - \frac{1}{2}h^{(1)}_{\| }{''} \right) \right]  T^{(1)}  \nonumber \\
 && + 2 \Bigg\{  \p_{\perp i}\frac{\ud}{\ud \bar \chi} \left(2 A^{(1)} - B^{(1)}_{\| }\right) + \p_{\perp i}   \left(A^{(1)}{'} - B^{(1)}_{\| }{'} - \frac{1}{2}h^{(1)}_{\| }{'} \right) -  \frac{1}{\bar \chi} \bigg[\p_{\perp i} \left(2 A^{(1)} - B^{(1)}_{\| }\right) + \left(  \frac{1}{\bar \chi} B_{\perp i}^{(1)} - \frac{\ud}{\ud \bar \chi} B_{\perp i}^{(1)} \right) \nonumber \\
 &&   -    \left( B_{\perp i}^{(1)}{'} +  n^j h_{ j k}^{(1)}{'} \Perp_i^k \right) \bigg]  \Bigg\}        \int_0^{\bar \chi} \ud \tilde \chi \left( B^{i (1)}_{\perp }+ n^k h_{k}^{j(1)} \Perp^i_j  - 2S_{\perp}^{i(1)} \right)  \;.  \nonumber
 \end{eqnarray}
 
Then we obtain
\begin{eqnarray} 
\label{dDx_||-2_3}
&& \p_\| \Delta x_{\parallel}^{(2)}=\left(A^{(1)}_o\right)^2 -2 A^{(1)}_o B^{(1)}_{\| \, o}+ \left(B^{(1)}_{\| \, o}\right)^2 + 2 A^{(1)}_o v^{(1)}_{\| \, o} - 2 B^{(1)}_{\| \, o}v^{(1)}_{\| \, o}-  v^{(1)}_{\| \, o} h^{(1)}_{\| \, o} -\frac{1}{4} \left(h^{(1)}_{\| \, o}\right)^2   -  v^{(1)}_{\perp k \, o} v^{k (1)}_{\perp \, o}\nonumber \\
&& + n^i h_{ik\,o}^{(1)} \Perp^k_j v^{j (1)}_o -  n^i h_{ik\,o}^{(1)} \Perp^k_j B^{j (1)}_o - \frac{1}{4} n^i h_{ij\,o}^{(1)}\,  \Perp^j_k \, h^{k (1)}_{p \, o} n^p   -  B_{\perp \, o}^{i(1)}B_{\perp i \, o}^{(1)}+ 2 v_{\perp \, o}^{i(1)}B_{\perp i \, o}^{(1)}+2 \left(A^{(1)}_o-v^{(1)}_{\| \, o}\right)\nonumber \\
&& \times \bigg\{\left(-\frac{\cH' }{\cH^3}+ \frac{1}{\cH}\right) \left[\frac{\ud \,}{\ud \bar \chi}\left(  A^{(1)} - v^{(1)}_{\| } \right)+\left(A^{(1)}{'} - B^{(1)}_{\| }{'} - \frac{1}{2}h^{(1)}_{\| }{'} \right)\right] - \frac{\bar \chi}{\cH} \frac{\ud \,}{\ud \bar \chi}\left[ \frac{\ud \,}{\ud \bar \chi} \left( 2 A^{(1)} - B^{(1)}_{\| } \right) \right. \nonumber \\
&& \left. + \left(A^{(1)}{'} - B^{(1)}_{\| }{'} - \frac{1}{2}h^{(1)}_{\| }{'} \right)\right]+\left (  A^{(1)} - B^{(1)}_{\| } - \frac{1}{2} h^{(1)}_{\| } \right) -\bar \chi \frac{\ud}{\ud \bar \chi}\left( A^{(1)} - B^{(1)}_{\| } - \frac{1}{2}h^{(1)}_{\| } \right) - \frac{1}{\cH^2}  \frac{\ud \,}{\ud \bar \chi} \bigg[ \frac{\ud \,}{\ud \bar \chi}\left(  A^{(1)} - v^{(1)}_{\| } \right) \nonumber \\
&& + \left(A^{(1)}{'} - B^{(1)}_{\| }{'} - \frac{1}{2}h^{(1)}_{\| }{'} \right)\bigg]\bigg\} + 2  \left(B_{\perp \, o}^{i(1)} -v_{\perp \, o}^{i(1)}+\frac{1}{2}n^k h_{k\,o}^{j(1)} \Perp^i_j  \right)\bigg[ -\frac{1}{\cH} \frac{\ud \,}{\ud \bar \chi}  B_{\perp i}^{(1)}   +\frac{1}{\cH}  \p_{\perp i} \left(A^{(1)} -  v^{(1)}_{\|} \right) \nonumber \\
&& -\frac{1}{\cH}\left(B_{\perp i}^{(1)}{'}+ n^j h_{ j k}^{(1)}{'} \Perp_i^k  \right) -2\frac{\bar \chi}{\cH}  \frac{\ud \,}{\ud \bar \chi} \p_{\perp i} \left(A^{(1)} +  v^{(1)}_{\|} -B^{(1)}_{\|}\right) +\frac{1}{\cH \bar \chi} B_{\perp i}^{(1)} -4  \delta_{il} S_{\perp }^{l (1)}  \bigg]+ A^{(2)}- B^{(2)}_{\| } -\frac{1}{2}h_\|^{(2)} \nonumber \\
&&+ \frac{1}{\cH}\bigg[  \frac{\ud \,}{\ud \bar \chi}\left( A^{(2)} -  v^{(2)}_{\|} \right)+ \left(A^{(2)}{'} - B^{(2)}_{\| }{'} - \frac{1}{2}h^{(2)}_{\| }{'} \right)\bigg] -2\left(2A^{(1)}+h^{(1)}_{\| } \right)\left(  A^{(1)} - B^{(1)}_{\| } - \frac{1}{2} h^{(1)}_{\| }\right)-\frac{1}{\cH}   \frac{\ud \,}{\ud \bar \chi} \bigg[ 7 \left(A^{(1)}\right)^2 \nonumber \\ 
&&+  \left(B^{(1)}_{\|}\right)^2-4A^{(1)} B^{(1)}_{\|} - 2A^{(1)} v^{(1)}_{\| } +  \left(v^{(1)}_{\|}\right)^2 + v^{(1)}_{\|} h^{(1)}_{\| } + v_{\perp i}^{(1)} v_{\perp}^{i(1)}  - 2 v_{\perp i}^{(1)} B_{\perp}^{i(1)} \bigg]  +  \frac{2}{\cH^2}\left( A^{(1)} - v^{(1)}_{\| }\right)\nonumber \\ 
&&\times \frac{\ud \,}{\ud \bar \chi}\left[\frac{\ud \,}{\ud \bar \chi} \left(A^{(1)} -  v^{(1)}_{\|} \right) + \left(A^{(1)}{'} - B^{(1)}_{\| }{'} - \frac{1}{2}h^{(1)}_{\| }{'} \right)\right]  -\frac{2}{\cH} \bigg[\left( 2 A^{(1)} +  h^{(1)}_{\| }  \right)+ \left(A^{(1)} -  v^{(1)}_{\|}\right) \bigg] \nonumber \\ 
&&\times \left(A^{(1)}{'} - B^{(1)}_{\| }{'} - \frac{1}{2}h^{(1)}_{\| }{'} \right) + \frac{2}{\cH}   \left(2 A^{(1)} - B^{(1)}_{\| } \right)\bigg[ \frac{\ud \,}{\ud \bar \chi}\left( 2 A^{(1)} - B^{(1)}_{\| } \right) +  \left(A^{(1)}{'} - B^{(1)}_{\| }{'} - \frac{1}{2}h^{(1)}_{\| }{'} \right)\bigg]  \nonumber \\
&&  +2\bigg\{  \frac{1}{\cH} \left( A^{(1)} - B^{(1)}_{\| } - \frac{1}{2}h^{(1)}_{\| } \right)+\frac{\cH'}{\cH^3}\left( A^{(1)} - v^{(1)}_{\| }\right)+  \frac{1}{\cH^2}\bigg[\frac{\ud \,}{\ud \bar \chi}\left( A^{(1)} - v^{(1)}_{\| }\right) + \left(A^{(1)}{'} - B^{(1)}_{\| }{'} - \frac{1}{2}h^{(1)}_{\| }{'} \right) \bigg]\bigg\}\nonumber \\ 
&& \times  \bigg[\frac{\ud \,}{\ud \bar \chi} \left(A^{(1)} -  v^{(1)}_{\|} \right) + \left(A^{(1)}{'} - B^{(1)}_{\| }{'} - \frac{1}{2}h^{(1)}_{\| }{'} \right)\bigg]   +\frac{2}{\cH}v_{\perp i}^{(1)} \p^i_\perp \left( A^{(1)} - B^{(1)}_{\| } - \frac{1}{2}h^{(1)}_{\| } \right)+ \frac{2}{\cH}  \left(A^{(1)} - B^{(1)}_{\| } - \frac{1}{2}h^{(1)}_{\| } \right) \nonumber \\ 
&& \times \left[ \frac{\ud \,}{\ud \bar \chi}\left( 2 A^{(1)} - B^{(1)}_{\| }\right)  -  \p_\| \left(A^{(1)} +  v^{(1)}_{\|} -B^{(1)}_{\|}\right)\right]    +\frac{2}{\cH} \frac{\ud \,}{\ud \bar \chi} \p_\| \left(A^{(1)} +  v^{(1)}_{\|} -B^{(1)}_{\|}\right) T^{(1)}+ 2 \left(B_{\perp}^{i(1)}+ n^j h_{ j k}^{(1)} \Perp^{ik}  \right) \nonumber \\ 
&& \times \left[\frac{1}{2}\left(B_{i \perp}^{(1)}+ n^p h_{p m}^{(1)} \Perp^m_i  \right) -\frac{1}{\cH}\p_{\perp i}\left(  A^{(1)} - v^{(1)}_{\| } \right)+\frac{1}{\cH} \left(B_{\perp i}^{(1)}{'} + n^j h_{ j k}^{(1)}{'} \Perp_i^k  \right)  \right]  +4 \bigg\{ \left(-\frac{\cH' }{\cH^3} +\frac{1}{\cH}\right) \bigg[\frac{\ud \,}{\ud \bar \chi}\left(A^{(1)} - v^{(1)}_{\| } \right)  \nonumber \\ 
&&  +\left(A^{(1)}{'} - B^{(1)}_{\| }{'} - \frac{1}{2}h^{(1)}_{\| }{'} \right)\bigg]- \frac{1}{\cH^2} \frac{\ud \,}{\ud \bar \chi}\bigg[\frac{\ud \,}{\ud \bar \chi}\left(A^{(1)} - v^{(1)}_{\| } \right)  +\bigg(A^{(1)}{'} - B^{(1)}_{\| }{'} - \frac{1}{2}h^{(1)}_{\| }{'} \bigg) \bigg]+\left(  A^{(1)} - B^{(1)}_{\| } - \frac{1}{2} h^{(1)}_{\| }\right)\bigg\} I^{(1)}   \nonumber \\
 &&+2 \bigg\{ \frac{\ud}{\ud \bar \chi}\left( A^{(1)} - B^{(1)}_{\| } - \frac{1}{2}h^{(1)}_{\| } \right)   + \frac{1}{\cH} \frac{\ud \,}{\ud \bar \chi} \bigg[ \frac{\ud \,}{\ud \bar \chi} \left( 2 A^{(1)} - B^{(1)}_{\| } \right)  + \left(A^{(1)}{'} - B^{(1)}_{\| }{'} - \frac{1}{2}h^{(1)}_{\| }{'} \right)\bigg]  \bigg\}  \int_0^{\bar \chi} \ud \tilde \chi \;\bigg[ 2 A^{(1)} - B^{(1)}_{\| } \nonumber \\
 && + \left(\bar \chi-\tilde \chi\right)\left(A^{(1)}{'} - B^{(1)}_{\| }{'} - \frac{1}{2}h^{(1)}_{\| }{'} \right) \bigg]+ 4 \bigg[\frac{1}{\cH}\p_{\perp i}\left(  A^{(1)} - v^{(1)}_{\| } \right)-\frac{1}{\cH} \left(B_{\perp i}^{(1)}{'} + n^j h_{ j k}^{(1)}{'} \Perp_i^k  \right)  + \frac{1}{\cH \bar \chi} v_{\perp i}^{(1)}  -\frac{1}{\cH}\frac{\ud \,}{\ud \bar \chi} v_{\perp i}^{(1)}\nonumber \\
 &&-2S_{\perp }^{j(1)} \delta_{ij}  \bigg]    S_{\perp }^{i(1)}  + \frac{2}{\cH} \frac{\ud \,}{\ud \bar \chi} \bigg[\p_{\perp i}\left(A^{(1)} +  v^{(1)}_{\|} -B^{(1)}_{\|}\right)  - \frac{1}{\bar \chi} \left( v^{(1)}_{\perp i}- B^{ (1)}_{\perp i }\right)\bigg] \int_0^{\bar \chi} \ud \tilde \chi \bigg\{ \left( B^{i (1)}_{\perp }+ n^k h_{k}^{j(1)} \Perp^i_j\right)\nonumber \\
 && + \left(\bar \chi-\tilde \chi\right)\left[ \tilde \p^i_\perp \left( A^{(1)} - B^{(1)}_{\| } - \frac{1}{2}h^{(1)}_{\| } \right) + \frac{1}{\tilde \chi} \left(B^{i (1)}_{\perp }+  n^k h_{kj}^{(1)} \Perp^{ij}  \right)\right] \bigg\} -  \frac{\cH'}{\cH^2} \Delta \ln a^{(2)} +2 \bigg\{-\frac{\cH' }{\cH^2} \left( A^{(1)} - B^{(1)}_{\| } - \frac{1}{2}h^{(1)}_{\| } \right)\nonumber \\
&&- \frac{1}{\cH} \frac{\ud}{\ud \bar \chi}\left( A^{(1)} - B^{(1)}_{\| } - \frac{1}{2}h^{(1)}_{\| } \right)- \left(\frac{\cH' }{\cH^3} +\frac{1}{\cH} \right)\left[\frac{\ud \,}{\ud \bar \chi} \left( A^{(1)}-  v_\|^{(1)}\right) + \left(A^{(1)}{'} - B^{(1)}_{\| }{'} - \frac{1}{2}h^{(1)}_{\| }{'} \right)\right] \bigg\} \Delta \ln a^{(1)} \nonumber \\
&&  + 2 \int_0^{\bar \chi} \ud \tilde \chi \bigg\{ \left(A^{(1)} - B^{(1)}_{\| } - \frac{1}{2}h^{(1)}_{\| } \right) \left[2\left(A^{(1)}{'} - B^{(1)}_{\| }{'} - \frac{1}{2}h^{(1)}_{\| }{'} \right)+  \frac{\ud}{\ud \tilde \chi}\left( 2 A^{(1)} - B^{(1)}_{\| }\right) \right]   - \left(B_{\perp}^{i(1)} + n^k h_{k}^{j(1)} \Perp^i_j  \right)\nonumber \\
&&\times\left[ \tilde \p_{\perp i}\left (  A^{(1)} - B^{(1)}_{\| } - \frac{1}{2} h^{(1)}_{\| } \right)  + \frac{1}{\tilde \chi} \left(B_{\perp i}^{(1)}+   n^m h_{mp}^{(1)} \Perp^p_i  \right)\right]\bigg\}+\left[-\frac{\cH'' }{\cH^3}+3\left( \frac{\cH' }{\cH^2} \right)^2 + \frac{\cH' }{\cH^2} \right] \left( \Delta \ln a^{(1)} \right)^2  \nonumber \\
 &&+ \p_\| \Delta x_{\|  \rm post-Born}^{(2)}\;,
\end{eqnarray}

where
  \begin{eqnarray} 
 &&  \p_\| \Delta x_{\|  \rm post-Born}^{(2)}= -2  \left( A^{(1)}_o -  B^{(1)}_{\| o} - \frac{1}{2} h^{(1)}_{\| o}\right) \left(3 v_{\| \, o}^{(1)} + 2 h^{(1)}_{\| \, o} - \frac{1}{2} h^{j (1)}_{j \, o} \right)  +2 \left(A^{(1)}_o-v^{(1)}_{\| \, o} \right) \Bigg\{- \left(A^{(1)}   -  B^{(1)}_{\| } - \frac{1}{2}  h^{(1)}_{\| }\right)   \nonumber \\
 && + \frac{ \bar \chi}{\cH}  \frac{\ud}{\ud \bar \chi} \bigg[\frac{\ud}{\ud \bar \chi}\left( 2 A^{(1)} - B^{(1)}_{\| }\right)    + \left(A^{(1)}{'} - B^{(1)}_{\| }{'} - \frac{1}{2}h^{(1)}_{\| }{'} \right) \bigg] \Bigg\}+ 2  \bar \chi \left(A^{(1)}_o-v^{(1)}_{\| \, o}\right) \frac{\ud}{\ud \bar \chi}\left(A^{(1)}  - B^{(1)}_{\| } - \frac{1}{2}  h^{(1)}_{\| }\right) \nonumber \\
   && + 2 \left(B^{i (1)}_{\perp \, o }-v^{i (1)}_{\perp \, o }+ \frac{1}{2} n^k h_{k\,o}^{j(1)} \Perp^i_j\right)   \Bigg\{   \bar \chi  \p_{\perp i}\left(A^{(1)} - B^{(1)}_{\| } - \frac{1}{2}  h^{(1)}_{\| } \right) +  \left(B^{(1)}_{\perp i}+  n^j h_{ j}^{k(1)} \Perp_{ik}  \right)   - \frac{ 1}{\cH} \p_{\perp i} \left(2 A^{(1)} - B^{(1)}_{\| }\right)  \nonumber \\
  &&+ \frac{ \bar \chi}{\cH} \p_{\perp i}\frac{\ud}{\ud \bar \chi} \left(2 A^{(1)} - B^{(1)}_{\| }\right)    + \frac{ \bar \chi}{\cH} \p_{\perp i}   \left(A^{(1)}{'} - B^{(1)}_{\| }{'} - \frac{1}{2}h^{(1)}_{\| }{'} \right) - \frac{ 1}{\bar \chi \cH}  B_{\perp i}^{(1)} + \frac{ 1}{\cH}  \frac{\ud}{\ud \bar \chi} B_{\perp i}^{(1)}   \nonumber \\
 &&  +  \frac{ 1}{\cH}    \left( B_{\perp i}^{(1)}{'} +  n^j h_{ j k}^{(1)}{'} \Perp_i^k \right)+ 2 S_{\perp}^{j(1)}    \Bigg\}     
     - 4 \left(- B^{(1)}_{\| o} + v_{\| \, o}^{(1)} - \frac34 h^{(1)}_{\| \, o} + \frac14 h^{m (1)}_{m \, o} \right) \int_0^{\bar \chi}   \frac{\ud \tilde{\chi}}{\tilde \chi} \left( A^{(1)} - B^{(1)}_{\| } - \frac{1}{2} h^{(1)}_{\|} \right)     \nonumber \\
 &&   - 2 \Bigg\{ \left(A^{(1)}{'} - B^{(1)}_{\| }{'}  - \frac{1}{2}  h^{(1)}_{\| }{'} \right)  +\frac{1}{\cH}  \left[ \frac{\ud}{\ud \bar \chi}\left(2 A^{(1)}{'} - B^{(1)}_{\| }{'}\right)   + \left(A^{(1)}{''} - B^{(1)}_{\| }{''} - \frac{1}{2}h^{(1)}_{\| }{''} \right) \right]  \Bigg\} T^{(1)}  \nonumber \\
  && - 2 \Bigg\{ \frac{\ud}{\ud \bar \chi}\left(A^{(1)}  - B^{(1)}_{\| } - \frac{1}{2}  h^{(1)}_{\| }\right)  + \frac{1}{\cH} \frac{\ud}{\ud \bar \chi} \bigg[\frac{\ud}{\ud \bar \chi}\left( 2 A^{(1)} - B^{(1)}_{\| }\right)    + \left(A^{(1)}{'} - B^{(1)}_{\| }{'} - \frac{1}{2}h^{(1)}_{\| }{'} \right) \bigg] \Bigg\}  \nonumber \\
 && \times \int_0^{\bar \chi} \ud \tilde \chi \left(A^{(1)}+ \frac{1}{2}  h^{(1)}_{\| } -2I^{(1)} \right) + 2 \left(A^{(1)}   -  B^{(1)}_{\| } - \frac{1}{2}  h^{(1)}_{\| }\right) \left(A^{(1)} + \frac{1}{2}  h^{(1)}_{\| }- 2 I^{(1)}\right)  \nonumber \\
  && - 4 \left(A^{(1)}   -  B^{(1)}_{\| } - \frac{1}{2}  h^{(1)}_{\| }\right)\kappa^{(1)}  - 2  \Bigg\{   \p_{\perp i}\left(A^{(1)} - B^{(1)}_{\| } - \frac{1}{2}  h^{(1)}_{\| } \right) + \frac{1}{\bar \chi} \left(B^{(1)}_{\perp i}+  n^j h_{ j}^{k(1)} \Perp_{ik}  \right)  \nonumber \\
 && + \frac{1}{\cH} \p_{\perp i}\frac{\ud}{\ud \bar \chi} \left(2 A^{(1)} - B^{(1)}_{\| }\right) + \frac{1}{\cH} \p_{\perp i}   \left(A^{(1)}{'} - B^{(1)}_{\| }{'} - \frac{1}{2}h^{(1)}_{\| }{'} \right) -  \frac{1}{\bar \chi \cH} \bigg[\p_{\perp i} \left(2 A^{(1)} - B^{(1)}_{\| }\right)\nonumber \\
 &&   + \left(  \frac{1}{\bar \chi} B_{\perp i}^{(1)} - \frac{\ud}{\ud \bar \chi} B_{\perp i}^{(1)} \right)   -    \left( B_{\perp i}^{(1)}{'} +  n^j h_{ j k}^{(1)}{'} \Perp_i^k \right) \bigg]  \Bigg\}        \int_0^{\bar \chi} \ud \tilde \chi \left( B^{i (1)}_{\perp }+ n^k h_{k}^{j(1)} \Perp^i_j  - 2S_{\perp}^{i(1)} \right)\nonumber \\
  && +2 \int_0^{\bar \chi}  \ud \tilde{\chi} \Bigg\{   - \left( A^{(1)} - B^{(1)}_{\| } - \frac{1}{2} h^{(1)}_{\|} \right)  \left[\frac{\ud}{\ud \tilde \chi}\left(A^{(1)} + \frac{1}{2}  h^{(1)}_{\| }\right) + 2 \left(A^{(1)}{'} - B^{(1)}_{\| }{'} - \frac{1}{2}h^{(1)}_{\| }{'} \right) \right]   \nonumber \\
 &&-  \left[ \tilde \p_{\perp j}\left( A^{(1)} - B^{(1)}_{\| } - \frac{1}{2} h^{(1)}_{\|} \right) +  \frac{1}{\bar \chi} \left(B^{(1)}_{\perp j}+  n^m h_{ m}^{k(1)} \Perp_{jk}  \right) \right] \left[   - \left( B^{j (1)}_{\perp }+ n^l h_{l}^{p(1)} \Perp^j_p\right) + 2S_{\perp}^{j(1)} \right]   \nonumber \\
 &&  + \left( A^{(1)} - B^{(1)}_{\| } - \frac{1}{2} h^{(1)}_{\|} \right)  \left[   \tilde \p_{\perp m} \left(B^{m (1)} + n^l h_{l}^{m(1)} \right) - \frac{2}{\tilde \chi} \left( B^{(1)}_{\| } + h^{(1)}_{\|}\right) - 2  \tilde \p_{\perp m}S_{\perp}^{m(1)}  \right]  \nonumber \\
 && + 2 \bigg[ \frac{\ud}{\ud \bar \chi}\left( A^{(1)} - B^{(1)}_{\| } - \frac{1}{2} h^{(1)}_{\|} \right)    -  \frac{1}{\tilde\chi} \left( A^{(1)} - B^{(1)}_{\| } - \frac{1}{2} h^{(1)}_{\|} \right)  \bigg] \kappa^{(1)}  \Bigg\}   \;.   \nonumber\\
 \end{eqnarray}

From Eq. (\ref{Dx_perp-2_2}) and using Eqs. (\ref{kappa1-1}) (\ref{kappa2-1}) and (\ref{kappa3-1}) we obtain  the coordinate convergence lensing term   at second order

\begin{eqnarray}
 \label{kappa-2}
 \kappa^{(2)}=- \frac{1}{2}  \p_{\perp i} \Delta x_{\perp}^{i (2)}= \kappa_1^{(2)}+\kappa_2^{(2)}+\kappa_3^{(2)}+\kappa_4^{(2)}  + \kappa^{(2)}_{ \rm post-Born}\;,
  \end{eqnarray}
  
 where
 
  \begin{eqnarray}
 \label{kappa1-2}
 \kappa_1^{(2)} &=&  \frac{1}{2}  \int_0^{\bar \chi} \ud \tilde \chi  \left(\bar \chi-\tilde \chi\right) \frac{\tilde \chi}{ \bar \chi}   \tilde \nabla^2_\perp \left( A^{(2)} - B^{(2)}_{\| } - \frac{1}{2}h^{(2)}_{\| } \right)\;, \\
 \label{kappa2-2}
 \kappa_2^{(2)} &=& \frac{1}{2}  \int_0^{\bar \chi} \ud \tilde \chi \bigg[ \tilde \p_\perp^i B_i^{ (2)} - \frac{2}{\tilde\chi}B_\|^{ (2)}+ \Perp^{ij} n^k  \tilde \p_i h_{jk}^{ (2)}+ \frac{1}{\tilde \chi} \left(h_i^{i (2)}-3h_\|^{ (2)} \right)\bigg] \;, 
  \end{eqnarray}
 \begin{eqnarray}
 \label{kappa3-2}
 \kappa_3^{(2)} &=& \bigg\{- \frac{2}{\bar \chi} B_{\|}^{(1)} +  \p_\perp^i B_i^{ (1)} +  \frac{1}{\bar \chi}  \left(h_i^{i (1)}-3h_\|^{ (1)} \right)+ \Perp^{ij} n^k   \p_i h_{jk}^{ (1)}+ \frac{1}{\bar \chi}  \int_0^{\bar \chi} \ud \tilde \chi  \bigg[ \tilde \p_\perp^i B_i^{ (1)}+ \Perp^{ij} n^k  \tilde \p_i h_{jk}^{ (1)} \nonumber \\
  &&- \frac{1}{\tilde \chi} \left( 2B_{\|}^{(1)} + 3 h_\|^{ (1)} - h_i^{i (1)} \right) \bigg]+ \int_0^{\bar \chi} \ud \tilde \chi \; \frac{\tilde \chi}{ \bar \chi} \,  \tilde \nabla^2_\perp \left( A^{(1)} - B^{(1)}_{\| } - \frac{1}{2}h^{(1)}_{\| } \right)\bigg\} \times \bigg\{ \int_0^{\bar \chi} \ud \tilde \chi \bigg[ 2 A^{(1)} - B^{(1)}_{\| }  \nonumber \\
&&  + \left(\bar \chi-\tilde \chi\right) \left(A^{(1)}{'} - B^{(1)}_{\| }{'} - \frac{1}{2}h^{(1)}_{\| }{'} \right) \bigg]   -  \frac{1}{\cH}\Delta \ln a^{(1)} \bigg\}+\left( B^{i (1)}_{\perp }+ n^k h_{k}^{j(1)} \Perp^i_j-2S_{\perp}^{i(1)} \right)    \nonumber \\
 && \times   \bigg\{ \int_0^{\bar \chi} \ud \tilde \chi \,  \frac{\tilde \chi}{ \bar \chi}  \left[\tilde \p_{\perp i} \left(2 A^{(1)} - B^{(1)}_{\| }\right) + \left(\bar \chi-\tilde \chi\right) \tilde \p_{\perp i} \left(A^{(1)}{'} - B^{(1)}_{\| }{'} - \frac{1}{2}h^{(1)}_{\| }{'} \right) \right]   -  \frac{1}{\cH} \p_{\perp i}  \Delta \ln a^{(1)} \bigg\} \nonumber \\
 && + \int_0^{\bar \chi} \ud \tilde \chi  \left[ \frac{4}{\tilde \chi} A^{(1)}  B_{\|}^{(1)} - 4 A^{(1)} \tilde  \p_\perp^i B_i^{ (1)}  -  \frac{2}{\tilde \chi} \left( B_{\|}^{(1)} \right)^2 +B_{\|}^{(1)} \tilde\p_\perp^i B_i^{ (1)}-  \frac{3}{\tilde \chi}  A^{(1)}  h_i^{i (1)}+ \frac{9}{\tilde\chi}  A^{(1)} h_\|^{ (1)}  \right. \nonumber \\
&& - 3  A^{(1)}  \Perp^{ij} n^k  \tilde \p_i h_{jk}^{ (1)}  + \frac{2}{\tilde \chi}B_{\|}^{(1)}h_\|^{ (1)} -h_\|^{ (1)} \tilde \p_\perp^i B_i^{ (1)}- \frac{3}{2 \tilde \chi}  h_\|^{ (1)}  h_i^{i (1)}+ \frac{9}{2 \tilde \chi}\left( h_\|^{ (1)} \right)^2 -  \frac{3}{2} h_\|^{ (1)} \Perp^{ij} n^k  \tilde  \p_i h_{jk}^{ (1)} \nonumber \\
&& \left. -4 \left( \frac{2}{\tilde\chi} B_{\|}^{(1)} - \tilde \p_\perp^i B_i^{ (1)} -  \frac{1}{\tilde\chi} h_i^{i (1)} +  \frac{3}{\tilde\chi} h_\|^{ (1)} - \Perp^{ij} n^k  \tilde \p_i h_{jk}^{ (1)}  \right) I^{(1)} \right] + \int_0^{\bar \chi} \ud \tilde \chi \, \frac{\tilde \chi}{ \bar \chi}  \bigg[- \frac{4}{\tilde \chi} A^{(1)}  B_{\|}^{(1)} + 2 A^{(1)}\tilde \p_\perp^i B_i^{ (1)}  \nonumber \\
&& + \frac{2}{\tilde \chi} A^{(1)}  h_i^{i (1)} - \frac{6}{\tilde\chi} A^{(1)} h_\|^{ (1)} + 2 A^{(1)}  \Perp^{ij} n^k  \tilde \p_i h_{jk}^{ (1)} +  \frac{1}{\tilde\chi} B_{\|}^{(1)} h_i^{i (1)}  - \frac{3}{\tilde\chi} B_{\|}^{(1)} 
h_\|^{ (1)} -h_\|^{ (1)} \tilde \p_\perp^i B_i^{ (1)} - \frac{4}{\tilde\chi} \left( h_\|^{ (1)} \right)^2\nonumber \\
&&+ h_\|^{ (1)} \Perp^{ij} n^k  \tilde \p_i h_{jk}^{ (1)} - 2 \, \tilde \p_{\perp i}  A^{(1)} B^{i(1)}_{\perp} + \frac{1}{\tilde \chi}  B^{(1)}_{\perp i} B^{i(1)}_{\perp} +  n^j B^{i (1)}_{\perp}  \tilde \p_{\perp i}  B_j^{(1)} -  \tilde \p_{i} A^{(1)} n^j h_{ j k}^{(1)} \Perp^{ki} +\frac{2}{\tilde \chi} n^p h_{ p q}^{(1)} \Perp^{qj} h_{ j k}^{(1)} n^k  \nonumber \\
&& - \frac{1}{2} n^j h_{ j k}^{(1)} \Perp^{ki}    n^p n^q \tilde \p_{\perp i} h_{ p q}^{(1)} -   h_{l}^{k(1)}  \Perp^{l}_q \,  \tilde \p_{\perp k} B^{q (1)}- \frac{1}{\tilde \chi} \Perp^{m}_k \Perp^{l}_p h_{m}^{p(1)}  h_{l}^{k(1)}+ \frac{3}{\tilde \chi} n^l  B_{\perp k}^{(1)}   h_{l}^{k(1)}-B^{l(1)}_{\perp}   \tilde \p_{\perp k} h_l^{k(1)} \nonumber \\
&&+  \frac{2}{\tilde\chi} h_\|^{ (1)} h_i^{i (1)}- n^m  h_{m}^{p(1)}\Perp^l_p \tilde \p_{\perp k}h_{l}^{k(1)}  + 4  \left(B_{\perp}^{i(1)}+ n^j h_{ j k}^{(1)} \Perp^{ik}  \right) \tilde \p_{\perp i}  I^{(1)} + 2 \left(- \frac{3}{\tilde \chi}  n^l \Perp_{kj} h_l^{ k(1)} +  \Perp^l_j  \tilde \p_{\perp k} h_l^{ k (1)} \right) S_{\perp}^{j(1)} \nonumber \\
&&- \frac{2}{\tilde \chi}   \Perp^l_k   h_l^{ k (1)}S_{\|}^{(1)} +2 \, \Perp^l_m   h_l^{ k (1)}  \tilde \p_{\perp k} S^{m(1)}\bigg]  + \int_0^{\bar \chi} \ud \tilde \chi \left(\bar \chi-\tilde \chi\right)\frac{\tilde \chi}{ \bar \chi}  \bigg\{   \tilde \p_{\perp i}\left(A^{(1)}{'} - B^{(1)}_{\| }{'} - \frac{1}{2}h^{(1)}_{\| }{'} \right) \nonumber \\
&& \times \left(B_{\perp}^{i(1)}+ n^j h_{ j k}^{(1)} \Perp^{ik}  \right)  + \left(A^{(1)}{'} - B^{(1)}_{\| }{'} - \frac{1}{2}h^{(1)}_{\| }{'} \right) \left(-\frac{2}{\tilde \chi}  B_{\|}^{(1)}  + \tilde \p_\perp^i B_i^{ (1)} +   \frac{1}{\tilde \chi}  h_i^{i (1)}- \frac{3}{\tilde \chi}  h_\|^{ (1)} +  n^k  \tilde \p_{\perp}^{j(1)}  h_{jk}^{ (1)}\right)  \nonumber \\
&& - \left(2  \p_{\perp i} A^{(1)} -\frac{1}{\tilde \chi}  B_{\perp i}^{(1)} - n_k   \tilde \p_{\perp i} B_k^{ (1)} \right)  \frac{\ud}{\ud \tilde \chi} \left(B_{\perp}^{i(1)}+ n^j h_{ j k}^{(1)} \Perp^{ik}  \right)-  \left(2 A^{(1)} -B_\|^{(1)}\right)  \frac{\ud}{\ud \tilde \chi} \left(-\frac{2}{\tilde \chi}  B_{\|}^{(1)}  + \tilde \p_\perp^i B_i^{ (1)} \right. \nonumber \\
&& \left. +  \frac{1}{\tilde \chi}  h_i^{i (1)}- \frac{3}{\tilde \chi}  h_\|^{ (1)} + n^k  \tilde \p_{\perp}^{j(1)}  h_{jk}^{ (1)} \right) - 2 \left(- \Perp_{jq}  \tilde \p_{\perp i} B^{ q (1)}- n^k   \Perp_{jq}  \tilde \p_{\perp i}  h_k^{q (1)}+ 2\Perp_{jm} \tilde \p_{\perp j} S^{m(1)}  \right)   \nonumber \\
&&\times \left( \Perp^{l[j} \,  \p_\perp^{i]} B_l^{ (1)} + n^m  \Perp^{l[j} \, \p_\perp^{i]}  h_{lm}^{ (1)} \right)+  \left(B_{\perp j}^{(1)}+ n^p h_{p}^{k(1)} \Perp_{j k}  - 2 \delta_{jp}  S_{\perp}^{p(1)} \right) \left(\Perp^{jl}   \tilde \nabla^2_\perp B^{(1)}_l   -   \tilde \p_{\perp}^j \tilde \p_{\perp}^k B^{(1)}_k \right. \nonumber \\
&& \left.+  \Perp^{jl}  n^m  \tilde \nabla^2_\perp h^{(1)}_{lm} - n^m  \, \tilde \p_{\perp}^j \tilde \p_{\perp}^k h^{(1)}_{km}  +  \frac{1}{\tilde \chi} n^m n^k   \tilde \p_{\perp}^j h^{(1)}_{km}+  \frac{1}{\tilde \chi}   \Perp^{jl}  \tilde \p_{\perp}^m h^{(1)}_{lm}  - \frac{1}{\tilde \chi}\tilde \p_{\perp}^j  h_i^{i (1)} \right)  \nonumber \\
&&- \tilde \p_{\perp i} \left( A^{(1)} - B^{(1)}_{\| } - \frac{1}{2}h^{(1)}_{\| } \right) \bigg[ \tilde\p^i_\perp \left(B^{(1)}_{\| }+ h^{(1)}_{\| } \right) - \p_\| \left(B_{\perp}^{i (1)}+n^p h_{pq}^{(1)} \Perp^{iq} \right)   - \frac{1}{\tilde \chi} \left(B^{i (1)}_{\perp }+  2 n^p h_{pq}^{(1)} \Perp^{iq}  \right) \bigg] \nonumber \\ 
&& - \left( A^{(1)} - B^{(1)}_{\| } - \frac{1}{2}h^{(1)}_{\| } \right) \bigg[ \tilde \nabla_{\perp}^2 \left(B^{(1)}_{\| }+ h^{(1)}_{\| } \right)  - \tilde \p_\| \left(-\frac{2}{\tilde \chi}  B_{\|}^{(1)}  + \tilde \p_\perp^i B_i^{ (1)} +  \frac{1}{\tilde \chi}  h_i^{i (1)}- \frac{3}{\tilde \chi}  h_\|^{ (1)} + n^k  \tilde \p_{\perp}^{j(1)}  h_{jk}^{ (1)} \right) \bigg]   \nonumber \\ 
&& - 2 \left( 2\tilde \p_\perp^i A^{ (1)} - \frac{1}{\tilde \chi} B_{\perp j}^{(1)} -  n_k   \tilde \p_{\perp i} B_k^{ (1)}  - 2  \tilde  \p_\perp^i I^{ (1)} \right)\left[ \tilde \p_{\perp}^i \left (  A^{(1)} - B^{(1)}_{\| } - \frac{1}{2} h^{(1)}_{\| } \right)  + \frac{1}{\tilde \chi} \left(B_{\perp}^{i(1)}+   n^j h_{jk}^{(1)} \Perp^{k i}  \right)\right]\nonumber \\ 
&&  - 2 \left(2 A^{(1)} - B^{(1)}_{\| } - 2 I^{ (1)}  \right) \tilde \nabla_{\perp}^2 \left( A^{(1)} - B^{(1)}_{\| } - \frac{1}{2}h^{(1)}_{\| } \right) \bigg\} \;, 
  \end{eqnarray} 
 \begin{eqnarray}
 \label{kappa4-2}
  \kappa_4^{(2)} &=&  B_{\| \, o}^{(2)}- v_{\| \, o}^{(2)}+\frac{3}{4}h_{\| \, o}^{ (2)} - \frac{1}{4}h_{i \, o}^{i (2)}-2A_o^{(1)}  B_{\| \, o}^{(1)}+ 2A_o^{(1)}  v_{\| \, o}^{(1)} -\frac{1}{2}A_o^{(1)}  h_{i \, o}^{i(1)}+\frac{3}{2}A_o^{(1)}  h_{\| \, o}^{(1)} + \frac{1}{2} v^{(1)}_{\perp i \, o } v^{i (1)}_{\perp \, o}-\left(v_{\| \, o}^{(1)}\right)^2  \nonumber\\
  && +   v^{(1)}_{\perp i \, o } B^{i (1)}_{\perp \, o} -2 v_{\| \, o}^{(1)} B_{\| \, o}^{(1)} -\frac{1}{2} B^{(1)}_{\perp i \, o } B^{i (1)}_{\perp \, o} + \left(B_{\| \, o}^{(1)}\right)^2 - \frac{7}{2} v_{\| \, o}^{(1)}h_{\| \, o}^{(1)}
  +\frac{1}{2}  v_{\| \, o}^{(1)}h_{i \, o}^{i(1)} - \frac{3}{2}   v^{(1)}_{i \, o} n^k h_{k\, o}^{j(1)} \Perp^i_j   \nonumber\\
  &&+ \frac{1}{8} h_{k \, o}^{j (1)}  h_{j \, o}^{k(1)} - \frac{3}{8} n^i h_{i \, o}^{j (1)}  h_{j \, o}^{k(1)} n_k    +  v^{(1)}_{\perp i \, o} \bigg\{  \left(B^{i (1)}_{\perp }+  n^k h_{kj}^{(1)} \Perp^{ij}\right)- 2 S_{\perp}^{i(1)} - \int_0^{\bar \chi} \ud \tilde \chi \bigg[\frac{2}{\tilde \chi}\left(B_{\perp}^{i(1)}+ n^k h_{ k}^{j(1)} \Perp^{i}_j \right) \nonumber \\
 && + 2\left(1-\frac{\tilde \chi}{\bar \chi}\right)\tilde \p^i_\perp \left( A^{(1)} - B^{(1)}_{\| } - \frac{1}{2}h^{(1)}_{\| } \right) \bigg] \bigg\} + \left (A^{(1)}_o-v^{(1)}_{\| \, o}\right) \bigg\{2 B^{(1)}_{\|} - \bar \chi \p_{\perp i} B^{i (1)} + 3 h_{\|}^{(1)} - h_{i}^{i(1)}\nonumber \\
 && - \bar \chi n^k  \p_{\perp j} h^{j (1)}_k - \int_0^{\bar \chi} \ud \tilde \chi \bigg[\tilde \p_{\perp i} B^{i (1)} + n^k  \tilde \p_{\perp j} h^{j (1)}_k   - \frac{2}{\tilde\chi}B_\|^{ (1)}+ \frac{1}{\tilde \chi} \left(h_i^{i (1)}-3h_\|^{ (1)} \right)+ \tilde \chi   \tilde \nabla^2_\perp \left( A^{(1)} - B^{(1)}_{\| } - \frac{1}{2}h^{(1)}_{\| } \right) \bigg] \nonumber \\
 &&+ 4 \kappa_2^{(1)}+ 4 \kappa_1^{(1)} \bigg\}-\frac{1}{\bar \chi} \left(B_{\| \, o}^{(1)} - v_{\| \, o}^{(1)}+\frac{1}{2}h_{\| \, o}^{ (1)} \right) \int_0^{\bar \chi} \ud \tilde \chi \left(h_i^{i (1)}-h_\|^{ (1)} \right) + \frac{1}{2\bar \chi} \Perp^k_i  \, h^{l (1)}_{k \, o} \Perp_{jl}  \int_0^{\bar \chi} \ud \tilde \chi \left(\Perp^{jm}  \, h^{p (1)}_{m} \Perp_{p}^{i(1)}\right)  \nonumber \\
 &&+ \left( B^{ (1)}_{\perp j \, o }-v^{ (1)}_{\perp j \, o }+ \frac{1}{2} n^k \, h_{k\,o}^{l(1)} \Perp_{j l} \right) \int_0^{\bar \chi} \ud \tilde \chi \bigg\{ -  \Perp^{jl} \tilde \p_{\perp}^m h^{(1)}_{lm} +  \frac{\tilde \chi}{ \bar \chi} \bigg[ -\frac{3}{\tilde\chi} n_p \, h_{m}^{p(1)} \Perp^{mj}+2 \Perp^{jm} \p_{\perp p} h^{p(1)}_m  \nonumber \\
 && - \left(\bar \chi-\tilde \chi\right)  \left(\Perp^{jl}   \tilde \nabla^2_\perp B^{(1)}_l   -   \tilde \p_{\perp}^j \tilde \p_{\perp}^k B^{(1)}_k +  \Perp^{jl}  n^m  \tilde \nabla^2_\perp h^{(1)}_{lm} - n^m  \, \tilde \p_{\perp}^j \tilde \p_{\perp}^k h^{(1)}_{km}   \right)\nonumber \\
  &&- \left(\frac{\bar \chi}{\tilde \chi}-1\right)\left(n^m n^k \tilde \p_{\perp}^j h_{k m}^{(1)}- \tilde \p_{\perp}^j h_{k }^{k(1)}\right)\bigg] \bigg\} +\bigg[ \frac{2}{\bar \chi} \left(B_{\| \, o}^{(1)} - v_{\| \, o}^{(1)}\right) - \frac{1}{2 \bar \chi}h_{i \, o}^{i (1)}+\frac{3}{2 \bar \chi} h_{\| \, o}^{ (1)}\bigg]  \nonumber \\
  && \times  \bigg\{ \int_0^{\bar \chi} \ud \tilde \chi \left[ 2 A^{(1)} - B^{(1)}_{\| } + \left(\bar \chi-\tilde \chi\right) \left(A^{(1)}{'} - B^{(1)}_{\| }{'} - \frac{1}{2}h^{(1)}_{\| }{'} \right) \right]   -  \frac{1}{\cH}\Delta \ln a^{(1)} \bigg\} \nonumber \\
  && - \left(  B^{i (1)}_{\perp \, o }-v^{i (1)}_{\perp \, o }+ \frac{1}{2} n^k h_{k\,o}^{j(1)} \Perp^i_j\right)  \bigg\{ \int_0^{\bar \chi} \ud \tilde \chi   \frac{\tilde \chi}{ \bar \chi}  \left[\tilde \p_{\perp i} \left(2 A^{(1)} - B^{(1)}_{\| }\right) + \left(\bar \chi-\tilde \chi\right) \tilde \p_{\perp i} \left(A^{(1)}{'} - B^{(1)}_{\| }{'} - \frac{1}{2}h^{(1)}_{\| }{'} \right) \right]   \nonumber \\
  &&   -  \frac{1}{\cH} \p_{\perp i}  \Delta \ln a^{(1)} \bigg\}\;.
 \end{eqnarray}

Here
\begin{equation}
\kappa^{(2)}_{ \rm post-Born}=\kappa^{(2)}_{1 \rm post-Born}+ \kappa^{(2)}_{2 \rm post-Born}+ \kappa^{(2)}_{3 \rm post-Born}\;,
\end{equation}
where
        \begin{eqnarray}
  &&\kappa^{(2)}_{1 \rm post-Born}=    -   \left(B^{(1)i }_{\perp }+  n^j h_{ j}^{k(1)} \Perp^i_{k}  \right)  \int_0^{\bar \chi} \ud \tilde{\chi}    \frac{ \tilde \chi }{\bar \chi} \bigg[\tilde \p_{\perp i}  \left( A^{(1)} + \frac{1}{2}  h^{(1)}_{\| } \right) - 2 \tilde \p_{\perp i} I^{(1)} \bigg]   - \bigg[ \p_{\perp i} \left(B^{i (1)}  + n^k h_{k}^{i(1)} \right)\nonumber \\
  &&  -\frac{2}{\tilde \chi} \left( B^{(1)}_{\| } + h^{(1)}_{\|}\right)  \bigg]  \int_0^{\bar \chi} \ud \tilde \chi \left(A^{(1)} + \frac{1}{2}  h^{(1)}_{\| }- 2 I^{(1)}\right)   + \int_0^{\bar \chi} \ud \tilde{\chi}  \frac{ \tilde \chi }{\bar \chi} \Bigg\{  2   \left[   \tilde \p_{\perp m} \left(B^{m (1)} + n^l h_{l}^{m(1)} \right) - \frac{2}{\tilde \chi} \left( B^{(1)}_{\| } + h^{(1)}_{\|}\right)  \right]    \nonumber \\
 && \times \left( A^{(1)} +  \frac{1}{2}  h^{(1)}_{\| } - 2I^{(1)} \right) + 2  \left(B^{(1)i }_{\perp }+  n^j h_{ j}^{k(1)} \Perp^i_{k}  \right)   \bigg[\tilde \p_{\perp i}  \left( A^{(1)} + \frac{1}{2}  h^{(1)}_{\| } \right) - 2 \tilde \p_{\perp i} I^{(1)} \bigg]     \nonumber \\
 && + \bigg[ \tilde \nabla^2_{\perp} \left(A^{(1)} - B^{(1)}_{\| } - \frac{1}{2}h^{(1)}_{\| } \right) + \frac{1}{\tilde \chi}  \tilde \p_{\perp m} \left(B^{m (1)} + n^l h_{l}^{m(1)} \right) - \frac{2}{\tilde \chi^2 } \left( B^{(1)}_{\| } + h^{(1)}_{\|}\right)   \bigg] \delta x_{\|}^{(1)} \nonumber \\
 &&+ \bigg[ \p_{\perp i}  \left(A^{(1)} - B^{(1)}_{\| } - \frac{1}{2}h^{(1)}_{\| } \right) +  \frac{1}{\bar \chi} \left(B^{(1)}_{\perp i}+ n^k h_{k}^{j(1)} \Perp_{ij} \right) \bigg] \p^i_{\perp } \delta  x_{\|}^{(1)}  - \bigg[   \tilde \p_{\perp m} \left(B^{m (1)}{'} + n^l h_{l}^{m(1)}{'} \right)  \nonumber\\
 &&  - \frac{2}{\tilde \chi} \left( B^{(1)}_{\| }{'} + h^{(1)}_{\|}{'}\right)  \bigg] T^{(1)}  -  \left( B_{\perp i}^{(1)}{'} +  n^j h_{ j k}^{(1)}{'} \Perp_i^k \right) \tilde \p^i_{\perp }T^{(1)}  + \bigg[ - \frac{1}{\tilde \chi} \tilde \p_{\perp i}\left( B^{(1)}_{\| } + h^{(1)}_{\|}\right)   + \frac{1}{\tilde \chi^2}\left(B^{ (1)}_{\perp i}+ n^k h_{k}^{j(1)} \Perp_{ij} \right)    \nonumber\\
 &&  + \frac{2}{\tilde \chi^2} n^k h_{ki}^{(1)}   + \tilde \p_{\perp i} \tilde \p_{\perp m} \left(B^{m (1)} + n^k h_{k}^{m(1)} \right) - \frac{1}{\tilde \chi} \tilde \p_{\perp m} h_{i}^{m(1)} + \frac{1}{\tilde \chi} \tilde \p_{\perp i} \left(A^{(1)} - B^{(1)}_{\| } - \frac{1}{2}h^{(1)}_{\| }  \right)  \bigg] \delta x_{\perp}^{i (1)}  \nonumber\\
 &&  - \frac{2}{\tilde \chi} \left(A^{(1)} - B^{(1)}_{\| } - \frac{1}{2}h^{(1)}_{\| }  \right)  \kappa^{(1)} +  \frac{4}{\tilde \chi} \left(B^{(1)}_{\| } +  h^{(1)}_{\| } \right) \kappa^{(1)}  - 2 \tilde \p_{\perp m} \left(B^{m (1)} + n^k h_{k}^{m(1)} \right) \kappa^{(1)}  \nonumber\\
 &&  - 2 \left(B^{m (1)} + n^l h_{l}^{m(1)}\right)  \tilde \p_{\perp m} \kappa^{(1)} -   \bigg[ \p^j_{\perp } \left(B^{m (1)}  + n^k h_{k}^{m(1)} \right)  -  \frac{1}{\tilde\chi} h^{jm(1)} \bigg]  \left(\gamma_{mj}^{(1)}+  \Perp_{jm} \kappa^{(1)} + \theta_{jm}^{(1)}\right) \Bigg\}\;, \nonumber
    \end{eqnarray}
    
     \begin{eqnarray}
  &&  \kappa^{(2)}_{2 \rm post-Born}  = \int_0^{\bar \chi} \ud \tilde{\chi} ~ (\bar \chi - \tilde \chi) \frac{ \tilde \chi }{\bar \chi} \Bigg\{-\bigg[ \tilde \nabla^2_{\perp} \left(A^{(1)}{'}  - B^{(1)}_{\| }{'}  - \frac{1}{2}h^{(1)}_{\| }{'}  \right) + \frac{1}{\tilde \chi^2}  \tilde \p_{\perp i}   \left(B^{i (1)} + n^l h_{l}^{i(1)} \right) - \frac{2}{\tilde \chi^3} \left( B^{(1)}_{\| } + h^{(1)}_{\|}\right)  \nonumber\\
 && + \frac{1}{\tilde\chi} \tilde \p_{\perp i}   \left(B^{i (1)}{'} + n^l h_{l}^{i(1)}{'} \right) -   \frac{2}{\tilde \chi^2} \left( B^{(1)}_{\| }{'} + h^{(1)}_{\|}{'} \right) \bigg]  T^{(1)}   - \left( B_{\perp i}^{(1)}{'} +  n^j h_{ j k}^{(1)}{'} \Perp_i^k \right) \tilde \p^i_{\perp } \left(A^{(1)}-B^{(1)}_{\| }- \frac{1}{2} h^{(1)}_{\| }\right)   \nonumber\\
  &&  - \bigg[ \tilde \p_{\perp }^i    \left(A^{(1)}{'} - B^{(1)}_{\| }{'} - \frac{1}{2}h^{(1)}_{\| }{'} \right)   +  \frac{1}{\tilde\chi^2} \left(B^{i (1)}_{\perp }+ n^k h_{k}^{j(1)} \Perp^i_j \right)      +  \frac{1}{\tilde\chi}  \left( B_{\perp }^{i(1)}{'} +  n^j h_{ j }^{k(1)}{'} \Perp_k^i \right) \bigg] \tilde \p_{\perp i} T^{(1)}   \nonumber\\
&& - \bigg[ \tilde \p_{\perp m} \left(B^{m (1)}{'} + n^l h_{l}^{m(1)}{'} \right) - \frac{2}{\tilde \chi} \left( B^{(1)}_{\| }{'} + h^{(1)}_{\|}{'}\right) \bigg]\left(A^{(1)}-B^{(1)}_{\| }- \frac{1}{2} h^{(1)}_{\| }\right)  +\bigg[ \tilde \nabla^2_{\perp} \left(A^{(1)} - B^{(1)}_{\| } - \frac{1}{2}h^{(1)}_{\| } \right)  \nonumber\\
&&+ \frac{1}{\tilde \chi}  \tilde \p_{\perp m} \left(B^{m (1)} + n^l h_{l}^{m(1)} \right) - \frac{2}{\tilde \chi^2 } \left( B^{(1)}_{\| } + h^{(1)}_{\|}\right)   \bigg] \left( A^{(1)} +  \frac{1}{2}  h^{(1)}_{\| } - 2I^{(1)} \right) + \bigg[ \p_{\perp i}  \left(A^{(1)} - B^{(1)}_{\| } - \frac{1}{2}h^{(1)}_{\| } \right)  \nonumber\\
&&+  \frac{1}{\bar \chi} \left(B^{(1)}_{\perp i}+ n^k h_{k}^{j(1)} \Perp_{ij} \right) \bigg]\bigg[\tilde \p^i_{\perp }  \left( A^{(1)} + \frac{1}{2}  h^{(1)}_{\| } \right) - 2 \tilde \p^i_{\perp} I^{(1)} \bigg]  - \bigg[ \tilde \p_{\perp m} \left(B^{m (1)} + n^l h_{l}^{m(1)} \right) - \frac{2}{\tilde \chi} \left( B^{(1)}_{\| } + h^{(1)}_{\|}\right) \bigg]    \nonumber\\
&& \times  \left[\frac{\ud}{\ud \bar \chi}\left(A^{(1)} + \frac{1}{2}  h^{(1)}_{\| }\right)+\left(A^{(1)}{'} - B^{(1)}_{\| }{'} - \frac{1}{2}h^{(1)}_{\| }{'} \right) \right]  -  \left(B^{i (1)}_{\perp }+ n^k h_{k}^{j(1)} \Perp^i_j \right) \tilde \p_{\perp i} \bigg[\frac{\ud}{\ud \bar \chi}\left(A^{(1)} + \frac{1}{2}  h^{(1)}_{\| }\right) \nonumber\\
&& +\left(A^{(1)}{'} - B^{(1)}_{\| }{'} - \frac{1}{2}h^{(1)}_{\| }{'} \right) \bigg] +2  \frac{\ud}{\ud \tilde \chi}  \bigg[ \tilde \p_{\perp m} \left(B^{m (1)} + n^l h_{l}^{m(1)} \right) - \frac{2}{\tilde \chi} \left( B^{(1)}_{\| } + h^{(1)}_{\|}\right) \bigg]  \kappa^{(1)}  \nonumber\\
&&    +  \frac{2}{\tilde\chi}  \bigg[-     \frac{1}{\tilde\chi}  \left(A^{(1)} - B^{(1)}_{\| } - \frac{1}{2}h^{(1)}_{\| } \right) +  \frac{\ud}{\ud \tilde \chi}  \left(A^{(1)} - B^{(1)}_{\| } - \frac{1}{2}h^{(1)}_{\| } \right) \bigg]  \kappa^{(1)}   + 2 \bigg[  \frac{\ud}{\ud \tilde \chi}\left(B^{i (1)}_{\perp }+ n^k h_{k}^{j(1)} \Perp^i_j \right)     \nonumber\\
&&   -  \frac{1}{\tilde\chi} \left(B^{i (1)}_{\perp }+ n^k h_{k}^{j(1)} \Perp^i_j \right)  \bigg] \tilde \p_{\perp i} \kappa^{(1)}  -\bigg[- \tilde \p_{\perp i} \tilde \p_{\perp m} \left(B^{m (1)} + n^k h_{k}^{m(1)} \right) -\frac{1}{\tilde\chi} \tilde \p_{\perp i}  \left(A^{(1)} - B^{(1)}_{\| } - \frac{1}{2}h^{(1)}_{\| } \right) \nonumber\\
&& +   \frac{1}{\tilde\chi} \tilde \p_{\perp i} \left( B^{(1)}_{\| } + h^{(1)}_{\|}\right) +  \frac{1}{\tilde\chi^2}   \left(B^{(1)}_{\perp i}+ n^k h_{k}^{j(1)} \Perp_{ij} \right) + \frac{1}{\tilde\chi}\tilde \p_{\perp m} h_{i}^{m(1)}  \bigg] \left( B^{i (1)}_{\perp }+ n^k h_{k}^{j(1)} \Perp^i_j - 2S_{\perp}^{i(1)} \right)  \nonumber\\
&&     +  \bigg[ \tilde \p^j_{\perp } \left(B^{m (1)}  + n^k h_{k}^{m(1)} \right) -  \frac{1}{\tilde\chi} h^{jm(1)} \bigg]  \tilde \p_{\perp m} \left( B^{ (1)}_{j \perp }+ n^l h_{l}^{p(1)} \Perp_{pj}  - 2S_{\perp j}^{(1)} \right) - \left(B^{(1)}_{\perp i}+ n^k h_{k}^{j(1)} \Perp_{ij} \right) \nonumber\\
&& \times \bigg[  - \tilde \p^i_{\perp }  \tilde \p_{\perp m} \left(B^{m (1)} + n^l h_{l}^{m(1)} \right) + \frac{2}{\tilde \chi}\tilde \p^i_{\perp } \left( B^{(1)}_{\| } + h^{(1)}_{\|}\right)  + 2\tilde \p^i_{\perp }  \tilde \p_{\perp m}S_{\perp}^{m(1)}  \bigg] + \bigg[ - \tilde \p_{\perp m} \left(B^{m (1)} + n^l h_{l}^{m(1)} \right)  \nonumber\\
&& - \frac{1}{\tilde\chi}   \left(A^{(1)} - B^{(1)}_{\| } - \frac{1}{2}h^{(1)}_{\| } \right) + \frac{2}{\tilde \chi} \left( B^{(1)}_{\| } + h^{(1)}_{\|}\right) \bigg]  \bigg[ -  \tilde \p_{\perp m} \left(B^{m (1)} + n^l h_{l}^{m(1)} \right)  + \frac{2}{\tilde \chi} \left( B^{(1)}_{\| } + h^{(1)}_{\|}\right) + 2  \tilde \p_{\perp m}S_{\perp}^{m(1)}  \bigg]  \nonumber\\
&&  + \bigg[   +  \tilde \p_{\perp i} \tilde \nabla^2_{\perp} \left(A^{(1)} - B^{(1)}_{\| } - \frac{1}{2}h^{(1)}_{\| } \right)    +   \frac{1}{\tilde\chi} \tilde \p_{\perp i}    \left(A^{(1)}{'} - B^{(1)}_{\| }{'} - \frac{1}{2}h^{(1)}_{\| }{'} \right) +  \frac{2}{\tilde\chi^3} \left(B^{(1)}_{\perp i}+ n^k h_{k}^{j(1)} \Perp_{ij} \right)   \nonumber\\
&& - \frac{3}{\tilde \chi^2}  \tilde \p_{\perp i} \left( B^{(1)}_{\| } + h^{(1)}_{\|}\right)  + \frac{1}{\tilde\chi} \tilde \p_{\perp i} \tilde \p_{\perp m} \left(B^{m (1)} + n^k h_{k}^{m(1)} \right)   + \frac{2}{\tilde \chi^2}  \tilde \p_{\perp i}   \left(A^{(1)} - B^{(1)}_{\| } - \frac{1}{2}h^{(1)}_{\| } \right)   - \frac{1}{\tilde \chi^2}  \tilde \p_{\perp k}  h_{i}^{k(1)} \nonumber\\
&&   + \frac{1}{\tilde \chi}   \Perp^l_i \tilde \nabla^2_{\perp} \left(B_l^{(1)} + n^m h_{lm}^{(1)}\right)  + \frac{2}{\tilde\chi^3}  n^k h_{k}^{j(1)} \Perp_{ij}  \bigg] \delta x_{\perp}^{i (1)}  +   \bigg[- \tilde \p^{(j}_{\perp} \tilde \p^{m)}_{\perp} \left(A^{(1)} - B^{(1)}_{\| } - \frac{1}{2}h^{(1)}_{\| } \right) \nonumber\\
 && - \frac{1}{\tilde \chi}  \Perp^{jm} \left(A^{(1)}{'} - B^{(1)}_{\| }{'} - \frac{1}{2}h^{(1)}_{\| }{'} \right)  - \frac{1}{\tilde \chi^2}\Perp^{jm}  \left(A^{(1)} - B^{(1)}_{\| } - \frac{1}{2}h^{(1)}_{\| } \right)       - \frac{1}{\tilde \chi} \tilde \p^j_{\perp} \left(B^{m (1)}  + n^k h_{k}^{l(1)}  \Perp^m_{l} \right)
 \nonumber \\
&&  -\frac{1}{\tilde \chi} n^{[ j} \p^{m]}_\perp  \left(A^{(1)} - B^{(1)}_{\| } - \frac{1}{2}h^{(1)}_{\| } \right)  -  \frac{1}{\tilde \chi} \Perp^{kj} \tilde \p^m_{\perp}\left( B_k^{(1)} + n^l h_{kl}^{(1)}  \right) +   \frac{1}{\tilde \chi^2} \Perp^m_{l}   h^{lj(1)} \bigg] \left(\gamma_{mj}^{(1)}+  \Perp_{mj} \kappa^{(1)} + \theta_{mj}^{(1)}\right) \Bigg\}\nonumber   
  \end{eqnarray} 

and

   \begin{eqnarray}
 &&\kappa^{(2)}_{3 \rm post-Born}= - \left( A^{(1)}_o -  B^{(1)}_{\| o} - \frac{1}{2} h^{(1)}_{\| o}\right) \left(2v_{\| \, o}^{(1)} + \frac32 h^{(1)}_{\| \, o} - \frac12 h^{j (1)}_{j \, o} \right)   +  \left(2B^{(1)}_{\| o}+3  h^{(1)}_{\| \, o} - h^{j (1)}_{j \, o} \right)  \left(3 v_{\| \, o}^{(1)} + 2 h^{(1)}_{\| \, o}    - \frac{1}{2} h^{j (1)}_{j \, o} \right) \nonumber\\
 && - \left(   B^{i (1)}_{\perp \, o } +  n^k h_{k\,o}^{j(1)} \Perp^i_j \right) \left( 4 v^{ (1)}_{\perp i \, o }+ \frac{9}{2} n^l h_{l\,o}^{m(1)} \Perp_{mi} \right) + \left(B^{(1)}_{\| o} -v^{(1)}_{\| \, o} + \frac{1}{2}h^{(1)}_{\| \, o}\right)  \int_0^{\bar \chi} \ud \tilde{\chi} ~  \frac{(\bar \chi - \tilde \chi) }{\bar \chi}    \nonumber \\
 &&   \times   \left[ \tilde \p^m_{\perp } \left(B^{m (1)}  + n^k h_{k}^{m(1)} \right) -   \frac{1}{\tilde\chi}  \Perp_{mj}  h^{jm(1)} \right]   + 2 \left(- B^{(1)}_{\| o} + v_{\| \, o}^{(1)} - \frac34 h^{(1)}_{\| \, o} + \frac14 h^{m (1)}_{m \, o} \right)  \int_0^{\bar \chi} \ud \tilde{\chi} ~ \frac{(\bar \chi - \tilde \chi)  }{\bar \chi}    \nonumber\\ 
  &&  \times \bigg[ - \tilde \p_{\perp m} \left(B^{m (1)} + n^l h_{l}^{m(1)} \right) - \frac{1}{\tilde\chi}   \left(A^{(1)} - B^{(1)}_{\| } - \frac{1}{2}h^{(1)}_{\| } \right)  + \frac{2}{\tilde \chi} \left( B^{(1)}_{\| } + h^{(1)}_{\|}\right) \bigg]  + \left(A^{(1)}_o-v^{(1)}_{\| \, o}\right) \nonumber \\
 && \times  \Bigg\{ \bar \chi  \left[ \p_{\perp i} \left(B^{i (1)}  + n^k h_{k}^{i(1)} \right) -\frac{2}{\tilde \chi} \left( B^{(1)}_{\| } + h^{(1)}_{\|}\right)  \right]     -2    \int_0^{\bar \chi} \ud \tilde{\chi}  \frac{ \tilde \chi }{\bar \chi}  \left[   \tilde \p_{\perp m} \left(B^{m (1)} + n^l h_{l}^{m(1)} \right) - \frac{2}{\tilde \chi} \left( B^{(1)}_{\| } + h^{(1)}_{\|}\right)  \right]   \nonumber \\
     &&  -   \int_0^{\bar \chi} \ud \tilde{\chi} ~ (\bar \chi - \tilde \chi) \frac{ \tilde \chi }{\bar \chi}   \bigg[ \tilde \nabla^2_{\perp} \left(A^{(1)} - B^{(1)}_{\| } - \frac{1}{2}h^{(1)}_{\| } \right) + \frac{1}{\tilde \chi}  \tilde \p_{\perp m} \left(B^{m (1)} + n^l h_{l}^{m(1)} \right) - \frac{2}{\tilde \chi^2 } \left( B^{(1)}_{\| } + h^{(1)}_{\|}\right)   \bigg] \Bigg\} \nonumber\\
     &&+ v^{ (1)}_{\perp i \, o } \Bigg\{  -    \left(B^{(1)i }_{\perp }+  n^j h_{ j}^{k(1)} \Perp^i_{k}  \right)    + \frac{ 2}{\bar \chi}  \int_0^{\bar \chi} \ud \tilde{\chi}   \left(B^{(1)i }_{\perp }+  n^j h_{ j}^{k(1)} \Perp^i_{k}  \right)  \nonumber \\
      &&   +\frac{ 1 }{\bar \chi}   \int_0^{\bar \chi} \ud \tilde{\chi} ~ (\bar \chi - \tilde \chi) \bigg[ \p^i_{\perp }  \left(A^{(1)} - B^{(1)}_{\| } - \frac{1}{2}h^{(1)}_{\| } \right) +  \frac{1}{\bar \chi} \left(B^{i(1)}_{\perp }+ n^k h_{k}^{j(1)} \Perp^i_{j} \right) \bigg]  \Bigg\} \nonumber\\
 && +  \left(B^{i (1)}_{\perp \, o }-v^{i (1)}_{\perp \, o }+ \frac{1}{2} n^k h_{k\,o}^{j(1)} \Perp^i_j \right)  \int_0^{\bar \chi} \ud \tilde{\chi} ~ (\bar \chi - \tilde \chi) \frac{ \tilde \chi }{\bar \chi}  \bigg[- \tilde \p_{\perp i} \tilde \p_{\perp m} \left(B^{m (1)} + n^k h_{k}^{m(1)} \right)  \nonumber\\
 &&   -\frac{1}{\tilde\chi} \tilde \p_{\perp i}  \left(A^{(1)} - B^{(1)}_{\| } - \frac{1}{2}h^{(1)}_{\| } \right) +   \frac{1}{\tilde\chi} \tilde \p_{\perp i} \left( B^{(1)}_{\| } + h^{(1)}_{\|}\right)   +  \frac{1}{\tilde\chi^2}   \left(B^{(1)}_{\perp i}+ n^k h_{k}^{j(1)} \Perp_{ij} \right) + \frac{1}{\tilde\chi}\tilde \p_{\perp m} h_{i}^{m(1)}  \bigg] \bigg]    \nonumber\\
  &&   -2\left(B^{i (1)}_{\perp \, o } - v^{i (1)}_{\perp \, o } - \frac32 n^k h_{k\,o}^{m(1)} \Perp^i_m  \right)  \int_0^{\bar \chi} \ud \tilde{\chi} ~\frac{ (\bar \chi - \tilde \chi)  }{\bar \chi \tilde \chi} \left(B^{(1)}_{\perp i}+ n^k h_{k}^{j(1)} \Perp_{ij} \right)    \nonumber \\
 &&  - \frac{1}{2}\Perp^k_m \Perp^l_{j} h_{lk\,o}^{(1)}   \int_0^{\bar \chi} \ud \tilde{\chi} ~  \frac{ (\bar \chi - \tilde \chi) }{\bar \chi} \left[ \tilde \p^j_{\perp } \left(B^{m (1)}  + n^k h_{k}^{m(1)} \right) -  \frac{1}{\tilde\chi} h^{jm(1)} \right]\;.   \nonumber
 \end{eqnarray}

    Finally, in order to obtain $\Delta_g^{(2)}$ [see Eq.\ (\ref{Deltag-2})],  we need the following 

\begin{equation}
\hat g_\mu^{\mu (2)} - \hat g_\mu^{\nu (1)}  \; \hat g_\nu^{\mu (1)}=2A^{(2)}+ h_i^{i (2)}  - 4\left(A^{(1)}\right)^2+2   B^{(1)}_i B^{i (1)}-  h_{i}^{k(1)} h_{k}^{i(1)}\;,
\end{equation}

\begin{equation}
  E_{\hat 0}^{0 (2)}   + E_{\hat 0 }^{\| (2)}  = - A^{(2)}+ v_{\|}^{(2)}+ 3 \left(A^{(1)}\right)^2+   v^{(1)}_i v^{i (1)}- 2  v^{(1)}_i B^{i (1)}\;,
\end{equation}

\begin{eqnarray}
 \frac{1}{ \cH} {\hat g_\mu^{\mu(1)}}{'} \Delta \ln a^{(1)} +   \left(\p_{\parallel}\hat g_\mu^{\mu (1)}\right) \Delta x_{\parallel}^{(1)} &=&- \frac{1}{ \cH} \Delta \ln a^{(1)}   \frac{\ud \,}{\ud \bar \chi}\left( \hat g_\mu^{\mu (1)} \right)-   \left(\p_{\parallel}\hat g_\mu^{\mu (1)}\right)  T^{(1)} \nonumber \\
 &=&- \frac{1}{ \cH} \frac{\ud \,}{\ud  \bar \chi}\left(2A^{(1)} + h_i^{i (1)}  \right)\Delta \ln a^{(1)} - \p_{\parallel}\left(2A^{(1)} + h_i^{i (1)}  \right) T^{(1)}\;,
\end{eqnarray}

\begin{eqnarray}
 \left(\p_{\perp i}\hat g_\mu^{\mu (1)}\right) \Delta x_{\perp}^{i (1)} &=& \p_{\perp i}\left(2A^{(1)} + h_i^{i (1)}  \right) \delta x_{\perp}^{i (1)} = \bar \chi \left(B^{i (1)}_{\perp \, o }-v^{i (1)}_{\perp \, o }+ \frac{1}{2} n^k h_{k\,o}^{j(1)} \Perp^i_j\right) \p_{\perp i}\left(2A^{(1)} + h_i^{i (1)}  \right)  \nonumber \\
&-& \p_{\perp i}\left(2A^{(1)} + h_i^{i (1)}  \right)\int_0^{\bar \chi} \ud \tilde \chi \left\{ \left( B^{i (1)}_{\perp }+ n^k h_{k}^{j(1)} \Perp^i_j\right) + \left(\bar \chi-\tilde \chi\right)  \right.  \nonumber \\
  &\times&  \left. \left[ \tilde \p^i_\perp \left( A^{(1)} - B^{(1)}_{\| } - \frac{1}{2}h^{(1)}_{\| } \right) + \frac{1}{\tilde \chi} \left(B^{i (1)}_{\perp }+  n^k h_{kj}^{(1)} \Perp^{ij}  \right)\right]\right\} \;,
\end{eqnarray}

\begin{equation}
 \frac{1}{\cH}  \delta_g^{(1)}{'} \Delta \ln a^{(1)}   +   \left(\p_{\parallel}\delta_g^{(1)}\right) \Delta x_{\parallel}^{(1)} =- \frac{1}{\cH}  \frac{\ud \,}{\ud  \bar \chi} \left(\delta_g^{(1)}\right) \Delta \ln a^{(1)}- \left(\p_{\parallel}\delta_g^{(1)}\right) T^{(1)}\;,
\end{equation}

\begin{eqnarray}
  \Delta x_{\perp i}^{(1)} \,  \p_{\perp}^{i}  \delta_g^{(1)}  &=& \delta x_{\perp i}^{(1)} \, \p_{\perp}^{i} \delta_g^{(1)}  = \bar \chi \left(B^{i (1)}_{\perp \, o }-v^{i (1)}_{\perp \, o }+ \frac{1}{2} n^k h_{k\,o}^{j(1)} \Perp^i_j\right) \p_{\perp i}  \delta_g^{(1)} - \left( \p_{\perp i} \delta_g^{(1)} \right) \int_0^{\bar \chi} \ud \tilde \chi \bigg\{ \left( B^{i (1)}_{\perp }+ n^k h_{k}^{j(1)} \Perp^i_j\right) \nonumber \\
&+& \left(\bar \chi-\tilde \chi\right) \left[ \tilde \p^i_\perp \left( A^{(1)} - B^{(1)}_{\| } - \frac{1}{2}h^{(1)}_{\| } \right) + \frac{1}{\tilde \chi} \left(B^{i (1)}_{\perp }+  n^k h_{kj}^{(1)} \Perp^{ij}  \right)\right]\bigg\} \;,  
\end{eqnarray}

\begin{equation}
- \frac{1}{\bar \chi^2} \left( \Delta x_{\parallel}^{(1)} \right)^2 =- \frac{1}{\bar \chi^2} \left( T^{(1)} \right)^2- \frac{1}{\bar \chi^2 \cH^2} \left( \Delta \ln a^{(1)} \right)^2- \frac{2}{\bar \chi^2 \cH}  \Delta \ln a^{(1)} T^{(1)}\;,
\end{equation}

\begin{equation}
 \frac{2}{\bar \chi} \Delta x_{\parallel}^{(1)}  \kappa^{(1)}  =- \frac{2}{\bar \chi \cH}  \Delta \ln a^{(1)}   \kappa^{(1)} -  \frac{2}{\bar \chi}  T^{(1)}  \kappa^{(1)}\;,
\end{equation}

\begin{eqnarray}
 &-&  \frac{1}{2} \left( \p_{\perp j}   \Delta x_{\perp}^{i (1)}  \right)  \left( \p_{\perp i}   \Delta x_{\perp}^{j (1)}  \right)= -\left(v^{(1)}_{\| \, o}\right)^2- \left(A^{(1)}_{\, o}\right)^2-\frac{1}{2} v^{(1)}_{\| \, o}h_{i \, o}^{i (1)}+ \frac{1}{2} v^{(1)}_{\| \, o} h^{(1)}_{\| \, o} -  \frac{1}{8} \Perp^m_p \Perp^k_n h_{m \, o}^{n (1)}  h_{k \, o}^{p(1)} \nonumber \\
 &+& 2 A^{(1)}_{\, o}v^{(1)}_{\| \, o}+ \frac{1}{2}A^{(1)}_{\, o}h_{i \, o}^{i (1)}-  \frac{1}{2}A^{(1)}_{\, o} h^{(1)}_{\| \, o} + \left(2A^{(1)}_{\, o} - 2  v^{(1)}_{\| \, o} -\frac{1}{2}  h_{i \, o}^{i (1)} +  \frac{1}{2}h^{(1)}_{\| \, o} \right)\left(A^{(1)} - B^{(1)}_{\| } - \frac{1}{2}h^{(1)}_{\| }  - 2 I^{(1)} \right)\nonumber \\
 &+& \left( \Perp^i_j A^{(1)}_{\, o} -  \Perp^i_j   v^{(1)}_{\| \, o}  - \frac{1}{2} \Perp^k_j \Perp^i_p h_{k \, o}^{p (1)} \right) \int_0^{\bar \chi} \ud \tilde \chi\bigg\{ \frac{1}{\tilde \chi} \Perp^j_i  B^{(1)}_{\| }  -\Perp^j_m \tilde \p_{ \perp i} B^{m (1)}-  \frac{1}{\tilde \chi}  \Perp^n_i \Perp^j_m   h_n^{m (1)} + \frac{1}{\tilde \chi}  \Perp^j_i h^{(1)}_{\| }   - n^m \Perp^j_n \tilde \p_{ \perp i}  h_m^{n (1)}\nonumber \\
 & -&  \frac{\tilde \chi}{ \bar \chi} \left[ \Perp^j_i \tilde \p_\| + \left(\bar \chi-\tilde \chi\right) \Perp^n_i \Perp^{jm}\tilde  \p_m \tilde \p_n \right] \left( A^{(1)} - B^{(1)}_{\| } - \frac{1}{2}h^{(1)}_{\| } \right) \bigg\} - \left(A^{(1)} - B^{(1)}_{\| } - \frac{1}{2}h^{(1)}_{\| }  - 2 I^{(1)} \right)^2 \nonumber \\
& -& \left(A^{(1)} - B^{(1)}_{\| } - \frac{1}{2}h^{(1)}_{\| }  - 2 I^{(1)} \right)  \int_0^{\bar \chi} \ud \tilde \chi\bigg\{\frac{2}{\tilde \chi} B^{(1)}_{\| }  - \tilde \p_{ \perp m} B^{m (1)}- \frac{1}{\tilde \chi}  h_i^{i (1)} + \frac{3}{\tilde \chi}   h^{(1)}_{\| }  - n^m  \tilde \p_{ \perp n}  h_m^{n (1)}\nonumber \\
 & -&  \frac{\tilde \chi}{ \bar \chi} \left[ 2 \tilde \p_\| + \left(\bar \chi-\tilde \chi\right)  \Perp^{mn} \tilde \p_m  \tilde \p_n \right] \left( A^{(1)} - B^{(1)}_{\| } - \frac{1}{2}h^{(1)}_{\| } \right) \bigg\} - \frac{1}{2} \int_0^{\bar \chi} \ud \tilde \chi\bigg\{   \frac{1}{\tilde \chi}  \Perp^i_j  B^{(1)}_{\| }  -\Perp^i_p \tilde \p_{ \perp j} B^{p (1)}- \frac{1}{\tilde \chi}   \Perp^p_j \Perp^i_q   h_p^{q (1)}\nonumber \\
 & +&  \frac{1}{\tilde \chi}  \Perp_j^i h^{(1)}_{\| }   - n^p \Perp^i_q \tilde \p_{ \perp j}  h_p^{q (1)}-  \frac{\tilde \chi}{ \bar \chi} \left[ \Perp_j^i \tilde \p_\| + \left(\bar \chi-\tilde \chi\right) \Perp^p_j \Perp^{iq} \tilde \p_q  \tilde \p_p \right] \left( A^{(1)} - B^{(1)}_{\| } - \frac{1}{2}h^{(1)}_{\| } \right) \bigg\}  \int_0^{\bar \chi} \ud \tilde \chi\bigg\{  \frac{1}{\tilde \chi}   \Perp^j_i  B^{(1)}_{\| }  \nonumber \\
 &-&\Perp^j_m \tilde \p_{ \perp i} B^{m (1)}- \frac{1}{\tilde \chi}  \Perp^n_i \Perp^j_m   h_n^{m (1)} +  \frac{1}{\tilde \chi} \Perp^j_i h^{(1)}_{\| }   - n^m \Perp^j_n \tilde \p_{ \perp i}  h_m^{n (1)}-  \frac{\tilde \chi}{ \bar \chi} \left[ \Perp^j_i \tilde \p_\| + \left(\bar \chi-\tilde \chi\right) \Perp^n_i \Perp^{jm}  \tilde \p_m   \tilde \p_n \right]  \nonumber \\
 &\times& \left( A^{(1)} - B^{(1)}_{\| } - \frac{1}{2}h^{(1)}_{\| } \right) \bigg\} \;,
   \end{eqnarray}

\begin{eqnarray}
 &&\left(\frac{1}{\bar \chi}  \Delta x_{\perp i}^{(1)}  -\p_{\perp i}   \Delta x_{\parallel}^{(1)}  \right)   \p_{\parallel}   \Delta x_{\perp}^{i (1)}=  \left(\frac{1}{\bar \chi}  \delta x_{\perp i}^{(1)}  -\p_{\perp i}   \Delta x_{\parallel}^{(1)}  \right)  \delta n_{\perp}^{i (1)}  
=  \left(B^{(1)}_{\perp i \, o }-v^{ (1)}_{\perp i \, o }+ \frac{1}{2} n^k h_{k\,o}^{j(1)} \Perp_{ij} \right)  \nonumber \\
&& \times \left(B^{i (1)}_{\perp \, o }-v^{i (1)}_{\perp \, o }+ \frac{1}{2} n^k h_{k\,o}^{j(1)} \Perp^i_j\right) +  \left(B^{(1)}_{\perp i \, o }-v^{ (1)}_{\perp i \, o }+ \frac{1}{2} n^k h_{k\,o}^{j(1)} \Perp_{ij} \right) \bigg\{-\left( B^{i (1)}_{\perp }+ n^k h_{k}^{j(1)} \Perp^i_j\right) + 2S_{\perp}^{i(1)} \nonumber \\
&&+ \p_{\perp i}   \left(\frac{1}{\cH}  \Delta \ln a^{(1)}+   T^{(1)}  \right) 
  - \frac{1}{\bar \chi} \int_0^{\bar \chi} \ud \tilde \chi \left[ \frac{ \bar \chi}{\tilde \chi}\left( B^{i (1)}_{\perp }+ n^k h_{k}^{j(1)} \Perp^i_j\right) + \left(\bar \chi-\tilde \chi\right) \tilde \p^i_\perp \left( A^{(1)} - B^{(1)}_{\| } - \frac{1}{2}h^{(1)}_{\| } \right)\right]\bigg\}\nonumber \\
 &&+ \left[-\left( B^{i (1)}_{\perp }+ n^k h_{k}^{j(1)} \Perp^i_j\right) + 2S_{\perp}^{i(1)} \right]  \bigg\{ - \frac{1}{\bar \chi}\int_0^{\bar \chi} \ud \tilde \chi \left[ \frac{ \bar \chi}{\tilde \chi}\left( B^{i (1)}_{\perp }+ n^k h_{k}^{j(1)} \Perp^i_j\right) + \left(\bar \chi-\tilde \chi\right) \tilde \p^i_\perp \left( A^{(1)} - B^{(1)}_{\| } - \frac{1}{2}h^{(1)}_{\| } \right)\right] \nonumber \\
&&+\p_{\perp i}   \left(\frac{1}{\cH}  \Delta \ln a^{(1)}+   T^{(1)}  \right) \bigg\}\;,
\end{eqnarray}

\begin{eqnarray}
&&\frac{1}{\cH} {\left( E_{\hat 0}^{0(1)}+ E_{\hat 0}^{\|(1)} \right)}' \Delta \ln a^{(1)}  +  \p_{\parallel} \left( E_{\hat 0}^{0(1)}+ E_{\hat 0}^{\|(1)} \right) \Delta x_{\parallel}^{(1)}= \frac{1}{\cH} \frac{\ud \,}{\ud \bar \chi}\left(A^{(1)}- v^{(1)}_{\| }  \right) \Delta \ln a^{(1)} +  \p_{\parallel} \left( A^{(1)}- v^{(1)}_{\| } \right)  T^{(1)} \nonumber \\
\end{eqnarray}

\begin{eqnarray}
&& \p_{\perp i}\left( E_{\hat 0}^{0(1)}+ E_{\hat 0}^{\|(1)} \right)\Delta x_{\perp}^{i (1)}=- \bar \chi \left(B^{i (1)}_{\perp \, o }-v^{i (1)}_{\perp \, o }+ \frac{1}{2} n^k h_{k\,o}^{j(1)} \Perp^i_j\right) \p_{\perp i}\left( A^{(1)}- v^{(1)}_{\| } \right) + \p_{\perp i}\left( A^{(1)}- v^{(1)}_{\| } \right) \nonumber \\
&& \times \int_0^{\bar \chi} \ud \tilde \chi \left\{ \left( B^{i (1)}_{\perp }+ n^k h_{k}^{j(1)} \Perp^i_j\right) + \left(\bar \chi-\tilde \chi\right)\left[\tilde \p^i_\perp \left( A^{(1)} - B^{(1)}_{\| } - \frac{1}{2}h^{(1)}_{\| } \right) + \frac{1}{\tilde \chi} \left(B^{i (1)}_{\perp }+  n^k h_{kj}^{(1)} \Perp^{ij}  \right)\right] \right\}\;,  \end{eqnarray}
 
 \begin{equation}
 -   \left(\delta n_\|^{ (1)} +  \delta\nu^{(1)} \right)  E_{\hat 0}^{\| (1)}=v^{(1)}_{\| }   \left(\frac{\ud T}{\ud \bar \chi}\right)^{(1)} =-v^{(1)}_{\| }  \left( A^{(1)}-B^{(1)}_{\| }- \frac{1}{2} h^{(1)}_{\| }\right)\;,
 \end{equation}
 
 \begin{equation}
  -  E_{\hat 0 }^{\perp i (1)} \p_{\perp i} \left(  \Delta x^{0(1)}+ \Delta x_{\parallel}^{(1)} \right)=-  E_{\hat 0 }^{\perp i (1)} \p_{\perp i} \left(  \delta x^{0(1)}+ \delta x_{\parallel}^{(1)} \right)=v^{i (1)}_{\perp } \p_{\perp i} T^{(1)}\;,
  \end{equation}
  
   \begin{eqnarray}
&&  -\frac{1}{2} \left(\p_\|  \Delta x_{\parallel}^{(1)} \right)^2 = -\frac{1}{2} \left( A^{(1)} - B^{(1)}_{\| }- \frac{1}{2}h^{(1)}_{\| }\right)^2  -\frac{1}{2\cH^2} \left[\frac{\ud \,}{\ud \bar \chi} \left( A^{(1)}-  v_\|^{(1)}\right)+ \left(A^{(1)}{'} - B^{(1)}_{\| }{'} - \frac{1}{2}h^{(1)}_{\| }{'} \right)\right]^2 \nonumber\\
 &&-\frac{1}{\cH}\left(A^{(1)} - B^{(1)}_{\| }- \frac{1}{2}h^{(1)}_{\| }\right)\left[\frac{\ud \,}{\ud \bar \chi} \left( A^{(1)}-  v_\|^{(1)}\right)+ \left(A^{(1)}{'} - B^{(1)}_{\| }{'} - \frac{1}{2}h^{(1)}_{\| }{'} \right)\right]
 + \frac{\cH'}{\cH^2} \left(A^{(1)} - B^{(1)}_{\| }- \frac{1}{2}h^{(1)}_{\| }\right)\Delta \ln a^{(1)} \nonumber\\
 && +\frac{\cH'}{\cH^3} \left[\frac{\ud \,}{\ud \bar \chi} \left( A^{(1)}-  v_\|^{(1)}\right)+ \left(A^{(1)}{'} - B^{(1)}_{\| }{'} - \frac{1}{2}h^{(1)}_{\| }{'} \right)\right]\Delta \ln a^{(1)} -\frac{1}{2} \left(\frac{\cH'}{\cH^2} \right)^2\left(\Delta \ln a^{(1)} \right)^2 \;,
     \end{eqnarray}
  
  \begin{equation}
 -\frac{1}{2}  \left(   E_{\hat 0}^{0(1)}+ E_{\hat 0}^{\|(1)} \right)^2=- \frac{1}{2} \left( A^{(1)}-  v_\|^{(1)}  \right)^2\;.
  \end{equation}
  
Putting together the above relations we obtain finally
  \begin{eqnarray}
&&  \Delta_g^{(2)}   =  \delta_g^{(2)}+   b_e \, \Delta \ln a^{(2)} +  \p_{\parallel} \Delta x_{\parallel}^{(2)}  + \frac{2}{\bar \chi} \Delta x_{\parallel}^{(2)} - 2\kappa^{(2)}  + A^{(2)}+ v_{\|}^{(2)}+ \frac{1}{2} h_i^{i (2)} +\left(\Delta_g^{(1)} \right)^2 - 3 \left(A^{(1)}\right)^2+6A^{(1)}B^{(1)}_{\| }\nonumber \\
 &&+3A^{(1)}h^{(1)}_{\| } -3B^{(1)}_{\| }h^{(1)}_{\| }+   v^{(1)}_{\perp i} v^{i (1)}_{\perp}- 2  v^{(1)}_{\perp i} B^{i (1)}_{\perp}  -2\left(B^{(1)}_{\| } \right)^2 +  B^{(1)}_{\perp i} B^{i (1)}_{\perp} - \frac{1}{2}h_{i}^{k(1)} h_{k}^{i(1)}- \frac{3}{4}\left(h^{(1)}_{\| } \right)^2-8\left( I^{(1)} \right)^2 \nonumber \\
 &&+8A^{(1)} I^{(1)}-8B^{(1)}_{\| } I^{(1)}-4h^{(1)}_{\| } I^{(1)}- \left(\delta_g^{(1)}\right)^2  + v^{(1)}_{\| } h^{(1)}_{\| } -\frac{1}{\cH^2} \left[\frac{\ud \,}{\ud \bar \chi} \left( A^{(1)}-  v_\|^{(1)}\right)+ \left(A^{(1)}{'} - B^{(1)}_{\| }{'} - \frac{1}{2}h^{(1)}_{\| }{'} \right)\right]^2 \nonumber \\
 &&-\frac{2}{\cH}\left(A^{(1)} - B^{(1)}_{\| }- \frac{1}{2}h^{(1)}_{\| }\right)\left[\frac{\ud \,}{\ud \bar \chi} \left( A^{(1)}-  v_\|^{(1)}\right)+ \left(A^{(1)}{'} - B^{(1)}_{\| }{'} - \frac{1}{2}h^{(1)}_{\| }{'} \right)\right]- \frac{1}{ \cH} \frac{\ud \,}{\ud \bar \chi}\left(2 v^{(1)}_{\| } + h_i^{i (1)}+ 2\delta_g^{(1)} \right)\Delta \ln a^{(1)} \nonumber \\
 && - \p_{\parallel}\left(2 v^{(1)}_{\| } + h_i^{i (1)} + 2\delta_g^{(1)} \right) T^{(1)}- \frac{4}{\bar \chi^2 \cH}  \Delta \ln a^{(1)} T^{(1)}+ 2v^{i (1)}_{\perp } \p_{\perp i} T^{(1)} - \frac{4}{\bar \chi \cH}  \Delta \ln a^{(1)}   \kappa^{(1)} -  \frac{4}{\bar \chi}  T^{(1)}  \kappa^{(1)}  \nonumber \\
  && + \left[- b_e +  \frac{\ud \ln b_e }{\ud  \ln \bar a} - \left(\frac{\cH'}{\cH^2} \right)^2- \frac{2}{\bar \chi^2 \cH^2}\right] \left( \Delta \ln a^{(1)}\right)^2- \frac{2}{\bar \chi^2} \left( T^{(1)} \right)^2  + 2\frac{\cH'}{\cH^2} \left(A^{(1)} - B^{(1)}_{\| }- \frac{1}{2}h^{(1)}_{\| }\right)\Delta \ln a^{(1)}\nonumber \\
   && +2\frac{\cH'}{\cH^3} \left[\frac{\ud \,}{\ud \bar \chi} \left( A^{(1)}-  v_\|^{(1)}\right)+ \left(A^{(1)}{'} - B^{(1)}_{\| }{'} - \frac{1}{2}h^{(1)}_{\| }{'} \right)\right]\Delta \ln a^{(1)}+2 \left[-\left( B^{i (1)}_{\perp }+ n^k h_{k}^{j(1)} \Perp^i_j\right) + 2S_{\perp}^{i(1)} \right] \nonumber\\
     && \times \p_{\perp i}   \left(\frac{1}{\cH}  \Delta \ln a^{(1)}+   T^{(1)}  \right) - \left[-\frac{2}{\bar \chi}\left( B^{i (1)}_{\perp }+ n^k h_{k}^{j(1)} \Perp^i_j\right)  + \frac{4}{\bar \chi} S_{\perp}^{i(1)}+ \p^i_{\perp}\left( 2v^{(1)}_{\| } + h_l^{l (1)}  + 2\delta_g^{(1)}\right)  \right]  \nonumber \\
 &&\times  \int_0^{\bar \chi} \ud \tilde \chi \bigg[ \frac{ \bar \chi}{\tilde \chi}\left( B^{(1)}_{\perp i}+ n^k h_{k}^{j(1)} \Perp_{ij}\right) + \left(\bar \chi-\tilde \chi\right) \tilde \p_{\perp i}  \left( A^{(1)} - B^{(1)}_{\| } - \frac{1}{2}h^{(1)}_{\| } \right)\bigg]\nonumber \\
& &- 2\left(A^{(1)} - B^{(1)}_{\| } - \frac{1}{2}h^{(1)}_{\| }  - 2 I^{(1)} \right)  \int_0^{\bar \chi} \ud \tilde \chi\bigg\{\frac{2}{\tilde \chi} B^{(1)}_{\| }  - \tilde \p_{ \perp m} B^{m (1)}- \frac{1}{\tilde \chi}  h_i^{i (1)} + \frac{3}{\tilde \chi}   h^{(1)}_{\| }  - n^m  \tilde \p_{ \perp n}  h_m^{n (1)}\nonumber \\
 & &-  \frac{\tilde \chi}{ \bar \chi} \left[ 2 \tilde \p_\| + \left(\bar \chi-\tilde \chi\right)  \Perp^{mn} \tilde \p_m  \tilde \p_n \right] \left( A^{(1)} - B^{(1)}_{\| } - \frac{1}{2}h^{(1)}_{\| } \right) \bigg\} - \int_0^{\bar \chi} \ud \tilde \chi\bigg\{   \frac{1}{\tilde \chi}  \Perp^i_j  B^{(1)}_{\| }  -\Perp^i_p \tilde \p_{ \perp j} B^{p (1)}- \frac{1}{\tilde \chi}   \Perp^p_j \Perp^i_q   h_p^{q (1)}+ \frac{1}{\tilde \chi}  \Perp_j^i h^{(1)}_{\| }  \nonumber \\
 &&  - n^p \Perp^i_q \tilde \p_{ \perp j}  h_p^{q (1)}-  \frac{\tilde \chi}{ \bar \chi} \left[ \Perp_j^i \tilde \p_\| + \left(\bar \chi-\tilde \chi\right) \Perp^p_j \Perp^{iq} \tilde \p_q  \tilde \p_p \right] \left( A^{(1)} - B^{(1)}_{\| } - \frac{1}{2}h^{(1)}_{\| } \right) \bigg\} \int_0^{\bar \chi} \ud \tilde \chi\bigg\{  \frac{1}{\tilde \chi}   \Perp^j_i  B^{(1)}_{\| } -\Perp^j_m \tilde \p_{ \perp i} B^{m (1)} \nonumber \\
  &&+  \frac{1}{\tilde \chi} \Perp^j_i h^{(1)}_{\| }- \frac{1}{\tilde \chi}  \Perp^n_i \Perp^j_m   h_n^{m (1)}    - n^m \Perp^j_n \tilde \p_{ \perp i}  h_m^{n (1)} -  \frac{\tilde \chi}{ \bar \chi} \left[ \Perp^j_i \tilde \p_\| + \left(\bar \chi-\tilde \chi\right) \Perp^n_i \Perp^{jm}  \tilde \p_m   \tilde \p_n \right] \left( A^{(1)} - B^{(1)}_{\| } - \frac{1}{2}h^{(1)}_{\| } \right) \bigg\} -2\left(v^{(1)}_{\| \, o}\right)^2\nonumber \\
 && +  v^{(1)}_{\| \, o} h^{(1)}_{\| \, o} +4 A^{(1)}_{\, o}v^{(1)}_{\| \, o}+ A^{(1)}_{\, o}h_{i \, o}^{i (1)}-  A^{(1)}_{\, o} h^{(1)}_{\| \, o} + 2\left(B^{(1)}_{\perp i \, o }-v^{ (1)}_{\perp i \, o }+ \frac{1}{2} n^k h_{k\,o}^{j(1)} \Perp_{ij} \right) \left(B^{i (1)}_{\perp \, o }-v^{i (1)}_{\perp \, o }+ \frac{1}{2} n^k h_{k\,o}^{j(1)} \Perp^i_j\right)\nonumber \\
 && -2 \left(A^{(1)}_{\, o}\right)^2 - v^{(1)}_{\| \, o}h_{i \, o}^{i (1)}  -  \frac{1}{4} \Perp^m_p \Perp^k_n h_{m \, o}^{n (1)}  h_{k \, o}^{p(1)}+ 2\left(2A^{(1)}_{\, o} - 2  v^{(1)}_{\| \, o} -\frac{1}{2}  h_{i \, o}^{i (1)} +  \frac{1}{2}h^{(1)}_{\| \, o} \right)\left(A^{(1)} - B^{(1)}_{\| } - \frac{1}{2}h^{(1)}_{\| }  - 2 I^{(1)} \right)\nonumber \\
 &&+2 \left( \Perp^i_j A^{(1)}_{\, o} -  \Perp^i_j   v^{(1)}_{\| \, o}  - \frac{1}{2} \Perp^k_j \Perp^i_p h_{k \, o}^{p (1)} \right) \int_0^{\bar \chi} \ud \tilde \chi\bigg\{ \frac{1}{\tilde \chi} \Perp^j_i  B^{(1)}_{\| }  -\Perp^j_m \tilde \p_{ \perp i} B^{m (1)}-  \frac{1}{\tilde \chi}  \Perp^n_i \Perp^j_m   h_n^{m (1)} + \frac{1}{\tilde \chi}  \Perp^j_i h^{(1)}_{\| } \nonumber \\
 & &  - n^m \Perp^j_n \tilde \p_{ \perp i}  h_m^{n (1)}-  \frac{\tilde \chi}{ \bar \chi} \left[ \Perp^j_i \tilde \p_\| + \left(\bar \chi-\tilde \chi\right) \Perp^n_i \Perp^{jm}\tilde  \p_m \tilde \p_n \right] \left( A^{(1)} - B^{(1)}_{\| } - \frac{1}{2}h^{(1)}_{\| } \right) \bigg\} \nonumber \\
&& +  \left(B^{(1)}_{\perp i \, o }-v^{ (1)}_{\perp i \, o }+ \frac{1}{2} n^k h_{k\,o}^{j(1)} \Perp_{ij} \right) \bigg\{-2\left( B^{i (1)}_{\perp }+ n^m h_m^{l(1)} \Perp^i_l\right) + 4S_{\perp}^{i(1)} + \bar \chi \p^i_{\perp}\left(2v^{(1)}_{\| } + h_l^{l (1)} + 2  \delta_g^{(1)} \right)\nonumber \\
&&+ 2 \p_{\perp i}   \left(\frac{1}{\cH}  \Delta \ln a^{(1)}+   T^{(1)}  \right)  -2 \frac{1}{\bar \chi} \int_0^{\bar \chi} \ud \tilde \chi \left[ \frac{ \bar \chi}{\tilde \chi}\left( B^{i (1)}_{\perp }+ n^k h_{k}^{j(1)} \Perp^i_j\right) + \left(\bar \chi-\tilde \chi\right) \tilde \p^i_\perp \left( A^{(1)} - B^{(1)}_{\| } - \frac{1}{2}h^{(1)}_{\| } \right)\right]\bigg\}\;, \label{maindel}
 \end{eqnarray}
where for $\Delta \ln a^{(2)}/2$ see Eq.\ (\ref{Deltalna-2}), for $\Delta x_{\parallel}^{(2)}$ see Eq.\ (\ref{Dx_||-2_2}),   for $\p_{\parallel} \Delta x_{\parallel}^{(2)}/2$ see Eq.\ (\ref{dDx_||-2_3}) and  for $\kappa^{(2)}$ see Eq.\ (\ref{kappa-2}).

Equation \eqref{maindel} is the main result of this paper -- giving the number counts at second order in a general gauge, valid for any dark energy model and also in metric theories of modified gravity.

We can simplify the result by explicitly introducing the weak lensing shear and rotation, defined in Eqs. (\ref{volumedistortion1})--(\ref{shear}). This leads to
   \begin{eqnarray}
&&  \Delta_g^{(2)}   =  \delta_g^{(2)}+   b_e \, \Delta \ln a^{(2)} +  \p_{\parallel} \Delta x_{\parallel}^{(2)}  + \frac{2}{\bar \chi} \Delta x_{\parallel}^{(2)} - 2\kappa^{(2)}  + A^{(2)}+ v_{\|}^{(2)}+ \frac{1}{2} h_i^{i (2)} +\left(\Delta_g^{(1)} \right)^2 -  \left(A^{(1)}\right)^2+2A^{(1)}B^{(1)}_{\| }\nonumber \\
 &&+A^{(1)}h^{(1)}_{\| } -B^{(1)}_{\| }h^{(1)}_{\| }+   v^{(1)}_{\perp i} v^{i (1)}_{\perp}- 2  v^{(1)}_{\perp i} B^{i (1)}_{\perp}   +  B^{(1)}_{\perp i} B^{i (1)}_{\perp} - \frac{1}{2}h_{i}^{k(1)} h_{k}^{i(1)}- \frac{1}{4}\left(h^{(1)}_{\| } \right)^2- \left(\delta_g^{(1)}\right)^2  + v^{(1)}_{\| } h^{(1)}_{\| }\nonumber \\
 && -2\big|\gamma^{(1)}\big|^2-2\left(\kappa^{(1)}\right)^2+\vartheta_{ij}^{(1)}\vartheta^{ij(1)}  -\frac{1}{\cH^2} \left[\frac{\ud \,}{\ud \bar \chi} \left( A^{(1)}-  v_\|^{(1)}\right)+ \left(A^{(1)}{'} - B^{(1)}_{\| }{'} - \frac{1}{2}h^{(1)}_{\| }{'} \right)\right]^2 \nonumber \\
 &&-\frac{2}{\cH}\left(A^{(1)} - B^{(1)}_{\| }- \frac{1}{2}h^{(1)}_{\| }\right)\left[\frac{\ud \,}{\ud \bar \chi} \left( A^{(1)}-  v_\|^{(1)}\right)+ \left(A^{(1)}{'} - B^{(1)}_{\| }{'} - \frac{1}{2}h^{(1)}_{\| }{'} \right)\right]- \frac{1}{ \cH} \frac{\ud \,}{\ud \bar \chi}\left(2 v^{(1)}_{\| } + h_i^{i (1)}+ 2\delta_g^{(1)} \right)\Delta \ln a^{(1)} \nonumber \\
 && - \p_{\parallel}\left(2 v^{(1)}_{\| } + h_i^{i (1)} + 2\delta_g^{(1)} \right) T^{(1)}- \frac{4}{\bar \chi^2 \cH}  \Delta \ln a^{(1)} T^{(1)}+ 2v^{i (1)}_{\perp } \p_{\perp i} T^{(1)} - \frac{4}{\bar \chi \cH}  \Delta \ln a^{(1)}   \kappa^{(1)} -  \frac{4}{\bar \chi}  T^{(1)}  \kappa^{(1)}  \nonumber \\
  && + \left[- b_e +  \frac{\ud \ln b_e }{\ud  \ln \bar a} - \left(\frac{\cH'}{\cH^2} \right)^2- \frac{2}{\bar \chi^2 \cH^2}\right] \left( \Delta \ln a^{(1)}\right)^2- \frac{2}{\bar \chi^2} \left( T^{(1)} \right)^2  + 2\frac{\cH'}{\cH^2} \left(A^{(1)} - B^{(1)}_{\| }- \frac{1}{2}h^{(1)}_{\| }\right)\Delta \ln a^{(1)}\nonumber \\
   && +2\frac{\cH'}{\cH^3} \left[\frac{\ud \,}{\ud \bar \chi} \left( A^{(1)}-  v_\|^{(1)}\right)+ \left(A^{(1)}{'} - B^{(1)}_{\| }{'} - \frac{1}{2}h^{(1)}_{\| }{'} \right)\right]\Delta \ln a^{(1)}+2 \left[-\left( B^{i (1)}_{\perp }+ n^k h_{k}^{j(1)} \Perp^i_j\right) + 2S_{\perp}^{i(1)} \right] \nonumber\\
     && \times \p_{\perp i}   \left(\frac{1}{\cH}  \Delta \ln a^{(1)}+   T^{(1)}  \right) - \left[-\frac{2}{\bar \chi}\left( B^{i (1)}_{\perp }+ n^k h_{k}^{j(1)} \Perp^i_j\right)  + \frac{4}{\bar \chi} S_{\perp}^{i(1)}+ \p^i_{\perp}\left( 2v^{(1)}_{\| } + h_l^{l (1)}  + 2\delta_g^{(1)}\right)  \right]  \nonumber \\
 &&\times  \int_0^{\bar \chi} \ud \tilde \chi \bigg[ \frac{ \bar \chi}{\tilde \chi}\left( B^{(1)}_{\perp i}+ n^k h_{k}^{j(1)} \Perp_{ij}\right) + \left(\bar \chi-\tilde \chi\right) \tilde \p_{\perp i}  \left( A^{(1)} - B^{(1)}_{\| } - \frac{1}{2}h^{(1)}_{\| } \right)\bigg]\nonumber \\
&& + 2\left(B^{(1)}_{\perp i \, o }-v^{ (1)}_{\perp i \, o }+ \frac{1}{2} n^k h_{k\,o}^{j(1)} \Perp_{ij} \right) \left(B^{i (1)}_{\perp \, o }-v^{i (1)}_{\perp \, o }+ \frac{1}{2} n^k h_{k\,o}^{j(1)} \Perp^i_j\right)\nonumber \\
&& +  \left(B^{(1)}_{\perp i \, o }-v^{ (1)}_{\perp i \, o }+ \frac{1}{2} n^k h_{k\,o}^{j(1)} \Perp_{ij} \right) \bigg\{-2\left( B^{i (1)}_{\perp }+ n^m h_m^{l(1)} \Perp^i_l\right) + 4S_{\perp}^{i(1)} + \bar \chi \p^i_{\perp}\left(2v^{(1)}_{\| } + h_l^{l (1)} + 2  \delta_g^{(1)} \right)\nonumber \\
&&+ 2 \p_{\perp i}   \left(\frac{1}{\cH}  \Delta \ln a^{(1)}+   T^{(1)}  \right)  -2 \frac{1}{\bar \chi} \int_0^{\bar \chi} \ud \tilde \chi \left[ \frac{ \bar \chi}{\tilde \chi}\left( B^{i (1)}_{\perp }+ n^k h_{k}^{j(1)} \Perp^i_j\right) + \left(\bar \chi-\tilde \chi\right) \tilde \p^i_\perp \left( A^{(1)} - B^{(1)}_{\| } - \frac{1}{2}h^{(1)}_{\| } \right)\right]\bigg\}\;. \label{maindel2}
 \end{eqnarray}
 
\subsection*{Assuming no velocity bias}

Up to now, we have made no assumption about the CDM velocity $u_m^\mu$ or about the conservation equations. If we define
\begin{equation}\label{cure}
{\cal E}_m^\mu = \nabla_\nu T_m^{\mu\nu}, ~~ T_m^{\mu\nu}=\rho_m u_m^\mu u_m^\nu,
\end{equation}
then ${\cal E}_m^\mu = 0$ in General Relativity, in the absence of interactions between CDM and dark energy. However, to include interacting CDM and some modified gravity models, we allow for nonzero ${\cal E}_m^\mu$.

We now assume that galaxy velocities follow the matter velocity field (at first and second order), i.e. $u_m^\mu=u^\mu$.  Then the expressions for ${\cal E}_m^{\mu(n)}$ are given in Eqs. (A11)--(A13).

At first order
\begin{eqnarray}
\label{partial_||Dx||2-1}
\p_\| \Delta x^{(1)}_\| &=& A^{(1)}  - v^{(1)}_{\| }- \frac{1}{2} h^{(1)}_{\| }- \frac{1}{\cH} \p_\| v_\|^{(1)} - \frac{1}{2\cH} h^{(1)}_{\| }{'} + \frac{\bar a^2}{\cH \bar \rho_m} \left( \mathcal{E}_m^{\| (1)}- \mathcal{E}_m^{0 (0)} v_\|^{(1)} \right)- \frac{\cH'}{\cH^2}\Delta \ln a^{(1)}\;, 
\end{eqnarray}
where 
\begin{eqnarray}
\label{Em||-1}
 \mathcal{E}_m^{\| (1)}&=&n_i  \mathcal{E}_m^{i (1)}=\frac{\bar \rho_m}{a^2} \left(v_\|^{(1)}{'} - B_\|^{(1)}{'} +\cH v_\|^{(1)}  -\cH B_\|^{(1)} + \p_\| A^{(1)} \right)+  \mathcal{E}_m^{0 (0)} v_\|^{(1)},\\
\label{Em0-0}
\mathcal{E}_m^{0 (0)} &=&  \frac{1}{\bar a^2} \left(\bar \rho_m'+3 \cH \bar \rho_m \right)= \frac{\cH \bar \rho_m}{\bar a^2} b_m.
\end{eqnarray}
Here $b_m = \ud (a^3 \bar \rho_m)/\ud \ln \bar a$.
Then
\begin{eqnarray}
\label{Deltag-1_3}
\Delta_g^{(1)}
 &=& \delta_g^{(1)} + \left( b_e  - \frac{\cH'}{\cH^2} - \frac{2}{\bar \chi \cH}\right)\Delta \ln a^{(1)}- \frac{1}{\cH} \p_\| v_\|^{(1)}- \frac{1}{2\cH} h^{(1)}_{\| }{'} +  A^{(1)}- \frac{1}{2}h^{(1)}_{\| }+\frac{1}{2} h^{i(1)}_i  - b_m v_\|^{(1)} \nonumber \\
 &&+ \frac{a^2}{\cH \bar \rho_m}  \mathcal{E}_m^{\| (1)}- \frac{2}{\bar \chi} T^{(1)}- 2 \kappa^{(1)}\;.
\end{eqnarray}
 
 This relation generalizes for any gauge the results previously obtained in Refs.\cite{Yoo:2009au, Yoo:2010ni, Bonvin:2011bg, Challinor:2011bk,Jeong:2011as}  and can be apply to general dark energy models, including those where dark energy interacts non-gravitationally with dark matter, and to metric theories of modified gravity as an alternative to dark energy. 
 
At  second order, using Eqs.\ (\ref{ConEqu||-1}) we have
\begin{eqnarray}
\label{Deltalna-2}
 \Delta\ln a^{(2)}&=& + A^{(2)}_o- v^{(2)}_{\| \, o}-\left(A^{(1)}_o\right)^2+2A^{(1)}_o B^{(1)}_{\| \, o}-\left(B^{(1)}_{\| \, o}\right)^2-6A^{(1)}_o v^{(1)}_{\| \, o}+2B^{(1)}_{\| \, o}v^{(1)}_{\| \, o} +  v^{(1)}_{k\, o} v^{k (1)}_o \nonumber \\
&&-n^i h_{ij\,o}^{(1)} v^{j (1)}_o  + 2\left(A^{(1)}_o-v^{(1)}_{\| \, o}\right) \bigg[-3 A^{(1)}+2 B^{(1)}_{\| }- \frac{1}{\cH} \p_\| v_\|^{(1)} - \frac{1}{2\cH} h^{(1)}_{\| }{'} - b_m v_\|^{(1)} \nonumber \\
&&+ \bar \chi \p_\|\left( 2 A^{(1)} - B^{(1)}_{\| } \right)-\bar \chi \left(A^{(1)}{'} + \frac{1}{2}h^{(1)}_{\| }{'} \right) +\frac{a^2}{\cH \bar \rho_m} \mathcal{E}_m^{\| (1)} + 4 I^{(1)} \bigg] + 2 \left(B_{\perp \, o}^{i(1)} -v_{\perp \, o}^{i(1)}+\frac{1}{2}n^k h_{k\,o}^{j(1)} \Perp^i_j  \right) \nonumber \\
&& \times \bigg\{ B_{\perp i}^{(1)}+ \bar \chi \, \p_{\perp i} \left(A^{(1)} +  v^{(1)}_{\|} -B^{(1)}_{\|}\right) -  \left. \int_0^{\bar \chi} \ud \tilde \chi \left[\tilde\p_{\perp i}\left( 2 A^{(1)} - B^{(1)}_{\| }\right)- \left(B_{\perp i}^{(1)}{'}+ n^j h_{ j k}^{(1)}{'} \Perp_i^k  \right) +\frac{1}{\tilde \chi} B_{\perp i}^{(1)}\right] \right\}\nonumber \\
&& - A^{(2)}+  v^{(2)}_{\|} +  7 \left(A^{(1)}\right)^2  +    \left(B^{(1)}_{\|}\right)^2-6A^{(1)} B^{(1)}_{\|}- 2 \left(\frac{1}{2}+b_m \right) \left(v^{(1)}_{\|}\right)^2+v_{\perp i}^{(1)} v_{\perp}^{i(1)}+v^{(1)}_{\|} h^{(1)}_{\| } \nonumber \\
&&  +2 b_m A^{(1)} v^{(1)}_{\|} + 2 v^{(1)}_{\|}B^{(1)}_{\|} - 2 v_{\perp i}^{(1)} B_{\perp}^{i(1)}  + \frac{2}{\cH}\left( A^{(1)} - v^{(1)}_{\| }\right) \left( \frac{1}{2}h^{(1)}_{\| }{'} + \p_\|  v^{(1)}_{\|}  -  \frac{a^2}{\bar \rho_m} \mathcal{E}_m^{\| (1)} \right)  \nonumber \\
&& + 4 \bigg[ -3 A^{(1)}+2B^{(1)}_{\| } -  \frac{1}{\cH} \p_\| v_\|^{(1)}  - \frac{1}{2\cH} h^{(1)}_{\| }{'} - b_m v_\|^{(1)}+\frac{\bar a^2}{\cH \bar \rho_m} \mathcal{E}_m^{\| (1)} \bigg] I^{(1)} + 4 v_{\perp i}^{(1)}S_{\perp}^{i(1)}   \nonumber \\
&&-2 \p_\| \left(A^{(1)} +  v^{(1)}_{\|} -B^{(1)}_{\|}\right) T^{(1)}-2 \bigg[ \p_\| \left( 2 A^{(1)} - B^{(1)}_{\| } \right) - \left(A^{(1)}{'} +  \frac{1}{2}h^{(1)}_{\| }{'} \right)\bigg] \int_0^{\bar \chi} \ud \tilde \chi\bigg[ 2 A^{(1)} - B^{(1)}_{\| } \nonumber \\
&&+ \left(\bar \chi-\tilde \chi\right) \left(A^{(1)}{'} - B^{(1)}_{\| }{'} - \frac{1}{2}h^{(1)}_{\| }{'} \right) \bigg]  - 2 \bigg[\p_{\perp i}\left(A^{(1)} +  v^{(1)}_{\|} -B^{(1)}_{\|}\right)   - \frac{1}{\bar \chi} \left( v^{(1)}_{\perp i}- B^{ (1)}_{\perp i }\right)\bigg]\nonumber \\
&&   \int_0^{\bar \chi} \ud \tilde \chi \left\{ \left( B^{i (1)}_{\perp }+ n^k h_{k}^{j(1)} \Perp^i_j\right) + \left(\bar \chi-\tilde \chi\right)\left[ \tilde\p^i_\perp \left( A^{(1)} - B^{(1)}_{\| } - \frac{1}{2}h^{(1)}_{\| } \right)+ \frac{1}{\tilde \chi} \left(B^{i (1)}_{\perp }+  n^k h_{kj}^{(1)} \Perp^{ij}  \right)\right] \right\}  + 2I^{(2)} \nonumber \\
&&+2 \int_0^{\bar \chi} \ud \tilde \chi  \left\{ \left( B^{(1)}_{\| } +  h^{(1)}_{\| } - 4 I^{(1)}\right) \left(A^{(1)}{'} - B^{(1)}_{\| }{'} - \frac{1}{2}h^{(1)}_{\| }{'} \right) - \left(A^{(1)} - B^{(1)}_{\| } - \frac{1}{2}h^{(1)}_{\| } \right) \frac{\ud}{\ud \tilde \chi}\left( 2 A^{(1)} - B^{(1)}_{\| }\right)  \right. \nonumber \\
 &&  + \left. \left[ \left(B_{\perp}^{i(1)}+ n^j h_{ j k}^{(1)} \Perp^{ik}  \right) - 2 S_{\perp }^{i(1)}  \right] \left[\tilde \p_{\perp i}\left( 2 A^{(1)} - B^{(1)}_{\| } \right) -  \left(B_{\perp i}^{(1)}{'} + n^j h_{ j k}^{(1)}{'} \Perp_i^k  \right) + \frac{1}{\tilde \chi} B_{\perp i}^{(1)}\right]  \right\}\nonumber \\
 && + \Delta\ln a^{(2)}_{\rm post-Born}\;,
\end{eqnarray}

\begin{eqnarray}
\label{Dx0-2_3}
\Delta x^{0(2)}&=& + \frac{1}{\cH} A^{(2)}_o-\frac{1}{\cH} v^{(2)}_{\| \, o}-\left(\frac{\cH'}{\cH^3}+\frac{2}{\cH}\right)\left(A^{(1)}_o\right)^2-\frac{1}{\cH} \left(B^{(1)}_{\| \, o}\right)^2
+\frac{2}{\cH}A^{(1)}_o B^{(1)}_{\| \, o}+2\left( \frac{\cH'}{\cH^3}-\frac{2}{\cH}\right)A^{(1)}_o v^{(1)}_{\| \, o}\nonumber \\
&& +\frac{2}{\cH}B^{(1)}_{\| \, o}v^{(1)}_{\| \, o}- \frac{\cH'}{\cH^3}\left(v^{(1)}_{\| \, o}\right)^2+ \frac{1}{\cH}v^{(1)}_{\perp i \, o}v^{i (1)}_{\perp \, o}-\frac{1}{\cH}n^i h_{ij\,o}^{(1)} v^{j (1)}_o + 2 \left(A^{(1)}_o-v^{(1)}_{\| \, o}\right) \nonumber \\
&& \times  \bigg\{\left( \frac{\cH'}{\cH^3}- \frac{2}{\cH}\right) A^{(1)}-\left[ \frac{\cH'}{\cH^3}+ \frac{1}{\cH}\left(1+b_m\right)\right] v^{(1)}_{\|}+ \frac{2}{\cH} B^{(1)}_{\|} -\frac{\bar \chi}{\cH}A^{(1)}{'} - \frac{1}{\cH^2} \left(\p_\|  v^{(1)}_{\|}  - \frac{\bar a^2}{\bar \rho_m} \mathcal{E}^{\| (1)}\right)   \nonumber \\  
 &&- \frac{1}{2} \left(\frac{\bar \chi}{\cH} + \frac{1}{\cH^2}\right) h^{(1)}_{\|}{'}+ \frac{\bar \chi}{\cH} \p_\| \left( 2 A^{(1)} - B^{(1)}_{\| } \right)-  2 \left( \frac{\cH'}{\cH^3}-\frac{1}{\cH}\right) I^{(1)} \bigg\} + 2 \left(B_{\perp \, o}^{i(1)} -v_{\perp \, o}^{i(1)}+\frac{1}{2}n^k h_{k\,o}^{j(1)} \Perp^i_j  \right) \nonumber \\  
 && \times  \bigg\{\frac{1}{\cH} B_{\perp i}^{(1)}+ \frac{\bar \chi}{\cH}  \p_{\perp i} \left(A^{(1)} +  v^{(1)}_{\|} -B^{(1)}_{\|}\right)  -  \frac{1}{\cH} \int_0^{\bar \chi} \ud \tilde \chi \left[ \tilde \p_{\perp i}\left( 2 A^{(1)} - B^{(1)}_{\| }\right)- \left(B_{\perp i}^{(1)}{'}+ n^j h_{ j k}^{(1)}{'} \Perp_i^k  \right) \right.   \nonumber \\  
&& \left. \left.   + \frac{1}{\tilde \chi} B_{\perp i}^{(1)}\right] \right\}  -\frac{1}{\cH} A^{(2)} + \frac{1}{\cH}  v^{(2)}_{\|} +   \left(- \frac{\cH'}{\cH^3}+\frac{6}{\cH}\right) \left(A^{(1)}\right)^2    +  \frac{1}{\cH}  \left(B^{(1)}_{\|}\right)^2  -\frac{6}{\cH}A^{(1)} B^{(1)}_{\|}  + \frac{2}{\cH} v^{(1)}_{\| } B^{(1)}_{\|}   \nonumber \\
&&- \left[ \frac{\cH'}{\cH^3} + \frac{2}{\cH}\left(1+ b_m \right) \right]\left(v^{(1)}_{\|}\right)^2+  \frac{1}{\cH} v_{\perp i}^{(1)} v_{\perp}^{i(1)} + \frac{1}{\cH}  v^{(1)}_{\|} h^{(1)}_{\| } +2 \left[ \frac{\cH'}{\cH^3}+ \frac{1}{\cH}\left(1+ b_m\right)\right]A^{(1)} v^{(1)}_{\| }  -  \frac{2}{\cH} v_{\perp i}^{(1)} B_{\perp}^{i(1)}\nonumber \\
&&  +  \frac{2}{\cH^2} \left( A^{(1)} - v^{(1)}_{\| }\right) \left(  \frac{1}{2}h^{(1)}_{\| }{'} + \p_\|  v^{(1)}_{\|}  - \frac{\bar a^2}{\bar \rho_m} \mathcal{E}_m^{\| (1)} \right) + 4 \bigg\{ \left( \frac{\cH'}{\cH^3}- \frac{2}{\cH}\right) A^{(1)}-\left[ \frac{\cH'}{\cH^3}+ \frac{1}{\cH}\left(1+b_m \right)\right] v^{(1)}_{\|} \nonumber \\
&& + \frac{2}{\cH} B^{(1)}_{\|} -  \frac{1}{\cH^2}  \left( \frac{1}{2}h^{(1)}_{\| }{'} + \p_\|  v^{(1)}_{\|}  -  \frac{\bar a^2}{\bar \rho_m}\mathcal{E}_m^{\| (1)} \right) - \left( \frac{\cH'}{\cH^3}+ \frac{1}{\cH}\right)I^{(1)} \bigg\} I^{(1)}  + \frac{4}{\cH} v_{\perp i}^{(1)}S_{\perp}^{i(1)} \nonumber \\
&& -  \frac{2}{\cH} \p_\| \left(A^{(1)} +  v^{(1)}_{\|} -B^{(1)}_{\|}\right) T^{(1)}- \frac{2}{\cH}  \bigg[ \p_\| \left( 2 A^{(1)} - B^{(1)}_{\| } \right) - \left(A^{(1)}{'} +  \frac{1}{2}h^{(1)}_{\| }{'} \right)\bigg]  \nonumber \\
&& \times   \int_0^{\bar \chi} \ud \tilde \chi\left[ 2 A^{(1)} - B^{(1)}_{\| } + \left(\bar \chi-\tilde \chi\right) \left(A^{(1)}{'} - B^{(1)}_{\| }{'} - \frac{1}{2}h^{(1)}_{\| }{'} \right) \right]  - \frac{2}{\cH} \left[\p_{\perp i}\left(A^{(1)} +  v^{(1)}_{\|} -B^{(1)}_{\|}\right)  \right.\nonumber \\
&& \left. - \frac{1}{\bar \chi} \left( v^{(1)}_{\perp i}- B^{ (1)}_{\perp i }\right)\right] \int_0^{\bar \chi} \ud \tilde \chi \left\{ \left( B^{i (1)}_{\perp }+ n^k h_{k}^{j(1)} \Perp^i_j\right)+\left(\bar \chi-\tilde \chi\right)\left[ \tilde\p^i_\perp \left( A^{(1)} - B^{(1)}_{\| } - \frac{1}{2}h^{(1)}_{\| } \right)  \right. \right. \nonumber \\
&&\left. \left. + \frac{1}{\tilde \chi} \left(B^{i (1)}_{\perp }+  n^k h_{kj}^{(1)} \Perp^{ij}  \right)\right] \right\} +  \frac{2}{\cH} I^{(2)}  + \frac{2}{\cH} \int_0^{\bar \chi} \ud \tilde \chi  \left\{ \left( B^{(1)}_{\| } +  h^{(1)}_{\| } - 4 I^{(1)}\right) \left(A^{(1)}{'} - B^{(1)}_{\| }{'} - \frac{1}{2}h^{(1)}_{\| }{'} \right) \right.\nonumber \\
&& \left. - \left(A^{(1)} - B^{(1)}_{\| } - \frac{1}{2}h^{(1)}_{\| } \right)  \frac{\ud}{\ud \tilde \chi}\left( 2 A^{(1)} - B^{(1)}_{\| }\right) + \left[ \left(B_{\perp}^{i(1)}+ n^j h_{ j k}^{(1)} \Perp^{ik}  \right) - 2 S_{\perp }^{i(1)}  \right]  \right. \nonumber \\
&& \left. \times \left[\tilde\p_{\perp i}\left( 2 A^{(1)} - B^{(1)}_{\| } \right)- \left(B_{\perp i}^{(1)}{'} + n^j h_{ j k}^{(1)}{'} \Perp_i^k  \right) + \frac{1}{\tilde \chi} B_{\perp i}^{(1)}\right]  \right\}+\Delta x^{0(2)}_{\rm post-Born}\;,
\end{eqnarray}

\begin{eqnarray} 
\label{dDx_||-2_4}
&& \p_\| \Delta x_{\parallel}^{(2)}=\left(A^{(1)}_o\right)^2 - 2A^{(1)}_o B^{(1)}_{\| \, o}+\left(B^{(1)}_{\| \, o}\right)^2 + 2A^{(1)}_o v^{(1)}_{\| \, o} -2B^{(1)}_{\| \, o}v^{(1)}_{\| \, o}-   v^{(1)}_{\| \, o} h^{(1)}_{\| \, o} -\frac{1}{4} \left(h^{(1)}_{\| \, o}\right)^2   -   v^{(1)}_{\perp k \, o} v^{k (1)}_{\perp \, o}\nonumber \\
&& +  n^i h_{ik\,o}^{(1)} \Perp^k_j v^{j (1)}_o - n^i h_{ik\,o}^{(1)} \Perp^k_j B^{j (1)}_o - \frac{1}{4} n^i h_{ij\,o}^{(1)}\,  \Perp^j_k \, h^{k (1)}_{p \, o} n^p   -  B_{\perp \, o}^{i(1)}B_{\perp i \, o}^{(1)}+2 v_{\perp \, o}^{i(1)}B_{\perp i \, o}^{(1)}+2 \left(A^{(1)}_o-v^{(1)}_{\| \, o}\right)\nonumber \\
&& \times \Bigg\{\left(-\frac{\cH' }{\cH^3}+ \frac{1}{\cH}\right) \left[\frac{\bar a^2}{\bar \rho_m} \Em^{\| (1)}-\p_\| v_\|^{(1)}-\cH \left( b_m  + 1\right) v_\|^{(1)}+\cH B_\|^{(1)} - \frac{1}{2}h^{(1)}_{\| }{'} \right] - \frac{\bar \chi}{\cH} \frac{\ud \,}{\ud \bar \chi}\bigg[ \p_\| \left( 2A^{(1)}-B_\|^{(1)}\right) \nonumber \\
&& \left. -\left(A^{(1)}{'}+\frac{1}{2}h^{(1)}_{\| }{'} \right)  \right]+\left (  A^{(1)} - B^{(1)}_{\| } - \frac{1}{2} h^{(1)}_{\| } \right) -\bar \chi \frac{\ud}{\ud \bar \chi}\left( A^{(1)} - B^{(1)}_{\| } - \frac{1}{2}h^{(1)}_{\| } \right) - \frac{1}{\cH^2}  \frac{\ud \,}{\ud \bar \chi} \bigg[\frac{\bar a^2}{\bar \rho_m} \Em^{\| (1)}-\p_\| v_\|^{(1)}\nonumber \\
&&-\cH \left( b_m  + 1\right) v_\|^{(1)}  +\cH B_\|^{(1)} - \frac{1}{2}h^{(1)}_{\| }{'} \bigg]\Bigg\} + 2 \left(B_{\perp \, o}^{i(1)} -v_{\perp \, o}^{i(1)}+\frac{1}{2}n^k h_{k\,o}^{j(1)} \Perp^i_j  \right) \bigg[ -\frac{1}{\cH} \frac{\ud \,}{\ud \bar \chi}  B_{\perp i}^{(1)}   +\frac{1}{\cH}  \p_{\perp i} \left(A^{(1)} -  v^{(1)}_{\|} \right)\nonumber \\
&&  -\frac{1}{\cH}\left(B_{\perp i}^{(1)}{'}+ n^j h_{ j k}^{(1)}{'} \Perp_i^k  \right) -\frac{\bar \chi}{\cH}  \frac{\ud \,}{\ud \bar \chi} \p_{\perp i} \left(A^{(1)} +  v^{(1)}_{\|} -B^{(1)}_{\|}\right) +\frac{1}{\cH \bar \chi} B_{\perp i}^{(1)} -4  \delta_{il} S_{\perp }^{l (1)} \bigg]+A^{(2)}- B^{(2)}_{\| } -\frac{1}{2}h_\|^{(2)}-  \frac{1}{2 \cH} h^{(2)}_{\| }{'}\nonumber \\
&&  +2\frac{\bar a^2}{\cH \bar \rho_m} \left[ \frac{1}{2} \Em^{\| (2)} -\Em^{0 (1)}  v_\|^{(1)}-  \Em^{\| (1)} \left(\delta_m^{(1)} - A^{(1)} \right) \right] -  2 b_m \left(\frac{1}{2} v_\|^{(2)}-\delta_m^{(1)}  v_\|^{(1)} +2 A^{(1)} v_\|^{(1)} \right) -\frac{2}{\cH}\bigg[ \frac{1}{2}\p_\| v_\|^{(2)}- v_\|^{(1)} \p_\| v_\|^{(1)} \nonumber \\
&&  + \cH  \left(\frac{1}{2} v_\|^{(2)}- \frac{1}{2} B_\|^{(2)}\right)    -\frac{2}{\bar \chi} \left( v_\|^{(1)} \right)^2 - v_\|^{(1)}  \p_{\perp j}  v_\perp^{j(1)} + A^{(1)} B_\|^{(1)}{'} + A^{(1)}{'} B_\|^{(1)}+ \cH A^{(1)} B_\|^{(1)} + v_\|^{(1)} h_\|^{(1)}{'}  +   \cH  B_\|^{(1)}   h_\|^{(1)} \nonumber \\
&&+  B_\|^{(1)}{'} h_\|^{(1)}  - A^{(1)} \p_\| A^{(1)} -\p_\| A^{(1)} h_\|^{(1)} + v_{\perp k}^{(1)} \p_\|  B_{\perp}^{k(1)} - v_{\perp}^{j (1)} \p_{\perp j}  B_\|^{(1)}  + \frac{1}{\bar \chi} v_{\perp}^{j (1)}  B_{\perp j}^{(1)}+\cH  B_k^{(1)} \Perp^k_j  h^{ij (1)} n_i \nonumber \\ 
&& +v_k^{(1)} \Perp^k_j  h^{ij (1)}{'} n_i +   B_k^{(1)}{'} \Perp^k_j  h^{ij (1)} n_i  - \p_k A^{(1)}\Perp^k_j  h^{ij (1)} n_i \bigg] -2 \left(2A^{(1)}+h^{(1)}_{\| } \right)\left(  A^{(1)} - B^{(1)}_{\| } - \frac{1}{2} h^{(1)}_{\| }\right)\nonumber \\ 
&& -\frac{2}{\cH}   \frac{\ud \,}{\ud \bar \chi} \bigg[ \frac{7}{2} \left(A^{(1)}\right)^2 + \frac{1}{2}  \left(B^{(1)}_{\|}\right)^2-2A^{(1)} B^{(1)}_{\|} - A^{(1)} v^{(1)}_{\| } + \frac{1}{2} \left(v^{(1)}_{\|}\right)^2 + \frac{1}{2} v^{(1)}_{\|} h^{(1)}_{\| } +\frac{1}{2} v_{\perp i}^{(1)} v_{\perp}^{i(1)}  - v_{\perp i}^{(1)} B_{\perp}^{i(1)} \bigg]\nonumber \\ 
&& +  \frac{2}{\cH^2}\left( A^{(1)} - v^{(1)}_{\| }\right) \frac{\ud \,}{\ud \bar \chi}\left[\frac{\bar a^2}{\bar \rho_m} \Em^{\| (1)}-\p_\| v_\|^{(1)}-\cH \left( b_m  + 1\right) v_\|^{(1)}+\cH B_\|^{(1)} - \frac{1}{2}h^{(1)}_{\| }{'} \right]  -\frac{2}{\cH} \bigg( 3 A^{(1)} +  h^{(1)}_{\| }  -  v^{(1)}_{\|} \bigg)\nonumber \\ 
&& \times \left(A^{(1)}{'} - B^{(1)}_{\| }{'} - \frac{1}{2}h^{(1)}_{\| }{'} \right)  + \frac{2}{\cH} \left(2 A^{(1)} - B^{(1)}_{\| } \right) \bigg[ \p_\| \left( 2A^{(1)}-B_\|^{(1)}\right)-\left(A^{(1)}{'}+\frac{1}{2}h^{(1)}_{\| }{'} \right)\bigg] + \bigg\{ 2 \frac{\cH'}{\cH^3}\left( A^{(1)} - v^{(1)}_{\| }\right) \nonumber \\ 
&&  +  \frac{2}{\cH} \left( A^{(1)} - B^{(1)}_{\| } - \frac{1}{2}h^{(1)}_{\| } \right)   +  \frac{2}{\cH^2}\left[\frac{\bar a^2}{\bar \rho_m} \Em^{\| (1)}-\p_\| v_\|^{(1)}-\cH \left( b_m  + 1\right) v_\|^{(1)}+\cH B_\|^{(1)}  - \frac{1}{2}h^{(1)}_{\| }{'}  \right]\bigg\} \bigg[\frac{\bar a^2}{\bar \rho_m} \Em^{\| (1)}  -\p_\| v_\|^{(1)}  \nonumber \\
&&   -\cH \left( b_m  + 1\right) v_\|^{(1)}+\cH B_\|^{(1)}  - \frac{1}{2}h^{(1)}_{\| }{'} \bigg] +  \frac{2}{\cH} \left[ \frac{\ud \,}{\ud  \bar \chi}\left( 2 A^{(1)} - B^{(1)}_{\| }\right)  -  \p_\| \left(A^{(1)} +  v^{(1)}_{\|} -B^{(1)}_{\|}\right)\right] \left(A^{(1)} - B^{(1)}_{\| } - \frac{1}{2}h^{(1)}_{\| } \right) \nonumber \\ 
&&+2 \left(B_{\perp}^{i(1)}+ n^j h_{ j k}^{(1)} \Perp^{ik}  \right) \bigg[\frac{1}{2}\left(B_{i \perp}^{(1)}+ n^p h_{p m}^{(1)} \Perp^m_i  \right) -\frac{1}{\cH}\p_{\perp i}\left(  A^{(1)} - v^{(1)}_{\| } \right) +\frac{1}{\cH} \left(B_{\perp i}^{(1)}{'} + n^j h_{ j k}^{(1)}{'} \Perp_i^k  \right)  \bigg]   +\frac{2}{\cH}v_{\perp i}^{(1)}\nonumber \\ 
&& \times  \p^i_\perp \bigg( A^{(1)} - B^{(1)}_{\| } - \frac{1}{2}h^{(1)}_{\| } \bigg)     +\frac{2}{\cH} \frac{\ud \,}{\ud \bar \chi} \p_\| \left(A^{(1)} +  v^{(1)}_{\|} -B^{(1)}_{\|}\right) T^{(1)}  + 4 \bigg\{ \left(-\frac{\cH' }{\cH^3} +\frac{1}{\cH}\right)  \bigg[\frac{\bar a^2}{\bar \rho_m} \Em^{\| (1)} -\p_\| v_\|^{(1)}+\cH B_\|^{(1)}  \nonumber \\ 
&& -\cH \left( b_m  + 1\right) v_\|^{(1)}- \frac{1}{2}h^{(1)}_{\| }{'} \bigg]    - \frac{1}{\cH^2} \frac{\ud \,}{\ud \bar \chi}\bigg[\frac{\bar a^2}{\bar \rho_m} \Em^{\| (1)}-\p_\| v_\|^{(1)} -\cH \left( b_m  + 1\right) v_\|^{(1)}+\cH B_\|^{(1)} - \frac{1}{2}h^{(1)}_{\| }{'}  \bigg] +\bigg(  A^{(1)} - B^{(1)}_{\| }\nonumber \\ 
&&  - \frac{1}{2} h^{(1)}_{\| }\bigg)\bigg\} I^{(1)}+ 2 \bigg\{ \frac{\ud}{\ud \bar \chi}\left( A^{(1)} - B^{(1)}_{\| } - \frac{1}{2}h^{(1)}_{\| } \right)  + \frac{1}{\cH} \frac{\ud \,}{\ud \bar \chi} \bigg[\p_\| \left( 2A^{(1)}-B_\|^{(1)}\right)-\left(A^{(1)}{'}+\frac{1}{2}h^{(1)}_{\| }{'} \right)\bigg]  \bigg\} \int_0^{\bar \chi} \ud \tilde \chi \;\bigg[ 2 A^{(1)}  \nonumber \\
 && - B^{(1)}_{\| } + \left(\bar \chi-\tilde \chi\right)\left(A^{(1)}{'} - B^{(1)}_{\| }{'} - \frac{1}{2}h^{(1)}_{\| }{'} \right) \bigg] +  4\bigg[\frac{1}{\cH}\p_{\perp i}\left(  A^{(1)} - v^{(1)}_{\| } \right) -\frac{1}{\cH} \left(B_{\perp i}^{(1)}{'} + n^j h_{ j k}^{(1)}{'} \Perp_i^k  \right)    + \frac{1}{\cH \bar \chi} v_{\perp i}^{(1)}  \nonumber \\
 &&-\frac{1}{\cH}\frac{\ud \,}{\ud \bar \chi} v_{\perp i}^{(1)} -2S_{\perp }^{j(1)} \delta_{ij}  \bigg]   S_{\perp }^{i(1)} + \frac{2}{\cH} \frac{\ud \,}{\ud \bar \chi} \left[\p_{\perp i}\left(A^{(1)} +  v^{(1)}_{\|} -B^{(1)}_{\|}\right)  - \frac{1}{\bar \chi} \left( v^{(1)}_{\perp i}- B^{ (1)}_{\perp i }\right)\right] \int_0^{\bar \chi} \ud \tilde \chi \bigg\{ \left( B^{i (1)}_{\perp }+ n^k h_{k}^{j(1)} \Perp^i_j\right) \nonumber \\
 && + \left(\bar \chi-\tilde \chi\right)  \left[ \tilde \p^i_\perp \left( A^{(1)} - B^{(1)}_{\| } - \frac{1}{2}h^{(1)}_{\| } \right) + \frac{1}{\tilde \chi} \left(B^{i (1)}_{\perp }+  n^k h_{kj}^{(1)} \Perp^{ij}  \right)\right] \bigg\}  -   \frac{\cH'}{\cH^2} \Delta \ln a^{(2)} + \bigg\{-2\frac{\cH' }{\cH^2} \left( A^{(1)} - B^{(1)}_{\| } - \frac{1}{2}h^{(1)}_{\| } \right)\nonumber \\
&&   - \frac{2}{\cH} \frac{\ud}{\ud \bar \chi}\left( A^{(1)} - B^{(1)}_{\| } - \frac{1}{2}h^{(1)}_{\| } \right)- 2\left(\frac{\cH' }{\cH^3} +\frac{1}{\cH} \right) \bigg[\frac{\bar a^2}{\bar \rho_m} \Em^{\| (1)}-\p_\| v_\|^{(1)} +\cH B_\|^{(1)} -\cH \left( b_m  + 1\right) v_\|^{(1)} - \frac{1}{2}h^{(1)}_{\| }{'} \bigg] \bigg\} \Delta \ln a^{(1)}\nonumber \\
&&+\left[-\frac{\cH'' }{\cH^3} +3\left( \frac{\cH' }{\cH^2} \right)^2 + \frac{\cH' }{\cH^2} \right] \left( \Delta \ln a^{(1)} \right)^2 +2 \int_0^{\bar \chi} \ud \tilde \chi \bigg\{ \left(A^{(1)} - B^{(1)}_{\| } - \frac{1}{2}h^{(1)}_{\| } \right)  \bigg[2\left(A^{(1)}{'} - B^{(1)}_{\| }{'} - \frac{1}{2}h^{(1)}_{\| }{'} \right)+  \frac{\ud}{\ud \tilde \chi}\bigg( 2 A^{(1)}  \nonumber \\
&&  - B^{(1)}_{\| }\bigg) \bigg]  -2 \left(B_{\perp}^{i(1)} + n^k h_{k}^{j(1)} \Perp^i_j  \right)\left[ \tilde \p_{\perp i}\left (  A^{(1)} - B^{(1)}_{\| } - \frac{1}{2} h^{(1)}_{\| } \right)  + \frac{1}{\tilde \chi} \left(B_{\perp i}^{(1)}+   n^m h_{mp}^{(1)} \Perp^p_i  \right)\right]\bigg\} + \p_\| \Delta x_{\|  \rm post-Born}^{(2)}\;.
\end{eqnarray}

We can see immediately that $ \p_\| \Delta x_{\parallel}^{(2)}$ can be further expanded. We apply again Eq.\  (\ref{Em})  in the following terms
 \begin{eqnarray}
 && -\frac{1}{\cH}\frac{\ud \,}{\ud \bar \chi} v_{\perp i}^{(1)} =   \frac{1}{\cH} \frac{\bar a^2}{\bar \rho_m} \Perp_{ij}  \Em^{j (1)} - \frac{1}{\cH}\p_{\|} v_{\perp i}^{(1)}- \left( b_m  + 1\right) v^{(1)}_{\perp i}+ B^{(1)}_{\perp i} -  \frac{1}{\cH}\p_{\perp i} A^{(1)} + \frac{1}{\cH} B^{(1)}_{\perp i}{'}\;, \\
\nonumber \\
 \nonumber \\
 && +\frac{2}{\cH} \frac{\ud \,}{\ud \bar \chi} \p_\| \left(A^{(1)} +  v^{(1)}_{\|} -B^{(1)}_{\|}\right) T^{(1)} = +\frac{2}{\cH} \bigg[    -\frac{\bar a^2}{\bar \rho_m} \p_\| \Em^{\| (1)}+ \p_\|^2 v_\|^{(1)}+\cH \left( b_m  + 1\right) \p_\| v_\|^{(1)} -\cH \p_\| B_\|^{(1)}-\p_\| A^{(1)}{'}\nonumber \\
&& +2 \p_\|^2 A^{(1)}- \p_\|^2 B^{(1)}_{\|}\bigg] T^{(1)}\;,  \\
 \nonumber \\
\nonumber \\
&&-\frac{\bar \chi}{\cH}  \frac{\ud \,}{\ud \bar \chi} \p_{\perp i} \left(A^{(1)} +  v^{(1)}_{\|} -B^{(1)}_{\|}\right)= -\frac{\bar \chi}{\cH}\p_{\perp i}  \bigg[- A^{(1)}{'}  -\frac{\bar a^2}{\bar \rho_m} \Em^{\| (1)}+\p_\| v_\|^{(1)}+\cH \left( b_m  + 1\right) v_\|^{(1)} -\cH B_\|^{(1)}  + 2\p_\| A^{(1)} \nonumber \\
 &&- \p_\| B^{(1)}_{\|}\bigg] +\frac{1}{\cH}\p_{\perp i} \left(A^{(1)} +  v^{(1)}_{\|} -B^{(1)}_{\|}\right)\;, \\
 \nonumber \\
\nonumber \\
&&+ \frac{2}{\cH} \frac{\ud \,}{\ud \bar \chi} \left[\p_{\perp i}\left(A^{(1)} +  v^{(1)}_{\|} -B^{(1)}_{\|}\right)  - \frac{1}{\bar \chi} \left( v^{(1)}_{\perp i}- B^{ (1)}_{\perp i }\right)\right] =\frac{2}{\cH}\bigg[\p_{\perp i}\bigg( -A^{(1)}{'} -\frac{\bar a^2}{\bar \rho_m} \Em^{\| (1)}+\p_\| v_\|^{(1)}+\cH \left( b_m  + 1\right) v_\|^{(1)} \nonumber \\
&&-\cH B_\|^{(1)} + 2 \p_\| A^{(1)} - \p_\| B^{(1)}_{\|}\bigg)  +\frac{1}{\bar \chi} \left(-2\p_{\perp i}A^{(1)} - \p_{\perp i} v^{(1)}_{\|} - \p_{\|} v^{(1)}_{\perp i} +\p_{\perp i}B^{(1)}_{\|} +\frac{\bar a^2}{\bar \rho_m} \Perp_{ij} \Em^{j (1)}-\cH \left( b_m  + 1\right) v^{(1)}_{\perp i}\right.\nonumber \\
&&\left. +\cH B^{(1)}_{\perp i} +\p_\| B_{\perp i}^{ (1)}\right)+\frac{1}{\bar \chi^2}\left( v^{(1)}_{\perp i}- B^{ (1)}_{\perp i }\right)\bigg] \;, \\
 \nonumber \\
\nonumber \\
&&  \frac{\ud \,}{\ud \bar \chi} \bigg[\frac{\bar a^2}{\bar \rho_m} \Em^{\| (1)}-\p_\| v_\|^{(1)}-\cH \left( b_m  + 1\right) v_\|^{(1)}  +\cH B_\|^{(1)} - \frac{1}{2}h^{(1)}_{\| }{'} \bigg]=-{\left(\frac{\bar a^2}{\bar \rho_m} \Em^{\| (1)}\right)}'+\frac{\bar a^2 \cH}{\bar \rho_m}  \left( b_m  + 1\right)  \Em^{\| (1)} \nonumber \\
&& + \cH^2 \frac{\ud b_m}{\ud \ln a}  v_\|^{(1)} + b_m \bigg[ -\cH^2 \left( b_m  + 1\right) v_\|^{(1)}  +\cH B_\|^{(1)}{'} - 2 \cH \p_\| v_\|^{(1)} +\left(\cH'-\cH^2\right) v_\|^{(1)} + \cH^2  B_\|^{(1)}  - \cH \p_\| A^{(1)} \bigg] \nonumber \\
&&-2 \cH \p_\| v_\|^{(1)} - \p_\|^2 A^{(1)} + \p_\|  B_\|^{(1)}{'} -\p_\|^2 v^{(1)}_{\|} + 2 \cH \p_\| B_\|^{(1)}+ \left(\cH'-\cH^2\right) \left(v_\|^{(1)}-B_\|^{(1)}\right)- \cH  \p_\| A^{(1)}  - \frac{1}{2} \frac{\ud \,}{\ud \bar \chi}h^{(1)}_{\| }{'}\;,\\
\nonumber \\
\nonumber \\
&&\frac{\ud \,}{\ud \bar \chi} \bigg[ \frac{7}{2} \left(A^{(1)}\right)^2 + \frac{1}{2}  \left(B^{(1)}_{\|}\right)^2-2A^{(1)} B^{(1)}_{\|} - A^{(1)} v^{(1)}_{\| } + \frac{1}{2} \left(v^{(1)}_{\|}\right)^2 + \frac{1}{2} v^{(1)}_{\|} h^{(1)}_{\| } +\frac{1}{2} v_{\perp i}^{(1)} v_{\perp}^{i(1)}  - v_{\perp i}^{(1)} B_{\perp}^{i(1)} \bigg] \nonumber \\ 
&&=6A^{(1)}\frac{\ud \,}{\ud \bar \chi}A^{(1)}-A^{(1)}A^{(1)}{'}+ B^{(1)}_{\|}\frac{\ud \,}{\ud \bar \chi}B^{(1)}_{\|}-2\frac{\ud \,}{\ud \bar \chi}\left(A^{(1)} B^{(1)}_{\|}\right)+A^{(1)}B^{(1)}_{\|}{'}+  v^{(1)}_{\| } A^{(1)}{'}+\frac{1}{2}v^{(1)}_{\| }\frac{\ud \,}{\ud \bar \chi}h^{(1)}_{\| }\nonumber \\ 
&&-A^{(1)}\p_\|v^{(1)}_{\| }-\cH A^{(1)}v^{(1)}_{\| }+\cH A^{(1)}B^{(1)}_{\| }-v^{(1)}_{\| }B^{(1)}_{\| }{'}+ v^{(1)}_{\| } \p_\| v^{(1)}_{\| }+\cH \left(v^{(1)}_{\| }\right)^2-\cH v^{(1)}_{\| }B^{(1)}_{\|}+\frac{1}{2} h^{(1)}_{\| } \p_\| A^{(1)} -\frac{1}{2} h^{(1)}_{\| } B^{(1)}_{\|}{'} \nonumber \\ 
&&+ \frac{1}{2} h^{(1)}_{\| } \p_\| v^{(1)}_{\| }+\frac{\cH}{2}v^{(1)}_{\| }h^{(1)}_{\| }-\frac{\cH}{2}B^{(1)}_{\| }h^{(1)}_{\| }+\left( B_{\perp i}^{(1)}  - v_{\perp i}^{(1)} \right) \left(-\p_\|  v_{\perp}^{i(1)}+ B_{\perp}^{i(1)}{'}-\cH  v_{\perp}^{i(1)}+\cH B_{\perp}^{i(1)}-\p_{\perp}^i A^{(1)} \right)\nonumber \\ 
&& -\left(A^{(1)}-v^{(1)}_{\| }-\frac{1}{2} h^{(1)}_{\| }\right)\left(\frac{\bar a^2}{\bar \rho_m} \Em^{\| (1)}- \cH b_m v^{(1)}_{\| } \right)+\left( B_{\perp i}^{(1)}  - v_{\perp i}^{(1)} \right)\left(\frac{\bar a^2}{\bar \rho_m}\Perp^i_j \Em^{j (1)}- \cH b_m v^{i(1)}_{\perp }\right)\;.
   \end{eqnarray}

Finally, we get
  \begin{eqnarray}
&&  \Delta_g^{(2)}   =  \delta_g^{(2)}+   b_e \, \Delta \ln a^{(2)} +  \p_{\parallel} \Delta x_{\parallel}^{(2)}  + \frac{2}{\bar \chi} \Delta x_{\parallel}^{(2)} - 2\kappa^{(2)}  + A^{(2)}+ v_{\|}^{(2)}+ \frac{1}{2} h_i^{i (2)} +\left(\Delta_g^{(1)} \right)^2 - 3 \left(A^{(1)}\right)^2+4A^{(1)}B^{(1)}_{\| }\nonumber \\
 &&+3A^{(1)}h^{(1)}_{\| } -2B^{(1)}_{\| }h^{(1)}_{\| }  -\left(v^{(1)}_{\| } \right)^2 -\left(B^{(1)}_{\| } \right)^2  - \frac{1}{2}h_{i}^{k(1)} h_{k}^{i(1)}- \frac{3}{4}\left(h^{(1)}_{\| } \right)^2 +2A^{(1)} v^{(1)}_{\| }-\frac{1}{\cH^2}\left(\p_\|v^{(1)}_{\| } \right)^2  +\frac{1}{\cH}A^{(1)}h^{(1)}_{\| }{'} \nonumber \\
 &&- \frac{1}{4\cH^2}\left(h^{(1)}_{\| }{'} \right)^2  -\frac{1}{\cH} v^{(1)}_{\| }h^{(1)}_{\| }{'}-\frac{1}{2\cH} h^{(1)}_{\| }h^{(1)}_{\| }{'} +\frac{2}{\cH}A^{(1)}\p_\|v^{(1)}_{\| }-\frac{2}{\cH}v^{(1)}_{\| }\p_\|v^{(1)}_{\| }-\frac{1}{\cH}h^{(1)}_{\| }\p_\|v^{(1)}_{\| }  -\frac{1}{\cH^2}h^{(1)}_{\| }{'}\p_\|v^{(1)}_{\| }\nonumber \\
&&+  B^{(1)}_{\perp i} B^{i (1)}_{\perp}+   v^{(1)}_{\perp i} v^{i (1)}_{\perp}- 2  v^{(1)}_{\perp i} B^{i (1)}_{\perp} -8\left( I^{(1)} \right)^2+8A^{(1)} I^{(1)}-8B^{(1)}_{\| } I^{(1)}-4h^{(1)}_{\| } I^{(1)} - \left(\delta_g^{(1)}\right)^2 + 2v^{i (1)}_{\perp } \p_{\perp i} T^{(1)} \nonumber \\
 &&  + \frac{2}{ \cH} \left(-\p_\| A^{(1)}+ B_\|^{(1)}{'} -\p_\| v_\|^{(1)}-\cH  v_\|^{(1)}+\cH B_\|^{(1)} \right)\Delta \ln a^{(1)} - \frac{1}{ \cH} \frac{\ud \,}{\ud \bar \chi}\left(  h_i^{i (1)}+ 2\delta_g^{(1)} \right)\Delta \ln a^{(1)}- \frac{4}{\bar \chi^2 \cH}  \Delta \ln a^{(1)} T^{(1)}\nonumber \\
 &&  + 2\frac{\cH'}{ \cH^2} \left(A^{(1)}-  v^{(1)}_{\| } - \frac{1}{2}h^{(1)}_{\| }- \frac{1}{\cH}\p_\|v^{(1)}_{\| }-\frac{1}{2\cH}h^{(1)}_{\| }{'}\right)\Delta \ln a^{(1)} - \p_{\parallel}\left(2 v^{(1)}_{\| } + h_i^{i (1)} + 2\delta_g^{(1)} \right) T^{(1)}- \frac{4}{\bar \chi \cH}  \Delta \ln a^{(1)}   \kappa^{(1)}\nonumber \\
 &&  -\frac{4}{\bar \chi}  T^{(1)}  \kappa^{(1)} + \left[- b_e +  \frac{\ud \ln b_e }{\ud  \ln \bar a} - \left(\frac{\cH'}{\cH^2} \right)^2- \frac{2}{\bar \chi^2 \cH^2}\right] \left( \Delta \ln a^{(1)}\right)^2- \frac{2}{\bar \chi^2} \left( T^{(1)} \right)^2  +2 \left[-\left( B^{i (1)}_{\perp }+ n^k h_{k}^{j(1)} \Perp^i_j\right) + 2S_{\perp}^{i(1)} \right] \nonumber\\
  && \times \p_{\perp i}   \left(\frac{1}{\cH}  \Delta \ln a^{(1)}+   T^{(1)}  \right) - \left[-\frac{2}{\bar \chi}\left( B^{i (1)}_{\perp }+ n^k h_{k}^{j(1)} \Perp^i_j\right)  + \frac{4}{\bar \chi} S_{\perp}^{i(1)}+ \p^i_{\perp}\left( 2v^{(1)}_{\| } + h_l^{l (1)}  + 2\delta_g^{(1)}\right)  \right]  \nonumber \\
 &&\times  \int_0^{\bar \chi} \ud \tilde \chi \bigg[ \frac{ \bar \chi}{\tilde \chi}\left( B^{(1)}_{\perp i}+ n^k h_{k}^{j(1)} \Perp_{ij}\right) + \left(\bar \chi-\tilde \chi\right) \tilde \p_{\perp i}  \left( A^{(1)} - B^{(1)}_{\| } - \frac{1}{2}h^{(1)}_{\| } \right)\bigg]\nonumber \\
& &- 2\left(A^{(1)} - B^{(1)}_{\| } - \frac{1}{2}h^{(1)}_{\| }  - 2 I^{(1)} \right) \int_0^{\bar \chi} \ud \tilde \chi\bigg\{\frac{2}{\tilde \chi} B^{(1)}_{\| }  - \tilde \p_{ \perp m} B^{m (1)}- \frac{1}{\tilde \chi}  h_i^{i (1)} + \frac{3}{\tilde \chi}   h^{(1)}_{\| }  - n^m  \tilde \p_{ \perp n}  h_m^{n (1)}\nonumber \\
 & &-  \frac{\tilde \chi}{ \bar \chi} \left[ 2 \tilde \p_\| + \left(\bar \chi-\tilde \chi\right)  \Perp^{mn} \tilde \p_m  \tilde \p_n \right] \left( A^{(1)} - B^{(1)}_{\| } - \frac{1}{2}h^{(1)}_{\| } \right) \bigg\} - \int_0^{\bar \chi} \ud \tilde \chi\bigg\{   \frac{1}{\tilde \chi}  \Perp^i_j  B^{(1)}_{\| }  -\Perp^i_p \tilde \p_{ \perp j} B^{p (1)}- \frac{1}{\tilde \chi}   \Perp^p_j \Perp^i_q   h_p^{q (1)}  \nonumber \\
 && + \frac{1}{\tilde \chi}  \Perp_j^i h^{(1)}_{\| } - n^p \Perp^i_q \tilde \p_{ \perp j}  h_p^{q (1)}-  \frac{\tilde \chi}{ \bar \chi} \left[ \Perp_j^i \tilde \p_\| + \left(\bar \chi-\tilde \chi\right) \Perp^p_j \Perp^{iq} \tilde \p_q  \tilde \p_p \right] \left( A^{(1)} - B^{(1)}_{\| } - \frac{1}{2}h^{(1)}_{\| } \right) \bigg\}\nonumber \\
 &&  \int_0^{\bar \chi} \ud \tilde \chi\bigg\{  \frac{1}{\tilde \chi}   \Perp^j_i  B^{(1)}_{\| } -\Perp^j_m \tilde \p_{ \perp i} B^{m (1)} - \frac{1}{\tilde \chi}  \Perp^n_i \Perp^j_m   h_n^{m (1)} +  \frac{1}{\tilde \chi} \Perp^j_i h^{(1)}_{\| }   - n^m \Perp^j_n \tilde \p_{ \perp i}  h_m^{n (1)} \nonumber \\
  && -  \frac{\tilde \chi}{ \bar \chi} \left[ \Perp^j_i \tilde \p_\| + \left(\bar \chi-\tilde \chi\right) \Perp^n_i \Perp^{jm}  \tilde \p_m   \tilde \p_n \right] \left( A^{(1)} - B^{(1)}_{\| } - \frac{1}{2}h^{(1)}_{\| } \right) \bigg\}-2\left(v^{(1)}_{\| \, o}\right)^2-2 \left(A^{(1)}_{\, o}\right)^2- v^{(1)}_{\| \, o}h_{i \, o}^{i (1)}+  v^{(1)}_{\| \, o} h^{(1)}_{\| \, o}   \nonumber \\
 &&+4 A^{(1)}_{\, o}v^{(1)}_{\| \, o}+ A^{(1)}_{\, o}h_{i \, o}^{i (1)}-  A^{(1)}_{\, o} h^{(1)}_{\| \, o}+ 2\left(B^{(1)}_{\perp i \, o }-v^{ (1)}_{\perp i \, o }+ \frac{1}{2} n^k h_{k\,o}^{j(1)} \Perp_{ij} \right) \left(B^{i (1)}_{\perp \, o }-v^{i (1)}_{\perp \, o }+ \frac{1}{2} n^k h_{k\,o}^{j(1)} \Perp^i_j\right)\nonumber \\
 && -  \frac{1}{4} \Perp^m_p \Perp^k_n h_{m \, o}^{n (1)}  h_{k \, o}^{p(1)}+ 2\left(2A^{(1)}_{\, o} - 2  v^{(1)}_{\| \, o} -\frac{1}{2}  h_{i \, o}^{i (1)} +  \frac{1}{2}h^{(1)}_{\| \, o} \right)\left(A^{(1)} - B^{(1)}_{\| } - \frac{1}{2}h^{(1)}_{\| }  - 2 I^{(1)} \right)\nonumber \\
 &&+2 \left( \Perp^i_j A^{(1)}_{\, o} -  \Perp^i_j   v^{(1)}_{\| \, o}  - \frac{1}{2} \Perp^k_j \Perp^i_p h_{k \, o}^{p (1)} \right) \int_0^{\bar \chi} \ud \tilde \chi\bigg\{ \frac{1}{\tilde \chi} \Perp^j_i  B^{(1)}_{\| }  -\Perp^j_m \tilde \p_{ \perp i} B^{m (1)}-  \frac{1}{\tilde \chi}  \Perp^n_i \Perp^j_m   h_n^{m (1)} + \frac{1}{\tilde \chi}  \Perp^j_i h^{(1)}_{\| }  \nonumber \\
 & & - n^m \Perp^j_n \tilde \p_{ \perp i}  h_m^{n (1)}-  \frac{\tilde \chi}{ \bar \chi} \left[ \Perp^j_i \tilde \p_\| + \left(\bar \chi-\tilde \chi\right) \Perp^n_i \Perp^{jm}\tilde  \p_m \tilde \p_n \right] \left( A^{(1)} - B^{(1)}_{\| } - \frac{1}{2}h^{(1)}_{\| } \right) \bigg\} \nonumber \\
&& +  \left(B^{(1)}_{\perp i \, o }-v^{ (1)}_{\perp i \, o }+ \frac{1}{2} n^k h_{k\,o}^{j(1)} \Perp_{ij} \right) \bigg\{-2\left( B^{i (1)}_{\perp }+ n^m h_m^{l(1)} \Perp^i_l\right) + 4S_{\perp}^{i(1)} + \bar \chi \p^i_{\perp}\left(2v^{(1)}_{\| } + h_l^{l (1)} + 2  \delta_g^{(1)} \right)\nonumber \\
&&+ 2 \p_{\perp i}   \left(\frac{1}{\cH}  \Delta \ln a^{(1)}+   T^{(1)}  \right)  -2 \frac{1}{\bar \chi} \int_0^{\bar \chi} \ud \tilde \chi \left[ \frac{ \bar \chi}{\tilde \chi}\left( B^{i (1)}_{\perp }+ n^k h_{k}^{j(1)} \Perp^i_j\right) + \left(\bar \chi-\tilde \chi\right) \tilde \p^i_\perp \left( A^{(1)} - B^{(1)}_{\| } - \frac{1}{2}h^{(1)}_{\| } \right)\right]\bigg\}\nonumber \\
&& - \left( \frac{\bar a^2}{\bar \rho_m \cH} \Em^{\| (1)}- b_m v_\|^{(1)} \right)^2-2 \left[A^{(1)}-  v^{(1)}_{\| } - \frac{1}{2}h^{(1)}_{\| }- \frac{1}{\cH}\p_\|v^{(1)}_{\| }-\frac{1}{2\cH}h^{(1)}_{\| }{'}-\left(1+\frac{\cH'}{ \cH^2}\right)\Delta \ln a^{(1)}\right] \nonumber \\
&&\times \left( \frac{\bar a^2}{\bar \rho_m \cH} \Em^{\| (1)}- b_m v_\|^{(1)} \right) \;. 
 \end{eqnarray}

This is the main result in a general gauge with no velocity bias. If we explicitly identify the weak lensing shear and rotation contributions, we arrive at Eq. \eqref{Poiss-Deltag-5}.


\section{Perturbations in the Poisson gauge}\label{Sec:PoissonGauge}

The Poisson gauge \cite{Ma:1995ey} is defined by $\p^i {B_i}^{(n)}=\p^j  {h_{ij}}^{(n)}=0$.  
In this case, one  scalar degree of freedom is eliminated from $g_{0i}$ and one scalar and two vector degrees of freedom from $g_{ij}$.  In addition, here we neglect vector and tensor perturbations at  first order. First-order vector perturbations have a decreasing amplitude and are not generated by inflation. The first-order tensor perturbations give a negligible contribution to second-order perturbations. 
Then the metric is 
 \begin{eqnarray} 
 \label{Poiss-metric}
 \ud s^2 = a(\eta)^2\left\{-\left(1 + 2\Phi^{(1)}+\Phi^{(2)}\right)\ud\eta^2+2\omega_{i}^{(2)}\ud\eta \, \ud x^i+\left[\delta_{ij} \left(1 -2\Psi^{(1)}-\Psi^{(2)}\right)+\frac{1}{2}\hat h_{ij}^{(2)}\right]\ud x^i\ud x^j\right\} \;,
\end{eqnarray}
where $A^{(n)}=\Phi^{(n)}$, $B_i^{(1)}=0$,  $B^{(2)}=0$, $\hat B_i^{(2)}=- 2 \omega_i^{(2)}$, $D^{(n)}=-\Psi^{(n)}$, $F^{(n)}=0$,  $\hat F^{(n)}_j=0$ (i.e. $h_{ij}^{(1)}=-2 \delta_{ij} \Psi^{(1)}$ and 
$h_{ij}^{(2)}=-2 \delta_{ij}\Psi^{(2)}+\hat h_{ij}^{(2)}$)\;. 

For the geodesic equation we obtain, at first order,
\begin{eqnarray} 
\label{Poiss-dkmu}
\frac{\ud}{\ud\bar \chi} \left(\delta \nu^{(1)} - 2 \Phi^{(1)}\right) = \Phi^{(1)}{'} +\Psi^{(1)}{'}\;, \quad \quad \quad \quad \quad \frac{\ud}{\ud\bar \chi} \left( \delta n^{i(1)} -2 \Psi^{(1)} n^i \right) = - \p^i \left(\Phi^{(1)} +  \Psi^{(1)}\right), 
\end{eqnarray}
and, at second order,
\begin{eqnarray} 
\label{Poiss-dnu-2}
&&\frac{\ud}{\ud\bar \chi} \left(\delta \nu^{(2)} - 2 \Phi^{(2)} -2\omega_{\|}^{(2)}+4\Phi^{(1)} \delta \nu^{(1)}\right) = \Phi^{(2)}{'} +2\omega_{\|}^{(2)}{'}+ \Psi^{(2)}{'} - \frac{1}{2} \hat h_{\|}^{(2)}{'}+
4\delta n^{i(1)}\p_i  \Phi^{(1)}+4\delta n^{(1)}_\| \Psi^{(1)}{'}   \nonumber \\
 &&+2 \left[2\frac{\ud}{\ud \bar \chi} \Phi^{(1)}{'} +  \left(\Phi^{(1)}{''}  + \Psi^{(1)}{''} \right)    \right] \left(\delta x^{0(1)} + \delta x_{\|}^{(1)}\right)+2  \frac{\ud}{\ud \bar \chi}  \left[2\frac{\ud}{\ud \bar \chi} \Phi^{(1)}+\left(\Phi^{(1)}{'}+ \Psi^{(1)}{'} \right) \right] \delta x_{\|}^{(1)}
 \nonumber \\
&&  + 2 \Bigg\{\p_{\perp i}\left[2\frac{\ud}{\ud \bar \chi} \Phi^{(1)}+\left(\Phi^{(1)}{'} + \Psi^{(1)}{'} \right) \right] - \frac{2}{\bar \chi} \p_{\perp i} \Phi^{(1)} \Bigg\} \delta x_{\perp}^{i (1)}\;,
\end{eqnarray}
\begin{eqnarray} 
&&\frac{\ud}{\ud\bar \chi} \left( \delta n^{i(2)} - 2 \omega^{i(2)}-2\Psi^{(2)} n^i+ \hat h_{j}^{i(2)} n^j- 4 \delta n^{i(1)} \Psi^{(1)} \right) = 
-  \p^i \left(\Phi^{(2)} + \Psi^{(2)} \right) -2 \p^i\omega_{\|}^{(2)}+ \frac{2}{\bar \chi} \omega_{\perp}^{i(2)}  + \frac{1}{2} \p^i \hat h_{\|}^{(2)}  \nonumber \\
&& - \frac{1}{\bar \chi} \Perp^{ij} \hat h_{jk}^{(2)} n^k + 4 \delta \nu^{(1)}\left(\p^i \Phi^{(1)}  +n^i \Psi^{(1)} {'} \right)- 4 \delta n^{(1)}_\| \p^i \Psi^{(1)} + 4 \delta n^{j(1)} n^i  \p_j \Psi^{(1)}  \nonumber \\
 && -2   {\left[\p^i\left( \Phi^{(1)}+ \Psi^{(1)} \right)   -2  n^i \frac{\ud}{\ud \bar \chi} \Psi^{(1)} \right]}'  \left(\delta x^{0(1)} + \delta x_{\|}^{(1)}\right)  -2 \frac{\ud}{\ud \bar \chi} \left[\p^i\left( \Phi^{(1)} + \Psi^{(1)} \right)   -2  n^i \frac{\ud}{\ud \bar \chi} \Psi^{(1)}  \right] \delta x_{\|}^{(1)}
 \nonumber \\
 && -2 \Bigg\{ \p_{\perp l} \left[\p^i\left( \Phi^{(1)}+ \Psi^{(1)} \right)  -2 n^i  \frac{\ud}{\ud \bar \chi}  \Psi^{(1)}\right]   + \frac{2}{\bar \chi} \Perp_l^j \left(  \delta_j^i\frac{\ud}{\ud \bar \chi}   \Psi^{(1)}  + n^i  \p_j  \Psi^{(1)}\right) \Bigg\}  \delta x_{\perp}^{l (1)} \;.
\end{eqnarray}

Using the constraints
\begin{eqnarray}
\label{Poiss-dnude-o}
\delta\nu^{(1)}_o&=&\Phi^{(1)}_o+v^{(1)}_{\| o}\;, \quad \quad \quad \quad \quad \delta n^{\hat a (1)}_o=-v^{\hat a (1)}_o + n^{\hat a} \Psi^{(1)}_{ \, o}  \;, \nonumber \\
\delta\nu^{(2)}_o&=&  \Phi^{(2)}_o+ v^{(2)}_{\| o}+ 2 \omega^{(2)}_{\| o}- 3 \left(\Phi^{(1)}_o\right)^2 -2 v^{(1)}_{\| o} \Phi^{(1)}_o -  v_{k \, o}^{(1)} v_o^{k (1)} - 2 \Psi^{(1)}_{ \, o}  v^{(1)}_{\| \, o} \nonumber \\
 \delta n^{\hat a (2)}_o&=&- v^{\hat a (2)}_o+ n^{\hat a} \Psi^{(2)}_{\, o} -\frac{1}{2}  n^i \hat h_{i \,o}^{\hat a (2)} +  v_o^{\hat a (1)}  v_{\| \, o}^{(1)}+  3 \, n^{\hat a}  \left(\Psi^{(1)}_{ \, o}\right)^2   \;,
\end{eqnarray}
we obtain at first order
\begin{eqnarray}
\label{Poiss-dnude-1}
\delta\nu^{(1)}&=&- \left (\Phi^{(1)}_o-v^{(1)}_{\| \, o}\right)+ 2 \Phi^{(1)}  + \int_0^{\bar \chi} \ud \tilde \chi \left(\Phi^{(1)}{'} +  \Psi^{(1)}{'} \right) = - \left (\Phi^{(1)}_o-v^{(1)}_{\| \, o}\right)+ 2 \Phi^{(1)} - 2I^{(1)} \;, \\
\delta n^{i (1)}&=& -v^{i (1)}_o- n^i \Psi^{(1)}_{\, o}  +2 n^i \Psi^{(1)} - \int_0^{\bar \chi} \ud \tilde \chi\,  \tilde \p^i\left( \Phi^{(1)} +  \Psi^{(1)}  \right) = n^i \delta n_\|^{ (1)}+\delta n_\perp^{i (1)}\;,
\end{eqnarray}
where
\begin{eqnarray}
\label{Poiss-dnue-||perp-1}
\delta n_\|^{ (1)}=\Phi^{(1)}_o-v^{(1)}_{\| \, o}-\Phi^{(1)}+ \Psi^{(1)}+2I^{(1)}\;,   \quad \quad \quad \delta n_\perp^{i (1)}=  -v^{i (1)}_{\perp \, o }+ 2S_{\perp}^{i(1)}  \;,
\end{eqnarray}
\begin{eqnarray}
\label{Poiss-iota}
I^{(1)}& =& -\frac{1}{2} \int_0^{\bar \chi} \ud \tilde \chi\left(\Phi^{(1)}{'} +  \Psi^{(1)}{'} \right) \;,\\
\label{Poiss-varsigma}
S^{i(1)} &=& -\frac{1}{2} \int_0^{\bar \chi} \ud \tilde \chi \left[ \tilde\p^i \left( \Phi^{(1)} +\Psi^{(1)}\right) -\frac{2}{\tilde \chi} n^i\Psi^{(1)} \right]\;, \\
S_{\perp}^{i(1)} &=& -\frac{1}{2} \int_0^{\bar \chi} \ud \tilde \chi \, \tilde\p^i_\perp \left( \Phi^{(1)}  + \Psi^{(1)}\right)\;,\\
S_{\|}^{(1)} &=& n_i S^{i(1)} = \frac{1}{2} \left( \Phi^{(1)}_o  +\Psi^{(1)}_{\| \, o} \right)- \frac{1}{2} \left( \Phi^{(1)} + \Psi^{(1)} \right)+ I^{(1)} + \int_0^{\bar \chi} \ud \tilde \chi \frac{\Psi^{(1)}}{\tilde \chi}.
\end{eqnarray}
Note the following useful relation 
\begin{eqnarray}
\label{Poiss-du+de}
\delta n_\|^{ (1)} +  \delta\nu^{(1)}&=&\Phi^{(1)}+ \Psi^{(1)}\;.
\end{eqnarray}

At second order we find
\begin{eqnarray}
\label{Poiss-dnu2-2}
 \delta\nu^{(2)}&=&- \Phi^{(2)}_o+ v^{(2)}_{\| \, o}+\left(\Phi^{(1)}_o\right)^2+ 6\Phi^{(1)}_o v^{(1)}_{\| \, o}- v^{(1)}_{k\, o} v^{k (1)}_o -2 \Psi^{(1)}_{\, o} v^{(1)}_{\| \, o} +4 \left(\Phi^{(1)}_o-v^{(1)}_{\| \, o}\right) \left( 2 \Phi^{(1)}  -2 I^{(1)}\right)\nonumber \\
&&- 4\, v_{\perp \, o}^{i(1)}  \int_0^{\bar \chi} \, \ud \tilde \chi \left( \tilde\p_{\perp i}\Phi^{(1)} \right)+2\Phi^{(2)}+ 2 \omega^{(2)}_{\| }-12 \left(\Phi^{(1)} \right)^2 +16 \, \Phi^{(1)} I^{(1)} - 2 I^{(2)} \nonumber \\
&& + 4 \int_0^{\bar \chi} \ud \tilde \chi \bigg\{\left(\Phi^{(1)} + \Psi^{(1)} \right)  \frac{\ud}{\ud \tilde \chi} \Phi^{(1)}  + \left(\Phi^{(1)}{'}  +  \Psi^{(1)}{'} \right) \left( \Psi^{(1)} + 2 I^{(1)} \right)  +  2 \tilde\p_{\perp i}\left(\Phi^{(1)}\right) S_{\perp }^{i(1)} \bigg\}\nonumber \\
&&+  \delta\nu^{(2)}_{\rm post-Born}\;,
\end{eqnarray}
where
    \begin{eqnarray}
   && \delta\nu^{(2)}_{\rm post-Born}= -4 \Phi^{(1)}_o  \left(3 v_{\| \, o}^{(1)} - \Psi^{(1)}_o\right) - 4 \left(\Phi^{(1)}_o-v^{(1)}_{\| \, o}\right)\left(  \Phi^{(1)} -   I^{(1)} \right) - 8 v_{\| \, o}^{(1)} \int_0^{\bar \chi}   \frac{\ud \tilde{\chi}}{\tilde \chi} \Phi^{(1)}   +4 v^{i (1)}_{\perp \, o }   \int_0^{\bar \chi}  \ud \tilde{\chi}   \tilde \p_{\perp i} \Phi^{(1)}  \nonumber\\
          &&  +4 \Phi^{(1)}  \left( \Phi^{(1)} - \Psi^{(1)} -2I^{(1)} -2\kappa^{(1)} \right)  + 2\left[2\frac{\ud}{\ud \bar \chi} \Phi^{(1)} +\left(\Phi^{(1)}{'} + \Psi^{(1)}{'} \right) \right] \delta x_{\|}^{(1)}  +4 \Phi^{(1)}{'} \left(\delta x^{0(1)} + \delta x_{\|}^{(1)}\right)  \nonumber\\
    &&+4  \p_{\perp i} \Phi^{(1)} \delta x_{\perp}^{i (1)}   +2 \int_0^{\bar \chi}  \ud \tilde{\chi} \Bigg\{   \left(\Phi^{(1)}{''} + \Psi^{(1)}{''} \right)   \left(\delta x^{0(1)} + \delta x_{\|}^{(1)}\right)  -2 \Phi^{(1)}{'}  \left(\Phi^{(1)} + \Psi^{(1)}\right)   \nonumber\\
      &&   +  \left(\Phi^{(1)}{'} + \Psi^{(1)}{'} \right)   \left(\Phi^{(1)}- \Psi^{(1)} -2I^{(1)}\right)  - 2 \Phi^{(1)}   \left[\frac{\ud}{\ud \tilde \chi}\left(\Phi^{(1)}- \Psi^{(1)}\right)+\left(\Phi^{(1)}{'} + \Psi^{(1)}{'} \right) \right] \nonumber \\
 &&    - 4  \tilde \p_{\perp i}  \Phi^{(1)}  S_{\perp}^{i(1)} -  4 \Phi^{(1)}   \tilde \p_{\perp j}S_{\perp}^{j(1)}   + 4 \bigg( \frac{\ud}{\ud \tilde \chi} \Phi^{(1)}    -  \frac{1}{\tilde\chi} \Phi^{(1)}   \bigg) \kappa^{(1)} + \bigg[  \tilde \p_{\perp i}   \left(\Phi^{(1)}{'} +\Psi^{(1)}{'} \right)  \bigg]   \delta x_{\perp}^{i (1)} \Bigg\} \;.  \nonumber
 \end{eqnarray}
 
and, splitting $\delta n^{i(2)}= n^i \delta n_\|^{ (2)}+\delta n_\perp^{i (2)}$, we obtain
\begin{eqnarray}
\label{Poiss-de_||-2}  
 \delta n_\|^{(2)}&=&\Phi^{(2)}_o- v^{(2)}_{\| \, o} +  \left(v^{(1)}_{\| \, o}\right)^2- \left(\Psi^{(1)}_{\, o}\right)^2-4\Phi^{(1)}_o v^{(1)}_{\| \, o}+4 v^{(1)}_{\| \, o} \Psi^{(1)}_{\, o} - 4\left (\Phi^{(1)}_o-v^{(1)}_{\| \, o}\right)\left( \Phi^{(1)} - \Psi^{(1)} \right) + 8  \left (\Phi^{(1)}_o-v^{(1)}_{\| \, o}\right) I^{(1)} \nonumber \\
&-&4v_{\perp \, o}^{i(1)} \int_0^{\bar \chi} \ud \tilde \chi \left( \p_{\perp i} \Psi^{(1)} \right)  -\Phi^{(2)}+ \Psi^{(2)} -\frac{1}{2}\hat h_{\|}^{(2)} + 4 \left(\Psi^{(1)}\right)^2  +4 \left( \Phi^{(1)} \right)^2 - 4\Phi^{(1)} \Psi^{(1)}   -8 \left (  \Phi^{(1)} - \Psi^{(1)} \right) I^{(1)}\nonumber \\
 &+& 2 I^{(2)} + 4 \int_0^{\bar \chi} \ud \tilde \chi \left[ \left( \Phi^{(1)}  - 2 I^{(1)}\right) \left(\Phi^{(1)}{'}  +  \Psi^{(1)}{'} \right)+2  \,  \tilde \p_{\perp i} \Psi^{(1)}   S_{\perp }^{i(1)} \right]+  \delta n^{(2)}_{\| \rm post-Born} \;, 
\end{eqnarray}
where
    \begin{eqnarray}
  &&  \delta n^{(2)}_{\| \rm post-Born} =  2 \left( \Phi^{(1)}_o - \Psi^{(1)}_o\right) \left(3 v_{\| \, o}^{(1)} - \Psi^{(1)}_o \right) -  \left(\Phi^{(1)}_o - \Psi^{(1)}_{o}\right)^2  +2  \left(\Phi^{(1)}_o-v^{(1)}_{\| \, o} \right) \left(\Phi^{(1)} - \Psi^{(1)}-2  I^{(1)} \right) \nonumber\\
 && +  4   v_{\| \, o}^{(1)}  \int_0^{\bar \chi}   \frac{\ud \tilde{\chi} }{\tilde \chi} \left( \Phi^{(1)} - \Psi^{(1}\right)    - 2 v^{i (1)}_{\perp \, o }  \int_0^{\bar \chi}  \ud \tilde{\chi} \left[ \tilde \p_{\perp i} \left(\Phi^{(1)} - \Psi^{(1)} \right)\right]   
- 2\bigg[\frac{\ud}{\ud \bar \chi}\left(\Phi^{(1)}- \Psi^{(1)}\right) +\left(\Phi^{(1)}{'} + \Psi^{(1)}{'} \right) \bigg] \delta x_{\|}^{(1)}\nonumber\\
&&  -2 \left(\Phi^{(1)}{'} - \Psi^{(1)}{'} \right)\left(\delta x^{0(1)} + \delta x_{\|}^{(1)}\right)- \left(\Phi^{(1)} -\Psi^{(1)}\right) \left(\Phi^{(1)} - \Psi^{(1)}-4I^{(1)}\right) 
+ 4  \left(\Phi^{(1)}-\Psi^{(1)}\right)  \kappa^{(1)} \nonumber\\
    && -2  \left[\p_{\perp i}\left(\Phi^{(1)} - \Psi^{(1)} \right) \right] \delta x_{\perp}^{i (1)}  +2 \int_0^{\bar \chi}  \ud \tilde{\chi} \Bigg\{ - \left(\Phi^{(1)}{''} + \Psi^{(1)}{''} \right)   \left(\delta x^{0(1)} + \delta x_{\|}^{(1)}\right)  \nonumber \\
 && + \left(\Phi^{(1)}{'} -\Psi^{(1)}{'}\right) \left(\Phi^{(1)} + \Psi^{(1)}\right)  +2  \left(\Phi^{(1)}{'} +  \Psi^{(1)}{'} \right) I^{(1)}      + 2 S_{\perp}^{i(1)} \tilde \p_{\perp i} \left(\Phi^{(1)} -\Psi^{(1)} \right) \nonumber \\
 &&     + 2 \left( \Phi^{(1)} - \Psi^{(1)}\right)    \tilde \p_{\perp j}S_{\perp}^{j(1)}  + 2 \bigg[ - \frac{\ud}{\ud \tilde \chi}\left(\Phi^{(1)} - \Psi^{(1)} \right)  +  \frac{1}{\tilde\chi} \left(\Phi^{(1)} - \Psi^{(1)} \right)  \bigg] \kappa^{(1)}  -   \tilde \p_{\perp i}   \left(\Phi^{(1)}{'}+  \Psi^{(1)}{'} \right)    \delta x_{\perp}^{i (1)} \Bigg\} \;,  \nonumber
 \end{eqnarray} 
and
\begin{eqnarray}
\label{Poiss-de_perp-2}  
&& \delta n_\perp^{i(2)}= - 2 \omega^{i(2)}_{\perp \, o}-  v^{i(2)}_{\perp \, o} + \frac{1}{2} n^j \hat h_{ j k\, o}^{(2)} \Perp^{ki}+   v^{(1)}_{\| \, o}v^{i(1)}_{\perp \, o}+ 4 v^{i (1)}_{\perp o}  \Psi^{(1)}_{\, o}  +8 \left(\Phi^{(1)}_o-v^{(1)}_{\| \, o}\right)  S_{\perp}^{i(1)} - 4 v^{i (1)}_{\perp \, o }   \Psi^{(1)} +2\omega^{i(2)}_{\perp}\nonumber \\
&& -   n^j \hat h_{jk}^{(2)} \Perp^{ki} +8 \, \Psi^{(1)} S_{\perp}^{i(1)}+2 S_{\perp }^{i(2)} + 4 \int_0^{\bar \chi} \ud \tilde \chi \left\{  +  2 \left( \Phi^{(1)}  -  I^{(1)}\right) \tilde\p^i_\perp \left( \Phi^{(1)} +  \Psi^{(1)} \right) -  \left( \Phi^{(1)}  + \Psi^{(1)} \right) \tilde\p^i_\perp \Psi^{(1)} \right\}   \nonumber\\
&&+  \delta n^{(2)}_{\perp \rm post-Born}\;,\nonumber \\
\end{eqnarray}
where
\begin{eqnarray}
 && +  \delta n^{(2)}_{\perp \rm post-Born} = - 2 v_{\perp o}^{i (1)}  \left( \Phi^{(1)}_o + \Psi^{(1)}_o\right)  -4  \left(\Phi^{(1)}_o-v^{(1)}_{\| \, o} \right) S_{\perp}^{i(1)}   - 2 v_{\perp o}^{i (1)}  \int_0^{\bar \chi} \frac{ \ud \tilde{\chi}}{\tilde \chi}  \left(\Phi^{(1)} + \Psi^{(1)} \right)      \nonumber\\
&& -2 \p^i_{\perp }  \left(\Phi^{(1)}+ \Psi^{(1)} \right)  \delta x_{\|}^{(1)}  -  \frac{2}{\bar \chi}  \left(\Phi^{(1)} + \Psi^{(1)} \right)  \delta x_{\perp}^{i (1)}    +2 \int_0^{\bar \chi}  \ud \tilde{\chi} \Bigg\{ -  \tilde \p_{\perp }^i    \left(\Phi^{(1)}{'} + \Psi^{(1)}{'} \right)    \left(\delta x^{0(1)} + \delta x_{\|}^{(1)}\right) \nonumber\\
&&    -  \tilde \p_{\perp }^i    \left(\Phi^{(1)}+ \Psi^{(1)}  \right)  \left(\Phi^{(1)}-\Psi^{(1)} -2 I^{(1)}\right) +   \frac{2}{\tilde \chi} \left(\Phi^{(1)}+ \Psi^{(1)} \right) S_{\perp}^{i(1)}
\nonumber\\
&&
  -   \bigg[ \Perp^{im} \tilde \p_{\perp j}   \p_{\perp m}    \left(\Phi^{(1)}+ \Psi^{(1)} \right) +  \frac{1}{\tilde\chi} \Perp^i_j  \left(\Phi^{(1)}{'} + \Psi^{(1)}{'}  \right) +   \frac{1}{\tilde\chi^2} \Perp^i_j \left(\Phi^{(1)} + \Psi^{(1)}  \right) \bigg]  \delta x_{\perp}^{j (1)}      
      \Bigg\}   \;.
 \nonumber
 \end{eqnarray}
 
Here
\begin{eqnarray}
\label{Poiss-iota}
I^{(2)}& =& -\frac{1}{2} \int_0^{\bar \chi} \ud \tilde \chi \left(\Phi^{(2)}{'} +2  \omega^{(2)}_{\| }{'} + \Psi^{(2)}{'}- \frac{1}{2} \hat h^{(2)}_{\| }{'} \right)\;,
\\
S_{\perp}^{i(2)} &=& -\frac{1}{2} \int_0^{\bar \chi} \ud \tilde \chi \left[ \tilde\p^i_\perp \left( \Phi^{(2)} +2  \omega^{(2)}_{\| } + \Psi^{(2)}- \frac{1}{2} \hat h^{(2)}_{\| }\right) + \frac{1}{\tilde \chi} \left(-2 \omega^{i (2)}_{\perp }+  n^k \hat h_{kj}^{(2)} \Perp^{ij}  \right)\right]\;.
\label{Poiss-varsigma}
\end{eqnarray}

Combining Eqs.\  (\ref{dnu-2}) and (\ref{de_||-2}),  we  obtain 
\begin{eqnarray} 
\label{Poiss-dnu+de_||-2}
 \delta\nu^{(2)} +  \delta n_\|^{(2)}&=&+\left(\Phi^{(1)}_o\right)^2 + 2\Phi^{(1)}_o v^{(1)}_{\| \, o}+2 v^{(1)}_{\| \, o} \Psi^{(1)}_{\, o} - \left( \Psi^{(1)}_{\, o}\right)^2  -   v^{(1)}_{\perp k \, o} v^{k (1)}_{\perp \, o}+4 \left (\Phi^{(1)}_o-v^{(1)}_{\| \, o}\right) \left (  \Phi^{(1)} + \Psi^{(1)} \right)  \nonumber \\
&& + 8 \, v_{\perp i \, o}^{(1)} \, S_{\perp }^{i (1)} + \Phi^{(2)}+ 2 \omega^{(2)}_{\| } +  \Psi^{(2)} -\frac{1}{2}\hat h_{\|}^{(2)}-8\left( \Phi^{(1)} \right)^2 - 4   \Psi^{(1)}  \left(  \Phi^{(1)} - \Psi^{(1)} \right) +8 \left(\Phi^{(1)}+ \Psi^{(1)}\right)  I^{(1)}   \nonumber \\
&&  -8 S_{\perp }^{i(1)}S_{\perp }^{j(1)} \delta_{ij}  + 4 \int_0^{\bar \chi} \ud \tilde \chi \bigg\{ \left(\Phi^{(1)}+ \Psi^{(1)} \right) \left[\left(\Phi^{(1)}{'} + \Psi^{(1)}{'} \right)+ \frac{\ud}{\ud \tilde \chi} \Phi^{(1)} \right] \bigg\} \nonumber \\
&&+\left(\delta\nu^{(2)} +  \delta n_\|^{(2)}\right)_{ \rm post-Born}\;,
\end{eqnarray}
where
 \begin{eqnarray}
&&  \left(\delta\nu^{(2)} +  \delta n_\|^{(2)}\right)_{ \rm post-Born}= -2  \left( \Phi^{(1)}_o + \Psi^{(1)}_o\right) \left(3 v_{\| \, o}^{(1)} - \Psi^{(1)}_o \right)  -2 \left(\Phi^{(1)}_o-v^{(1)}_{\| \, o} \right)  \left(\Phi^{(1)} +  \Psi^{(1)} \right)   - 4 v^{j (1)}_{\perp \, o }    S_{\perp}^{j(1)}    - 4  v_{\| \, o}^{(1)}  \int_0^{\bar \chi}   \frac{\ud \tilde{\chi}}{\tilde \chi} \left( \Phi^{(1)} + \Psi^{(1)} \right)     \nonumber \\
 && + 2 \frac{\ud}{\ud \bar \chi}\left(\Phi^{(1)}  + \Psi^{(1)} \right) \delta x_{\|}^{(1)}  + 2 \left(\Phi^{(1)}{'} + \Psi^{(1)}{'} \right)\left(\delta x^{0(1)} + \delta x_{\|}^{(1)}\right)  + 2 \left(\Phi^{(1)}  + \Psi^{(1)}\right) \left(\Phi^{(1)} - \Psi^{(1)} - 2 I^{(1)}\right)  \nonumber \\
 &&   +2  \p_{\perp i}\left(\Phi^{(1)} + \Psi^{(1)} \right)   \delta x_{\perp}^{i (1)}   - 4 \left(\Phi^{(1)}  + \Psi^{(1)}\right)\kappa^{(1)}  +2 \int_0^{\bar \chi}  \ud \tilde{\chi} \Bigg\{   - \left( \Phi^{(1)} + \Psi^{(1)} \right)  \bigg[\frac{\ud}{\ud \tilde \chi}\left(\Phi^{(1)} - \Psi^{(1)}\right)  \nonumber \\
 && + 2 \left(\Phi^{(1)}{'} + \Psi^{(1)}{'} \right) \bigg]    - 2 \tilde \p_{\perp j}\left( \Phi^{(1)}+  \Psi^{(1)} \right)     S_{\perp}^{j(1)}    - 2  \left( \Phi^{(1)} + \Psi^{(1)} \right)   \tilde \p_{\perp m}S_{\perp}^{m(1)}   \nonumber \\
 && + 2 \bigg[ \frac{\ud}{\ud \tilde \chi}\left( \Phi^{(1)} + \Psi^{(1)} \right)    -  \frac{1}{\tilde\chi} \left( \Phi^{(1)} + \Psi^{(1)} \right)  \bigg] \kappa^{(1)}  \Bigg\} \;.   \nonumber
\end{eqnarray}

From Eqs (\ref{deltax0-n}) and (\ref{deltaxi-n}), we find
\begin{eqnarray}
\label{Poiss-dx0-1}
\delta x^{0(1)}&=& -\bar \chi \left (\Phi^{(1)}_o-v^{(1)}_{\| \, o}\right)+ \int_0^{\bar \chi} \ud \tilde \chi \left[ 2 \Phi^{(1)} + \left(\bar \chi-\tilde \chi\right) \left(\Phi^{(1)}{'}+ \Psi^{(1)}{'} \right) \right] \; \\
\label{Poiss-dx||-1}
\delta x_{\|}^{(1)}&=& \bar \chi \left(\Phi^{(1)}_o-v^{(1)}_{\| \, o}\right)-\int_0^{\bar \chi} \ud \tilde \chi \left[ \left(\Phi^{(1)}- \Psi^{(1)}\right) +  \left(\bar \chi-\tilde \chi\right) \left(\Phi^{(1)}{'} +\Psi^{(1)}{'} \right) \right] \;,   \\
\label{Poiss-dxperp-1}
\delta x_{\perp}^{i (1)}&=& - \bar \chi \, v^{i (1)}_{\perp \, o }-\int_0^{\bar \chi} \ud \tilde \chi \left[ \left(\bar \chi-\tilde \chi\right) \tilde \p^i_\perp \left( \Phi^{(1)} + \Psi^{(1)} \right) \right] \;,
\end{eqnarray}
to first order. Using Eq.\ (\ref{du+de}), we note that
\begin{eqnarray}
\label{Poiss-s-1}
 T^{(1)}=-\left( \delta x^{0(1)} + \delta x_{\|}^{(1)}\right) = -  \int_0^{\bar \chi} \ud \tilde \chi \left(\Phi^{(1)} + \Psi^{(1)}\right)\;.
\end{eqnarray}
At second order,
\begin{eqnarray}
\label{Poiss-dx0-2}
 \delta x^{0(2)}&=& \bar \chi\left[-\Phi^{(2)}_o+ v^{(2)}_{\| \, o}+\left(\Phi^{(1)}_o\right)^2 + 6\Phi^{(1)}_o v^{(1)}_{\| \, o}- v^{(1)}_{k\, o} v^{k (1)}_o  -2 \Psi^{(1)}_{\, o} v^{(1)}_{\| \,o}\right] \nonumber \\
&+& 4  \left (\Phi^{(1)}_o-v^{(1)}_{\| \, o}\right)  \int_0^{\bar \chi} \ud \tilde \chi \left[ 2 \Phi^{(1)}  + \left(\bar \chi-\tilde \chi\right) \left(\Phi^{(1)}{'}+ \Psi^{(1)}{'} \right) \right]\nonumber \\
&-&4 v_{\perp \, o}^{i(1)}\int_0^{\bar \chi} \ud \tilde \chi \left(\bar \chi-\tilde \chi\right) \tilde \p_{\perp i}\Phi^{(1)} +2 \int_0^{\bar \chi} \ud \tilde \chi \bigg[ \Phi^{(2)}+  \omega^{(2)}_{\| }-6\left(\Phi^{(1)}\right)^2+ 8 \Phi^{(1)}  I^{(1)} \bigg]  \nonumber \\
&+&  \int_0^{\bar \chi} \ud \tilde \chi\left(\bar \chi-\tilde \chi\right) \left\{\Phi^{(2)}{'} + 2 \omega^{(2)}_{\| }{'} + \Psi^{(2)}- \frac{1}{2} \hat h^{(2)}_{\| }{'} + 4 \left(\Phi^{(1)}+ \Psi^{(1)} \right)  \frac{\ud}{\ud \tilde \chi}\Phi^{(1)} \right. \nonumber \\
&+& \left.   4 \left(\Phi^{(1)}{'} + \Psi^{(1)}{'} \right) \left(\Psi^{(1)} + 2 I^{(1)} \right)   + 8 S_{\perp }^{i(1)} \tilde \p_{\perp i} \Phi^{(1)}    \right\}+ \delta x^{0(2)}_{\rm post-Born}\;,
\end{eqnarray}
where
    \begin{eqnarray}
    && \delta x^{0(2)}_{\rm post-Born} = -4\bar \chi  \Phi^{(1)}_o  \left(3 v_{\| \, o}^{(1)} - \Psi^{(1)}_o \right) + 4  \left(\Phi^{(1)}_o-v^{(1)}_{\| \, o}\right) \bigg[ \bar \chi \Phi^{(1)}   -   \int_0^{\bar \chi} \ud \tilde \chi  \left(2 \Phi^{(1)} - I^{(1)} \right)    \bigg]    - 8 v_{\| \, o}^{(1)}  \int_0^{\bar \chi}  \ud \tilde{\chi} ~ \frac{ (\bar \chi - \tilde \chi)}{\tilde \chi}  \Phi^{(1)}   \nonumber\\
      && + 4 v^{i (1)}_{\perp \, o }  \int_0^{\bar \chi}  \ud \tilde{\chi} ~ (\bar \chi - \tilde \chi)   \tilde \p_{\perp i} \Phi^{(1)}  - 4 \Phi^{(1)}  \int_0^{\bar \chi} \ud \tilde \chi  \left(\Phi^{(1)}  -  \Psi^{(1)} - 2I^{(1)} \right)    
+2\int_0^{\bar \chi} \ud \tilde{\chi} \Bigg\{  \left(\Phi^{(1)}{'} +  \Psi^{(1)}{'} \right) \delta x_{\|}^{(1)}  \nonumber \\
 &&- 2  \Phi^{(1)}{'} T^{(1)}  + 4 \Phi^{(1)}   \left(\Phi^{(1)} - \Psi^{(1)} - 2I^{(1)} - \kappa^{(1)}\right) + 2 \tilde \p_{\perp i} \Phi^{(1)} \delta x_{\perp}^{ i (1)} \Bigg\}  +2 \int_0^{\bar \chi}  \ud \tilde{\chi} ~ (\bar \chi - \tilde \chi) \Bigg\{  - \left(\Phi^{(1)}{''} +  \Psi^{(1)}{''} \right)   T^{(1)} \nonumber\\
    &&     -  2 \Phi^{(1)}{'}  \left(\Phi^{(1)}+  \Psi^{(1)} \right)  + \left(\Phi^{(1)}{'} +  \Psi^{(1)}{'} \right)  \left( \Phi^{(1)}  - \Psi^{(1)} - 2I^{(1)}\right)  -  2  \Phi^{(1)}   \bigg[\frac{\ud}{\ud \tilde \chi}\left(\Phi^{(1)} - \Psi^{(1)} \right)  +\left(\Phi^{(1)}{'} + \Psi^{(1)}{'} \right) \bigg]   \nonumber \\
 &&  -4 \tilde \p_{\perp i}  \Phi^{(1)} S_{\perp}^{i(1)}  - 4 \Phi^{(1)}    \tilde \p_{\perp m}S_{\perp}^{m(1)}  + 4  \left( \frac{\ud}{\ud \tilde \chi} \Phi^{(1)}     -  \frac{1}{\tilde\chi} \Phi^{(1)} \right)  \kappa^{(1)}  + \tilde \p_{\perp i}  \left(\Phi^{(1)}{'} + \Psi^{(1)}{'} \right) \delta x_{\perp}^{i (1)}\Bigg\} \;,  \nonumber
 \end{eqnarray}

\begin{eqnarray}
\label{Poiss-dx_||-2}  
\delta x_\|^{(2)}&=&\bar \chi \left[ \Phi^{(2)}_o-  v^{(2)}_{\| \, o} + \left(v^{(1)}_{\| \, o}\right)^2- \left(\Psi^{(1)}_{\, o}\right)^2-4\Phi^{(1)}_o v^{(1)}_{\| \, o}+ 4v^{(1)}_{\| \, o} \Psi^{(1)}_{\, o} \right] \nonumber \\
&-& 4 \left (\Phi^{(1)}_o-v^{(1)}_{\| \, o}\right)  \int_0^{\bar \chi} \ud \tilde \chi \left[\left ( \Phi^{(1)} - \Psi^{(1)} \right)+ \left(\bar \chi-\tilde \chi\right) \left(\Phi^{(1)}{'}+ \Psi^{(1)}{'} \right) \right]-4  v_{\perp \, o}^{i(1)}  \int_0^{\bar \chi} \ud \tilde \chi  \left(\bar \chi-\tilde \chi\right)  \tilde \p_{\perp i}\Psi^{(1)} \nonumber \\
&+&  \int_0^{\bar \chi} \ud \tilde \chi \bigg\{- \Phi^{(2)} + \Psi^{(2)}-\frac{1}{2}\hat h_{\|}^{(2)}-4 \Psi^{(1)}\Phi^{(1)}+ 4 \left(\Psi^{(1)}\right)^2 + 4 \left(\Phi^{(1)}\right)^2 -8 \left ( \Phi^{(1)} - \Psi^{(1)} \right) I^{(1)}\bigg\} \nonumber \\
 &+&\int_0^{\bar \chi} \ud \tilde \chi  \left(\bar \chi-\tilde \chi\right)  \left\{- \left(\Phi^{(2)}{'}+2 \omega^{(2)}_{\| }{'} + \Psi^{(2)}{'}  - \frac{1}{2} \hat h^{(2)}_{\| }{'} \right)+ 4 \left( \Phi^{(1)}  - 2 I^{(1)}\right)  \left(\Phi^{(1)}{'}  + \Psi^{(1)}{'} \right)    + 8  S_{\perp }^{i(1)} \tilde \p_{\perp i} \Psi^{(1)}  \right\}\nonumber \\
 &+& \delta x^{(2)}_{\| \rm post-Born} \;,  
 \end{eqnarray}
where
\begin{eqnarray}
 &&\delta x^{(2)}_{\| \rm post-Born} = 2\bar \chi \left( \Phi^{(1)}_o - \Psi^{(1)}_o\right) \left(3 v_{\| \, o}^{(1)} -  \Psi^{(1)}_o \right)    - \bar \chi \left(\Phi^{(1)}_o - \Psi^{(1)}_o \right)^2  + 4\left(\Phi^{(1)}_o-v^{(1)}_{\| \, o} \right) \bigg[-\frac{\bar \chi}{2}\left(\Phi^{(1)} - \Psi^{(1)} \right) 
  \nonumber\\
    && +  \int_0^{\bar \chi} \ud \tilde{\chi}   \left(\Phi^{(1)} - \Psi^{(1)} - I^{(1)} \right) \bigg]
 + 4 v_{\| \, o}^{(1)}   \int_0^{\bar \chi}   \ud \tilde{\chi} ~ \frac{(\bar \chi - \tilde \chi)  }{\tilde \chi}\left( \Phi^{(1)} - \Psi^{(1)}\right)    \nonumber\\
 &&  - 2  v^{i (1)}_{\perp \, o }  \int_0^{\bar \chi}   \ud \tilde{\chi} \left[ (\bar \chi - \tilde \chi)  \tilde \p_{\perp i} \left(\Phi^{(1)}  -  \Psi^{(1)} \right)\right]   +2\left(\Phi^{(1)} -  \Psi^{(1)}\right) \int_0^{\bar \chi} \ud \tilde \chi \left(\Phi^{(1)}  -  \Psi^{(1)}  - 2I^{(1)} \right) \nonumber\\
    &&   +2\int_0^{\bar \chi} \ud \tilde{\chi} \Bigg\{    -\left(\Phi^{(1)}{'} + \Psi^{(1)}{'} \right)  \delta x_{\|}^{(1)}  - \left(\Phi^{(1)}-\Psi^{(1)} \right) \left[ \frac{3}{2}\left(\Phi^{(1)} - \Psi^{(1)} \right)  -4I^{(1)}\right] + \left(\Phi^{(1)}{'} - \Psi^{(1)}{'} \right)  T^{(1)}   \nonumber\\
         &&  +2 \left(\Phi^{(1)} -  \Psi^{(1)} \right)   \kappa^{(1)} -   \tilde \p_{\perp j}  \left(\Phi^{(1)} - \Psi^{(1)} \right)  \delta x_{\perp}^{j (1)} \Bigg\} +2 \int_0^{\bar \chi}   \ud \tilde{\chi} ~ (\bar \chi - \tilde \chi)\Bigg\{  \left(\Phi^{(1)}{'} - \Psi^{(1)}{'}\right) \left(\Phi^{(1)} + \Psi^{(1)} \right) \nonumber\\
&&    
+  \left(\Phi^{(1)}{''} +  \Psi^{(1)}{''} \right)  T^{(1)}  +2\left(\Phi^{(1)}{'} +  \Psi^{(1)}{'} \right) I^{(1)}    + 2 \tilde \p_{\perp i} \left(\Phi^{(1)} - \Psi^{(1)} \right) S_{\perp}^{i(1)}  \nonumber\\
    && + 2  \left( \Phi^{(1)} - \Psi^{(1)} \right)    \tilde \p_{\perp m}S_{\perp}^{m(1)} + 2 \bigg[ - \frac{\ud}{\ud \tilde \chi}\left(\Phi^{(1)} - \Psi^{(1)} \right)    +  \frac{1}{\tilde\chi} \left(\Phi^{(1)} - \Psi^{(1)} \right)  \bigg] \kappa^{(1)}  - \bigg[  \tilde \p_{\perp i}  \left(\Phi^{(1)}{'} + \Psi^{(1)}{'} \right) \bigg]   \delta x_{\perp}^{i (1)} \Bigg\} \;,  \nonumber
 \end{eqnarray}
 
and

\begin{eqnarray}
\label{Poiss-dx_perp-2}  
 \delta x_\perp^{i(2)}&=& \bar \chi \bigg[  -2 \omega^{i(2)}_{\perp \, o}-  v^{i(2)}_{\perp \, o} + \frac{1}{2} n^j \hat h_{ j k\, o}^{(2)} \Perp^{ki}+   v^{(1)}_{\| \, o}v^{i(1)}_{\perp \, o}   +4 v^{i (1)}_{\perp o} \Psi^{(1)}_{\, o}  \bigg] \nonumber \\
&-&  4  \left (\Phi^{(1)}_o-v^{(1)}_{\| \, o}\right) \int_0^{\bar \chi} \ud \tilde \chi \left(\bar \chi-\tilde \chi\right)\tilde \p^i_\perp \left( \Phi^{(1)} + \Psi^{(1)} \right)  - 4 v^{i (1)}_{\perp \, o } \int_0^{\bar \chi} \ud \tilde \chi \Psi^{(1)}  \nonumber \\
&+&  \int_0^{\bar \chi} \ud \tilde \chi \left( 2\omega^{i(2)}_{\perp} -  n^j \hat h_{jk}^{(2)} \Perp^{ki} + 8\,\Psi^{(1)} S_{\perp}^{i(1)}\right) + \int_0^{\bar \chi} \ud \tilde \chi  \left(\bar \chi-\tilde \chi\right) \bigg\{-\bigg[\tilde \p^i_\perp \left( \Phi^{(2)} + 2 \omega^{(2)}_{\| }+ \Psi^{(2)} - \frac{1}{2} \hat h^{(2)}_{\| } \right)  \nonumber \\
&+& \frac{1}{\tilde \chi} \left(- 2 \omega^{i (2)}_{\perp }+  n^k \hat h_{kj}^{(2)} \Perp^{ij}  \right)\bigg]    + 8 \left( \Phi^{(1)} -  I^{(1)}\right)  \tilde\p^i_\perp \left( \Phi^{(1)}+ \Psi^{(1)} \right) -4\left( \Phi^{(1)}+\Psi^{(1)} \right) \tilde\p^i_\perp  \Psi^{(1)}\bigg] \bigg\}\nonumber \\
&+&  \delta x_{\perp  \rm post-Born}^{i(2)}\;,
\end{eqnarray}
where
  \begin{eqnarray}
 && \delta x_{\perp  \rm post-Born}^{i(2)}= - 2 \bar \chi v_{\perp o}^{i (1)}  \left( \Phi^{(1)}_o + \Psi^{(1)}_o\right)  - 4 \left(\Phi^{(1)}_o-v^{(1)}_{\| \, o}  \right) \int_0^{\bar \chi} \ud \tilde{\chi}   S_{\perp}^{i(1)}  - 2 v_{\perp o}^{i (1)}   \int_0^{\bar \chi} \ud \tilde{\chi} ~ (\bar \chi - \tilde \chi)   \frac{1}{\tilde \chi} \left(\Phi^{(1)} + \Psi^{(1)} \right)  \nonumber\\
&&    - 2 \int_0^{\bar \chi}  \ud \tilde{\chi}\left[  \tilde \p^i_{\perp }  \left(\Phi^{(1)} + \Psi^{(1)} \right)  \delta x_{\|}^{(1)} + \frac{1}{\tilde \chi}  \left(\Phi^{(1)} + \Psi^{(1)} \right) \delta x_{\perp}^{i (1)} \right]     +2 \int_0^{\bar \chi} \ud \tilde{\chi} ~ (\bar \chi - \tilde \chi)\Bigg\{ -  \tilde \p_{\perp }^i    \left(\Phi^{(1)}{'}+  \Psi^{(1)}{'} \right)~ T^{(1)}      \nonumber\\
&&   -   \tilde \p_{\perp }^i    \left(\Phi^{(1)} + \Psi^{(1)}  \right)   \left(\Phi^{(1)} - \Psi^{(1)}  - 2I^{(1)} \right)  +  \frac{2}{\tilde \chi}  \left(\Phi^{(1)} + \Psi^{(1)} \right) S_{\perp}^{i(1)}  \nonumber\\
&&   -   \bigg[ \Perp^{im} \tilde \p_{\perp j}   \p_{\perp m}    \left(\Phi^{(1)}+ \Psi^{(1)} \right) +  \frac{1}{\tilde\chi} \Perp^i_j  \left(\Phi^{(1)}{'} + \Psi^{(1)}{'}  \right) +   \frac{1}{\tilde\chi^2} \Perp^i_j \left(\Phi^{(1)} + \Psi^{(1)}  \right) \bigg]  \delta x_{\perp}^{j (1)}    \Bigg\}\;.
 \nonumber
 \end{eqnarray}

Combining Eqs.\  (\ref{dx0-2}) and (\ref{dx_||-2}) we have
\begin{eqnarray} 
\label{Poiss-dx0+dx_||-2}
 \delta x^{0 (2)} + \delta x_\|^{(2)}&=&
 \bar \chi \bigg [\left(\Phi^{(1)}_o\right)^2 +2  \Phi^{(1)}_o v^{(1)}_{\| \, o} +2   v^{(1)}_{\| \, o}  \Psi^{(1)}_{\, o} - \left( \Psi^{(1)}_{\, o}\right)^2   -  v^{(1)}_{\perp k \, o} v^{k (1)}_{\perp \, o} \bigg]-4 \left (\Phi^{(1)}_o-v^{(1)}_{\| \, o}\right)T^{(1)} \nonumber \\
&& -4 \, v_{\perp \, o}^{i(1)} \int_0^{\bar \chi} \ud \tilde \chi\left(\bar \chi-\tilde\chi\right)\left[\tilde \p^i_\perp \left( \Phi^{(1)} + \Psi^{(1)} \right) \right]  - T^{(2)}+ 4 \int_0^{\bar \chi} \ud \tilde \chi \bigg[-2\left(  \Phi^{(1)}  \right)^2\nonumber \\
&&- \Psi^{(1)}  \left(  \Phi^{(1)} - \Psi^{(1)} \right)   +2 \left(\Phi^{(1)} +\Psi^{(1)}\right)  I^{(1)}  -2S_{\perp }^{i(1)}S_{\perp }^{j(1)} \delta_{ij}  \bigg]   \nonumber \\
&&+ 4 \int_0^{\bar \chi} \ud \tilde \chi\left(\bar \chi-\tilde\chi\right) \bigg\{ \left(\Phi^{(1)} + \Psi^{(1)} \right) \left[\left(\Phi^{(1)}{'} + \Psi^{(1)}{'} \right)+  \frac{\ud}{\ud \tilde \chi}\Phi^{(1)}  \right]\bigg\}\nonumber \\
&&+ \left(\delta x^{0 (2)} + \delta x_\|^{(2)}\right)_{\rm post-Born}\;,
\end{eqnarray}
where
\begin{eqnarray}
 && \left(\delta x^{0 (2)} + \delta x_\|^{(2)}\right)_{\rm post-Born}= -2 \bar \chi \left( \Phi^{(1)}_o+  \Psi^{(1)}_o \right) \left(3 v_{\| \, o}^{(1)} -  \Psi^{(1)}_o \right) +  4 \left(\Phi^{(1)}_o-v^{(1)}_{\| \, o}\right) \left[ \frac{\bar \chi }{2} \left(\Phi^{(1)}  +  \Psi^{(1)}\right) + T^{(1)} \right]  \nonumber \\
 &&  - 4  v_{\| \, o}^{(1)} \int_0^{\bar \chi}  \ud \tilde{\chi} ~  \frac{(\bar \chi - \tilde \chi)}{\tilde \chi}   \left( \Phi^{(1)} +  \Psi^{(1)} \right)    - 4 v^{i (1)}_{\perp \, o }  \int_0^{\bar \chi}  \ud \tilde{\chi} S_{\perp i}^{(1)}  - 2  \left(\Phi^{(1)} + \Psi^{(1)}\right) \int_0^{\bar \chi} \ud \tilde \chi  \left(\Phi^{(1)}  - \Psi^{(1)} - 2I^{(1)} \right)  \nonumber \\
 &&   +2 \int_0^{\bar \chi}  \ud \tilde{\chi} \Bigg\{ -  \left(\Phi^{(1)}{'} +  \Psi^{(1)}{'} \right)T^{(1)} +2 \left(\Phi^{(1)}  +  \Psi^{(1)})\right) \left(  \Phi^{(1)} - \Psi^{(1)} -2I^{(1)} - \kappa^{(1)}  \right)  \nonumber \\
 && +  \left[ \tilde \p_{\perp i}\left(\Phi^{(1)}+ \Psi^{(1)} \right) \right] \delta x_{\perp}^{i (1)}  \Bigg\}  +2 \int_0^{\bar \chi}  \ud \tilde{\chi} ~ (\bar \chi - \tilde \chi) \Bigg\{  - \left( \Phi^{(1)}+ \Psi^{(1)} \right)    \bigg[\frac{\ud}{\ud \tilde \chi}\left(\Phi^{(1)} - \Psi^{(1)} \right)  + 2 \left(\Phi^{(1)}{'} +  \Psi^{(1)}){'} \right) \bigg]     \nonumber \\
 && - 2 \left[ \tilde \p_{\perp j}\left( \Phi^{(1)} + \Psi^{(1)} \right) \right]  S_{\perp}^{i(1)}  - 2 \left( \Phi^{(1)}+ \Psi^{(1)} \right)    \tilde \p_{\perp m}S_{\perp}^{m(1)}     + 2 \bigg[ \frac{\ud}{\ud \tilde \chi}\left( \Phi^{(1)} + \Psi^{(1)} \right)    -  \frac{1}{\tilde\chi} \left( \Phi^{(1)} + \Psi^{(1)} \right)  \bigg] \kappa^{(1)}  \nonumber \Bigg\}\;.
\end{eqnarray}

To obtain all the second order terms we require $\Delta \ln a^{(1)}$ (or $\Delta x^{0(1)}$),  $\delta \chi^{(1)}$,  $\Delta x^{0(1)}$, $ \Delta x^{(1)}_\|$,  $\Delta x_{\perp}^{i (1)}$ and $\Delta_g^{(1)}$. From Eqs.  (\ref{Ek-1}) and (\ref{chi_1}) we have
\begin{eqnarray}
\label{Poiss-Deltalna-1}
\Delta \ln a^{(1)}&=&\left (\Phi^{(1)}_o-v^{(1)}_{\| \, o}\right) - \Phi^{(1)}+ v_\|^{(1)}+ 2I^{(1)} =\left (\Phi^{(1)}_o-v^{(1)}_{\| \, o}\right) - \Phi^{(1)}+ v_\|^{(1)}- \int_0^{\bar \chi} \ud \tilde \chi \left(\Phi^{(1)}{'} + \Psi^{(1)}{'} \right)\;, \\
\delta \chi^{(1)} &=&-\left(\bar \chi+\frac{1}{\cH}\right)\left(\Phi^{(1)}_o-v^{(1)}_{\| \, o}\right)+  \frac{1}{\cH}\left(\Phi^{(1)}- v_\|^{(1)}\right)  + \int_0^{\bar \chi} \ud \tilde \chi \left[ 2 \Phi^{(1)} +\left(\bar \chi-\tilde \chi\right) \left(\Phi^{(1)}{'} + \Psi^{(1)}{'} \right) \right] -\frac{2}{\cH}I^{(1)} \;.\nonumber \\
\end{eqnarray}
Then, from Eq. (\ref{Dx0_1})
\begin{eqnarray}
\label{Poiss-Dx^0-1}
\Delta x^{0(1)}&=&\frac{1}{\cH}\left[\left (\Phi^{(1)}_o-v^{(1)}_{\| \, o}\right) - \Phi^{(1)}+ v_\|^{(1)}+ 2I^{(1)}\right] =  \frac{1}{\cH}\left[\left (\Phi^{(1)}_o-v^{(1)}_{\| \, o}\right) - \Phi^{(1)}+ v_\|^{(1)}- \int_0^{\bar \chi} \ud \tilde \chi \left(\Phi^{(1)}{'}+ \Psi^{(1)}{'} \right) \right]\;, \nonumber \\
\end{eqnarray}
and from Eqs.\ (\ref{Dx_||-1}) and (\ref{s-1})
\begin{eqnarray}
\label{Poiss-Dx||-1}
\Delta x^{(1)}_\| &=&- T^{(1)} - \frac{1}{\cH}\left[\left (\Phi^{(1)}_o-v^{(1)}_{\| \, o}\right) - \Phi^{(1)}+ v_\|^{(1)}+ 2I^{(1)}\right]\nonumber \\
&=& \int_0^{\bar \chi} \ud \tilde \chi \left(\Phi^{(1)} +\Psi^{(1)}\right)- \frac{1}{\cH}\left[\left (\Phi^{(1)}_o-v^{(1)}_{\| \, o}\right) - \Phi^{(1)}+ v_\|^{(1)}- \int_0^{\bar \chi} \ud \tilde \chi \left(\Phi^{(1)}{'}  + \Psi^{(1)}{'} \right) \right].
 \end{eqnarray}
 Using Eq. (\ref{Dx_perp-1}), we have
 \begin{eqnarray}
 \Delta x_{\perp}^{i (1)}&=& - \bar \chi \, v^{i (1)}_{\perp \, o }-\int_0^{\bar \chi} \ud \tilde \chi  \left(\bar \chi-\tilde \chi\right)  \tilde \p^i_\perp \left( \Phi^{(1)} + \Psi^{(1)} \right)  \;.
\end{eqnarray}
In Eq.\ (\ref{Dx^0-1}) there is an ISW contribution and in Eq.\ (\ref{Dx||-1}) we have both time-delay and ISW contributions.

Now we can obtain $\Delta_g^{(1)} $. Using Eq.  (\ref{partialparallep}) for $\Delta x^{(1)}_\| $, we find
\begin{eqnarray}
\label{Poiss-partial_||Dx||-1}
\p_\| \Delta x^{(1)}_\| &=& \left(\Phi^{(1)} + \Psi^{(1)}\right) - \frac{\cH'}{\cH^2}\Delta \ln a^{(1)}
- \frac{1}{\cH}\left(\frac{\ud \, \Delta \ln a}{\ud \bar \chi}\right)^{(1)} \nonumber \\
&=&\left(\Phi^{(1)}+ \Psi^{(1)}\right)+\frac{1}{\cH}\left[\frac{\ud \,}{\ud \bar \chi} \left( \Phi^{(1)}-  v_\|^{(1)}\right)+ \left(\Phi^{(1)}{'} + \Psi^{(1)}{'} \right)\right]- \frac{\cH'}{\cH^2}\Delta \ln a^{(1)}.
\end{eqnarray}
With Eqs.\ (\ref{Deltalna-1}), (\ref{Dx||-1}) and  (\ref{partial_||Dx||-1}), Eq. (\ref{Deltag-1}) becomes 
\begin{eqnarray}
\label{Poiss-Deltag-1_2}
\Delta_g^{(1)}
 &=&  \delta_g^{(1)} + \left( b_e  - \frac{\cH'}{\cH^2} - \frac{2}{\bar \chi \cH}\right)\Delta \ln a^{(1)}  +\frac{1}{\cH}\left[\frac{\ud \,}{\ud \bar \chi} \left( \Phi^{(1)}-  v_\|^{(1)}\right)+ \left(\Phi^{(1)}{'} + \Psi^{(1)}{'} \right)\right] - \frac{2}{\bar \chi} T^{(1)}  - 2 \kappa^{(1)} \nonumber \\
&&   +   \Phi^{(1)}+v_\|^{(1)}-2 \Psi^{(1)} \;,
\end{eqnarray}
in agreement with \cite{Bonvin:2011bg, Challinor:2011bk}.
At first order,  the coordinate convergence lensing term  defined in Eq. (\ref{kappa-n}) yields
 \begin{eqnarray}
 \label{Poiss-kappa-1}
 \kappa^{(1)}=- \frac{1}{2}  \p_{\perp i} \Delta x_{\perp}^{i (1)}=  \frac{1}{2}  \int_0^{\bar \chi} \ud \tilde \chi  \left(\bar \chi-\tilde \chi\right) \frac{\tilde \chi}{ \bar \chi} \,   \tilde \nabla^2_\perp \left( \Phi^{(1)} + \Psi^{(1)} \right)- v_{\| \, o}^{(1)}.  
 \end{eqnarray}

At this point, we can finally compute $\Delta \ln a^{(2)}$,  $\Delta x^{0(2)}$, $ \Delta x^{(2)}_\|$ and  $\Delta x_{\perp}^{i (2)}$.
From Eq. (\ref{Ek-2}) we find
\begin{eqnarray}
\label{Poiss-Deltalna-2}
 \Delta\ln a^{(2)}&=&   \Phi^{(2)}_{\, o}- v^{(2)}_{\| \, o}-\left(\Phi^{(1)}_{\, o}\right)^2-6\Phi^{(1)}_{\, o} v^{(1)}_{\| \, o} +  v^{(1)}_{k\, o} v^{k (1)}_{\, o} +2 \Psi^{(1)}_{\, o} v^{(1)}_{\| \, o} +2 \left(\Phi^{(1)}_o-v^{(1)}_{\| \, o}\right) \bigg[-3\Phi^{(1)} +  v^{(1)}_{\|}   \nonumber \\
&&+\left(2 \bar \chi + \frac{1}{\cH}\right)\frac{\ud \,}{\ud \bar \chi} \Phi^{(1)} - \frac{1}{\cH}\frac{\ud \,}{\ud \bar \chi}   v^{(1)}_{\|}  +\left(\bar \chi + \frac{1}{\cH}\right)\left(\Phi^{(1)}{'} + \Psi^{(1)}{'} \right)   + 4 I^{(1)} \bigg] \nonumber \\
&&-2 \, v_{\perp \, o}^{i(1)}   \bigg[ \bar \chi \, \p_{\perp i} \left(\Phi^{(1)} +  v^{(1)}_{\|} \right) - 2 \int_0^{\bar \chi} \ud \tilde \chi \, \tilde\p_{\perp i} \Phi^{(1)} \bigg]-\Phi^{(2)}
+   v^{(2)}_{\|} +  7 \left(\Phi^{(1)}\right)^2 + v_{i}^{(1)} v^{i(1)}  \nonumber \\
&& -2v^{(1)}_{\|} \left( \Psi^{(1)} + \Phi^{(1)} \right) -  \frac{2}{\cH}\left( \Phi^{(1)} - v^{(1)}_{\| }\right) \left[\frac{\ud \,}{\ud \bar \chi} \left(\Phi^{(1)} -  v^{(1)}_{\|} \right) + \left(\Phi^{(1)}{'} + \Psi^{(1)}{'} \right)\right] \nonumber \\
&&-4 \bigg[ 3 \Phi^{(1)}   -   v^{(1)}_{\|} -  \frac{1}{\cH}\frac{\ud \,}{\ud \bar \chi} \left(\Phi^{(1)} - v^{(1)}_{\|} \right)  - \frac{1}{\cH}\left(\Phi^{(1)}{'} +  \Psi^{(1)}{'} \right) \bigg] I^{(1)}+ 4 v_{\perp i}^{(1)}S_{\perp}^{i(1)} \nonumber \\
&&- 2 \p_\| \left(\Phi^{(1)} +  v^{(1)}_{\|} \right) T^{(1)}- 2 \bigg[ 2\frac{\ud \,}{\ud \bar \chi}\Phi^{(1)}  +  \left(\Phi^{(1)}{'}+\Psi^{(1)}{'} \right)\bigg] \int_0^{\bar \chi} \ud \tilde \chi\left[ 2 \Phi^{(1)}  + \left(\bar \chi-\tilde \chi\right) \left(\Phi^{(1)}{'} + \Psi^{(1)}{'} \right) \right]   \nonumber \\
&& - 2 \left[\p_{\perp i}\left(\Phi^{(1)} +  v^{(1)}_{\|} \right)  - \frac{1}{\bar \chi} v^{(1)}_{\perp i}\right]\int_0^{\bar \chi} \ud \tilde \chi \left[  \left(\bar \chi-\tilde \chi\right) \tilde \p^i_\perp \left( \Phi^{(1)} +  \Psi^{(1)} \right) \right]  + 2 I^{(2)} \nonumber \\
&& -4 \int_0^{\bar \chi} \ud \tilde \chi  \bigg\{ \left( \Psi^{(1)} +2 I^{(1)}\right) \left(\Phi^{(1)}{'} +\Psi^{(1)}{'} \right) + \left(\Phi^{(1)} + \Psi^{(1)} \right)   \frac{\ud}{\ud \tilde \chi}\Phi^{(1)}  +2 S_{\perp }^{i(1)} \tilde\p_{\perp i} \Phi^{(1)}  \bigg\} \nonumber \\
&&+ \Delta\ln a^{(2)}_{\rm post-Born}  \;,
 \end{eqnarray}
 where
    \begin{eqnarray}
   &&
  \Delta\ln a^{(2)}_{\rm post-Born}= 4 \Phi^{(1)}_o   \left(3 v_{\| \, o}^{(1)} - \Psi^{(1)}_o \right) + 2\left(\Phi^{(1)}_o-v^{(1)}_{\| \, o}\right)  \Bigg\{    2 \Phi^{(1)}   -2   I^{(1)} -  \bar \chi   \left[ 2 \frac{\ud}{\ud \bar \chi}\Phi^{(1)}+\left(\Phi^{(1)}{'} + \Psi^{(1)}{'} \right) \right]  \Bigg\}  \nonumber\\
         &&    + 8 v_{\| \, o}^{(1)}    \int_0^{\bar \chi}   \frac{\ud \tilde{\chi}}{\tilde \chi}  \Phi^{(1)}    - 4    v^{i (1)}_{\perp \, o }  \Bigg( -   \bar \chi \p_{\perp i} \Phi^{(1)}   +   \int_0^{\bar \chi}  \ud \tilde{\chi}   \tilde \p_{\perp i}  \Phi^{(1)} \Bigg) 
          - 4 \Phi^{(1)} \left( \Phi^{(1)} - \Psi^{(1)} -2I^{(1)} -2\kappa^{(1)} \right) +  4  \Phi^{(1)}{'} T^{(1)}     \nonumber\\
  && + 2  \left[2 \frac{\ud}{\ud \bar \chi} \Phi^{(1)} +\left(\Phi^{(1)}{'} + \Psi^{(1)}{'} \right) \right] \int_0^{\bar \chi} \ud \tilde \chi     
  \left(\Phi^{(1)} - \Psi^{(1)}  - 2I^{(1)} \right)      -8   \p_{\perp i}\Phi^{(1)}   \int_0^{\bar \chi} \ud \tilde \chi S_{\perp}^{i(1)}   \nonumber\\
    && +2 \int_0^{\bar \chi}  \ud \tilde{\chi} \Bigg\{  + \left(\Phi^{(1)}{''}+ \Psi^{(1)}{''} \right)  T^{(1)} +  2  \Phi^{(1)}{'}  \left(\Phi^{(1)} + \Psi^{(1)} \right)  -  \left(\Phi^{(1)}{'} +  \Psi^{(1)}{'} \right)   \left(\Phi^{(1)} - \Psi^{(1)} -2 I^{(1)}\right)   \nonumber\\
      &&   + 2  \Phi^{(1)}   \bigg[\frac{\ud}{\ud \tilde \chi}\left(\Phi^{(1)} -  \Psi^{(1)}\right)     + \left(\Phi^{(1)}{'} + \Psi^{(1)}{'} \right) \bigg] + 4 S_{\perp}^{i(1)}  \tilde \p_{\perp i}  \Phi^{(1)}    + 4 \Phi^{(1)}    \tilde \p_{\perp j}S_{\perp}^{j(1)}   - 4 \bigg( \frac{\ud}{\ud \tilde \chi} \Phi^{(1)}    -  \frac{1}{\tilde\chi}  \Phi^{(1)}  \bigg) \kappa^{(1)}  \nonumber \\
 && - \bigg[  \tilde \p_{\perp i}   \left(\Phi^{(1)}{'} +  \Psi^{(1)}{'} \right) \bigg]   \delta x_{\perp}^{i (1)} \Bigg\} \;.  \nonumber
 \end{eqnarray}
Using Eqs\ (\ref{Deltalna-1}) and (\ref{Deltalna-2}), Eq.\ (\ref{Dx0_2}) yields
\begin{eqnarray}
\label{Poiss-Dx0_2-2}
\Delta x^{0(2)}&=& + \frac{1}{\cH} \Phi^{(2)}_o-\frac{1}{\cH} v^{(2)}_{\| \, o}-\left(\frac{\cH'}{\cH^3}+\frac{2}{\cH}\right)\left(\Phi^{(1)}_o\right)^2+2\left( \frac{\cH'}{\cH^3}-\frac{2}{\cH}\right)\Phi^{(1)}_o v^{(1)}_{\| \, o}  - \frac{\cH'}{\cH^3}\left(v^{(1)}_{\| \, o}\right)^2+\frac{2}{\cH}  \Psi^{(1)}_{\, o} v^{(1)}_{\| \, o} \nonumber \\
&&+ \frac{1}{\cH}v^{(1)}_{\perp i \, o}v^{i (1)}_{\perp \, o}+ 2 \left(\Phi^{(1)}_o-v^{(1)}_{\| \, o}\right) \bigg\{\left( \frac{\cH'}{\cH^3}- \frac{2}{\cH}\right) \Phi^{(1)}  - \frac{\cH'}{\cH^3} v^{(1)}_{\|}  +\left(2\frac{\bar \chi}{\cH} + \frac{1}{\cH^2}\right) \frac{\ud \,}{\ud \bar \chi}\Phi^{(1)} - \frac{1}{\cH^2} \frac{\ud \,}{\ud \bar \chi} v^{(1)}_{\|} \nonumber \\
&&+\left(\frac{\bar \chi}{\cH} + \frac{1}{\cH^2}\right) \left(\Phi^{(1)}{'}+ \Psi^{(1)}{'} \right)  -  2 \left( \frac{\cH'}{\cH^3}-\frac{1}{\cH}\right) I^{(1)} \bigg\}  -  2 v_{\perp \, o}^{i(1)} \bigg[+ \frac{\bar \chi}{\cH}  \p_{\perp i} \left(\Phi^{(1)} +  v^{(1)}_{\|} \right)  -  \frac{2}{\cH}  \int_0^{\bar \chi} \ud \tilde \chi\,   \tilde  \p_{\perp i} \Phi^{(1)} \bigg]\nonumber \\ 
&&  -\frac{1}{\cH} \Phi^{(2)} + \frac{1}{\cH}  v^{(2)}_{\|} +   \left(- \frac{\cH'}{\cH^3}+\frac{6}{\cH}\right) \left(\Phi^{(1)}\right)^2   -\frac{\cH'}{ \cH^3} \left(v^{(1)}_{\|}\right)^2+  \frac{1}{\cH} v_{\perp i}^{(1)} v_{\perp}^{i(1)}  -\frac{2}{\cH}  v^{(1)}_{\|} \Psi^{(1)}  + 2\frac{\cH'}{\cH^3}\Phi^{(1)} v^{(1)}_{\| } \nonumber \\
&&- \frac{2}{\cH^2} \left( \Phi^{(1)} - v^{(1)}_{\| }\right)  \left[\frac{\ud \,}{\ud \bar \chi} \left(\Phi^{(1)} -  v^{(1)}_{\|} \right) + \left(\Phi^{(1)}{'} + \Psi^{(1)}{'} \right)\right] -4 \bigg[  \left( \frac{2}{\cH}- \frac{\cH'}{\cH^3}\right) \Phi^{(1)}  +  \frac{\cH'}{\cH^3} v^{(1)}_{\|} \nonumber \\
&&   - \frac{1}{\cH^2}\frac{\ud \,}{\ud \bar \chi}\left( \Phi^{(1)} - v^{(1)}_{\| } \right)   - \frac{1}{\cH^2}\left(\Phi^{(1)}{'} + \Psi^{(1)}{'} \right)+ \left( \frac{\cH'}{\cH^3}+ \frac{1}{\cH}\right)I^{(1)} \bigg] I^{(1)}+ \frac{4}{\cH} v_{\perp i}^{(1)}S_{\perp}^{i(1)} -  \frac{2}{\cH} \p_\| \left(\Phi^{(1)} +  v^{(1)}_{\|}\right) T^{(1)} \nonumber \\
&& - \frac{2}{\cH} \left[ 2 \frac{\ud \,}{\ud \bar \chi} \Phi^{(1)}  +  \left(\Phi^{(1)}{'} + \Psi^{(1)}{'} \right)\right]   \int_0^{\bar \chi} \ud \tilde \chi\left[ 2 \Phi^{(1)} + \left(\bar \chi-\tilde \chi\right) \left(\Phi^{(1)}{'} + \Psi^{(1)}{'} \right) \right]  -  \frac{2}{\cH} \left[\p_{\perp i}\left(\Phi^{(1)} +  v^{(1)}_{\|} \right)  \right.\nonumber \\
&& \left. - \frac{1}{\bar \chi}  v^{(1)}_{\perp i}\right] \int_0^{\bar \chi} \ud \tilde \chi \left[  \left(\bar \chi-\tilde \chi\right) \tilde \p^i_\perp \left( \Phi^{(1)} +  \Psi^{(1)} \right) \right] +  \frac{2}{\cH} I^{(2)}  - \frac{4}{\cH} \int_0^{\bar \chi} \ud \tilde \chi  \left[ \left(   \Psi^{(1)} +2  I^{(1)}\right) \left(\Phi^{(1)}{'} + \Psi^{(1)}{'} \right) \right. \nonumber \\
&&+ \left(\Phi^{(1)} + \Psi^{(1)} \right)  \frac{\ud}{\ud \tilde \chi} \Phi^{(1)} + 2 S_{\perp }^{i(1)} \tilde\p_{\perp i}\Phi^{(1)} \bigg]+\Delta x^{0(2)}_{\rm post-Born} \;,
\end{eqnarray}
where
 \begin{eqnarray} 
 &&\Delta x^{0(2)}_{\rm post-Born}=
 + \frac{4} {\cH} \Phi^{(1)}_o  \left(3 v_{\| \, o}^{(1)} -  \Psi^{(1)}_o \right) + \frac{2}{\cH}\left(\Phi^{(1)}_o-v^{(1)}_{\| \, o}\right)  \Bigg\{    2 \Phi^{(1)}   -2   I^{(1)}  -  \bar \chi   \left[ 2 \frac{\ud}{\ud \bar \chi} \Phi^{(1)}  +\left(\Phi^{(1)}{'} + \Psi^{(1)}{'} \right) \right]  \Bigg\}     \nonumber\\
        && + \frac{8}{\cH} v_{\| \, o}^{(1)}  \int_0^{\bar \chi}   \frac{\ud \tilde{\chi}}{\tilde \chi}  \Phi^{(1)}   
          - \frac{4 \bar \chi}{\cH}  v^{i (1)}_{\perp \, o }  \Bigg\{ -     \p_{\perp i} \Phi^{(1)}   +  \int_0^{\bar \chi}  \ud \tilde{\chi}  \left( \tilde \p_{\perp i} \Phi^{(1)} \right)   \Bigg\}  
         - \frac{4}{\cH} \Phi^{(1)} \left( \Phi^{(1)} - \Psi^{(1)} -2I^{(1)} -2\kappa^{(1)} \right)    \nonumber\\
 && + \frac{2}{\cH}  \left[ 2 \frac{\ud}{\ud \bar \chi} \Phi^{(1)} +\left(\Phi^{(1)}{'} +  \Psi^{(1)}{'} \right) \right] \int_0^{\bar \chi} \ud \tilde \chi     
  \left(\Phi^{(1)} -  \Psi^{(1)}  -2I^{(1)} \right)  +  \frac{4}{\cH}  \Phi^{(1)}{'} T^{(1)}    - \frac{8} {\cH}   \p_{\perp i}  \Phi^{(1)}  \int_0^{\bar \chi} \ud \tilde \chi   S_{\perp}^{i(1)}   \nonumber\\
    && +\frac{2}{\cH}  \int_0^{\bar \chi}  \ud \tilde{\chi} \Bigg\{   \left(\Phi^{(1)}{''} +  \Psi^{(1)}{''} \right)  T^{(1)}
     + 2 \Phi^{(1)}{'} \left(\Phi^{(1)} +  \Psi^{(1)}\right)   -  \left(\Phi^{(1)}{'} +  \Psi^{(1)}{'} \right)   \left(\Phi^{(1)} - \Psi^{(1)} -2I^{(1)}\right)   \nonumber\\
          &&   + 2 \Phi^{(1)}   \bigg[\frac{\ud}{\ud \bar \chi}\left(\Phi^{(1)} - \Psi^{(1)}\right)     + \left(\Phi^{(1)}{'} + \Psi^{(1)}{'} \right) \bigg]  +  4S_{\perp}^{i(1)}  \tilde \p_{\perp i}  \Phi^{(1)}    +  4 \Phi^{(1)}   \tilde \p_{\perp j}S_{\perp}^{j(1)}  
          - 4 \bigg( \frac{\ud}{\ud \tilde \chi}\Phi^{(1)}   -  \frac{1}{\tilde\chi}  \Phi^{(1)}   \bigg) \kappa^{(1)}   \nonumber \\
 && 
  - \tilde \p_{\perp i}   \left(\Phi^{(1)}{'} +\Psi^{(1)} {'} \right)   \delta x_{\perp}^{i (1)} \Bigg\} \;.  \nonumber
 \end{eqnarray}

From Eqs.\  (\ref{Dx_||-2}) and   (\ref{dx0+dx_||-2}) we deduce
\begin{eqnarray} 
\label{Poiss-Dx_||-2_2}
\Delta x_{\parallel}^{(2)}&=&\bar \chi \bigg [\left(\Phi^{(1)}_o\right)^2  +2 \Phi^{(1)}_o v^{(1)}_{\| \, o} +2  v^{(1)}_{\| \, o} \Psi^{(1)}_{\, o} - \left(\Psi^{(1)}_{\, o}\right)^2   -   v^{(1)}_{\perp k \, o} v^{k (1)}_{\perp \, o}\bigg]- 2 \left (\Phi^{(1)}_o-v^{(1)}_{\| \, o}\right) \left[ \bar \chi \left (  \Phi^{(1)} +  \Psi^{(1)} \right) + 2T^{(1)}\right]  \nonumber \\
&& - 4 v_{\perp \, o}^{i(1)}  \int_0^{\bar \chi} \ud \tilde \chi\left(\bar \chi-\tilde\chi\right)\left[ \tilde\p^i_\perp \left( \Phi^{(1)}+ \Psi^{(1)} \right) \right]  + 2 \left (  \Phi^{(1)}+ \Psi^{(1)} \right)  \int_0^{\bar \chi} \ud \tilde \chi \left[ 2 \Phi^{(1)} + \left(\bar \chi-\tilde \chi\right) \left(\Phi^{(1)}{'} +  \Psi^{(1)}{'} \right) \right]  \nonumber \\ 
&& -  T^{(2)}+4 \int_0^{\bar \chi} \ud \tilde \chi \bigg[ -2 \left( \Phi^{(1)} \right)^2- \Psi^{(1)} \left( \Phi^{(1)} - \Psi^{(1)} \right)+2 \left(\Phi^{(1)} + \Psi^{(1)}\right)  I^{(1)}-2 S_{\perp }^{i(1)}S_{\perp }^{j(1)} \delta_{ij}  \bigg]   \nonumber \\
&&+ 4 \int_0^{\bar \chi} \ud \tilde \chi \bigg\{ \left(\bar \chi-\tilde\chi\right) \left(\Phi^{(1)}+ \Psi^{(1)} \right) \left[\left(\Phi^{(1)}{'} + \Psi^{(1)}{'} \right)+  \frac{\ud}{\ud \tilde \chi} \Phi^{(1)} \right]  \bigg\} - \frac{2}{\cH} \left (  \Phi^{(1)} + \Psi^{(1)} \right) \Delta \ln a^{(1)} \nonumber \\
&& -\frac{1}{ \cH} \Delta \ln a^{(2)}+ \left(\frac{\cH' }{\cH^3} +\frac{1}{\cH} \right)\left( \Delta \ln a^{(1)} \right)^2+ \left(\delta x^{0 (2)} + \delta x_\|^{(2)}\right)_{\rm post-Born} \;,
\end{eqnarray}
and from Eqs.\  (\ref{Dx_perp-2}) and   (\ref{dx_perp-2}) we find
\begin{eqnarray} 
\label{Poiss-Dx_perp-2_2}
\Delta x_{\perp}^{i(2)}&=&\bar \chi \bigg[ -2\omega^{i(2)}_{\perp \, o}-  v^{i(2)}_{\perp \, o} + \frac{1}{2} n^j \hat h_{ j k\, o}^{(2)} \Perp^{ki}+ 2 \Phi^{(1)}_o v^{i (1)}_{\perp \, o } -  v^{(1)}_{\| \, o}v^{i(1)}_{\perp \, o}  + 4 v^{i (1)}_{\perp o}    \Psi^{(1)}_{\, o}\bigg] -4 \bar \chi \left(\Phi^{(1)}_o-v^{(1)}_{\| \, o}\right) S_{\perp}^{i(1)} \nonumber \\
&& - 4  \left (\Phi^{(1)}_o-v^{(1)}_{\| \, o}\right) \int_0^{\bar \chi} \ud \tilde \chi \bigg[ \left(\bar \chi-\tilde \chi\right) \tilde \p^i_\perp \left( \Phi^{(1)}+  \Psi^{(1)} \right) \bigg]  - 2 \, v^{i (1)}_{\perp \, o } \bigg\{ \int_0^{\bar \chi} \ud \tilde \chi \left[ 2 \left( \Phi^{(1)}  +  \Psi^{(1)} \right) \right. \nonumber \\
&& \left. + \left(\bar \chi-\tilde \chi\right) \left(\Phi^{(1)}{'}+ \Psi^{(1)}{'} \right) \right] -  \frac{1}{\cH}\Delta \ln a^{(1)} \bigg\} -  \frac{4}{\cH}  S_{\perp}^{i(1)} \Delta \ln a^{(1)} + 4 \, S_{\perp}^{i(1)} \int_0^{\bar \chi} \ud \tilde \chi \bigg[ 2 \Phi^{(1)} \nonumber \\
&&+ \left(\bar \chi-\tilde \chi\right) \left(\Phi^{(1)}{'} + \Psi^{(1)}{'} \right) \bigg]  + \int_0^{\bar \chi} \ud \tilde \chi \bigg\{  2\omega^{i(2)}_{\perp} - n^j \hat h_{jk}^{(2)} \Perp^{ki} + 8 \, \Psi^{(1)} S_{\perp}^{i(1)} \bigg\} \nonumber \\
&& + \int_0^{\bar \chi} \ud \tilde \chi \left(\bar \chi-\tilde \chi\right) \left\{-\left[ \tilde \p^i_\perp \left( \Phi^{(2)} + 2 \omega^{(2)}_{\| } + \Psi^{(2)} - \frac{1}{2} \hat h^{(2)}_{\| } \right) + \frac{1}{\tilde \chi} \left(-2\, \omega^{i(2)}_{\perp} +  n^k \hat h_{kj}^{(2)} \Perp^{ij}  \right)\right]   \right. \nonumber \\
&&  \left.  +  8 \left( \Phi^{(1)} -  I^{(1)}\right) \tilde \p^i_\perp \left( \Phi^{(1)}+ \Psi^{(1)} \right) - 4 \left( \Phi^{(1)} + \Psi^{(1)} \right)  \tilde\p^i_\perp  \Psi^{(1)} \right\}+  \delta x_{\perp  \rm post-Born}^{i(2)}\; .
\end{eqnarray}

To obtain explicitly Eq.\ (\ref{Dx_||-2_2}) we need
\begin{eqnarray}
&& \left(\frac{\ud \, \Delta \ln a}{\ud \bar \chi}\right)^{(2)} = +2 \left(\Phi^{(1)}_o-v^{(1)}_{\| \, o}\right) \bigg\{\left(\frac{\cH' }{\cH^2}-1\right) \left[\frac{\ud \,}{\ud \bar \chi}\left(  \Phi^{(1)} - v^{(1)}_{\| } \right)+\left(\Phi^{(1)}{'} + \Psi^{(1)}{'} \right)\right] +   \frac{1}{\cH}  \frac{\ud \,}{\ud \bar \chi} \bigg[ \frac{\ud \,}{\ud \bar \chi}\left(  \Phi^{(1)} - v^{(1)}_{\| } \right)  \nonumber \\
&& + \left(\Phi^{(1)}{'}+ \Psi^{(1)}{'} \right)\bigg]+\bar \chi \frac{\ud \,}{\ud \bar \chi}\left[ 2 \frac{\ud \,}{\ud \bar \chi} \Phi^{(1)}  + \left(\Phi^{(1)}{'} + \Psi^{(1)}{'} \right)\right]\bigg\}  - 2 \, v_{\perp \, o}^{i(1)} \bigg[    -  \p_{\perp i} \left(\Phi^{(1)} -  v^{(1)}_{\|} \right)+\bar \chi  \frac{\ud \,}{\ud \bar \chi} \p_{\perp i} \left(\Phi^{(1)} +  v^{(1)}_{\|} \right) \bigg] \nonumber \\
&&   +  \frac{\ud \,}{\ud \bar \chi}\left(- \Phi^{(2)}+   v^{(2)}_{\|} \right) - \left(\Phi^{(2)}{'} +2 \, \omega^{(2)}_{\| }{'} +\Psi^{(2)}{'}  - \frac{1}{2} \hat h^{(2)}_{\| }{'} \right) +  \frac{\ud \,}{\ud \bar \chi}\ \bigg[ 7 \left(\Phi^{(1)}\right)^2 - 2\Phi^{(1)} v^{(1)}_{\| }  + \left(v^{(1)}_{\|}\right)^2-2 v^{(1)}_{\|}  \Psi^{(1)}  \bigg]   \nonumber \\
&& - 4 \bigg[ 2 \frac{\ud \,}{\ud \bar \chi}\Phi^{(1)}    + \left. \left(\Phi^{(1)}{'} + \Psi^{(1)}{'} \right)\right] \Phi^{(1)}   -2\bigg\{ \frac{\cH'}{\cH^2}\left( \Phi^{(1)} - v^{(1)}_{\| }\right) +  \frac{1}{\cH}\left[\frac{\ud \,}{\ud \bar \chi}\left( \Phi^{(1)} - v^{(1)}_{\| }\right) + \left(\Phi^{(1)}{'}+ \Psi^{(1)}{'} \right) \right]\bigg\} \nonumber \\
&& \times \left[\frac{\ud \,}{\ud \bar \chi} \left(\Phi^{(1)} -  v^{(1)}_{\|} \right) + \left(\Phi^{(1)}{'} +  \Psi^{(1)}{'} \right)\right] +2 \bigg[3 \Phi^{(1)} - 2\Psi^{(1)} -  v^{(1)}_{\|} \bigg] \left(\Phi^{(1)}{'} +  \Psi^{(1)}{'} \right)\nonumber \\
&& -  \frac{2}{\cH}\left( \Phi^{(1)} - v^{(1)}_{\| }\right)\frac{\ud \,}{\ud \bar \chi}\left[\frac{\ud \,}{\ud \bar \chi} \left(\Phi^{(1)} -  v^{(1)}_{\|} \right) + \left(\Phi^{(1)}{'}+ \Psi^{(1)}{'} \right)\right]  +2 \left(\Phi^{(1)} + \Psi^{(1)} \right) \left[-2 \frac{\ud \,}{\ud \bar \chi} \Phi^{(1)}   + \p_\| \left(\Phi^{(1)} +  v^{(1)}_{\|} \right)\right]  \nonumber \\
&&-2v_{\perp i}^{(1)} \p^i_\perp \left( \Phi^{(1)}+ \Psi^{(1)} \right) +  2 v_{\perp i}^{(1)} \frac{\ud \,}{\ud \bar \chi}  v_{\perp}^{i(1)}  + 4 \bigg\{ \left(\frac{\cH' }{\cH^2} -1\right) \left[\frac{\ud \,}{\ud \bar \chi}\left(\Phi^{(1)} - v^{(1)}_{\| } \right)   +\left(\Phi^{(1)}{'} + \Psi^{(1)}{'} \right)\right]\nonumber \\
&&+ \frac{1}{\cH} \frac{\ud \,}{\ud \bar \chi}\left[\frac{\ud \,}{\ud \bar \chi}\left(\Phi^{(1)} - v^{(1)}_{\| } \right) +\left(\Phi^{(1)}{'} + \Psi^{(1)}{'} \right) \right]\bigg\} I^{(1)}  - 2 \left[\frac{\ud \,}{\ud \bar \chi} \p_\| \left(\Phi^{(1)} +  v^{(1)}_{\|} \right) \right] T^{(1)}  - 2 \frac{\ud \,}{\ud \bar \chi} \bigg[2 \frac{\ud \,}{\ud \bar \chi}  \Phi^{(1)}  \nonumber \\
&& + \left(\Phi^{(1)}{'} +\Psi^{(1)}{'} \right)\bigg]  \int_0^{\bar \chi} \ud \tilde \chi\left[ 2 \Phi^{(1)}  + \left(\bar \chi-\tilde \chi\right) \left(\Phi^{(1)}{'} + \Psi^{(1)}{'} \right) \right]   -  4 \left[\p_{\perp i}\left(  \Phi^{(1)} - v^{(1)}_{\| } \right) + \frac{1}{\tilde \chi} v_{\perp i}^{(1)}  -\frac{\ud \,}{\ud \bar \chi} v_{\perp i}^{(1)} \right]   S_{\perp }^{i(1)} \nonumber \\
&&- 2 \frac{\ud \,}{\ud \bar \chi} \bigg[\p_{\perp i}\left(\Phi^{(1)} +  v^{(1)}_{\|} \right)  - \frac{1}{\bar \chi}  v^{(1)}_{\perp i}\bigg]\int_0^{\bar \chi} \ud \tilde \chi \left[ \left(\bar \chi-\tilde \chi\right) \tilde \p^i_\perp \left( \Phi^{(1)}+ \Psi^{(1)} \right)\right] + \left(\frac{\ud \, \Delta \ln a}{\ud \bar \chi}\right)^{(2)}_{ \rm post-Born}\,,
 \end{eqnarray}
 where
\begin{eqnarray} 
\left(\frac{\ud \, \Delta \ln a}{\ud \bar \chi}\right)^{(2)}_{ \rm post-Born}&=& -2 \bar \chi \left(\Phi^{(1)}_o-v^{(1)}_{\| \, o}\right)  \frac{\ud}{\ud \bar \chi} \bigg[2 \frac{\ud}{\ud \bar \chi} \Phi^{(1)}
    + \left(\Phi^{(1)}{'} +  \Psi^{(1)}{'} \right) \bigg]  \nonumber \\
&&+ 2  \bar \chi v^{i (1)}_{\perp \, o }  \Bigg[2  \p_{\perp i}\frac{\ud}{\ud \bar \chi}  \Phi^{(1)} + \p_{\perp i}   \left(\Phi^{(1)}{'} +  \Psi^{(1)}{'} \right)   -  \frac{2}{\bar \chi} \p_{\perp i}  \Phi^{(1)}   \Bigg]  \nonumber \\
&& + 2 \frac{\ud}{\ud \bar \chi} \bigg[2 \frac{\ud}{\ud \bar \chi} \Phi^{(1)}    + \left(\Phi^{(1)}{'} + \Psi^{(1)}{'} \right) \bigg] \int_0^{\bar \chi} \ud \tilde \chi \left(\Phi^{(1)} - \Psi^{(1)} -2I^{(1)} \right) \nonumber \\
&&+ 2 \left[2 \frac{\ud}{\ud \bar \chi} \Phi^{(1)}{'}    + \left(\Phi^{(1)}{''} + \Psi^{(1)}{''} \right) \right]  T^{(1)}  \nonumber \\
 && -4 \Bigg[ 2 \p_{\perp i}\frac{\ud}{\ud \bar \chi} \Phi^{(1)}  + \p_{\perp i}   \left(\Phi^{(1)}{'} + \Psi^{(1)}{'} \right) 
 -  \frac{2}{\bar \chi} \p_{\perp i} \Phi^{(1)}   \Bigg]       \int_0^{\bar \chi} \ud \tilde \chi S_{\perp}^{i(1)}   \;.  \nonumber
 \end{eqnarray}
Then we obtain
\begin{eqnarray} 
\label{Poiss-dDx_||-2_3}
&& \p_\| \Delta x_{\parallel}^{(2)}=\left(\Phi^{(1)}_o\right)^2 + 2 \Phi^{(1)}_o v^{(1)}_{\| \, o} +2 v^{(1)}_{\| \, o} \Psi^{(1)}_{\, o} - \left(\Psi^{(1)}_{\, o}\right)^2   -  v^{(1)}_{\perp k \, o} v^{k (1)}_{\perp \, o}+2 \left(\Phi^{(1)}_o-v^{(1)}_{\| \, o}\right)\nonumber \\
&& \times \bigg\{\left(-\frac{\cH' }{\cH^3}+ \frac{1}{\cH}\right) \bigg[\frac{\ud \,}{\ud \bar \chi}\left(  \Phi^{(1)} - v^{(1)}_{\| } \right)+\left(\Phi^{(1)}{'} + \Psi^{(1)}{'} \right)\bigg] - \frac{\bar \chi}{\cH} \frac{\ud \,}{\ud \bar \chi}\left[ 2\frac{\ud \,}{\ud \bar \chi}  \Phi^{(1)} + \left(\Phi^{(1)}{'} + \Psi^{(1)}{'} \right)\right] \nonumber \\
&& +\left (  \Phi^{(1)} + \Psi^{(1)} \right) -\bar \chi \frac{\ud}{\ud \bar \chi}\left( \Phi^{(1)}+ \Psi^{(1)} \right) -   \frac{1}{\cH^2}  \frac{\ud \,}{\ud \bar \chi} \left[ \frac{\ud \,}{\ud \bar \chi}\left(  \Phi^{(1)} - v^{(1)}_{\| } \right) + \left(\Phi^{(1)}{'}+\Psi^{(1)}{'} \right)\right]\bigg\} - 2  v_{\perp \, o}^{i(1)}  \nonumber \\
&& \times \bigg[   +\frac{1}{\cH}  \p_{\perp i} \left(\Phi^{(1)} -  v^{(1)}_{\|} \right)  -2\frac{\bar \chi}{\cH}  \frac{\ud \,}{\ud \bar \chi} \p_{\perp i} \left(\Phi^{(1)} +  v^{(1)}_{\|}\right)  -4 \delta_{il} S_{\perp }^{l (1)} \bigg]+ \Phi^{(2)} +2 \omega^{(2)}_{\| } + \Psi^{(2)}-\frac{1}{2}\hat h_{\|}^{(2)} \nonumber \\
&&+ \frac{1}{\cH}\bigg[  \frac{\ud \,}{\ud \bar \chi}\left( \Phi^{(2)} -  v^{(2)}_{\|} \right)+ \left(\Phi^{(2)}{'} + 2 \omega^{(2)}_{\| }{'} + \Psi^{(2)}{'}  - \frac{1}{2} \hat h^{(2)}_{\| }{'} \right)\bigg]  -4\left(\Phi^{(1)}- \Psi^{(1)} \right)\left(  \Phi^{(1)}+ \Psi^{(1)}\right)-\frac{1}{\cH}   \frac{\ud \,}{\ud \bar \chi} \bigg[ 7 \left(\Phi^{(1)}\right)^2\nonumber \\ 
&&- 2\Phi^{(1)} v^{(1)}_{\| } +  \left(v^{(1)}_{\|}\right)^2 -2 v^{(1)}_{\|} \Psi^{(1)} + v_{\perp i}^{(1)} v_{\perp}^{i(1)} \bigg]  +  \frac{2}{\cH^2}\left( \Phi^{(1)} - v^{(1)}_{\| }\right)\frac{\ud \,}{\ud \bar \chi}\left[\frac{\ud \,}{\ud \bar \chi} \left(\Phi^{(1)} -  v^{(1)}_{\|} \right) + \left(\Phi^{(1)}{'} + \Psi^{(1)}{'} \right)\right]  \nonumber \\ 
&& -\frac{4}{\cH} \bigg[\left(  \Phi^{(1)}- \Psi^{(1)}  \right)+ \left(\Phi^{(1)} -  v^{(1)}_{\|}\right) \bigg]\left(\Phi^{(1)}{'} + \Psi^{(1)}{'} \right) + \frac{4}{\cH}  \Phi^{(1)}  \bigg[ 2 \frac{\ud \,}{\ud \bar \chi} \Phi^{(1)} + \left(\Phi^{(1)}{'} + \Psi^{(1)}{'} \right)\bigg]   \nonumber \\
&&+2\bigg\{  \frac{1}{\cH} \left( \Phi^{(1)}+ \Psi^{(1)} \right)+\frac{\cH'}{\cH^3}\left( \Phi^{(1)} - v^{(1)}_{\| }\right) +  \frac{1}{\cH^2}\left[\frac{\ud \,}{\ud \bar \chi}\left( \Phi^{(1)} - v^{(1)}_{\| }\right) + \left(\Phi^{(1)}{'} + \Psi^{(1)}{'} \right) \right]\bigg\}\nonumber \\ 
&&\times  \bigg[\frac{\ud \,}{\ud \bar \chi} \left(\Phi^{(1)} -  v^{(1)}_{\|} \right) + \left(\Phi^{(1)}{'} +\Psi^{(1)}{'} \right)\bigg]  + \frac{2}{\cH} \left[ 2 \frac{\ud \,}{\ud \bar \chi}\Phi^{(1)} -  \p_\| \left(\Phi^{(1)} +  v^{(1)}_{\|} \right)\right]  \left(\Phi^{(1)} +  \Psi^{(1)} \right)  \nonumber \\ 
&&  +\frac{2}{\cH}v_{\perp i}^{(1)} \p^i_\perp \left( \Phi^{(1)} + \Psi^{(1)} \right)  +\frac{2}{\cH} \frac{\ud \,}{\ud \bar \chi} \p_\| \left(\Phi^{(1)} +  v^{(1)}_{\|} \right) T^{(1)}  +4 \bigg\{ \left(-\frac{\cH' }{\cH^3} +\frac{1}{\cH}\right) \bigg[\frac{\ud \,}{\ud \bar \chi}\left(\Phi^{(1)} - v^{(1)}_{\| } \right)   +\left(\Phi^{(1)}{'} + \Psi^{(1)}{'} \right)\bigg] \nonumber \\
 && - \frac{1}{\cH^2} \frac{\ud \,}{\ud \bar \chi}\bigg[\frac{\ud \,}{\ud \bar \chi}\left(\Phi^{(1)} - v^{(1)}_{\| } \right)  +\left(\Phi^{(1)}{'} + \Psi^{(1)}{'} \right) \bigg]+\left(  \Phi^{(1)} + \Psi^{(1)}\right)\bigg\} I^{(1)} +2 \bigg\{ \frac{\ud}{\ud \bar \chi}\left( \Phi^{(1)}+ \Psi^{(1)} \right) \nonumber \\
&&   + \frac{1}{\cH} \frac{\ud \,}{\ud \bar \chi} \bigg[ 2 \frac{\ud \,}{\ud \bar \chi}  \Phi^{(1)}  + \left(\Phi^{(1)}{'} + \Psi^{(1)}{'} \right)\bigg]  \bigg\}  \int_0^{\bar \chi} \ud \tilde \chi \;\bigg[ 2 \Phi^{(1)} + \left(\bar \chi-\tilde \chi\right)\left(\Phi^{(1)}{'} + \Psi^{(1)}{'} \right) \bigg]+ 4 \bigg[\frac{1}{\cH}\p_{\perp i}\left(  \Phi^{(1)} - v^{(1)}_{\| } \right) \nonumber \\
 && + \frac{1}{\cH \bar \chi} v_{\perp i}^{(1)}  -\frac{1}{\cH}\frac{\ud \,}{\ud \bar \chi} v_{\perp i}^{(1)}-2S_{\perp }^{j(1)} \delta_{ij}  \bigg]    S_{\perp }^{i(1)}  + \frac{2}{\cH} \frac{\ud \,}{\ud \bar \chi} \left[\p_{\perp i}\left(\Phi^{(1)} +  v^{(1)}_{\|} \right)  - \frac{1}{\bar \chi}  v^{(1)}_{\perp i}\right] \int_0^{\bar \chi} \ud \tilde \chi \bigg[ \left(\bar \chi-\tilde \chi\right)\ \tilde \p^i_\perp \left( \Phi^{(1)}+ \Psi^{(1)} \right)\bigg] \nonumber \\
&& -  \frac{\cH'}{\cH^2} \Delta \ln a^{(2)}+\left[-\frac{\cH'' }{\cH^3} +3\left( \frac{\cH' }{\cH^2} \right)^2 + \frac{\cH' }{\cH^2} \right] \left( \Delta \ln a^{(1)} \right)^2   + 4 \int_0^{\bar \chi} \ud \tilde \chi \bigg[ \left(\Phi^{(1)} + \Psi^{(1)} \right) \left(\Phi^{(1)}{'} + \Psi^{(1)}{'} +  \frac{\ud}{\ud \tilde \chi} \Phi^{(1)}  \right) \bigg]\nonumber \\
&&+2 \bigg\{-\frac{\cH' }{\cH^2} \left( \Phi^{(1)} + \Psi^{(1)} \right)- \frac{1}{\cH} \frac{\ud}{\ud \bar \chi}\left( \Phi^{(1)}+ \Psi^{(1)} \right)- \left(\frac{\cH' }{\cH^3} +\frac{1}{\cH} \right)\left[\frac{\ud \,}{\ud \bar \chi} \left( \Phi^{(1)}-  v_\|^{(1)}\right) + \left(\Phi^{(1)}{'}+ \Psi^{(1)}{'} \right)\right] \bigg\} \Delta \ln a^{(1)} \nonumber \\
 &&+ \p_\| \Delta x_{\|  \rm post-Born}^{(2)}\;,
\end{eqnarray}
where
\begin{eqnarray} 
 &&\p_\| \Delta x_{\|  \rm post-Born}^{(2)}= -2  \left( \Phi^{(1)}_o + \Psi^{(1)}_o\right) \left(3 v_{\| \, o}^{(1)} - \Psi^{(1)}_o \right)  +2 \left(\Phi^{(1)}_o-v^{(1)}_{\| \, o} \right) \Bigg\{- \left(\Phi^{(1)} + \Psi^{(1)}\right) + \frac{ \bar \chi}{\cH}  \frac{\ud}{\ud \bar \chi} \bigg[ 2 \frac{\ud}{\ud \bar \chi} \Phi^{(1)}     \nonumber \\
 &&  + \Phi^{(1)}{'} + \Psi^{(1)}{'}  \bigg] \Bigg\}
 + 2  \bar \chi \left(\Phi^{(1)}_o-v^{(1)}_{\| \, o}\right) \frac{\ud}{\ud \bar \chi}\left(\Phi^{(1)}  +\Psi^{(1)}\right)  - 2 v^{i (1)}_{\perp \, o }  \Bigg\{   \bar \chi  \p_{\perp i}\left(\Phi^{(1)} + \Psi^{(1)} \right)   - \frac{ 2}{\cH} \p_{\perp i} \Phi^{(1)} +2 \frac{ \bar \chi}{\cH} \p_{\perp i}\frac{\ud}{\ud \bar \chi}  \Phi^{(1)}    \nonumber \\
  && + \frac{ \bar \chi}{\cH} \p_{\perp i}   \left(\Phi^{(1)}{'}  + \Psi^{(1)}{'} \right)    + 2 S_{\perp}^{j(1)}    \Bigg\}
-  4  v_{\| \, o}^{(1)} \int_0^{\bar \chi}   \frac{\ud \tilde{\chi}}{\tilde \chi} \left( \Phi^{(1)}+ \Psi^{(1)} \right)      - 2 \Bigg[ \left(\Phi^{(1)}{'} +  \Psi^{(1)}{'} \right)  \nonumber \\
 &&  +\frac{1}{\cH}  \left( 2\frac{\ud}{\ud \bar \chi} \Phi^{(1)}{'}   + \Phi^{(1)}{''} + \Psi^{(1)}{''}  \right)  \Bigg] T^{(1)}  - 2 \Bigg\{ \frac{\ud}{\ud \bar \chi}\left(\Phi^{(1)} + \Psi^{(1)} \right)  + \frac{1}{\cH} \frac{\ud}{\ud \bar \chi} \bigg[ 2 \frac{\ud}{\ud \bar \chi}\Phi^{(1)}   + \left(\Phi^{(1)}{'} + \Psi^{(1)}{'} \right) \bigg] \Bigg\}  \nonumber \\
 && \times \int_0^{\bar \chi} \ud \tilde \chi \left(\Phi^{(1)} - \Psi^{(1)} -2I^{(1)} \right) + 2 \left(\Phi^{(1)} + \Psi^{(1)} \right) \left(\Phi^{(1)} - \Psi^{(1)} - 2 I^{(1)} -2 \kappa^{(1)} \right)  \nonumber \\
  && +4  \Bigg[   \p_{\perp i}\left(\Phi^{(1)}+  \Psi^{(1)} \right)  + \frac{2}{\cH} \p_{\perp i}\frac{\ud}{\ud \bar \chi} \Phi^{(1)} + \frac{1}{\cH} \p_{\perp i}   \left(\Phi^{(1)}{'}  + \Psi^{(1)}{'} \right) -  \frac{2}{\bar \chi \cH}  \p_{\perp i}  \Phi^{(1)}  \Bigg]        \int_0^{\bar \chi} \ud \tilde \chi S_{\perp}^{i(1)}  \nonumber \\
  && +2 \int_0^{\bar \chi}  \ud \tilde{\chi} \Bigg\{   - \left( \Phi^{(1)} +  \Psi^{(1)} \right)  \left[\frac{\ud}{\ud \tilde \chi}\left(\Phi^{(1)} -  \Psi^{(1)}\right) + 2 \left(\Phi^{(1)}{'} + \Psi^{(1)}{'} \right) \right]  - 2 \tilde \p_{\perp j}\left( \Phi^{(1)} +\Psi^{(1)} \right) S_{\perp}^{j(1)}   \nonumber \\
 &&  - 2 \left( \Phi^{(1)} + \Psi^{(1)} \right)    \tilde \p_{\perp m}S_{\perp}^{m(1)}   + 2 \bigg[ \frac{\ud}{\ud \bar \chi}\left( \Phi^{(1)} + \Psi^{(1)} \right)    -  \frac{1}{\tilde\chi} \left( \Phi^{(1)} + \Psi^{(1)} \right)  \bigg] \kappa^{(1)}  \Bigg\}   \;.   \nonumber
 \end{eqnarray}

From Eq. (\ref{Dx_perp-2_2}) and using Eqs. (\ref{kappa1-1}), (\ref{kappa2-1}) and (\ref{kappa3-1}), we obtain  the coordinate convergence lensing term  at second order
\begin{eqnarray}
 \label{Poiss-kappa-2}
 \kappa^{(2)}=- \frac{1}{2}  \p_{\perp i} \Delta x_{\perp}^{i (2)}= \kappa_1^{(2)}+\kappa_2^{(2)}+\kappa_3^{(2)}+\kappa_4^{(2)}  + \kappa^{(2)}_{ \rm post-Born}\;,
  \end{eqnarray}
 where
  \begin{eqnarray}
 \label{Poiss-kappa1-2}
 \kappa_1^{(2)} &=&  \frac{1}{2}  \int_0^{\bar \chi} \ud \tilde \chi  \left(\bar \chi-\tilde \chi\right) \frac{\tilde \chi}{ \bar \chi}   \tilde \nabla^2_\perp \left( \Phi^{(2)} + 2 \omega^{(2)}_{\| }+\Psi^{(2)} - \frac{1}{2} \hat h^{(2)}_{\| } \right)\;, \\
 \label{Poiss-kappa2-2}
 \kappa_2^{(2)} &=& \frac{1}{2}  \int_0^{\bar \chi} \ud \tilde \chi \left(-2  \tilde \p_\perp^i \omega_i^{ (2)} + \frac{4}{\tilde\chi} \omega_\|^{ (2)}+ \Perp^{ij} n^k  \tilde \p_i  \hat h_{jk}^{ (2)}- \frac{3}{\tilde \chi}  \hat h_\|^{ (2)} \right) \;, 
  \end{eqnarray}
 \begin{eqnarray}
 \label{Poiss-kappa3-2}
 \kappa_3^{(2)} &=& \bigg[ \int_0^{\bar \chi} \ud \tilde \chi \; \frac{\tilde \chi}{ \bar \chi} \,  \tilde \nabla^2_\perp \left( \Phi^{(1)}+ \Psi^{(1)} \right)\bigg] \bigg\{ \int_0^{\bar \chi} \ud \tilde \chi \bigg[ 2 \Phi^{(1)}  + \left(\bar \chi-\tilde \chi\right) \left(\Phi^{(1)}{'}+ \Psi^{(1)}{'} \right) \bigg]   -  \frac{1}{\cH}\Delta \ln a^{(1)} \bigg\}    \nonumber \\
 &&-2S_{\perp}^{i(1)}   \bigg\{ \int_0^{\bar \chi} \ud \tilde \chi \,  \frac{\tilde \chi}{ \bar \chi}  \left[2\;  \tilde \p_{\perp i}  \Phi^{(1)} + \left(\bar \chi-\tilde \chi\right) \tilde \p_{\perp i} \left(\Phi^{(1)}{'} +\Psi^{(1)}{'} \right) \right]   -  \frac{1}{\cH} \p_{\perp i}  \Delta \ln a^{(1)} \bigg\} \nonumber \\
 && + \int_0^{\bar \chi} \ud \tilde \chi \, \frac{\tilde \chi}{ \bar \chi}  \bigg[  - 4  \,  \tilde \p_{\perp j} \Psi^{(1)} S_{\perp}^{j(1)}+ \frac{8}{\tilde \chi}   \Psi^{(1)} S_{\|}^{(1)} - 4\, \Psi^{(1)}  \tilde \p_{\perp m} S^{m(1)}\bigg]   \nonumber \\
&&+ 2 \int_0^{\bar \chi} \ud \tilde \chi \bigg\{ \left(\bar \chi-\tilde \chi\right)\frac{\tilde \chi}{ \bar \chi}  \; \bigg[  \tilde\p^i_\perp  \Psi^{(1)} \tilde \p_{\perp i} \left( \Phi^{(1)} + \Psi^{(1)} \right)  +  \left( \Phi^{(1)} + \Psi^{(1)} \right)    \tilde \nabla^2_\perp \Psi^{(1)}\nonumber \\ 
&&  -2 \tilde \p_{\perp i} \left( \Phi^{(1)} - I^{(1)}\right)  \tilde\p^i_\perp  \left( \Phi^{(1)}+ \Psi^{(1)} \right) -2  \left( \Phi^{(1)} - I^{(1)}\right) \tilde \nabla^2_\perp \left( \Phi^{(1)}+ \Psi^{(1)} \right) \bigg] \bigg\} \;,
  \end{eqnarray} 
 \begin{eqnarray}
 \label{Poiss-kappa4-2}
   \kappa_4^{(2)} &=&  - 2 \omega_{\| \, o}^{(2)}- v_{\| \, o}^{(2)}+\frac{3}{4}\hat h_{\| \, o}^{ (2)} - \frac{1}{4} \hat h_{i \, o}^{i (2)} + 2 \Phi_{\, o}^{(1)}  v_{\| \, o}^{(1)} + \frac{1}{2} v^{(1)}_{\perp i \, o } v^{i (1)}_{\perp \, o}-\left(v_{\| \, o}^{(1)}\right)^2  + 4 v_{\| \, o}^{(1)}\Psi_{\, o}^{(1)}  \nonumber\\
  && + \left (\Phi^{(1)}_o-v^{(1)}_{\| \, o}\right) \bigg\{ - \int_0^{\bar \chi} \ud \tilde \chi \bigg[  \tilde \chi   \tilde \nabla^2_\perp \left( \Phi^{(1)}+ \Phi^{(1)} \right) \bigg]+ 4 \kappa_2^{(1)}+ 4 \kappa_1^{(1)} \bigg\}\nonumber \\
 && -\frac{2}{\bar \chi}v_{\| \, o}^{(1)} \bigg\{ \int_0^{\bar \chi} \ud \tilde \chi \left[ 2 \left(\Phi^{(1)}+\Psi^{(1)}\right) + \left(\bar \chi-\tilde \chi\right) \left(\Phi^{(1)}{'} + \Psi^{(1)}{'} \right) \right]   -  \frac{1}{\cH}\Delta \ln a^{(1)} \bigg\} \nonumber \\
    && + v^{(1)}_{\perp i \, o} \bigg\{- 2 S_{\perp}^{i(1)} + \int_0^{\bar \chi} \ud \tilde \chi \bigg[-2\left(1-2\frac{\tilde \chi}{\bar \chi}\right)\tilde \p^i_\perp  \left(\Phi^{(1)}+\Psi^{(1)}\right)+\frac{\tilde \chi}{ \bar \chi}  \left(\bar \chi-\tilde \chi\right) \tilde \p^i_{\perp} \left(\Phi^{(1)}{'} + \Psi^{(1)}{'} \right)  \bigg]  \nonumber \\
  &&  -  \frac{1}{\cH} \p^i_{\perp}  \Delta \ln a^{(1)} \bigg\},
 \end{eqnarray}
and

        \begin{eqnarray}
  &&  \kappa^{(2)}_{ \rm post-Born}  = \int_0^{\bar \chi} \ud \tilde{\chi}  \frac{ \tilde \chi }{\bar \chi} \Bigg\{ \tilde \nabla^2_{\perp} \left(\Phi^{(1)} + \Psi^{(1)} \right)  \delta x_{\|}^{(1)} + \p_{\perp i}  \left(\Phi^{(1)} + \Psi^{(1)} \right)  \p^i_{\perp } \delta  x_{\|}^{(1)}  + \frac{1}{\tilde \chi} \tilde \p_{\perp i} \left(\Phi^{(1)} + \Psi^{(1)}  \right)   \delta x_{\perp}^{i (1)} \nonumber\\
 &&  - \frac{2}{\tilde \chi} \left(\Phi^{(1)} + \Psi^{(1)}  \right)  \kappa^{(1)}  \Bigg\}  + \int_0^{\bar \chi} \ud \tilde{\chi} ~ (\bar \chi - \tilde \chi) \frac{ \tilde \chi }{\bar \chi} \Bigg\{- \bigg[ \tilde \nabla^2_{\perp} \left(\Phi^{(1)}{'} + \Psi^{(1)}{'} \right)  \bigg]  T^{(1)}    -  \tilde \p_{\perp }^i    \left(\Phi^{(1)}{'}  + \Psi^{(1)}{'} \right) \tilde \p_{\perp i} T^{(1)}   \nonumber\\
&& +  \left( \Phi^{(1)} - \Psi^{(1)} - 2I^{(1)} \right)  \tilde \nabla^2_{\perp} \left(\Phi^{(1)} + \Psi^{(1)} \right)    +  \p_{\perp i}  \left(\Phi^{(1)} + \Psi^{(1)} \right) \bigg[\tilde \p^i_{\perp }  \left( \Phi^{(1)}- \Psi^{(1)} \right) - 2 \tilde \p^i_{\perp} I^{(1)} \bigg]      \nonumber\\
&&    +  \frac{2}{\tilde\chi}  \bigg[-     \frac{1}{\tilde\chi}  \left(\Phi^{(1)} + \Psi^{(1)} \right) +  \frac{\ud}{\ud \tilde \chi}  \left(\Phi^{(1)} + \Psi^{(1)} \right) \bigg]  \kappa^{(1)}   -  \frac{2}{\tilde\chi} \tilde \p_{\perp i}  \left(\Phi^{(1)} +  \Psi^{(1)} \right)  S_{\perp}^{i(1)}    - \frac{2}{\tilde\chi}   \left(\Phi^{(1)} + \Psi^{(1)} \right)     \tilde \p_{\perp m}S_{\perp}^{m(1)}  \nonumber\\
&&  + \bigg[  \tilde \p_{\perp i} \tilde \nabla^2_{\perp} \left(\Phi^{(1)} + \Psi^{(1)} \right) + \frac{1}{\tilde\chi} \tilde \p_{\perp i}    \left(\Phi^{(1)}{'} + \Psi^{(1)}{'} \right)    + \frac{2}{\tilde \chi^2}  \tilde \p_{\perp i}   \left(\Phi^{(1)} + \Psi^{(1)} \right) \bigg] \delta x_{\perp}^{i (1)}  -   \bigg[ \tilde \p^{(j}_{\perp} \tilde \p^{m)}_{\perp}  \left(\Phi^{(1)} + \Psi^{(1)} \right)\nonumber \\
&&+ \frac{1}{\tilde \chi} n^{[ j} \p^{m]}_\perp \left(\Phi^{(1)} + \Psi^{(1)} \right) + \frac{1}{\tilde \chi}  \Perp^{jm}  \left(\Phi^{(1)}{'} + \Psi^{(1)}{'} \right)  + \frac{1}{\tilde \chi^2}\Perp^{jm}  \left(\Phi^{(1)} + \Psi^{(1)} \right)    \bigg]\left(\gamma_{mj}^{(1)}+  \Perp_{mj} \kappa^{(1)} + \theta_{mj}^{(1)}\right) \Bigg\}\nonumber \\
 && - 2 \left( \Phi^{(1)}_o+  \Psi^{(1)}_o\right) v_{\| \, o}^{(1)}    - 2 v_{\| \, o}^{(1)} \left(\Phi^{(1)}_o-v^{(1)}_{\| \, o}\right)  - 2  v_{\| \, o}^{(1)}   \int_0^{\bar \chi} \ud \tilde{\chi} ~ \frac{(\bar \chi - \tilde \chi)}{\bar \chi \tilde\chi}  \left(\Phi^{(1)} +  \Psi^{(1)} \right)     -  2\left(\Phi^{(1)}_o-v^{(1)}_{\| \, o}\right) \kappa^{(1)}  \nonumber\\
     &&+  \frac{ 2 }{\bar \chi} v^{ (1)}_{\perp i \, o }   \int_0^{\bar \chi} \ud \tilde{\chi} ~ (\bar \chi - \tilde \chi)  \p^i_{\perp }  \left(\Phi^{(1)} + \Psi^{(1)} \right)  \;.  \nonumber
 \end{eqnarray} 

Finally, in order to obtain $\Delta_g^{(2)}$ [see Eq.\ (\ref{Deltag-2})],  we use 
\begin{equation}
\frac{1}{2} \hat g_\mu^{\mu (2)} -\frac{1}{2} \hat g_\mu^{\nu (1)}  \; \hat g_\nu^{\mu (1)}=\Phi^{(2)}-3  \Psi^{(2)}  - 2\left(\Phi^{(1)}\right)^2-  6 \left(  \Psi^{(1)} \right)^2\;,
\end{equation}

\begin{equation}
  E_{\hat 0}^{0 (2)}   + E_{\hat 0 }^{\| (2)}  = - \Phi^{(2)}+ v_{\|}^{(2)}+ 3 \left(\Phi^{(1)}\right)^2+   v^{(1)}_i v^{i (1)}\;,
\end{equation}

\begin{eqnarray}
 \frac{1}{ \cH} {\hat g_\mu^{\mu(1)}}{'} \Delta \ln a^{(1)} +   \left(\p_{\parallel}\hat g_\mu^{\mu (1)}\right) \Delta x_{\parallel}^{(1)} =- \frac{2}{ \cH} \frac{\ud \,}{\ud \bar \chi}\left(\Phi^{(1)} -3 \Psi^{(1)}  \right)\Delta \ln a^{(1)} - 2\p_{\parallel}\left(\Phi^{(1)} -3 \Psi^{(1)}  \right) T^{(1)}\;,
\end{eqnarray}

\begin{eqnarray}
 \left(\p_{\perp i}\hat g_\mu^{\mu (1)}\right) \Delta x_{\perp}^{i (1)} &=&-2 \bar \chi v^{i (1)}_{\perp \, o } \p_{\perp i}\left(\Phi^{(1)} -3  \Psi^{(1)}  \right) -2 \p_{\perp i}\left(\Phi^{(1)}-3 \Psi^{(1)}  \right)\int_0^{\bar \chi} \ud \tilde \chi \left\{+ \left(\bar \chi-\tilde \chi\right) \left[ \tilde \p^i_\perp \left( \Phi^{(1)}+ \Psi^{(1)} \right) \right]\right\} \;, \nonumber \\
\end{eqnarray}

\begin{equation}
 \frac{2}{\cH}  \delta_g^{(1)}{'} \Delta \ln a^{(1)}   +  2 \left(\p_{\parallel}\delta_g^{(1)}\right) \Delta x_{\parallel}^{(1)} =  - \frac{2}{\cH}  \frac{\ud \,}{\ud  \bar \chi} \left(\delta_g^{(1)}\right) \Delta \ln a^{(1)}- 2\left(\p_{\parallel}\delta_g^{(1)}\right) T^{(1)}\;,
\end{equation}

\begin{eqnarray}
  \Delta x_{\perp i}^{(1)} \,  \p_{\perp}^{i}  \delta_g^{(1)}  &=& \delta x_{\perp i}^{(1)} \, \p_{\perp}^{i} \delta_g^{(1)}  = - \bar \chi v^{i (1)}_{\perp \, o } \p_{\perp i}  \delta_g^{(1)} - \left( \p_{\perp i} \delta_g^{(1)} \right) \int_0^{\bar \chi} \ud \tilde \chi \bigg[ \left(\bar \chi-\tilde \chi\right)  \tilde \p^i_\perp \left( \Phi^{(1)}+ \Psi^{(1)} \right)\bigg] \;, 
  \end{eqnarray}

\begin{equation}
- \frac{2}{\bar \chi^2} \left( \Delta x_{\parallel}^{(1)} \right)^2 =- \frac{2}{\bar \chi^2} \left( T^{(1)} \right)^2- \frac{2}{\bar \chi^2 \cH^2} \left( \Delta \ln a^{(1)} \right)^2- \frac{4}{\bar \chi^2 \cH}  \Delta \ln a^{(1)} T^{(1)}\;,
\end{equation}

\begin{equation}
 \frac{4}{\bar \chi} \Delta x_{\parallel}^{(1)}  \kappa^{(1)}  =  - \frac{4}{\bar \chi \cH}  \Delta \ln a^{(1)}   \kappa^{(1)} -  \frac{4}{\bar \chi}  T^{(1)}  \kappa^{(1)}\;,
\end{equation}

\begin{eqnarray}
 &&-  \left( \p_{\perp j}   \Delta x_{\perp}^{i (1)}  \right)  \left( \p_{\perp i}   \Delta x_{\perp}^{j (1)}  \right)= -2\left(v^{(1)}_{\| \, o}\right)^2-2 \left(\Phi^{(1)}_{\, o}\right)^2+4 v^{(1)}_{\| \, o} \Psi^{(1)}_{\, o} -2 \left( \Psi^{(1)}_{\, o}  \right)^2+ 4 \Phi^{(1)}_{\, o}v^{(1)}_{\| \, o} -4\Phi^{(1)}_{\, o} \Psi^{(1)}_{\, o} \nonumber \\
 &&+ 2 \left(\Phi^{(1)}_{\, o} -  v^{(1)}_{\| \, o}+ \Psi^{(1)}_{\, o} \right)\bigg\{2\left(\Phi^{(1)} +\Psi^{(1)}  - 2 I^{(1)} \right)- \int_0^{\bar \chi} \ud \tilde \chi\bigg[   \frac{\tilde \chi}{ \bar \chi} \left(2\tilde \p_\| + \left(\bar \chi-\tilde \chi\right)\Perp^{mn} \tilde \p_m \tilde \p_n \right)\left( \Phi^{(1)}+ \Psi^{(1)} \right) \bigg] \bigg\} \nonumber \\
& &-  2\left(\Phi^{(1)} +\Psi^{(1)}  - 2 I^{(1)} \right)^2 +2\left(\Phi^{(1)}+ \Psi^{(1)}  - 2 I^{(1)} \right)  \int_0^{\bar \chi} \ud \tilde \chi\bigg[  \frac{\tilde \chi}{ \bar \chi} \left( 2 \tilde \p_\| + \left(\bar \chi-\tilde \chi\right)  \Perp^{mn} \tilde \p_m  \tilde \p_n \right) \left( \Phi^{(1)} + \Psi^{(1)} \right) \bigg]  \nonumber \\
&& -\bigg[  \int_0^{\bar \chi} \ud \tilde \chi  \frac{\tilde \chi}{ \bar \chi} \left( \Perp_j^i \tilde \p_\| + \left(\bar \chi-\tilde \chi\right) \Perp^p_j \Perp^{iq} \tilde \p_q  \tilde \p_p \right) \left( \Phi^{(1)} +\Psi^{(1)} \right) \bigg]  \nonumber \\
&& \times \bigg[\int_0^{\bar \chi} \ud \tilde \chi \frac{\tilde \chi}{ \bar \chi} \left( \Perp^j_i \tilde \p_\| + \left(\bar \chi-\tilde \chi\right) \Perp^n_i \Perp^{jm}  \tilde \p_m   \tilde \p_n \right)  \left( \Phi^{(1)} + \Psi^{(1)} \right) \bigg],
   \end{eqnarray}

\begin{eqnarray}
 &&2 \left(\frac{1}{\bar \chi}  \Delta x_{\perp i}^{(1)}  -\p_{\perp i}   \Delta x_{\parallel}^{(1)}  \right)   \p_{\parallel}   \Delta x_{\perp}^{i (1)}= 2 v^{ (1)}_{\perp i \, o }  v^{i (1)}_{\perp \, o } -2v^{ (1)}_{\perp i \, o } \bigg\{ + 2S_{\perp}^{i(1)}+ \p_{\perp i}   \left(\frac{1}{\cH}  \Delta \ln a^{(1)}+   T^{(1)}  \right)  \nonumber \\
 && - \frac{1}{\bar \chi} \int_0^{\bar \chi} \ud \tilde \chi \left[  \left(\bar \chi-\tilde \chi\right) \tilde \p^i_\perp \left( \Phi^{(1)} +\Psi^{(1)} \right)\right]\bigg\}+ 4S_{\perp}^{i(1)}  \bigg\{ - \frac{1}{\bar \chi}\int_0^{\bar \chi} \ud \tilde \chi \left[ \left(\bar \chi-\tilde \chi\right) \tilde \p^i_\perp \left( \Phi^{(1)}+ \Psi^{(1)} \right)\right] \nonumber \\
&&+\p_{\perp i}   \left(\frac{1}{\cH}  \Delta \ln a^{(1)}+   T^{(1)}  \right) \bigg\}\;,
\end{eqnarray}

\begin{eqnarray}
&&\frac{2}{\cH} {\left( E_{\hat 0}^{0(1)}+ E_{\hat 0}^{\|(1)} \right)}' \Delta \ln a^{(1)}  + 2 \p_{\parallel} \left( E_{\hat 0}^{0(1)}+ E_{\hat 0}^{\|(1)} \right) \Delta x_{\parallel}^{(1)}= \frac{2}{\cH} \frac{\ud \,}{\ud \bar \chi}\left(\Phi^{(1)}- v^{(1)}_{\| }  \right) \Delta \ln a^{(1)} +2  \p_{\parallel} \left( \Phi^{(1)}- v^{(1)}_{\| } \right)  T^{(1)} \nonumber \\
\end{eqnarray}

\begin{eqnarray}
 \p_{\perp i}\left( E_{\hat 0}^{0(1)}+ E_{\hat 0}^{\|(1)} \right)\Delta x_{\perp}^{i (1)}=\bar \chi v^{i (1)}_{\perp \, o } \p_{\perp i}\left( \Phi^{(1)}- v^{(1)}_{\| } \right) + \p_{\perp i}\left( \Phi^{(1)}- v^{(1)}_{\| } \right)   \int_0^{\bar \chi} \ud \tilde \chi \left[ \left(\bar \chi-\tilde \chi\right)\tilde \p^i_\perp \left( \Phi^{(1)} + \Psi^{(1)} \right) \right],
 \end{eqnarray}
 
 \begin{equation}
 -   2\left(\delta n_\|^{ (1)} +  \delta\nu^{(1)} \right)  E_{\hat 0}^{\| (1)} =  -2v^{(1)}_{\| }  \left( \Phi^{(1)}+\Psi^{(1)}\right)\;,
 \end{equation}
 
 \begin{equation}
  -  2E_{\hat 0 }^{\perp i (1)} \p_{\perp i} \left(  \Delta x^{0(1)}+ \Delta x_{\parallel}^{(1)} \right)=+2v^{i (1)}_{\perp } \p_{\perp i} T^{(1)}\;,
  \end{equation}
  
   \begin{eqnarray}
&&   - \left(\p_\|  \Delta x_{\parallel}^{(1)} \right)^2= - \left\{\left(\Phi^{(1)} + \Psi^{(1)}\right)+\frac{1}{\cH}\left[\frac{\ud \,}{\ud \bar \chi} \left( \Phi^{(1)}-  v_\|^{(1)}\right)+ \left(\Phi^{(1)}{'} +\Psi^{(1)}{'} \right)\right]- \frac{\cH'}{\cH^2}\Delta \ln a^{(1)}\right\}^2 \nonumber\\
 &=&  -  \left( \Phi^{(1)}+ \Psi^{(1)}\right)^2  -\frac{1}{\cH^2} \left[\frac{\ud \,}{\ud \bar \chi} \left( \Phi^{(1)}-  v_\|^{(1)}\right)+ \left(\Phi^{(1)}{'} + \Psi^{(1)}{'} \right)\right]^2  -\left(\frac{\cH'}{\cH^2} \right)^2\left(\Delta \ln a^{(1)} \right)^2 \nonumber\\
 &&-\frac{2}{\cH}\left(\Phi^{(1)}+ \Psi^{(1)}\right)\left[\frac{\ud \,}{\ud \bar \chi} \left( \Phi^{(1)}-  v_\|^{(1)}\right)+ \left(\Phi^{(1)}{'} + \Psi^{(1)}{'} \right)\right]
 +2 \frac{\cH'}{\cH^2} \left(\Phi^{(1)}+ \Psi^{(1)}\right)\Delta \ln a^{(1)} \nonumber\\
 && +2\frac{\cH'}{\cH^3} \left[\frac{\ud \,}{\ud \bar \chi} \left( \Phi^{(1)}-  v_\|^{(1)}\right)+ \left(\Phi^{(1)}{'} +\Psi^{(1)}{'} \right)\right]\Delta \ln a^{(1)} 
     \end{eqnarray}
  
  \begin{equation}
 - \left( E_{\hat 0}^{0(1)}+ E_{\hat 0}^{\|(1)} \right)^2= - \left( \Phi^{(1)}-  v_\|^{(1)}  \right)^2\;.
  \end{equation}

 Using $\Delta \ln a^{(2)}$ in Eq.\ (\ref{Poiss-Deltalna-1}) , $\Delta x_{\parallel}^{(2)}$ in Eq.\ (\ref{Poiss-Dx_||-2_2}) ,   $\p_{\parallel} \Delta x_{\parallel}^{(2)}$ in Eq.\ (\ref{Poiss-dDx_||-2_3}) and  $\kappa^{(2)}$ in Eq.\ (\ref{Poiss-kappa-2}), we finally obtain in Poisson gauge the number density fluctuations at second order:

 \begin{eqnarray}
\label{Poiss-Deltag-2}
 \Delta_g^{(2)}
  &=&  \delta_g^{(2)} +   b_e \, \Delta \ln a^{(2)} +  \p_{\parallel} \Delta x_{\parallel}^{(2)}  + \frac{2}{\bar \chi} \Delta x_{\parallel}^{(2)} - 2\kappa^{(2)}  + v_{\|}^{(2)}   -3  \Psi^{(2)}   +\left(\Delta_g^{(1)} \right)^2  \nonumber \\
 &&  - \left(\delta_g^{(1)}\right)^2 -3 \left(\Phi^{(1)}\right)^2-  9 \left(  \Psi^{(1)} \right)^2  +   v^{(1)}_{\perp i} v^{i (1)}_{\perp}  - 2v^{(1)}_{\| }\Psi^{(1)}-6 \Phi^{(1)} \Psi^{(1)}+8\Phi^{(1)} I^{(1)}+8\Psi^{(1)} I^{(1)} -8\left( I^{(1)}\right)^2 \nonumber \\
   &&  -\frac{1}{\cH^2} \left[\frac{\ud \,}{\ud \bar \chi} \left( \Phi^{(1)}-  v_\|^{(1)}\right)+ \left(\Phi^{(1)}{'} + \Psi^{(1)}{'} \right)\right]^2   -\frac{2}{\cH}\left(\Phi^{(1)}+ \Psi^{(1)}\right)\left[\frac{\ud \,}{\ud \bar \chi} \left( \Phi^{(1)}-  v_\|^{(1)}\right)+ \left(\Phi^{(1)}{'} + \Psi^{(1)}{'} \right)\right]\nonumber\\
 &&   \nonumber \\
 &&+ \bigg\{ - \frac{2}{ \cH} \frac{\ud \,}{\ud \bar \chi}\left( v^{(1)}_{\| }  -3 \Psi^{(1)}  + \delta_g^{(1)}\right)  +2 \frac{\cH'}{\cH^2} \left(\Phi^{(1)}+ \Psi^{(1)}\right)   +2\frac{\cH'}{\cH^3} \left[\frac{\ud \,}{\ud \bar \chi} \left( \Phi^{(1)}-  v_\|^{(1)}\right)+ \left(\Phi^{(1)}{'} +\Psi^{(1)}{'} \right)\right]\nonumber \\
  && - \frac{4}{\bar \chi \cH}     \kappa^{(1)} \bigg\} \Delta \ln a^{(1)} + \left[- b_e +  \frac{\ud \ln b_e }{\ud  \ln \bar a}  -\left(\frac{\cH'}{\cH^2} \right)^2  - \frac{2}{\bar \chi^2 \cH^2}  \right] \left( \Delta \ln a^{(1)}\right)^2 - \frac{4}{\bar \chi^2 \cH}  \Delta \ln a^{(1)} T^{(1)}+2v^{i (1)}_{\perp } \p_{\perp i} T^{(1)} \nonumber \\
   &&   - 2\p_{\parallel}\left( v^{(1)}_{\| }  -3 \Psi^{(1)} + \delta_g^{(1)} \right) T^{(1)}  -  \frac{4}{\bar \chi}  T^{(1)}  \kappa^{(1)}  - \frac{2}{\bar \chi^2} \left( T^{(1)} \right)^2  +4 \p_{\perp i}   \left(\frac{1}{\cH}  \Delta \ln a^{(1)}+   T^{(1)}  \right)S_{\perp}^{i(1)} \nonumber \\
    && +2\left(\Phi^{(1)}+ \Psi^{(1)}  - 2 I^{(1)} \right) \int_0^{\bar \chi} \ud \tilde \chi\bigg[  \frac{\tilde \chi}{ \bar \chi} \left( 2 \tilde \p_\| + \left(\bar \chi-\tilde \chi\right)  \Perp^{mn} \tilde \p_m  \tilde \p_n \right) \left( \Phi^{(1)} + \Psi^{(1)} \right) \bigg]  \nonumber \\
     &&  -2   \p_{\perp i} \left(  v_\|^{(1)}  -3  \Psi^{(1)}  + \delta_g^{(1)}   \right) \int_0^{\bar \chi} \ud \tilde \chi  \left(\bar \chi-\tilde \chi\right)  \tilde \p^i_\perp \left( \Phi^{(1)} + \Psi^{(1)} \right) - \frac{4}{\bar \chi}S_{\perp}^{i(1)}  \int_0^{\bar \chi} \ud \tilde \chi \left[ \left(\bar \chi-\tilde \chi\right) \tilde \p^i_\perp \left( \Phi^{(1)}+ \Psi^{(1)} \right)\right]   \nonumber \\
&& -\bigg[  \int_0^{\bar \chi} \ud \tilde \chi  \frac{\tilde \chi}{ \bar \chi} \left( \Perp_j^i \tilde \p_\| + \left(\bar \chi-\tilde \chi\right) \Perp^p_j \Perp^{iq} \tilde \p_q  \tilde \p_p \right) \left( \Phi^{(1)} +\Psi^{(1)} \right) \bigg]  \nonumber \\
&& \times \bigg[\int_0^{\bar \chi} \ud \tilde \chi \frac{\tilde \chi}{ \bar \chi} \left( \Perp^j_i \tilde \p_\| + \left(\bar \chi-\tilde \chi\right) \Perp^n_i \Perp^{jm}  \tilde \p_m   \tilde \p_n \right)  \left( \Phi^{(1)} + \Psi^{(1)} \right) \bigg]\nonumber \\
 &&  -2\left(v^{(1)}_{\| \, o}\right)^2-2 \left(\Phi^{(1)}_{\, o}\right)^2+4 v^{(1)}_{\| \, o} \Psi^{(1)}_{\, o} -2 \left( \Psi^{(1)}_{\, o}  \right)^2+ 4 \Phi^{(1)}_{\, o}v^{(1)}_{\| \, o} -4\Phi^{(1)}_{\, o} \Psi^{(1)}_{\, o}  + 2 v^{ (1)}_{\perp i \, o }  v^{i (1)}_{\perp \, o } \nonumber \\
 &&+ 2 \left(\Phi^{(1)}_{\, o} -  v^{(1)}_{\| \, o}+ \Psi^{(1)}_{\, o} \right)\bigg\{2\left(\Phi^{(1)} +\Psi^{(1)}  - 2 I^{(1)} \right)- \int_0^{\bar \chi} \ud \tilde \chi\bigg[   \frac{\tilde \chi}{ \bar \chi} \left(2\tilde \p_\| + \left(\bar \chi-\tilde \chi\right)\Perp^{mn} \tilde \p_m \tilde \p_n \right)\left( \Phi^{(1)}+ \Psi^{(1)} \right) \bigg] \bigg\} \nonumber \\
&& -2v^{ (1)}_{\perp i \, o } \bigg\{ \bar \chi   \p_{\perp}^i \left(  v_\|^{(1)}  -3  \Psi^{(1)}  + \delta_g^{(1)}   \right)  + 2S_{\perp}^{i(1)}+ \p_{\perp i}   \left(\frac{1}{\cH}  \Delta \ln a^{(1)}+   T^{(1)}  \right)   \nonumber \\
 && - \frac{1}{\bar \chi} \int_0^{\bar \chi} \ud \tilde \chi \left[  \left(\bar \chi-\tilde \chi\right) \tilde \p^i_\perp \left( \Phi^{(1)} +\Psi^{(1)} \right)\right]\bigg\}.
  \end{eqnarray}

This is the main result in Poisson gauge. If we explicitly identify the weak lensing shear and rotation contributions, it becomes
 \begin{eqnarray}
\label{Poiss-Deltag-3}
 \Delta_g^{(2)}
  &=&  \delta_g^{(2)} +   b_e \, \Delta \ln a^{(2)} +  \p_{\parallel} \Delta x_{\parallel}^{(2)}  + \frac{2}{\bar \chi} \Delta x_{\parallel}^{(2)} - 2\kappa^{(2)}  + v_{\|}^{(2)}   -3  \Psi^{(2)}   +\left(\Delta_g^{(1)} \right)^2  - \left(\delta_g^{(1)}\right)^2 - \left(\Phi^{(1)}\right)^2  -  7 \left(  \Psi^{(1)} \right)^2   \nonumber \\
 &&   +   v^{(1)}_{\perp i} v^{i (1)}_{\perp}  - 2v^{(1)}_{\| }\Psi^{(1)}-2 \Phi^{(1)} \Psi^{(1)}
 -2\left(\kappa^{(1)}\right)^2 -2\big|\gamma^{(1)}\big|^2+\vartheta_{ij}^{(1)}\vartheta^{ij(1)}  \nonumber \\
   &&    -\frac{1}{\cH^2} \left[\frac{\ud \,}{\ud \bar \chi} \left( \Phi^{(1)}-  v_\|^{(1)}\right)+ \left(\Phi^{(1)}{'} + \Psi^{(1)}{'} \right)\right]^2 -\frac{2}{\cH}\left(\Phi^{(1)}+ \Psi^{(1)}\right)\left[\frac{\ud \,}{\ud \bar \chi} \left( \Phi^{(1)}-  v_\|^{(1)}\right)+ \left(\Phi^{(1)}{'} + \Psi^{(1)}{'} \right)\right]\nonumber\\
 &&   \nonumber \\
 && +\bigg\{ - \frac{2}{ \cH} \frac{\ud \,}{\ud \bar \chi}\left( v^{(1)}_{\| }  -3 \Psi^{(1)}  + \delta_g^{(1)}\right)  +2 \frac{\cH'}{\cH^2} \left(\Phi^{(1)}+ \Psi^{(1)}\right)   +2\frac{\cH'}{\cH^3} \left[\frac{\ud \,}{\ud \bar \chi} \left( \Phi^{(1)}-  v_\|^{(1)}\right)+ \left(\Phi^{(1)}{'} +\Psi^{(1)}{'} \right)\right]\nonumber \\
  && - \frac{4}{\bar \chi \cH}     \kappa^{(1)} \bigg\} \Delta \ln a^{(1)} + \left[- b_e +  \frac{\ud \ln b_e }{\ud  \ln \bar a}  -\left(\frac{\cH'}{\cH^2} \right)^2  - \frac{2}{\bar \chi^2 \cH^2}  \right] \left( \Delta \ln a^{(1)}\right)^2 - \frac{4}{\bar \chi^2 \cH}  \Delta \ln a^{(1)} T^{(1)}+2v^{i (1)}_{\perp } \p_{\perp i} T^{(1)} \nonumber \\
   &&   - 2\p_{\parallel}\left( v^{(1)}_{\| }  -3 \Psi^{(1)} + \delta_g^{(1)} \right) T^{(1)}  -  \frac{4}{\bar \chi}  T^{(1)}  \kappa^{(1)}  - \frac{2}{\bar \chi^2} \left( T^{(1)} \right)^2  +4 \p_{\perp i}   \left(\frac{1}{\cH}  \Delta \ln a^{(1)}+   T^{(1)}  \right)S_{\perp}^{i(1)} \nonumber \\
     &&  -2   \p_{\perp i} \left(  v_\|^{(1)}  -3  \Psi^{(1)}  + \delta_g^{(1)}   \right) \int_0^{\bar \chi} \ud \tilde \chi  \left(\bar \chi-\tilde \chi\right)  \tilde \p^i_\perp \left( \Phi^{(1)} + \Psi^{(1)} \right) - \frac{4}{\bar \chi}S_{\perp}^{i(1)}  \int_0^{\bar \chi} \ud \tilde \chi \left[ \left(\bar \chi-\tilde \chi\right) \tilde \p^i_\perp \left( \Phi^{(1)}+ \Psi^{(1)} \right)\right]   \nonumber \\
 &&  + 2 v^{ (1)}_{\perp i \, o }  v^{i (1)}_{\perp \, o }  - 2v^{ (1)}_{\perp i \, o } \bigg\{ \bar \chi   \p_{\perp}^i \left(  v_\|^{(1)}  -3  \Psi^{(1)}  + \delta_g^{(1)}   \right)  + 2S_{\perp}^{i(1)}+ \p_{\perp i}   \left(\frac{1}{\cH}  \Delta \ln a^{(1)}+   T^{(1)}  \right)  \nonumber \\
 &&    - \frac{1}{\bar \chi} \int_0^{\bar \chi} \ud \tilde \chi \left[  \left(\bar \chi-\tilde \chi\right) \tilde \p^i_\perp \left( \Phi^{(1)} +\Psi^{(1)} \right)\right]\bigg\}, 
  \end{eqnarray}
where  $\gamma_{ij}^{(1)}$  and $\vartheta_{ij}^{(1)}\vartheta^{ij(1)}$ are given in Eqs. \eqref{gamp} and \eqref{varp}.

\subsection*{Assuming no velocity bias}

At first order, assuming galaxy velocities follow the matter velocity field,  then the comoving velocities are the same and we can write
\begin{eqnarray}
\label{Poiss-partial_||Dx||2-1}
\p_\| \Delta x^{(1)}_\| &=& \Phi^{(1)}  - v^{(1)}_{\| }+\Psi^{(1)}- \frac{1}{\cH} \p_\| v_\|^{(1)} +\frac{1}{\cH} \Psi^{(1)}{'} + \frac{\bar a^2}{\cH \bar \rho_m} \left( \mathcal{E}_m^{\| (1)}- \mathcal{E}_m^{0 (0)} v_\|^{(1)} \right)- \frac{\cH'}{\cH^2}\Delta \ln a^{(1)}.
\end{eqnarray}
Then
\begin{eqnarray}
\label{Poiss-Deltag-1_3}
\Delta_g^{(1)}
 &=& \delta_g^{(1)} + \left( b_e  - \frac{\cH'}{\cH^2} - \frac{2}{\bar \chi \cH}\right)\Delta \ln a^{(1)}- \frac{1}{\cH} \p_\| v_\|^{(1)}+ \frac{1}{\cH}  \Psi^{(1)}{'} +  \Phi^{(1)}-2 \Psi^{(1)}  - b_m v_\|^{(1)} - \frac{2}{\bar \chi} T^{(1)}- 2 \kappa^{(1)}\nonumber \\
 &+& \frac{a^2}{\cH \bar \rho_m}  \mathcal{E}_m^{\| (1)}\;.
\end{eqnarray}

At second order,  through Eq.\ (\ref{Poiss-ConEqu||-1}) we have
\begin{eqnarray}
\label{Poiss-Deltalna-21}
 \Delta\ln a^{(2)}&=& + \Phi^{(2)}_{\, o}- v^{(2)}_{\| \, o}-\left(\Phi^{(1)}_{\, o} \right)^2-6\Phi^{(1)}_{\, o} v^{(1)}_{\| \, o} +  v^{(1)}_{k\, o} v^{k (1)}_o +2  \Psi^{(1)}_{\, o} v^{(1)}_{\| \, o}  + 2\left(\Phi^{(1)}_o-v^{(1)}_{\| \, o}\right) \bigg[-3 \Phi^{(1)}- \frac{1}{\cH} \p_\| v_\|^{(1)} \nonumber \\
&& + \frac{1}{\cH}  \Psi^{(1)}{'} - b_m v_\|^{(1)} +2 \bar \chi \p_\| \Phi^{(1)} -\bar \chi \left(\Phi^{(1)}{'} - \Psi^{(1)}{'} \right) +\frac{a^2}{\cH \bar \rho_m} \mathcal{E}_m^{\| (1)} + 4 I^{(1)} \bigg] - 2 v_{\perp \, o}^{i(1)} \bigg[  \bar \chi \, \p_{\perp i} \left(\Phi^{(1)} +  v^{(1)}_{\|} \right)\nonumber \\
&& -  2\left. \int_0^{\bar \chi} \ud \tilde \chi \, \tilde\p_{\perp i}\Phi^{(1)} \right] - \Phi^{(2)}+  v^{(2)}_{\|} +  7 \left(\Phi^{(1)}\right)^2  -  \left(1+2 b_m \right) \left(v^{(1)}_{\|}\right)^2+v_{\perp i}^{(1)} v_{\perp}^{i(1)} 
-2 v^{(1)}_{\|}  \Psi^{(1)}  +2 b_m \Phi^{(1)} v^{(1)}_{\|}  \nonumber \\
&& - \frac{2}{\cH}\left( \Phi^{(1)} - v^{(1)}_{\| }\right) \left(  \Psi^{(1)}{'} - \p_\|  v^{(1)}_{\|}  +  \frac{a^2}{\bar \rho_m} \mathcal{E}_m^{\| (1)} \right)- 4 \bigg[ 3 \Phi^{(1)} +  \frac{1}{\cH} \p_\| v_\|^{(1)}  - \frac{1}{\cH}  \Psi^{(1)}{'} + b_m v_\|^{(1)}\nonumber \\
&&-\frac{\bar a^2}{\cH \bar \rho_m} \mathcal{E}_m^{\| (1)} \bigg] I^{(1)}  + 4 v_{\perp i}^{(1)}S_{\perp}^{i(1)}-2 \p_\| \left(\Phi^{(1)} +  v^{(1)}_{\|}\right) T^{(1)}-2 \bigg[ 2 \p_\|  \Phi^{(1)}  - \left(\Phi^{(1)}{'} - \Psi^{(1)}{'} \right)\bigg] \nonumber \\
&&\times \int_0^{\bar \chi} \ud \tilde \chi\left[ 2 \Phi^{(1)} + \left(\bar \chi-\tilde \chi\right) \left(\Phi^{(1)}{'} + \Psi^{(1)}{'} \right) \right]- 2 \bigg[\p_{\perp i}\left(\Phi^{(1)} +  v^{(1)}_{\|} \right)   - \frac{1}{\bar \chi} v^{(1)}_{\perp i}\bigg] \nonumber \\
&&     \times \int_0^{\bar \chi} \ud \tilde \chi \left[ \left(\bar \chi-\tilde \chi\right) \tilde\p^i_\perp \left( \Phi^{(1)} +  \Psi^{(1)} \right)\right]+ 2I^{(2)} -4 \int_0^{\bar \chi} \ud \tilde \chi  \bigg[ \left(  \Psi^{(1)} +2 I^{(1)}\right) \left(\Phi^{(1)}{'} + \Psi^{(1)}{'} \right)  \nonumber \\
&&+  \left(\Phi^{(1)} + \Psi^{(1)} \right) \frac{\ud}{\ud \tilde \chi} \Phi^{(1)}  +2  S_{\perp }^{i(1)}\tilde \p_{\perp i} \Phi^{(1)}  \bigg]+ \Delta\ln a^{(2)}_{\rm post-Born}\;, 
\end{eqnarray}

\begin{eqnarray}
\label{Poiss-Dx0-2_3}
\Delta x^{0(2)}&=& + \frac{1}{\cH} \Phi^{(2)}_o-\frac{1}{\cH} v^{(2)}_{\| \, o}-\left(\frac{\cH'}{\cH^3}+\frac{2}{\cH}\right)\left(\Phi^{(1)}_o\right)^2+2\left( \frac{\cH'}{\cH^3}-\frac{2}{\cH}\right)\Phi^{(1)}_o v^{(1)}_{\| \, o}- \frac{\cH'}{\cH^3}\left(v^{(1)}_{\| \, o}\right)^2 + \frac{1}{\cH}v^{(1)}_{\perp i \, o}v^{i (1)}_{\perp \, o}\nonumber \\
&&+\frac{2}{\cH}  \Psi^{(1)}_{\, o} v^{(1)}_{\| \, o}+ 2 \left(\Phi^{(1)}_o-v^{(1)}_{\| \, o}\right)   \bigg\{\left( \frac{\cH'}{\cH^3}- \frac{2}{\cH}\right) \Phi^{(1)}-\left[ \frac{\cH'}{\cH^3}+ \frac{1}{\cH}\left(1+b_m\right)\right] v^{(1)}_{\|} -\frac{\bar \chi}{\cH}\Phi^{(1)}{'}  \nonumber \\  
 && - \frac{1}{\cH^2} \left(\p_\|  v^{(1)}_{\|}  - \frac{\bar a^2}{\bar \rho_m} \mathcal{E}^{\| (1)}\right)  +\left(\frac{\bar \chi}{\cH} + \frac{1}{\cH^2}\right) \Psi^{(1)}{'}+2 \frac{\bar \chi}{\cH} \p_\| \Phi^{(1)} -  2 \left( \frac{\cH'}{\cH^3}-\frac{1}{\cH}\right) I^{(1)} \bigg\} \nonumber \\  
 &&  - 2 v_{\perp \, o}^{i(1)}  \bigg[\frac{\bar \chi}{\cH}  \p_{\perp i} \left(\Phi^{(1)} +  v^{(1)}_{\|} \right)  -  \frac{2}{\cH} \int_0^{\bar \chi} \ud \tilde \chi \, \tilde \p_{\perp i} \Phi^{(1)} \bigg]  -\frac{1}{\cH} \Phi^{(2)} + \frac{1}{\cH}  v^{(2)}_{\|} +   \left(- \frac{\cH'}{\cH^3}+\frac{6}{\cH}\right) \left(\Phi^{(1)}\right)^2      \nonumber \\
&&- \left[ \frac{\cH'}{\cH^3} + \frac{2}{\cH}\left(1+ b_m \right) \right]\left(v^{(1)}_{\|}\right)^2+  \frac{1}{\cH} v_{\perp i}^{(1)} v_{\perp}^{i(1)} - \frac{2}{\cH}  v^{(1)}_{\|}\Psi^{(1)} +2 \left[ \frac{\cH'}{\cH^3}+ \frac{1}{\cH}\left(1+ b_m\right)\right]\Phi^{(1)} v^{(1)}_{\| } \nonumber \\
&&  +  \frac{2}{\cH^2} \left( \Phi^{(1)} - v^{(1)}_{\| }\right) \left(  - \Psi^{(1)}{'} + \p_\|  v^{(1)}_{\|}  - \frac{\bar a^2}{\bar \rho_m} \mathcal{E}_m^{\| (1)} \right) + 4 \bigg\{ \left( \frac{\cH'}{\cH^3}- \frac{2}{\cH}\right) \Phi^{(1)}-\left[ \frac{\cH'}{\cH^3}+ \frac{1}{\cH}\left(1+b_m \right)\right] v^{(1)}_{\|} \nonumber \\
&& -  \frac{1}{\cH^2}  \left(- \Psi^{(1)}{'} + \p_\|  v^{(1)}_{\|}  -  \frac{\bar a^2}{\bar \rho_m}\mathcal{E}_m^{\| (1)} \right) - \left( \frac{\cH'}{\cH^3}+ \frac{1}{\cH}\right)I^{(1)} \bigg\} I^{(1)}  + \frac{4}{\cH} v_{\perp i}^{(1)}S_{\perp}^{i(1)}  -  \frac{2}{\cH} \p_\| \left(\Phi^{(1)} +  v^{(1)}_{\|} \right) T^{(1)} \nonumber \\
&&- \frac{2}{\cH}  \bigg[ 2 \p_\| \Phi^{(1)} - \left(\Phi^{(1)}{'}- \Psi^{(1)}{'} \right)\bigg]  \int_0^{\bar \chi} \ud \tilde \chi\left[ 2 \Phi^{(1)}+ \left(\bar \chi-\tilde \chi\right) \left(\Phi^{(1)}{'} +  \Psi^{(1)}{'} \right) \right]  - \frac{2}{\cH} \left[\p_{\perp i}\left(\Phi^{(1)} +  v^{(1)}_{\|} \right)  \right.\nonumber \\
&& \left. - \frac{1}{\bar \chi}  v^{(1)}_{\perp i}\right] \int_0^{\bar \chi} \ud \tilde \chi \left[\left(\bar \chi-\tilde \chi\right) \tilde\p^i_\perp \left( \Phi^{(1)}+ \Psi^{(1)} \right) \right]   +  \frac{2}{\cH} I^{(2)}  - \frac{4}{\cH} \int_0^{\bar \chi} \ud \tilde \chi  \left[ \left(   \Psi^{(1)}+2 I^{(1)}\right) \left(\Phi^{(1)}{'}+ \Psi^{(1)}{'} \right) \right.\nonumber \\
&& \left. +\left(\Phi^{(1)}+  \Psi^{(1)} \right)  \frac{\ud}{\ud \tilde \chi} \Phi^{(1)} +2 S_{\perp }^{i(1)} \tilde\p_{\perp i}\Phi^{(1)}   \right]+ \Delta x^{0(2)}_{\rm post-Born}\;,
\end{eqnarray}

\begin{eqnarray} 
\label{Poiss-dDx_||-2_4}
&& \p_\| \Delta x_{\parallel}^{(2)}=\left(\Phi^{(1)}_o\right)^2 + 2\Phi^{(1)}_o v^{(1)}_{\| \, o} + 2 v^{(1)}_{\| \, o} \Psi^{(1)}_{\, o} -  \left(\Psi^{(1)}_{\, o}\right)^2   -  v^{(1)}_{\perp k \, o} v^{k (1)}_{\perp \, o}+ 2\left(\Phi^{(1)}_o-v^{(1)}_{\| \, o}\right)\nonumber \\
&& \times \bigg\{\left(-\frac{\cH' }{\cH^3}+ \frac{1}{\cH}\right) \left[\frac{\bar a^2}{\bar \rho_m} \Em^{\| (1)}-\p_\| v_\|^{(1)}-\cH \left( b_m  + 1\right) v_\|^{(1)} +\Psi^{(1)}{'} \right] - \frac{\bar \chi}{\cH} \frac{\ud \,}{\ud \bar \chi}\left[ 2 \p_\| \Phi^{(1)}  -\left(\Phi^{(1)}{'}- \Psi^{(1)}{'} \right)  \right] \nonumber \\
&&+\left (  \Phi^{(1)}+ \Psi^{(1)} \right) -\bar \chi \frac{\ud}{\ud \bar \chi}\left( \Phi^{(1)} + \Psi^{(1)} \right) - \frac{1}{\cH^2}  \frac{\ud \,}{\ud \bar \chi} \bigg[\frac{\bar a^2}{\bar \rho_m} \Em^{\| (1)}-\p_\| v_\|^{(1)}-\cH \left( b_m  + 1\right) v_\|^{(1)}  +\Psi^{(1)}{'} \bigg]\bigg\} -2 v_{\perp \, o}^{i(1)} \nonumber \\
&& \times \bigg[\frac{1}{\cH}  \p_{\perp i} \left(\Phi^{(1)} -  v^{(1)}_{\|} \right) -\frac{\bar \chi}{\cH}  \frac{\ud \,}{\ud \bar \chi} \p_{\perp i} \left(\Phi^{(1)} +  v^{(1)}_{\|} \right)  -4  \delta_{il} S_{\perp }^{l (1)}  \bigg]+\Phi^{(2)} + \Psi^{(2)} -\frac{1}{2} \hat h_{\|}^{(2)}+  \frac{1}{ \cH} \Psi^{(2)}{'}-  \frac{1}{2 \cH} \hat h^{(2)}_{\| }{'}\nonumber \\
&&  +2\frac{\bar a^2}{ \cH  \bar \rho_m} \left[ \frac{1}{2} \Em^{\| (2)} -\Em^{0 (1)}  v_\|^{(1)}-  \Em^{\| (1)} \left(\delta_m^{(1)} - \Phi^{(1)} \right) \right] - 2 b_m \left(\frac{1}{2} v_\|^{(2)}-\delta_m^{(1)}  v_\|^{(1)} +2 \Phi^{(1)} v_\|^{(1)} \right) -\frac{2}{ \cH }\bigg[ \frac{1}{2}\p_\| v_\|^{(2)}- v_\|^{(1)} \p_\| v_\|^{(1)} \nonumber \\
&&  +\frac{\cH}{2} v_\|^{(2)}    -\frac{2}{\bar \chi} \left( v_\|^{(1)} \right)^2 - v_\|^{(1)}  \p_{\perp j}  v_\perp^{j(1)} -2 v_\|^{(1)}\Psi^{(1)}{'}    - \Phi^{(1)} \p_\| \Phi^{(1)} +2\Psi^{(1)} \p_\| \Phi^{(1)} \bigg]  -4\left(\Phi^{(1)}\right)^2+4 \left( \Psi^{(1)}\right)^2\nonumber \\ 
&& -\frac{2}{\cH}   \frac{\ud \,}{\ud \bar \chi} \bigg[ \frac{7}{2} \left(\Phi^{(1)}\right)^2 - \Phi^{(1)} v^{(1)}_{\| } + \frac{1}{2} \left(v^{(1)}_{\|}\right)^2 -  v^{(1)}_{\|}  \Psi^{(1)} +\frac{1}{2} v_{\perp i}^{(1)} v_{\perp}^{i(1)} \bigg] + \frac{4}{\cH}  \Phi^{(1)} \bigg[ 2 \p_\| \Phi^{(1)}-\left(\Phi^{(1)}{'}- \Psi^{(1)}{'} \right)\bigg] \nonumber \\ 
&& +  \frac{2}{\cH^2}\left( \Phi^{(1)} - v^{(1)}_{\| }\right) \frac{\ud \,}{\ud \bar \chi}\left[\frac{\bar a^2}{\bar \rho_m} \Em^{\| (1)}-\p_\| v_\|^{(1)}-\cH \left( b_m  + 1\right) v_\|^{(1)}+ \Psi^{(1)}{'} \right]  -\frac{2}{\cH} \bigg( 3 \Phi^{(1)} -2 \Psi^{(1)}  -  v^{(1)}_{\|} \bigg)\left(\Phi^{(1)}{'} + \Psi^{(1)}{'} \right)  \nonumber \\ 
&&+2\bigg\{  \frac{1}{\cH} \left( \Phi^{(1)} + \Psi^{(1)} \right)+\frac{\cH'}{\cH^3}\left( \Phi^{(1)} - v^{(1)}_{\| }\right)+  \frac{1}{\cH^2}\bigg[\frac{\bar a^2}{\bar \rho_m} \Em^{\| (1)}-\p_\| v_\|^{(1)}   -\cH \left( b_m  + 1\right) v_\|^{(1)}  + \Psi^{(1)}{'}  \bigg]\bigg\} \bigg[\frac{\bar a^2}{\bar \rho_m} \Em^{\| (1)}-\p_\| v_\|^{(1)}\nonumber \\
&& -\cH \left( b_m  + 1\right) v_\|^{(1)} + \Psi^{(1)}{'} \bigg] +  \frac{2}{\cH} \left[2 \frac{\ud \,}{\ud  \bar \chi} \Phi^{(1)}  -  \p_\| \left(\Phi^{(1)} +  v^{(1)}_{\|} \right)\right]  \left(\Phi^{(1)} + \Psi^{(1)} \right) +\frac{2}{\cH}v_{\perp i}^{(1)} \p^i_\perp \bigg( \Phi^{(1)}+\Psi^{(1)} \bigg)\nonumber \\ 
&&   +\frac{2}{\cH} \frac{\ud \,}{\ud \bar \chi} \p_\| \left(\Phi^{(1)} +  v^{(1)}_{\|} \right) T^{(1)}   + 4 \bigg\{ \left(-\frac{\cH' }{\cH^3} +\frac{1}{\cH}\right)  \bigg[\frac{\bar a^2}{\bar \rho_m} \Em^{\| (1)}-\p_\| v_\|^{(1)}-\cH \left( b_m  + 1\right) v_\|^{(1)}+\Psi^{(1)}{'} \bigg]\nonumber \\ 
&&     - \frac{1}{\cH^2} \frac{\ud \,}{\ud \bar \chi}\bigg[\frac{\bar a^2}{\bar \rho_m} \Em^{\| (1)}-\p_\| v_\|^{(1)}-\cH \left( b_m  + 1\right) v_\|^{(1)}+ \Psi^{(1)}{'}  \bigg]  + \Phi^{(1)} +\Psi^{(1)}\bigg\} I^{(1)} + 2\bigg\{ \frac{\ud}{\ud \bar \chi}\left( \Phi^{(1)} + \Psi^{(1)} \right) \nonumber \\ 
&&  + \frac{1}{\cH} \frac{\ud \,}{\ud \bar \chi} \bigg[2 \p_\| \Phi^{(1)}-\left(\Phi^{(1)}{'}-\Psi^{(1)}{'} \right)\bigg]  \bigg\}  \int_0^{\bar \chi} \ud \tilde \chi \;\bigg[ 2 \Phi^{(1)} + \left(\bar \chi-\tilde \chi\right)\left(\Phi^{(1)}{'} + \Psi^{(1)}{'} \right) \bigg]+  4\bigg[\frac{1}{\cH}\p_{\perp i}\left(  \Phi^{(1)} - v^{(1)}_{\| } \right) \nonumber \\
 &&+ \frac{1}{\cH \bar \chi} v_{\perp i}^{(1)}  -\frac{1}{\cH}\frac{\ud \,}{\ud \bar \chi} v_{\perp i}^{(1)}  -2S_{\perp }^{j(1)} \delta_{ij}  \bigg]   S_{\perp }^{i(1)} + \frac{2}{\cH} \frac{\ud \,}{\ud \bar \chi} \left[\p_{\perp i}\left(\Phi^{(1)} +  v^{(1)}_{\|} \right)  - \frac{1}{\bar \chi} v^{(1)}_{\perp i}\right]  \nonumber \\
 && \times \int_0^{\bar \chi} \ud \tilde \chi \left[ \left(\bar \chi-\tilde \chi\right)  \tilde \p^i_\perp \left( \Phi^{(1)} + \Psi^{(1)} \right) \right] -  \frac{\cH'}{\cH^2} \Delta \ln a^{(2)}+\left[-\frac{\cH'' }{\cH^3} +3\left( \frac{\cH' }{\cH^2} \right)^2 + \frac{\cH' }{\cH^2} \right] \left( \Delta \ln a^{(1)} \right)^2 \nonumber \\
&&+ 2\bigg\{-\frac{\cH' }{\cH^2} \left( \Phi^{(1)} + \Psi^{(1)} \right)- \frac{1}{\cH} \frac{\ud}{\ud \bar \chi}\left( \Phi^{(1)} + \Psi^{(1)} \right)- \left(\frac{\cH' }{\cH^3} +\frac{1}{\cH} \right)\bigg[\frac{\bar a^2}{\bar \rho_m} \Em^{\| (1)}-\p_\| v_\|^{(1)}-\cH \left( b_m  + 1\right) v_\|^{(1)}  \nonumber \\
&&+ \Psi^{(1)}{'} \bigg] \bigg\} \Delta \ln a^{(1)} +4  \int_0^{\bar \chi} \ud \tilde \chi \bigg\{ \left(\Phi^{(1)} +\Psi^{(1)} \right) \bigg[\left(\Phi^{(1)}{'} +  \Psi^{(1)}{'} \right) +  \frac{\ud}{\ud \tilde \chi}\Phi^{(1)} \bigg] \bigg\}+ \p_\| \Delta x_{\|  \rm post-Born}^{(2)}\;.
\end{eqnarray}

The last equation can be further expanded,  applying again Eq.\  (\ref{Poiss-ConEqu||-1}):
 \begin{eqnarray}
  && \frac{1}{\cH}\frac{\ud \,}{\ud \bar \chi} v_{\perp i}^{(1)} =   \frac{1}{\cH} \frac{\bar a^2}{\bar \rho_m} \Perp_{ij}  \Em^{j (1)} - \frac{1}{\cH}\p_{\|} v_{\perp i}^{(1)}  - \left( b_m  + 1\right) v^{(1)}_{\perp i} -  \frac{1}{\cH}\p_{\perp i} \Phi^{(1)}\;, \\
  && \frac{2}{\cH} \frac{\ud \,}{\ud \bar \chi} \p_\| \left(\Phi^{(1)} +  v^{(1)}_{\|} \right) T^{(1)} = +\frac{2}{\cH} \bigg[    -\frac{\bar a^2}{\bar \rho_m} \p_\| \Em^{\| (1)}+ \p_\|^2 v_\|^{(1)}+\cH \left( b_m  + 1\right) \p_\| v_\|^{(1)}-\p_\| \Phi^{(1)}{'} +2 \p_\|^2 \Phi^{(1)}\bigg] T^{(1)}\;, \\
&&\frac{\bar \chi}{\cH}  \frac{\ud \,}{\ud \bar \chi} \p_{\perp i} \left(\Phi^{(1)} +  v^{(1)}_{\|} \right)= -\frac{\bar \chi}{\cH}\p_{\perp i}  \left(-\Phi^{(1)}{'}  -\frac{\bar a^2}{\bar \rho_m} \Em^{\| (1)}+\p_\| v_\|^{(1)}+\cH \left( b_m  + 1\right) v_\|^{(1)}+ 2\p_\| \Phi^{(1)}\right)\nonumber \\
 &&+\frac{1}{\cH}\p_{\perp i} \left(\Phi^{(1)} +  v^{(1)}_{\|} \right)  \\
&& \frac{2}{\cH} \frac{\ud \,}{\ud \bar \chi} \left[\p_{\perp i}\left(\Phi^{(1)} +  v^{(1)}_{\|} \right)  - \frac{1}{\bar \chi}  v^{(1)}_{\perp i}\right] =\frac{2}{\cH}\bigg[\p_{\perp i}\bigg( -\Phi^{(1)}{'} -\frac{\bar a^2}{\bar \rho_m} \Em^{\| (1)}+\p_\| v_\|^{(1)}+\cH \left( b_m  + 1\right) v_\|^{(1)}+ 2 \p_\| \Phi^{(1)}\bigg) \nonumber \\
&&   +\frac{1}{\bar \chi} \left(-2\p_{\perp i} \Phi^{(1)} - \p_{\|} v^{(1)}_{\perp i} - \p_{\perp i} v^{(1)}_{\|} +\frac{\bar a^2}{\bar \rho_m} \Perp_{ij} \Em^{j (1)}-\cH \left( b_m  + 1\right) v^{(1)}_{\perp i} \right)+\frac{1}{\bar \chi^2} v^{(1)}_{\perp i}\bigg] \\
&&  \frac{\ud \,}{\ud \bar \chi} \bigg[\frac{\bar a^2}{\bar \rho_m} \Em^{\| (1)}-\p_\| v_\|^{(1)}-\cH \left( b_m  + 1\right) v_\|^{(1)} +\Psi^{(1)}{'}\bigg]= -{\left(\frac{\bar a^2}{\bar \rho_m} \Em^{\| (1)}\right)}'+\frac{\bar a^2 \cH}{\bar \rho_m}  \left( b_m  + 1\right)  \Em^{\| (1)}+ \cH^2 \frac{\ud b_m}{\ud \ln a}  v_\|^{(1)}  \nonumber \\
&& + b_m \bigg[ -\cH^2 \left( b_m  + 1\right) v_\|^{(1)} - 2 \cH \p_\| v_\|^{(1)} +\left(\cH'-\cH^2\right) v_\|^{(1)} - \cH \p_\| \Phi^{(1)} \bigg] -2 \cH \p_\| v_\|^{(1)} - \p_\|^2 \Phi^{(1)} -\p_\|^2 v^{(1)}_{\|}\nonumber \\
&&+ \left(\cH'-\cH^2\right) v_\|^{(1)} - \cH  \p_\| \Phi^{(1)}+\frac{\ud \,}{\ud \bar \chi}\Psi^{(1)}{'} \;,\\
&&\frac{\ud \,}{\ud \bar \chi} \bigg[ \frac{7}{2} \left(\Phi^{(1)}\right)^2  - \Phi^{(1)} v^{(1)}_{\| } + \frac{1}{2} \left(v^{(1)}_{\|}\right)^2 - v^{(1)}_{\|}  \Psi^{(1)} +\frac{1}{2} v_{\perp i}^{(1)} v_{\perp}^{i(1)}  \bigg]=6\Phi^{(1)}\frac{\ud \,}{\ud \bar \chi}\Phi^{(1)}-\Phi^{(1)}\Phi^{(1)}{'}+  v^{(1)}_{\| } \Phi^{(1)}{'}\nonumber \\ 
&&-v^{(1)}_{\| }\frac{\ud \,}{\ud \bar \chi}\Psi^{(1)}-\Phi^{(1)}\p_\|v^{(1)}_{\| }-\cH \Phi^{(1)}v^{(1)}_{\| }+ v^{(1)}_{\| } \p_\| v^{(1)}_{\| }+\cH \left(v^{(1)}_{\| }\right)^2- \Psi^{(1)} \p_\| \Phi^{(1)}  - \Psi^{(1)} \p_\| v^{(1)}_{\| }- \cH v^{(1)}_{\| } \Psi^{(1)}\nonumber \\ 
&&+ v_{\perp i}^{(1)}  \p_\|  v_{\perp}^{i(1)}+ \cH v_{\perp i}^{(1)}   v_{\perp}^{i(1)}+ v_{\perp i}^{(1)} \p_{\perp}^i \Phi^{(1)} -\left(\Phi^{(1)}-v^{(1)}_{\| }+\Psi^{(1)}\right)\left(\frac{\bar a^2}{\bar \rho_m} \Em^{\| (1)}- \cH b_m v^{(1)}_{\| } \right) \nonumber \\ 
&&- v_{\perp i}^{(1)} \left(\frac{\bar a^2}{\bar \rho_m}\Perp^i_j \Em^{j (1)}- \cH b_m v^{i(1)}_{\perp }\right)\;.
   \end{eqnarray}
 
 \begin{eqnarray} 
\label{Poiss-dDx_||-2_5}
&& \p_\| \Delta x_{\parallel}^{(2)}=\left(\Phi^{(1)}_o\right)^2 + 2\Phi^{(1)}_o v^{(1)}_{\| \, o} + 2 v^{(1)}_{\| \, o} \Psi^{(1)}_{\, o} -  \left(\Psi^{(1)}_{\, o}\right)^2   -  v^{(1)}_{\perp k \, o} v^{k (1)}_{\perp \, o}+ 2\left(\Phi^{(1)}_o-v^{(1)}_{\| \, o}\right)\bigg\{\left(-\frac{\cH' }{\cH^3}+ \frac{1}{\cH}\right) \bigg[\frac{\bar a^2}{\bar \rho_m} \Em^{\| (1)}\nonumber \\
&&  -\p_\| v_\|^{(1)}-\cH \left( b_m  + 1\right) v_\|^{(1)} +\Psi^{(1)}{'} \bigg] - \frac{\bar \chi}{\cH} \frac{\ud \,}{\ud \bar \chi}\left[ 2 \p_\| \Phi^{(1)}  -\left(\Phi^{(1)}{'}- \Psi^{(1)}{'} \right)  \right] +  \Phi^{(1)}+ \Psi^{(1)} -\bar \chi \frac{\ud}{\ud \bar \chi}\left( \Phi^{(1)} + \Psi^{(1)} \right)\nonumber \\
&&  - \frac{1}{\cH^2}  \bigg[ -2 \cH \p_\| v_\|^{(1)} - \p_\|^2 \Phi^{(1)} -\p_\|^2 v^{(1)}_{\|} + \left(\cH'-\cH^2\right) v_\|^{(1)} - \cH  \p_\| \Phi^{(1)}+\frac{\ud \,}{\ud \bar \chi}\Psi^{(1)}{'} \bigg]  - \frac{1}{\cH^2}  \bigg[-{\left(\frac{\bar a^2}{\bar \rho_m} \Em^{\| (1)}\right)}'\nonumber \\
&&+\frac{\bar a^2 \cH}{\bar \rho_m}  \left( b_m  + 1\right)  \Em^{\| (1)}+ \cH^2 \frac{\ud b_m}{\ud \ln a}  v_\|^{(1)}  - \cH^2  b_m \left( b_m  + 1\right) v_\|^{(1)} - 2 \cH  b_m \p_\| v_\|^{(1)} +  \left(\cH'-\cH^2\right)  b_m v_\|^{(1)}  
 - \cH  b_m \p_\| \Phi^{(1)} \bigg]  \bigg\}  \nonumber \\
&&-2 v_{\perp \, o}^{i(1)}  \bigg[ +\frac{2}{\cH}\p_{\perp i} \Phi^{(1)}  -\frac{\bar \chi}{\cH}\p_{\perp i}  \left(-\Phi^{(1)}{'}  -\frac{\bar a^2}{\bar \rho_m} \Em^{\| (1)}+\p_\| v_\|^{(1)}+\cH \left( b_m  + 1\right) v_\|^{(1)}+ 2\p_\| \Phi^{(1)}\right)  -4  \delta_{il} S_{\perp }^{l (1)}  \bigg] \nonumber \\
 && +\Phi^{(2)} + \Psi^{(2)} -\frac{1}{2} \hat h_{\|}^{(2)}+  \frac{1}{ \cH} \Psi^{(2)}{'}-  \frac{1}{2 \cH} \hat h^{(2)}_{\| }{'} -\frac{2}{ \cH }\bigg[ \frac{1}{2}\p_\| v_\|^{(2)}- v_\|^{(1)} \p_\| v_\|^{(1)}    +\frac{\cH}{2} v_\|^{(2)}    -\frac{2}{\bar \chi} \left( v_\|^{(1)} \right)^2 - v_\|^{(1)}  \p_{\perp j}  v_\perp^{j(1)}  \nonumber \\
&&   -2 v_\|^{(1)}\Psi^{(1)}{'}  - \Phi^{(1)} \p_\| \Phi^{(1)} +2\Psi^{(1)} \p_\| \Phi^{(1)} \bigg]   -4\left(\Phi^{(1)}\right)^2+4 \left( \Psi^{(1)}\right)^2  -\frac{2}{\cH}   \bigg[6\Phi^{(1)}\frac{\ud \,}{\ud \bar \chi}\Phi^{(1)} -\Phi^{(1)}\Phi^{(1)}{'} +  v^{(1)}_{\| } \Phi^{(1)}{'} \nonumber \\ 
&&-v^{(1)}_{\| }\frac{\ud \,}{\ud \bar \chi}\Psi^{(1)}-\Phi^{(1)}\p_\|v^{(1)}_{\| }-\cH \Phi^{(1)}v^{(1)}_{\| }+ v^{(1)}_{\| } \p_\| v^{(1)}_{\| }+\cH \left(v^{(1)}_{\| }\right)^2   - \Psi^{(1)} \p_\| \Phi^{(1)}  - \Psi^{(1)} \p_\| v^{(1)}_{\| }- \cH v^{(1)}_{\| } \Psi^{(1)} + v_{\perp i}^{(1)}  \p_\|  v_{\perp}^{i(1)} \nonumber \\ 
&& + \cH v_{\perp i}^{(1)}   v_{\perp}^{i(1)}+ v_{\perp i}^{(1)} \p_{\perp}^i \Phi^{(1)} -\left(\Phi^{(1)}-v^{(1)}_{\| }+\Psi^{(1)}\right)\left(\frac{\bar a^2}{\bar \rho_m} \Em^{\| (1)}- \cH b_m v^{(1)}_{\| } \right)  - v_{\perp i}^{(1)} \left(\frac{\bar a^2}{\bar \rho_m}\Perp^i_j \Em^{j (1)}   - \cH b_m v^{i(1)}_{\perp }\right)\bigg] \nonumber \\
&&  +2\frac{\bar a^2}{ \cH  \bar \rho_m} \left[ \frac{1}{2} \Em^{\| (2)} -\Em^{0 (1)}  v_\|^{(1)}-  \Em^{\| (1)} \left(\delta_m^{(1)} - \Phi^{(1)} \right) \right] - 2 b_m \left(\frac{1}{2} v_\|^{(2)}-\delta_m^{(1)}  v_\|^{(1)} +2 \Phi^{(1)} v_\|^{(1)} \right)\nonumber \\
&&  +  \frac{2}{\cH^2}\left( \Phi^{(1)} - v^{(1)}_{\| }\right)  \bigg[ -{\left(\frac{\bar a^2}{\bar \rho_m} \Em^{\| (1)}\right)}'+\frac{\bar a^2 \cH}{\bar \rho_m}  \left( b_m  + 1\right)  \Em^{\| (1)}+ \cH^2 \frac{\ud b_m}{\ud \ln a}  v_\|^{(1)}    -\cH^2 b_m \left( b_m  + 1\right) v_\|^{(1)} - 2 \cH b_m \p_\| v_\|^{(1)} \nonumber \\
&& +\left(\cH'-\cH^2\right)  b_m v_\|^{(1)} - \cH b_m \p_\| \Phi^{(1)}  -2 \cH \p_\| v_\|^{(1)} - \p_\|^2 \Phi^{(1)} -\p_\|^2 v^{(1)}_{\|}+ \left(\cH'-\cH^2\right) v_\|^{(1)} - \cH  \p_\| \Phi^{(1)}+\frac{\ud \,}{\ud \bar \chi}\Psi^{(1)}{'}\bigg] \nonumber \\ 
&&+ \frac{4}{\cH}  \Phi^{(1)} \bigg[ 2 \p_\| \Phi^{(1)}-\left(\Phi^{(1)}{'}- \Psi^{(1)}{'} \right)\bigg]    -\frac{2}{\cH} \bigg( 3 \Phi^{(1)} -2 \Psi^{(1)}  -  v^{(1)}_{\|} \bigg)\left(\Phi^{(1)}{'} + \Psi^{(1)}{'} \right) +2\bigg\{  \frac{1}{\cH} \left( \Phi^{(1)} + \Psi^{(1)} \right) \nonumber \\ 
&&+\frac{\cH'}{\cH^3}\left( \Phi^{(1)} - v^{(1)}_{\| }\right)+  \frac{1}{\cH^2}\bigg[\frac{\bar a^2}{\bar \rho_m} \Em^{\| (1)}-\p_\| v_\|^{(1)}   -\cH \left( b_m  + 1\right) v_\|^{(1)}  + \Psi^{(1)}{'}  \bigg]\bigg\} \bigg[\frac{\bar a^2}{\bar \rho_m} \Em^{\| (1)}-\p_\| v_\|^{(1)} -\cH \left( b_m  + 1\right) v_\|^{(1)}  \nonumber \\
&&  + \Psi^{(1)}{'} \bigg] +  \frac{2}{\cH} \left[2 \frac{\ud \,}{\ud  \bar \chi} \Phi^{(1)}  -  \p_\| \left(\Phi^{(1)} +  v^{(1)}_{\|} \right)\right]  \left(\Phi^{(1)} + \Psi^{(1)} \right) +\frac{2}{\cH}v_{\perp i}^{(1)} \p^i_\perp \bigg( \Phi^{(1)}+\Psi^{(1)} \bigg) +\frac{2}{\cH} \frac{\ud \,}{\ud \bar \chi} \p_\| \left(\Phi^{(1)} +  v^{(1)}_{\|} \right) T^{(1)} \nonumber \\ 
&&     + 4 \bigg\{ \left(-\frac{\cH' }{\cH^3} +\frac{1}{\cH}\right)  \bigg[\frac{\bar a^2}{\bar \rho_m} \Em^{\| (1)}-\p_\| v_\|^{(1)}-\cH \left( b_m  + 1\right) v_\|^{(1)}+\Psi^{(1)}{'} \bigg]     - \frac{1}{\cH^2}\bigg[ -{\left(\frac{\bar a^2}{\bar \rho_m} \Em^{\| (1)}\right)}'+\frac{\bar a^2 \cH}{\bar \rho_m}  \left( b_m  + 1\right)  \Em^{\| (1)}\nonumber \\ 
&& + \cH^2 \frac{\ud b_m}{\ud \ln a}  v_\|^{(1)}   + b_m \bigg[ -\cH^2 \left( b_m  + 1\right) v_\|^{(1)} - 2 \cH \p_\| v_\|^{(1)} +\left(\cH'-\cH^2\right) v_\|^{(1)} - \cH \p_\| \Phi^{(1)} \bigg] -2 \cH \p_\| v_\|^{(1)} - \p_\|^2 \Phi^{(1)} -\p_\|^2 v^{(1)}_{\|}\nonumber \\
&&+ \left(\cH'-\cH^2\right) v_\|^{(1)} - \cH  \p_\| \Phi^{(1)}+\frac{\ud \,}{\ud \bar \chi}\Psi^{(1)}{'}\bigg]  + \Phi^{(1)} +\Psi^{(1)}\bigg\} I^{(1)} + 2\bigg[ \frac{\ud}{\ud \bar \chi}\left( \Phi^{(1)} + \Psi^{(1)} \right)  + \frac{1}{\cH} \frac{\ud \,}{\ud \bar \chi} \bigg(2 \p_\| \Phi^{(1)}-\Phi^{(1)}{'}\nonumber \\ 
&&  +\Psi^{(1)}{'} \bigg)  \bigg] \int_0^{\bar \chi} \ud \tilde \chi \;\bigg[ 2 \Phi^{(1)} + \left(\bar \chi-\tilde \chi\right)\left(\Phi^{(1)}{'} + \Psi^{(1)}{'} \right) \bigg]  +  4\bigg[\frac{1}{\cH}\p_{\perp i}\left(  \Phi^{(1)} - v^{(1)}_{\| } \right) + \frac{1}{\cH \bar \chi} v_{\perp i}^{(1)} + \frac{1}{\cH} \frac{\bar a^2}{\bar \rho_m} \Perp_{ij}  \Em^{j (1)}  - \frac{1}{\cH}\p_{\|} v_{\perp i}^{(1)} \nonumber \\ 
&& - \left( b_m  + 1\right) v^{(1)}_{\perp i} -  \frac{1}{\cH}\p_{\perp i} \Phi^{(1)}  -2S_{\perp }^{j(1)} \delta_{ij}  \bigg]   S_{\perp }^{i(1)} +\frac{2}{\cH}\bigg[\p_{\perp i}\bigg( -\Phi^{(1)}{'} -\frac{\bar a^2}{\bar \rho_m} \Em^{\| (1)}+\p_\| v_\|^{(1)}+\cH \left( b_m  + 1\right) v_\|^{(1)}+ 2 \p_\| \Phi^{(1)}\bigg) \nonumber \\
&&   +\frac{1}{\bar \chi} \left(-2\p_{\perp i} \Phi^{(1)} - \p_{\perp i} v^{(1)}_{\|} - \p_{\|} v^{(1)}_{\perp i} +\frac{\bar a^2}{\bar \rho_m} \Perp_{ij} \Em^{j (1)}-\cH \left( b_m  + 1\right) v^{(1)}_{\perp i} \right)+\frac{1}{\bar \chi^2} v^{(1)}_{\perp i}\bigg]  \int_0^{\bar \chi} \ud \tilde \chi \left[ \left(\bar \chi-\tilde \chi\right)  \tilde \p^i_\perp \left( \Phi^{(1)} + \Psi^{(1)} \right) \right]\nonumber \\
 &&  -  \frac{\cH'}{\cH^2} \Delta \ln a^{(2)}+\left[-\frac{\cH'' }{\cH^3} +3\left( \frac{\cH' }{\cH^2} \right)^2 + \frac{\cH' }{\cH^2} \right] \left( \Delta \ln a^{(1)} \right)^2 + 2\bigg\{-\frac{\cH' }{\cH^2} \left( \Phi^{(1)} + \Psi^{(1)} \right)- \frac{1}{\cH} \frac{\ud}{\ud \bar \chi}\left( \Phi^{(1)} + \Psi^{(1)} \right)  \nonumber \\
&&- \left(\frac{\cH' }{\cH^3} +\frac{1}{\cH} \right)\bigg[\frac{\bar a^2}{\bar \rho_m} \Em^{\| (1)}-\p_\| v_\|^{(1)}-\cH \left( b_m  + 1\right) v_\|^{(1)}  + \Psi^{(1)}{'} \bigg] \bigg\}  \Delta \ln a^{(1)} +4  \int_0^{\bar \chi} \ud \tilde \chi \bigg\{ \left(\Phi^{(1)} +\Psi^{(1)} \right) \bigg[\left(\Phi^{(1)}{'} +  \Psi^{(1)}{'} \right)  \nonumber \\
&&+  \frac{\ud}{\ud \tilde \chi}\Phi^{(1)} \bigg] \bigg\}\;.
\end{eqnarray}

 Finally we obtain
 
\begin{eqnarray}
\label{Poiss-Deltag-21}
 \Delta_g^{(2)} &=&  \delta_g^{(2)}+v_{\|}^{(2)}-3  \Psi^{(2)}  +   b_e \, \Delta \ln a^{(2)} +  \p_{\parallel} \Delta x_{\parallel}^{(2)}  + \frac{2}{\bar \chi} \Delta x_{\parallel}^{(2)} - 2\kappa^{(2)} +\left(\Delta_g^{(1)} \right)^2  \nonumber \\
&& - \left(\delta_g^{(1)}\right)^2 -3 \left(\Phi^{(1)}\right)^2-\left(v^{(1)}_{\| }\right)^2+2\Phi^{(1)}v^{(1)}_{\| }-9\left(\Psi^{(1)}\right)^2-\frac{1}{\cH^2}\left( \p_\| v_\|^{(1)}\right)^2-\frac{1}{\cH^2}\left( \Psi^{(1)}{'} \right)^2-6\Phi^{(1)}\Psi^{(1)}\nonumber \\
&&-\frac{2}{\cH}\Phi^{(1)} \Psi^{(1)}{'} -\frac{2}{\cH}v_\|^{(1)}\p_\| v_\|^{(1)}+\frac{2}{\cH}v_\|^{(1)} \Psi^{(1)}{'} +\frac{2}{\cH}\Psi^{(1)}\p_\| v_\|^{(1)}-\frac{2}{\cH}\Psi^{(1)} \Psi^{(1)}{'} +\frac{2}{\cH^2} \Psi^{(1)}{'} \p_\| v_\|^{(1)}+\frac{2}{\cH}\Phi^{(1)} \p_\| v_\|^{(1)}\nonumber \\
&&\nonumber \\
&&+   v^{(1)}_{\perp i} v^{i (1)}_{\perp}-8\left( I^{(1)} \right)^2+8\Phi^{(1)}I^{(1)}+8\Psi^{(1)}I^{(1)}+2\p_{\parallel}\left( +3\Psi^{(1)} - v^{(1)}_{\| } -\delta_g^{(1)} \right)T^{(1)} -  \frac{4}{\bar \chi}  \kappa^{(1)}T^{(1)}- \frac{2}{\bar \chi^2} \left( T^{(1)} \right)^2 \nonumber \\
&&+ \frac{2}{\cH} \left(-\p_\| v_\|^{(1)}-\cH v_\|^{(1)} -\p_\| \Phi^{(1)}+3\frac{\ud \,}{\ud \tilde \chi} \Psi^{(1)} -\frac{\ud \,}{\ud  \bar \chi} \delta_g^{(1)} - \frac{2}{\bar \chi^2}   T^{(1)} - \frac{2}{\bar \chi}    \kappa^{(1)}\right) \Delta \ln a^{(1)} \nonumber \\
&&  +2  \frac{\cH'}{\cH^2}\left(\Phi^{(1)}  - v^{(1)}_{\| }+\Psi^{(1)}- \frac{1}{\cH} \p_\| v_\|^{(1)} +\frac{1}{\cH} \Psi^{(1)}{'} \right) \Delta \ln a^{(1)}+ \left(- b_e +  \frac{\ud \ln b_e}{\ud  \ln \bar a} -  \left(\frac{\cH'}{\cH^2} \right)^2 - \frac{2}{\bar \chi^2 \cH^2}\right) \left( \Delta \ln a^{(1)}\right)^2  \nonumber \\
 &&+2\left(\Phi^{(1)}+ \Psi^{(1)}  - 2 I^{(1)} \right)  \int_0^{\bar \chi} \ud \tilde \chi\bigg[  \frac{\tilde \chi}{ \bar \chi} \left( 2 \tilde \p_\| + \left(\bar \chi-\tilde \chi\right)  \Perp^{mn} \tilde \p_m  \tilde \p_n \right) \left( \Phi^{(1)} + \Psi^{(1)} \right) \bigg]  \nonumber \\
&& -\bigg[  \int_0^{\bar \chi} \ud \tilde \chi  \frac{\tilde \chi}{ \bar \chi} \left( \Perp_j^i \tilde \p_\| + \left(\bar \chi-\tilde \chi\right) \Perp^p_j \Perp^{iq} \tilde \p_q  \tilde \p_p \right) \left( \Phi^{(1)} +\Psi^{(1)} \right) \bigg]  \nonumber \\
&& \times\bigg[\int_0^{\bar \chi} \ud \tilde \chi \frac{\tilde \chi}{ \bar \chi} \left( \Perp^j_i \tilde \p_\| + \left(\bar \chi-\tilde \chi\right) \Perp^n_i \Perp^{jm}  \tilde \p_m   \tilde \p_n \right)  \left( \Phi^{(1)} + \Psi^{(1)} \right) \bigg] \nonumber \\
&&  -2 \p_{\perp i}\left(v^{(1)}_{\|}-3 \Psi^{(1)}  +\delta_g^{(1)} \right)\int_0^{\bar \chi} \ud \tilde \chi \left[ \left(\bar \chi-\tilde \chi\right)  \tilde \p^i_\perp \left( \Phi^{(1)}+ \Psi^{(1)} \right) \right]  + 2v^{i (1)}_{\perp } \p_{\perp i} T^{(1)}\nonumber \\
&&+ 4S_{\perp}^{i(1)}  \bigg\{ - \frac{1}{\bar \chi}\int_0^{\bar \chi} \ud \tilde \chi \left[ \left(\bar \chi-\tilde \chi\right) \tilde \p^i_\perp \left( \Phi^{(1)}+ \Psi^{(1)} \right)\right] +\p_{\perp i}   \left(\frac{1}{\cH}  \Delta \ln a^{(1)}+   T^{(1)}  \right) \bigg\}\nonumber \\
  & &+2v^{ (1)}_{\perp i \, o }  v^{i (1)}_{\perp \, o }  -2\left(v^{(1)}_{\| \, o}\right)^2- 2\left(\Phi^{(1)}_{\, o}\right)^2+4 v^{(1)}_{\| \, o} \Psi^{(1)}_{\, o} -2 \left( \Psi^{(1)}_{\, o}  \right)^2+ 4 \Phi^{(1)}_{\, o}v^{(1)}_{\| \, o}-4\Phi^{(1)}_{\, o} \Psi^{(1)}_{\, o}  -2v^{ (1)}_{\perp i \, o }\nonumber \\
   &&\times \bigg\{ \bar \chi \p^i_{\perp}\left(v^{(1)}_{\|} -3  \Psi^{(1)} + \delta_g^{(1)} \right)+ 2S_{\perp}^{i(1)}+ \p^i_{\perp}   \left(\frac{1}{\cH}  \Delta \ln a^{(1)}+   T^{(1)}  \right)  - \frac{1}{\bar \chi} \int_0^{\bar \chi} \ud \tilde \chi \left[  \left(\bar \chi-\tilde \chi\right) \tilde \p^i_\perp \left( \Phi^{(1)} +\Psi^{(1)} \right)\right]\bigg\} \nonumber \\
 &&+2 \left(\Phi^{(1)}_{\, o} -  v^{(1)}_{\| \, o}+ \Psi^{(1)}_{\, o} \right)\bigg\{2\left(\Phi^{(1)} +\Psi^{(1)}  - 2 I^{(1)} \right)- \int_0^{\bar \chi} \ud \tilde \chi\bigg[   \frac{\tilde \chi}{ \bar \chi} \left(2\tilde \p_\| + \left(\bar \chi-\tilde \chi\right)\Perp^{mn} \tilde \p_m \tilde \p_n \right)\left( \Phi^{(1)}+ \Psi^{(1)} \right) \bigg] \bigg\}\nonumber \\
 &&-2 \frac{\bar a^2}{\cH \bar \rho_m} \left( \mathcal{E}_m^{\| (1)}- \frac{\cH}{\bar a^2} \bar \rho_m b_m v_\|^{(1)} \right)\left[\Phi^{(1)}  - v^{(1)}_{\| }+\Psi^{(1)}- \frac{1}{\cH} \p_\| v_\|^{(1)} +\frac{1}{\cH} \Psi^{(1)}{'}- \left(1+\frac{\cH'}{\cH^2}\right)\Delta \ln a^{(1)}\right]\nonumber \\
&&-\left( \frac{\bar a^2}{\cH \bar \rho_m} \right)^2\left(\mathcal{E}_m^{\| (1)}- \frac{\cH}{\bar a^2} \bar \rho_m b_m v_\|^{(1)}\right)^2\;.
 \end{eqnarray}

This is the main result in Poisson gauge with no velocity bias. If we explicitly identify the weak lensing shear and rotation contributions, we arrive at Eq. \eqref{Poiss-Deltag-4} in Appendix B.


\section{Prescription for the galaxy bias}
\label{Sec:GalaxyBias}

We need to relate the fluctuations of galaxy number density to the underlying matter density fluctuation $\delta_m$, assuming scale-independent bias. 
In order to define correctly the bias  we have to choose an appropriate frame where the baryon velocity perturbations vanish. If we assume that both at first and second order  the baryon rest frame coincides with the rest frame of CDM, the most general gauge that meets these requirements is the comoving-time orthogonal  (CO) gauge (see e.g.\ \cite{Kodama:1985bj}), which becomes the usual  comoving-synchronous  gauge when the perturbations are dominated by pressure-free matter, for example in the $\Lambda$CDM model\footnote{An analogous approach is used in \ \cite{Bartolo:2010ec}.}. In this frame, galaxy and matter over-densities are gauge invariant \cite{Kodama:1985bj}. 
This gauge is defined by the conditions  $B^{i (n)}=v^{i (n)}=0$. Then
 \begin{eqnarray} 
 \label{Comoving-ortnogonal_metric}
 \ud s^2 = a(\eta)^2\left[-\left(1 + 2 \varphi^{(1)}+ \varphi^{(2)}\right)\ud\eta^2+\left(\delta_{ij} +h_{ij \, {\rm CO}}^{(1)}+\frac{1}{2}h_{ij\, {\rm CO}}^{(2)}\right)\ud x^i\ud x^j\right] \;,
\end{eqnarray}
where $A_{\rm CO}^{(n)}=\varphi^{(n)}$, $h^{(n)}_{ij\, {\rm CO}} =- 2 \psi^{(n)} \delta_{ij} + F_{ij \, {\rm CO}}^{(n)}$, with $F^{(n)}_{ij  \, {\rm CO}}= (\p_i\p_j- \delta_{ij}\nabla^2/3) \xi^{(n)}+\p_i \hat \xi^{(n)}_j+ \p_j \hat \xi^{(n)}_i+\hat h^{(n)}_{ij}$, $\p_i \hat \xi^{i(n)}=\p_i \hat h^{ij(n)}=0$. For simplicity, we neglect vector and tensor perturbations at  first order, i.e.  $\hat \xi^{(1)}_j=  \hat h^{(1)}_{ij}=0$\;.

In order to find $\delta_{g \, {\rm CO}}$,  we transform the metric perturbations from the Poisson  gauge to the comoving-time orthogonal gauge.
Using  \cite{Matarrese:1997ay}, we get, at first order,
 \begin{eqnarray} \label{delpc1}
\delta_{g \, {\rm P}}^{(1)}=\delta_{g \, {\rm CO}}^{(1)}- b_e \cH v_{\rm P}^{(1)}+ 3 \cH v_{\rm P}^{(1)} ,
\end{eqnarray}
 and, at second order,
\begin{eqnarray} 
\delta_{g \, {\rm P}}^{(2)} &=&  \delta_{g \, {\rm CO}}^{(2)}- b_e \cH v_{\rm P}^{(2)}+ 3 \cH v_{\rm P}^{(2)} + \left( b_e \cH'-3 \cH' + \cH^2  \frac{\ud b_e}{\ud  \ln \bar a} + b_e^2 \cH^2  -6  b_e  \cH^2 + 9 \cH^2 \right) \left( v_{\rm P}^{(1)} \right)^2 \nonumber\\
&& + \cH b_e  v_{\rm P}^{(1)}  {v_{\rm P}^{(1)}}' - 3 \cH   v_{\rm P}^{(1)}  {v_{\rm P}^{(1)}}' -2\cH b_e  v_{\rm P}^{(1)} \delta_{g \, {\rm CO}}^{(1)} + 6 \cH  v_{\rm P}^{(1)} \delta_{g \, {\rm CO}}^{(1)} - 2  v_{\rm P}^{(1)} {\delta_{g{\rm CO}}^{(1)}}' \nonumber\\
&& - \frac{1}{2} \p^i \xi^{(1)} \left(- b_e \cH \p_i v_{\rm P}^{(1)}+ 3 \cH \p_i v_{\rm P}^{(1)} + 2 \p_i \delta_{g \, {\rm CO}}^{(1)} \right) - \left(b_e-3\right) \cH \nabla^{-2} \Xi.\label{delpc2}
\end{eqnarray} 
Here
\begin{eqnarray} 
\Xi &=&+ v_{\rm P}^{(1)} \nabla^2 {v_{\rm P}^{(1)}}' - {v_{\rm P}^{(1)}}' \nabla^2 v_{\rm P}^{(1)} - 2 \p_i \Phi^{(1)} \p^i v_{\rm P}^{(1)} - 2 \Phi^{(1)} \nabla^2 v_{\rm P}^{(1)} - 4 \Psi^{(1)}\nabla^2 v_{\rm P}^{(1)}- 4 \p_i \Psi^{(1)} \p^i v_{\rm P}^{(1)}  \nonumber\\
&& + \frac{1}{2} \p_i \xi^{(1)} \p^i \nabla^2 v_{\rm P}^{(1)} + \frac{1}{2} \p_i v_{\rm P}^{(1)} \p^i \nabla^2 \xi^{(1)} + \p_i \p_j \xi^{(1)} \p^i \p^j v_{\rm P}^{(1)} \;.
\end{eqnarray} 
(Another useful relation is ${\xi^{(1)}}'/2=v_{\rm P}^{(1)}$.)

The scale-independent bias at first and second order is defined by\footnote{A typo in this equation has been corrected.}
\begin{eqnarray} \label{deltagCO-1-2}
\delta_{g {\rm CO}}^{(1)}+\frac{1}{2} \delta_{g  {\rm CO}}^{(2)} = b_{1}^L \delta_{m  {\rm CO}}^{(1)} +\frac{1}{2} b_{1}^L  \delta_{m  {\rm CO}}^{(2)} +\frac{1}{2} b_{2}^L \big( \delta_{m  {\rm CO}}^{(1)} \big)^2\;. \nonumber
\end{eqnarray} 
These are substituted into Eqs. \eqref{delpc1} and \eqref{delpc2}, and then we can replace the term $\delta_g^{(2)}$ in the expression for the observed number overdensity at second order in Poisson gauge, 
Eq. (\ref{Poiss-Deltag-2}). This allows us to relate the observed number counts to the underlying matter overdensity in a gauge invariant way.


\section{Conclusions}
\label{Sec:Conclusions}

We presented  for the first time a derivation of the observed galaxy number counts to second order on cosmological scales,  including all relativistic effects. 
Our results are given both in a general gauge and in Poisson gauge, and apply to general dark energy models, including those where dark energy interacts non-gravitationally with dark matter. Our results also apply to metric theories of modified gravity as an alternative to dark energy. The main, fully general, result for the galaxy number count fluctuations at second order is Eq. (\ref{maindel}), which is specialized to the Poisson gauge in Eq. (\ref{Poiss-Deltag-2}). We also gave these results in the form where the contribution from weak lensing shear and rotation is made explicit, in Eqs. \eqref{maindel2} and \eqref{Poiss-Deltag-3}.

We derived the expressions needed to relate the observed number over-density to the matter over-density via the bias in a gauge-invariant way, in Eqs. (\ref{delpc1})--(\ref{deltagCO-1-2}).

The second-order effects that we derive, especially those involving integrals along the line of sight, may make a non-negligible contribution to the observed number counts. This will be important for removing potential biases  on parameter estimation in precision cosmology with galaxy surveys. 
It will also be important for an accurate analysis of the `contamination' of primordial non-Gaussianity by relativistic projection effects. This is discussed in Paper I \cite{Bertacca:2014dra} and is the subject of ongoing work \cite{Bertacca:2014n}.

\[\]{\bf Acknowledgments:}

We thank Enea di Dio, Ruth Durrer, Giovanni Marozzi, Obinna Umeh for helpful discussions.
DB and RM  are supported by the South African Square Kilometre
Array Project. RM acknowledges support from the UK Science \& Technology Facilities Council (grant ST/K0090X/1). RM and CC are supported by the South African National Research Foundation. We thank Ruth Durrer for alerting us to the possibility of an error in our results.

\appendix


\section{Perturbation terms in general gauge}\label{A}

From Eq.\ (\ref{metric}) the perturbations of  $g_{\mu \nu}$ and $g^{\mu \nu}$ are
\begin{eqnarray}
\begin{array} {lll}
g_{00}=a^2 \hat g_{00}= - a^2 \left(1+ 2 A^{(1)}+ A^{(2)}\right),  & \quad \quad \quad& g^{00}=a^{-2} \hat g^{00}=- a^{-2} \left[1-2  A^{(1)} - A^{(2)} +4 \left(A^{(1)}\right)^2- B^{(1)}_i  B^{i (1)}\right], \\ \\
g_{0i}=a^2 \hat g_{0i}= a^2 \left(- B^{(1)}_i - B^{(2)}_i/2 \right),  & \quad \quad \quad&  g^{0i}=a^{-2} \hat g^{0i} =a^{-2}\left[-B^{i (1)}-B^{i (2)}/2+ 2 A^{(1)} B^{i (1)}+   B^{(1)}_k h^{ki (1)}  \right], \\ \\
g_{ij}=a^2 \hat g_{ij}= a^2 \left(\delta_{ij} + h^{(1)}_{ij} + h^{(2)}_{ij}/2 \right),  & \quad \quad \quad&   g^{ij}=a^{-2}\hat g^{ij}=a^{-2}  \left[ \delta^{ij}- h^{ij(1)} -h^{ij(2)}/2 + h^{ik(1)}h^{j(1)}_{k} -B^{i (1)}B^{j (1)} \right], \\
  \end{array}  
\end{eqnarray}
For Christoffel symbols $\Gamma^\mu_{\rho \sigma}=\Gamma^{\mu (0)}_{\rho \sigma}+\Gamma^{\mu (1)}_{\rho \sigma}+\Gamma^{\mu (2)}_{\rho \sigma}/2$ and $\hat \Gamma^\mu_{\rho \sigma}=\hat \Gamma^{\mu (0)}_{\rho \sigma}+\hat \Gamma^{\mu (1)}_{\rho \sigma}+\hat \Gamma^{\mu (2)}_{\rho \sigma}/2$ in comoving coordinates, we obtain

\begin{eqnarray}
\begin{array} {lll} 
\Gamma^{0 (0)}_{00}=\cH\,,   \quad   \quad  & \Gamma^{0 (0)}_{0i}=0\,,   \quad   \quad  &\Gamma^{i (0)}_{00}=0\,, \\ \\
\Gamma^{i (0)}_{00}=0\,,   \quad   \quad  & \Gamma^{i (0)}_{j0}=\cH \delta^i_j\,,  \quad    \quad &\Gamma^{i (0)}_{jk}=0\,, 
~~~~ \hat \Gamma^{\mu (0)}_{\rho \sigma} = 0\,,
\end{array}  
\end{eqnarray}

\begin{eqnarray}
\begin{array} {lll} 
\Gamma^{0 (1)}_{00}=\hat \Gamma^{0 (1)}_{00}  &  \quad \quad   & \Gamma^{0 (1)}_{0i}=\hat \Gamma^{0 (1)}_{0i} - \cH  B^{(1)}_i  \\ \\
\Gamma^{0 (1)}_{ij}=\hat \Gamma^{0 (1)}_{ij} + \cH \left(- 2 A^{(1)} \delta_{ij} + h_{ij}^{(1)}\right)&  \quad \quad  &\Gamma^{i (1)}_{00}=\hat \Gamma^{i (1)}_{00}-\cH B^{i(1)}\,  \\ \\
\Gamma^{i (1)}_{j0}=\hat \Gamma^{i (1)}_{j0}\,  &  \quad  \quad &  \Gamma^{i (1)}_{jk}=\hat \Gamma^{i (1)}_{jk} + \cH  B^{i(1)} \delta_{jk}\,, \\ \\
\hat \Gamma^{0 (1)}_{00} = {A^{(1)}}' \,,&  \quad  \quad & \hat \Gamma^{0 (1)}_{0i} = \p_i A^{(1)} \,,\\ \\
\hat \Gamma^{0 (1)}_{ij}  = \frac{1}{2}\p_{i} B_{j}^{(1)}+\frac{1}{2} \p_{j} B_{i}^{(1)}+\frac{1}{2} {h_{ij}^{(1)}}' \,,&  \quad  \quad & \hat \Gamma^{i (1)}_{00} = \p^i A^{(1)} - {B^{i(1)}}'\,, \\ \\
\hat \Gamma^{i (1)}_{j0} =  \frac{1}{2}\p^{i} B_{j}^{(1)}-\frac{1}{2} \p_{j} B^{i(1)}+ \frac{1}{2}  {h_{j}^{i(1)}}'\,, &  \quad  \quad & \hat \Gamma^{i (1)}_{jk} = \frac{1}{2} \p_j h_{k}^{i(1)}+ \frac{1}{2}\p_k h_{j}^{i(1)}-\frac{1}{2}\p^i h_{jk}^{(1)} \,,
\end{array}  
\end{eqnarray}

\begin{eqnarray}
\frac{1}{2}\Gamma^{0 (2)}_{00}&=&\frac{1}{2} \hat \Gamma^{0 (2)}_{00}+ \cH  B^{(1)}_k  B^{k (1)} \,,  \nonumber \\
 \frac{1}{2}\Gamma^{0 (2)}_{0i}&=&\frac{1}{2} \hat \Gamma^{0 (2)}_{0i} - \frac{\cH}{2}  B^{(2)}_i + 2\cH  A^{(1)}  B^{(1)}_i\,,  \nonumber \\
  \frac{1}{2}\Gamma^{i (2)}_{00}&=&\frac{1}{2} \hat \Gamma^{i (2)}_{00}-\frac{\cH}{2} B^{i(2)}+ \cH  B^{(1)}_k  h^{ik (1)}\,, \nonumber \\
\frac{1}{2}\Gamma^{i (2)}_{j0}&=&\frac{1}{2} \hat \Gamma^{i (2)}_{j0}-\cH  B^{i(1)}  B^{(1)}_j \,,    \nonumber \\
 \frac{1}{2}\Gamma^{0 (2)}_{ij}&=&\frac{1}{2} \hat \Gamma^{0 (2)}_{ij} +\cH\left\{ \left[ -A^{(2)} + 4 \left(A^{(1)}\right)^2- B^{(1)}_k  B^{k (1)} \right] \delta_{ij} + \frac{1}{2} h_{ij}^{(2)} - 2  A^{(1)}  h_{ij}^{(1)} \right\} \,,\nonumber \\
\frac{1}{2}\Gamma^{i (2)}_{jk}&=&\frac{1}{2} \hat \Gamma^{i (2)}_{jk} + \cH \left[ \left( \frac{1}{2}  B^{i(2)}  - 2 A^{(1)}B^{i(1)} -  B^{(1)}_l   h^{il(1)} \right) \delta_{jk}  - B^{i(1)} h_{jk}^{(1)} \right] \,,\nonumber \\
\frac{1}{2} \hat \Gamma^{0 (2)}_{00}&=&+\frac{1}{2} {A^{2)}}' -2 A^{(1)}  {A^{(1)}}' +  B^{(1)}_k  {B^{k (1)}}' - \p_k A^{(1)} B^{k (1)}  \,,\nonumber \\
\frac{1}{2} \hat \Gamma^{0 (2)}_{0i} &=&  \frac{1}{2} \p_i A^{(2)} -2A^{(1)}  \p_i A^{(1)} -\frac{1}{2} B^{k (1)} { h_{ik}^{(1)} }' + \frac{1}{2} B^{k (1)}  \left( \p_{i} B_{k}^{(1)}-\p_{k} B_i^{(1)}\right) \,, \nonumber \\
\frac{1}{2}  \hat \Gamma^{i (2)}_{00}&=& \frac{1}{2}\p^i A^{(2)} - \frac{1}{2} {B^{i(2)}}'  + {A^{(1)}}'  B^{i(1)} + {B_{k}^{(1)}}'   h^{ik(1)} -\p_k A^{(1)} h^{ik(1)} \,,\nonumber \\
\frac{1}{2} \hat \Gamma^{i (2)}_{j0} &=&  \frac{1}{4}\p^{i} B_{j}^{(2)}-\frac{1}{4} \p_{j} B^{i(2)}+ \frac{1}{4}  {h_{j}^{i(2)}}' -\frac{1}{2} h^{ik(1)} {h_{jk}^{(1)}}' +\p_jA^{(1)} B^{i(1)} -\frac{1}{2}h^{ik(1)} \left(\p_{k} B_{j}^{(1)}-\frac{1}{2} \p_{j} B^{(1)}_k\right)  \,,\nonumber \\
\frac{1}{2} \hat \Gamma^{0 (2)}_{ij}&=& \frac{1}{4}\p_{i} B_{j}^{(2)}+\frac{1}{4} \p_{j} B_{i}^{(2)}+\frac{1}{4} {h_{ij}^{(2)}}'  -A^{(1)}\left(\p_{i} B_{j}^{(1)}+ \p_{j} B_{i}^{(1)}\right) -  A^{(1)} {h_{ij}^{(1)}}' -\frac{1}{2} B^{k(1)}\left( \p_i h_{jk}^{(1)}+ \p_j h_{ik}^{(1)}-\p_k h_{ij}^{(1)}\right)\,, \nonumber \\
\frac{1}{2}\hat\Gamma^{i (2)}_{jk}&=&  \frac{1}{4} \p_j h_{k}^{i(2)}+ \frac{1}{4}\p_k h_{j}^{i(2)}-\frac{1}{4}\p^i h_{jk}^{(2)} + \frac{1}{2} B^{i(1)}\left(\p_{j} B_{k}^{(1)}+ \p_{k} B_{j}^{(1)}\right) +\frac{1}{2} B^{i(1)} {h_{jk}^{(1)}}'   \nonumber \\
 && -\frac{1}{2} h^{il(1)}\left( \p_j h_{kl}^{(1)}+ \p_k h_{jl}^{(1)}-\p_l h_{jk}^{(1)}\right) \,.
\end{eqnarray}

For four-velocity $u^\mu$ ($g_{\mu \nu} u^\mu u^\nu=-1$), we find
\begin{eqnarray}
\label{u0i}
u_0&=&-a\left[1+A^{(1)}+\frac{1}{2}A^{(2)}-\frac{1}{2}\left(A^{(1)}\right)^2+\frac{1}{2}v_k^{(1)}v^{k(1)}\right] ,\\ 
u_i&=&a\left[v_i^{(1)}-B_i^{(1)}+\frac{1}{2}\left(v_i^{(2)}-B_i^{(2)}\right)+ A^{(1)}B_i^{(1)}+h_{ik}^{(1)}v^{k(1)}\right], \\
u^0&=&\frac{1}{a}\left[1-A^{(1)}-\frac{1}{2}A^{(2)}+\frac{3}{2}\left(A^{(1)}\right)^2+\frac{1}{2}v_k^{(1)}v^{k(1)}-v_k^{(1)} B^{k(1)}\right] ,\\
u^i&=&\frac{1}{a}\left(v^{i(1)}+\frac{1}{2}v^{i(2)}\right)\;.
\end{eqnarray}
From Eqs.\ (\ref{E0mu}) and (\ref{u0i})  we obtain all components of $\Lambda_{\hat 0 \mu}^{(n)}$ and $E_{\hat 0 \mu}^{(n)}$. 
From Eq.\ (\ref{E0mu}), we have
\begin{equation}
\label{E0mu-2}
u_\mu= \Lambda_{\hat{0} \mu}=a \, E_{\hat{0} \mu} \quad \quad {\rm and}  \quad \quad u^\mu=\Lambda^\mu_{\hat 0}= E_{\hat{0}}^\mu/a\;,
\end{equation}
and, using  Eq.\ (\ref{LambdaE}) we can deduce all components of $\Lambda_{\hat a \mu}^{(n)}$ and $E_{\hat a \mu}^{(n)}$. We summarize as follows:

\begin{eqnarray} \label{LambdaE-1-2}
\begin{array} {lll}
\Lambda_{\hat 0 0}^{(1)}=a E_{\hat 0 0}^{(1)}= -a A^{(1)}\;, & \quad& \Lambda_{\hat 0 i}^{(1)}=a E_{\hat 0 i}^{(1)}= a \left(v_i^{(1)}-B_i^{(1)}\right)\;,  \\  \\
\Lambda_{\hat a 0}^{(1)}=a E_{\hat a 0}^{(1)}= -a v_{\hat a}^{(1)}\;, & \quad& \Lambda_{\hat a i}^{(1)}=a E_{\hat a i}^{(1)}=\frac{1}{2} a h_{\hat a i}^{(1)}\;,   \\  \\
 \frac{1}{2}\Lambda_{\hat 0 0}^{(2)}=\frac{1}{2} a E_{\hat 0 0}^{(2)} = a\left[ -\frac{1}{2}A^{(2)}+\frac{1}{2}\left(A^{(1)}\right)^2-\frac{1}{2}v_k^{(1)}v^{k(1)}  \right] \;,  \\  \\
  \frac{1}{2}\Lambda_{\hat 0 i}^{(2)}=\frac{1}{2} a E_{\hat 0 i}^{(2)}=  a\left[\frac{1}{2}\left(v_i^{(2)}-B_i^{(2)}\right)+ A^{(1)}B_i^{(1)}+h_{ik}^{(1)}v^{k(1)} \right] \;, \\  \\
 \frac{1}{2}\Lambda_{\hat a 0}^{(2)}=\frac{1}{2} a E_{\hat a 0}^{(2)}= a\left[-\frac{1}{2} v_{\hat a}^{(2)}-A^{(1)}v_{\hat a}^{(1)} -\frac{1}{2} v^{k(1)} h_{\hat a k}^{(1)}  \right]\;, \\  \\  
 \frac{1}{2} \Lambda_{\hat a i}^{(2)}=\frac{1}{2} a E_{\hat a i}^{(2)}= a\left[\frac{1}{4}  h_{\hat a j}^{(2)} +  \frac{1}{2} \left(v_i^{(1)}-B_i^{(1)}\right) \left(v_{\hat a}^{(1)}-B_{\hat a}^{ (1)}\right)- \frac{1}{8}  h_{j}^{k (1)}  h_{\hat a k}^{(1)}  \right] \;. \\  \\
\end{array} 
\end{eqnarray}

The four-vector $\Em^\nu$ defined in Eq. \eqref{cure} can be nonzero. For the background, first- and second-order perturbations we obtain
\begin{eqnarray}
\label{Em}
\Em^{0 (0)} &=&\frac{1}{a^2}\rho_m^{(0)}{'}+\frac{3}{a^2} \cH\rho_m^{(0)}\nonumber \\
\Em^{i (0)}&=&0\;, \nonumber \\  
\Em^{0 (1)}&=& \frac{1}{a^2}\rho_m^{(0)} \left( \delta_m^{(1)}{'} + \p_i v^{i(1)}+ \frac{1}{2} h^{i(1)}_i {'}\right)+ \Em^{0 (0)}\left(-2A^{(1)} + \delta_m^{(1)} \right)\;, \nonumber \\  
\Em^{i (1)}&=& \frac{1}{a^2}\rho_m^{(0)} \left[ \left(v^{i(1)}-B^{i(1)}\right){'}+ \cH  \left(v^{i(1)}-B^{i(1)}\right) + \p^i  A^{(1)} \right]+ \Em^{0 (0)}  v^{i(1)}\;, \nonumber \\ 
\frac{1}{2}\Em^{0 (2)}&=& \frac{1}{a^2}\rho_m^{(0)} \left[\frac{1}{2} \delta_m^{(2)}{'} + \frac{1}{2}\p_i v^{i(2)}+\frac{1}{4} h^{i(2)}_i {'}-  \cH  \left(v^{i(1)}-B^{i(1)}\right) \left(v^{(1)}_i-B^{(1)}_i\right)+ B^{(1)}_i \p^i  A^{(1)} + \left(A^{(1)} + \delta_m^{(1)} \right)\p_i v^{i(1)} \right. \nonumber \\
&&\left. +v^{i(1)}  \p_i \delta_m^{(1)}   +  \frac{1}{2}\delta_m^{(1)}  h^{i(1)}_i {'} +\frac{1}{2}v^{j(1)} \p_j  h^{i(1)}_i -\frac{1}{2} h^{ij(1)} h_{ij}^{(1)}{'} \right]+\Em^{0 (0)}\left(-A^{(2)} + \frac{1}{2}\delta_m^{(2)}-  v^{i(1)} v^{(1)}_i \right) \nonumber \\
 &&+2 \left(v^{i(1)} -B^{i(1)} \right) \Em^{i (1)}-2 A^{(1)}  \Em^{0 (1)}\;, \nonumber \\
\frac{1}{2}\Em^{i (2)}&=& \frac{1}{a^2}\rho_m^{(0)} \left[ \left( \frac{1}{2} v^{i(2)}- \frac{1}{2}B^{i(2)}\right){'}+ \cH  \left( \frac{1}{2} v^{i(2)}- \frac{1}{2} B^{i(2)}\right) + \frac{1}{2}  \p^i  A^{(2)} -  v^{i(1)} \p_j  v^{j(1)} + A^{(1)} B^{i(1)}{'}+\cH  A^{(1)} B^{i(1)} \right. \nonumber \\
&&  +  A^{(1)}{'} B^{i(1)} +\cH B^{(1)}_k h^{ik (1)} + B^{(1)}_k{'} h^{ik (1)}+ v^{k (1)} h_k^{i (1)}{'}  -   A^{(1)} \p^i  A^{(1)}  + \p^i B^{(1)}_k v^{k (1)}-v^{k (1)} \p_k B^{i(1)} \nonumber \\
&& - \p_k  A^{(1)}   h^{ik (1)}  \bigg] + \Em^{0 (0)} \bigg( \frac{1}{2} v^{i(2)}- \delta_m^{(1)} v^{i(1)} +2  A^{(1)}v^{i(1)}   \bigg) + \Em^{0 (1)}  v^{i(1)}  +\Em^{i (1)} \bigg(\delta_m^{(1)}  -  A^{(1)} \bigg)  \;. 
\end{eqnarray}
Here $ \delta_m^{(1)}=\rho_m^{(1)}/\rho_m^{(0)} -1$ is the CDM fractional overdensity.
If $\Em^\nu$ is evaluated in redshift-space, then $\rho_m^{(0)}=\bar \rho_m$, $a(\bar x^0)=\bar a$ and 
\begin{equation}
\Em^{0 (0)}=\frac{\cH}{\bar a^2} \bar \rho_m b_m\;,
\end{equation}
where $b_m = \ud (a^3 \bar \rho_m)/\ud \ln \bar a$.

$\Em^{\| (1)}$ and $\Em^{\| (2)}$, evaluated at $\bar x^\mu$, are given by
\begin{eqnarray}
\label{ConEqu||-1}
\Em^{\| (1)} &=&  \frac{1}{\bar a^2} \bar \rho_m \bigg[ \frac{\ud \,}{\ud \bar \chi}  \left(A^{(1)}-v_\|^{(1)}\right) +A^{(1)}{'} -  B_\|^{(1)}{'}+\p_\| v_\|^{(1)} + \cH  \left(v_\|^{(1)}-B_\|^{(1)}\right) \bigg] +\frac{\cH}{\bar a^2} \bar \rho_m b_m  v_\|^{(1)}\;, \nonumber \\ 
\label{ConEqu||-2}
 \frac{1}{2} \Em^{\| (2)} &=&  \frac{1}{\bar a^2} \bar \rho_m \bigg[  \frac{\ud \,}{\ud \bar \chi}  \left(\frac{1}{2}A^{(2)}-\frac{1}{2}v_\|^{(2)}\right) + \frac{1}{2} A^{(2)}{'} - \frac{1}{2} B_\|^{(2)}{'}+\frac{1}{2}\p_\| v_\|^{(2)} + \cH  \left(\frac{1}{2} v_\|^{(2)}- \frac{1}{2} B_\|^{(2)}\right)  - v_\|^{(1)} \p_\| v_\|^{(1)} \nonumber \\
&&  -\frac{2}{\bar \chi} \left( v_\|^{(1)} \right)^2 - v_\|^{(1)}  \p_{\perp j}  v_\perp^{j(1)} + A^{(1)} B_\|^{(1)}{'} + A^{(1)}{'} B_\|^{(1)}+ \cH A^{(1)} B_\|^{(1)} + v_\|^{(1)} h_\|^{(1)}{'}+   \cH  B_\|^{(1)}   h_\|^{(1)} +  B_\|^{(1)}{'} h_\|^{(1)} \nonumber \\ 
&& - A^{(1)} \p_\| A^{(1)} -\p_\| A^{(1)} h_\|^{(1)} + v_{\perp k}^{(1)} \p_\|  B_{\perp}^{k(1)} - v_{\perp}^{j (1)} \p_{\perp j}  B_\|^{(1)} + \frac{1}{\chi} v_{\perp}^{j (1)}  B_{\perp j}^{(1)} +\cH  B_k^{(1)} \Perp^k_j  h^{ij (1)} n_i \nonumber \\ 
&&+v_k^{(1)} \Perp^k_j  h^{ij (1)}{'} n_i +   B_k^{(1)}{'} \Perp^k_j  h^{ij (1)} n_i  - \p_k A^{(1)}\Perp^k_j  h^{ij (1)} n_i \bigg] + \frac{\cH}{\bar a^2} \bar \rho_m b_m \left(\frac{1}{2} v_\|^{(2)}-\delta_m^{(1)}  v_\|^{(1)} +2 A^{(1)} v_\|^{(1)} \right) \nonumber \\ 
&&+  \Em^{0 (1)}  v_\|^{(1)} + \Em^{\| (1)} \left(\delta_m^{(1)} - A^{(1)} \right)\;.
\end{eqnarray}

For the weak lensing shear and rotation:
 \begin{eqnarray}
 \label{p_perpDeltax_perp-1}
 \p_{\perp i}   \Delta x_{\perp j}^{(1)} & =& - \Perp_{ij} \left(B^{(1)}_{\|\, o}-v^{(1)}_{\|\, o}\right)+\frac{1}{2} \Perp_i^m \Perp_j^n h_{mn\,o}^{(1)}-\frac{1}{2}\Perp_{ij}  h^{(1)}_{\| \, o}-\left(B_{\perp i o}^{(1)}- v_{\perp i \, o}^{(1)}\right)n_{j} -\frac{1}{2}\Perp_{ip} n^k h_{k\, o}^{p (1)}n_{j} \nonumber\\\
 && +  \int_0^{\bar \chi} \ud \tilde \chi\bigg[    \frac{1}{\tilde \chi}  \Perp_{ij}  B^{(1)}_{\| }  -\Perp_{jp} \tilde \p_{ \perp i} B^{p (1)}- \frac{1}{\tilde \chi}   \Perp^p_i \Perp_{jq}   h_p^{q (1)}  +\frac{1}{\tilde \chi}  \Perp_{ij} h^{(1)}_{\| }                  - n^p  \Perp_{jq} \tilde \p_{ \perp i}   h_p^{q (1)} \nonumber \\
&&
  + \frac{1}{\tilde \chi}  \left( B^{(1)}_{\perp i}+ \Perp_{in} n^m h_{m}^{n(1)} \right)n_{j} \bigg]  +  \int_0^{\bar \chi} \ud \tilde \chi\left(\bar \chi-\tilde \chi\right) \frac{\tilde \chi}{ \bar \chi}  \tilde \p_{\perp i}  \tilde \p_{\perp j}\left( A^{(1)} - B^{(1)}_{\| } - \frac{1}{2}h^{(1)}_{\| } \right)\;,\\
 \gamma_{ij}^{(1)}&=&\Perp_{ij} \left(B^{(1)}_{\|\, o}-v^{(1)}_{\|\, o}\right)-\frac{1}{2} \Perp_{(i}^m \Perp_{j)}^n h_{mn\,o}^{(1)} + \frac{1}{2}\Perp_{ij}  h^{(1)}_{\| \, o} + n_{(j}\left(B_{\perp i) \, o}^{(1)}- v_{\perp i) \, o}^{(1)}\right) +\frac{1}{2}\Perp_{p(i} n_{j)} n^k h_{k\, o}^{p (1)}  \nonumber \\
 && -  \int_0^{\bar \chi} \ud \tilde \chi\bigg[    \frac{1}{\tilde \chi}  \Perp_{ij}  B^{(1)}_{\| }  -\Perp_{p(j} \tilde \p_{ \perp i)} B^{p (1)}- \frac{1}{\tilde \chi}   \Perp^p_{(i} \Perp_{j)q}   h_p^{q (1)}  +\frac{1}{\tilde \chi}  \Perp_{ij} h^{(1)}_{\| }                  - n^p  \Perp_{q(j} \tilde \p_{ \perp i)}   h_p^{q (1)} \nonumber \\
&& + \frac{1}{\tilde \chi}  \left( B^{(1)}_{\perp (i}+ n^m h_{m}^{n(1)} \Perp_{n(i}  \right)n_{j)} \bigg] -  \int_0^{\bar \chi} \ud \tilde \chi\left(\bar \chi-\tilde \chi\right) \frac{\tilde \chi}{ \bar \chi}  \tilde \p_{\perp (i}  \tilde \p_{\perp j)}\left( A^{(1)} - B^{(1)}_{\| } - \frac{1}{2}h^{(1)}_{\| } \right)-\Perp_{ij}\kappa^{(1)}\;,
\label{gamg} \\
 \vartheta_{ij}^{(1)}\vartheta^{ij(1)} &=&\frac{1}{2}B_{\perp i o}^{(1)}B_{\perp  o}^{i(1)}-B_{\perp i o}^{(1)}v_{\perp  o}^{i(1)} +\frac{1}{2}v_{\perp i o}^{(1)}v_{\perp  o}^{i(1)}+\frac{1}{2}\left(B_{\perp i o}^{(1)}- v_{\perp i \, o}^{(1)}\right) n^k h_{k\, o}^{i (1)}+\frac{1}{8}n^m \Perp^j_n   h_m^{n (1)} n^k  h_{jk}^{ (1)}\nonumber \\
 &&-\left(B_{\perp i o}^{(1)}- v_{\perp i \, o}^{(1)}+\frac{1}{2}\Perp_{ip} n^k  h^{p(1)}_{k \, o}\right)  \int_0^{\bar \chi} \ud \tilde \chi\bigg[    \frac{1}{\tilde \chi}  \left( B^{i(1)}_{\perp}+ \Perp^{i}_l n^m h_{m}^{l(1)} \right)+\left(\bar \chi-\tilde \chi\right) \frac{1}{ \bar \chi}  \tilde \p^i_{\perp } \left( A^{(1)} - B^{(1)}_{\| } - \frac{1}{2}h^{(1)}_{\| } \right)\bigg]\nonumber \\
  && +  \int_0^{\bar \chi} \ud \tilde \chi\bigg[  \Perp_{q[i} \tilde \p_{ \perp j]} B^{q (1)}  + n^k  \Perp_{q [i} \tilde \p_{ \perp j]}   h_k^{q (1)}  \bigg]   \int_0^{\bar \chi} \ud \tilde \chi\bigg[  \Perp_{p}^{[i} \tilde \p_{ \perp}^{j]} B^{p (1)}  + n^m  \Perp_{p}^{[i} \tilde \p_{ \perp}^{ j]}   h_m^{p (1)}  \bigg] \nonumber \\
  &&+\frac{1}{2}  \int_0^{\bar \chi} \ud \tilde \chi\bigg[  \frac{1}{\tilde \chi}  \left( B^{(1)}_{\perp i}+ \Perp_{in} n^m h_{m}^{n(1)} \right) +\left(\bar \chi-\tilde \chi\right) \frac{1}{ \bar \chi}  \tilde \p_{\perp i}  \left( A^{(1)} - B^{(1)}_{\| } - \frac{1}{2}h^{(1)}_{\| } \right)\bigg]\nonumber \\
&& \times  \int_0^{\bar \chi} \ud \tilde \chi\bigg[  \frac{1}{\tilde \chi}  \left( B^{i(1)}_{\perp }+ \Perp^i_{p} n^q h_{q}^{p(1)} \right) +\left(\bar \chi-\tilde \chi\right) \frac{1}{ \bar \chi}  \tilde \p^i_{\perp}  \left( A^{(1)} - B^{(1)}_{\| } - \frac{1}{2}h^{(1)}_{\| } \right)\bigg] \;.\label{varg}
 \end{eqnarray}

If we assume that galaxy velocities follow the matter velocity field,
 \begin{eqnarray}
 \label{Poiss-Deltag-5}
&&  \Delta_g^{(2)}   =  \delta_g^{(2)}+   b_e \, \Delta \ln a^{(2)} +  \p_{\parallel} \Delta x_{\parallel}^{(2)}  + \frac{2}{\bar \chi} \Delta x_{\parallel}^{(2)} - 2\kappa^{(2)}  + A^{(2)}+ v_{\|}^{(2)}+ \frac{1}{2} h_i^{i (2)} +\left(\Delta_g^{(1)} \right)^2 -  \left(A^{(1)}\right)^2  \nonumber \\
 &&+A^{(1)}h^{(1)}_{\| } -\left(v^{(1)}_{\| } \right)^2 +\left(B^{(1)}_{\| } \right)^2  - \frac{1}{2}h_{i}^{k(1)} h_{k}^{i(1)}- \frac{1}{4}\left(h^{(1)}_{\| } \right)^2 +2A^{(1)} v^{(1)}_{\| }-\frac{1}{\cH^2}\left(\p_\|v^{(1)}_{\| } \right)^2  +\frac{1}{\cH}A^{(1)}h^{(1)}_{\| }{'} \nonumber \\
 &&- \frac{1}{4\cH^2}\left(h^{(1)}_{\| }{'} \right)^2  -\frac{1}{\cH} v^{(1)}_{\| }h^{(1)}_{\| }{'}-\frac{1}{2\cH} h^{(1)}_{\| }h^{(1)}_{\| }{'} +\frac{2}{\cH}A^{(1)}\p_\|v^{(1)}_{\| }-\frac{2}{\cH}v^{(1)}_{\| }\p_\|v^{(1)}_{\| }-\frac{1}{\cH}h^{(1)}_{\| }\p_\|v^{(1)}_{\| }  -\frac{1}{\cH^2}h^{(1)}_{\| }{'}\p_\|v^{(1)}_{\| }\nonumber \\
&&  -2\big|\gamma^{(1)}\big|^2-2\left(\kappa^{(1)}\right)^2+\vartheta_{ij}^{(1)}\vartheta^{ij(1)}+  B^{(1)}_{\perp i} B^{i (1)}_{\perp}+   v^{(1)}_{\perp i} v^{i (1)}_{\perp}- 2  v^{(1)}_{\perp i} B^{i (1)}_{\perp} - \left(\delta_g^{(1)}\right)^2 + 2v^{i (1)}_{\perp } \p_{\perp i} T^{(1)} \nonumber \\
 &&  + \frac{2}{ \cH} \left(-\p_\| A^{(1)}+ B_\|^{(1)}{'} -\p_\| v_\|^{(1)}-\cH  v_\|^{(1)}+\cH B_\|^{(1)} \right)\Delta \ln a^{(1)} - \frac{1}{ \cH} \frac{\ud \,}{\ud \bar \chi}\left(  h_i^{i (1)}+ 2\delta_g^{(1)} \right)\Delta \ln a^{(1)}- \frac{4}{\bar \chi^2 \cH}  \Delta \ln a^{(1)} T^{(1)}\nonumber \\
 &&  + 2\frac{\cH'}{ \cH^2} \left(A^{(1)}-  v^{(1)}_{\| } - \frac{1}{2}h^{(1)}_{\| }- \frac{1}{\cH}\p_\|v^{(1)}_{\| }-\frac{1}{2\cH}h^{(1)}_{\| }{'}\right)\Delta \ln a^{(1)} - \p_{\parallel}\left(2 v^{(1)}_{\| } + h_i^{i (1)} + 2\delta_g^{(1)} \right) T^{(1)}- \frac{4}{\bar \chi \cH}  \Delta \ln a^{(1)}   \kappa^{(1)}\nonumber \\
 &&  -\frac{4}{\bar \chi}  T^{(1)}  \kappa^{(1)} + \left[- b_e +  \frac{\ud \ln b_e }{\ud  \ln \bar a} - \left(\frac{\cH'}{\cH^2} \right)^2- \frac{2}{\bar \chi^2 \cH^2}\right] \left( \Delta \ln a^{(1)}\right)^2- \frac{2}{\bar \chi^2} \left( T^{(1)} \right)^2  +2 \left[-\left( B^{i (1)}_{\perp }+ n^k h_{k}^{j(1)} \Perp^i_j\right) + 2S_{\perp}^{i(1)} \right] \nonumber\\
  && \times \p_{\perp i}   \left(\frac{1}{\cH}  \Delta \ln a^{(1)}+   T^{(1)}  \right) - \left[-\frac{2}{\bar \chi}\left( B^{i (1)}_{\perp }+ n^k h_{k}^{j(1)} \Perp^i_j\right)  + \frac{4}{\bar \chi} S_{\perp}^{i(1)}+ \p^i_{\perp}\left( 2v^{(1)}_{\| } + h_l^{l (1)}  + 2\delta_g^{(1)}\right)  \right]  \nonumber \\
 &&\times  \int_0^{\bar \chi} \ud \tilde \chi \bigg[ \frac{ \bar \chi}{\tilde \chi}\left( B^{(1)}_{\perp i}+ n^k h_{k}^{j(1)} \Perp_{ij}\right) + \left(\bar \chi-\tilde \chi\right) \tilde \p_{\perp i}  \left( A^{(1)} - B^{(1)}_{\| } - \frac{1}{2}h^{(1)}_{\| } \right)\bigg]\nonumber \\
 &&+ 2\left(B^{(1)}_{\perp i \, o }-v^{ (1)}_{\perp i \, o }+ \frac{1}{2} n^k h_{k\,o}^{j(1)} \Perp_{ij} \right) \left(B^{i (1)}_{\perp \, o }-v^{i (1)}_{\perp \, o }+ \frac{1}{2} n^k h_{k\,o}^{j(1)} \Perp^i_j\right)\nonumber \\
&& +  \left(B^{(1)}_{\perp i \, o }-v^{ (1)}_{\perp i \, o }+ \frac{1}{2} n^k h_{k\,o}^{j(1)} \Perp_{ij} \right) \bigg\{-2\left( B^{i (1)}_{\perp }+ n^m h_m^{l(1)} \Perp^i_l\right) + 4S_{\perp}^{i(1)} + \bar \chi \p^i_{\perp}\left(2v^{(1)}_{\| } + h_l^{l (1)} + 2  \delta_g^{(1)} \right)\nonumber \\
&&+ 2 \p_{\perp i}   \left(\frac{1}{\cH}  \Delta \ln a^{(1)}+   T^{(1)}  \right)  -2 \frac{1}{\bar \chi} \int_0^{\bar \chi} \ud \tilde \chi \left[ \frac{ \bar \chi}{\tilde \chi}\left( B^{i (1)}_{\perp }+ n^k h_{k}^{j(1)} \Perp^i_j\right) + \left(\bar \chi-\tilde \chi\right) \tilde \p^i_\perp \left( A^{(1)} - B^{(1)}_{\| } - \frac{1}{2}h^{(1)}_{\| } \right)\right]\bigg\}\nonumber \\
&& - \left( \frac{\bar a^2}{\bar \rho_m \cH} \Em^{\| (1)}- b_m v_\|^{(1)} \right)^2-2 \left[A^{(1)}-  v^{(1)}_{\| } - \frac{1}{2}h^{(1)}_{\| }- \frac{1}{\cH}\p_\|v^{(1)}_{\| }-\frac{1}{2\cH}h^{(1)}_{\| }{'}-\left(1+\frac{\cH'}{ \cH^2}\right)\Delta \ln a^{(1)}\right] \nonumber \\
&&\times \left( \frac{\bar a^2}{\bar \rho_m \cH} \Em^{\| (1)}- b_m v_\|^{(1)} \right) \;.  
  \end{eqnarray}

\section{Perturbation terms in Poisson Gauge}\label{Poiss-pert}

From Eq.\ (\ref{Poiss-metric}), the perturbations of $g_{\mu \nu}$ and $g^{\mu \nu}$ are
\begin{eqnarray}
\begin{array} {lll}
g_{00}=- a^2 \left(1+ 2 \Phi^{(1)}+ \Phi^{(2)}\right),  & \quad& g^{00}=- a^{-2} \left[1-2  \Phi^{(1)} - \Phi^{(2)} +4 \left(\Phi^{(1)}\right)^2\right], \\ \\
g_{0i}= a^2 \omega^{(2)}_i,  & \quad&  g^{0i}=a^{-2}\omega^{i (2)}, \\ \\
g_{ij}= a^2 \left(\delta_{ij} -2 \delta_{ij} \Psi^{(1)} -\delta_{ij} \Psi^{(2)}+\hat h^{(2)}_{ij}/2 \right),  & \quad&   g^{ij}=a^{-2}  \left[ \delta^{ij}+2 \delta^{ij} \Psi^{(1)} + \delta^{ij} \Psi^{(2)}-\hat h^{ij(2)}/2 +4\delta^{ij} (\Psi^{(1)})^2\right], \\
  \end{array} 
\end{eqnarray}

For four-velocity $u^\mu$, we find
\begin{eqnarray}
\label{Poiss-u0i}
u_0&=&-a\left[1+\Phi^{(1)}+\frac{1}{2}\Phi^{(2)}-\frac{1}{2}\left(\Phi^{(1)}\right)^2+\frac{1}{2}v_k^{(1)}v^{k(1)}\right] ,\\ 
u_i&=&a\left[v_i^{(1)}+\frac{1}{2}\left(v_i^{(2)}+2\omega_i^{(2)}\right)- 2 \Psi^{(1)}v_i^{(1)}\right], \\
u^0&=&\frac{1}{a}\left[1-\Phi^{(1)}-\frac{1}{2}\Phi^{(2)}+\frac{3}{2}\left(\Phi^{(1)}\right)^2+\frac{1}{2}v_k^{(1)}v^{k(1)}\right] ,\\
u^i&=&\frac{1}{a}\left(v^{i(1)}+\frac{1}{2}v^{i(2)}\right)\;.
\end{eqnarray}

For the tetrad:

\begin{eqnarray} \label{Poiss-LambdaE-1-2}
\begin{array} {lll}
\Lambda_{\hat 0 0}^{(1)}=a E_{\hat 0 0}^{(1)}= -a \Phi^{(1)}\;, & \quad& \Lambda_{\hat 0 i}^{(1)}=a E_{\hat 0 i}^{(1)}= a v_i^{(1)}\;,  \\  \\
\Lambda_{\hat a 0}^{(1)}=a E_{\hat a 0}^{(1)}= -a v_{\hat a}^{(1)}\;, & \quad& \Lambda_{\hat a i}^{(1)}=a E_{\hat a i}^{(1)}=-a \delta_{\hat a i}\Psi^{(1)}\;,   \\  \\
 \frac{1}{2}\Lambda_{\hat 0 0}^{(2)}=\frac{1}{2} a E_{\hat 0 0}^{(2)} = a\left[ -\frac{1}{2}\Phi^{(2)}+\frac{1}{2}\left(\Phi^{(1)}\right)^2-\frac{1}{2}v_k^{(1)}v^{k(1)}  \right] \;,  \\  \\
  \frac{1}{2}\Lambda_{\hat 0 i}^{(2)}=\frac{1}{2} a E_{\hat 0 i}^{(2)}=  a\left[\frac{1}{2}\left(v_i^{(2)}+2 \omega_i^{(2)}\right)-2 \Psi^{(1)}v_i^{(1)} \right] \;, \\  \\
 \frac{1}{2}\Lambda_{\hat a 0}^{(2)}=\frac{1}{2} a E_{\hat a 0}^{(2)}= a\left[-\frac{1}{2} v_{\hat a}^{(2)}-\Phi^{(1)}v_{\hat a}^{(1)} + v^{(1)}_{\hat a}\Psi^{(1)}  \right]\;, \\  \\  
 \frac{1}{2} \Lambda_{\hat a i}^{(2)}=\frac{1}{2} a E_{\hat a i}^{(2)}= a\left[-\frac{1}{2} \delta_{\hat a j} \Psi^{(2)}+\frac{1}{4}  \hat h_{\hat a j}^{(2)} +  \frac{1}{2} v_i^{(1)} v_{\hat a}^{(1)}- \frac{1}{2}\delta_{\hat a j}  \left(\Phi^{(1)}\right)^2\right] \;. \\  \\
\end{array} 
\end{eqnarray}

For the energy-momentum exchange four-vector $\Em^\nu$, defined in Eq. \eqref{cure}:
\begin{eqnarray}
\Em^{0 (1)}&=& \frac{1}{a^2}\rho_m^{(0)} \left( \delta_m^{(1)}{'} + \p_i v^{i(1)}- 3 \Psi^{(1)} {'}\right)+ \Em^{0 (0)}\left(-2\Phi^{(1)} + \delta_m^{(1)} \right)\;, \nonumber \\  
\Em^{i (1)}&=& \frac{1}{a^2}\rho_m^{(0)} \left[ v^{i(1)}{'}+ \cH  v^{i(1)}+ \p^i  \Phi^{(1)} \right]+ \Em^{0 (0)}  v^{i(1)}\;, \nonumber \\ 
\frac{1}{2}\Em^{0 (2)}&=& \frac{1}{a^2}\rho_m^{(0)} \left[\frac{1}{2} \delta_m^{(2)}{'} + \frac{1}{2}\p_i v^{i(2)}-\frac{3}{2} \Psi^{(2)}{'}+\frac{1}{4} \hat h^{i(2)}_i {'}-  \cH  v^{i(1)} v^{(1)}_i + \left(\Phi^{(1)} + \delta_m^{(1)} \right)\p_i v^{i(1)} \right. \nonumber \\
&&\left. +v^{i(1)}  \p_i \delta_m^{(1)}   -3 \delta_m^{(1)}  \Psi^{(1)}{'} -3 v^{j(1)} \p_j  \Psi^{(1)} - 6 \Psi^{(1)} \Psi^{(1)}{'} \right]+\Em^{0 (0)}\left(-\Phi^{(2)} + \frac{1}{2}\delta_m^{(2)}-  v^{i(1)} v^{(1)}_i \right) \nonumber \\
 &&+2 v^{i(1)} \Em^{i (1)}-2 \Phi^{(1)}  \Em^{0 (1)}\;, \nonumber \\
\frac{1}{2}\Em^{i (2)}&=& \frac{1}{a^2}\rho_m^{(0)} \left[ \left( \frac{1}{2} v^{i(2)} + \omega^{i(2)}\right){'}+ \cH  \left( \frac{1}{2} v^{i(2)}+ \omega^{i(2)} \right) + \frac{1}{2}  \p^i  \Phi^{(2)} -  v^{i(1)} \p_j  v^{j(1)}  -2 v^{i (1)} \Psi^{(1)}{'}  -   \Phi^{(1)} \p^i  \Phi^{(1)} \right. \nonumber \\
&&    +2 \p^i  \Phi^{(1)}   \Psi^{(1)}  \bigg] + \Em^{0 (0)} \bigg( \frac{1}{2} v^{i(2)}- \delta_m^{(1)} v^{i(1)} +2  \Phi^{(1)}v^{i(1)}   \bigg) + \Em^{0 (1)}  v^{i(1)}  +\Em^{i (1)} \bigg(\delta_m^{(1)}  -  \Phi^{(1)} \bigg)  \;,
\end{eqnarray}
or
\begin{eqnarray}
\label{Poiss-ConEqu||-1}
\Em^{\| (1)} &=&  \frac{1}{\bar a^2} \bar \rho_m \bigg[ \frac{\ud \,}{\ud \bar \chi}  \left(\Phi^{(1)}-v_\|^{(1)}\right) +\Phi^{(1)}{'}+\p_\| v_\|^{(1)} + \cH  v_\|^{(1)} \bigg] +\frac{\cH}{\bar a^2} \bar \rho_m b_m  v_\|^{(1)}\;, \nonumber \\ 
\label{Poiss-ConEqu||-2}
 \frac{1}{2} \Em^{\| (2)} &=&  \frac{1}{\bar a^2} \bar \rho_m \bigg[  \frac{\ud \,}{\ud \bar \chi}  \left(\frac{1}{2}\Phi^{(2)}-\frac{1}{2}v_\|^{(2)}\right) + \frac{1}{2} \Phi^{(2)}{'} +\omega_{\|}^{(2)}{'}+\frac{1}{2}\p_\| v_\|^{(2)} + \cH  \left(\frac{1}{2} v_\|^{(2)}+\omega_{\|}^{(2)}\right)  - v_\|^{(1)} \p_\| v_\|^{(1)}  -\frac{2}{\bar \chi} \left( v_\|^{(1)} \right)^2\nonumber \\
&&  - v_\|^{(1)}  \p_{\perp j}  v_\perp^{j(1)} -2 v_\|^{(1)}\Psi^{(1)}{'} - \Phi^{(1)} \p_\| \Phi^{(1)} +2\p_\| \Phi^{(1)} \Psi^{(1)}   \bigg]  + \frac{\cH}{\bar a^2} \bar \rho_m b_m \left(\frac{1}{2} v_\|^{(2)}-\delta_m^{(1)}  v_\|^{(1)} +2 \Phi^{(1)} v_\|^{(1)} \right)\nonumber \\ 
&&+  \Em^{0 (1)}  v_\|^{(1)} + \Em^{\| (1)} \left(\delta_m^{(1)} - \Phi^{(1)} \right)\;.
\end{eqnarray}

For the weak lensing shear and rotation: 
 \begin{eqnarray}
 \p_{\perp i}   \Delta x_{\perp j}^{(1)}  &=&  \Perp_{ij} v^{(1)}_{\|\, o}+ v_{\perp i \, o}^{(1)} n_{j}   +  \int_0^{\bar \chi} \ud \tilde \chi\left(\bar \chi-\tilde \chi\right) \frac{\tilde \chi}{ \bar \chi}  \tilde \p_{\perp i}  \tilde \p_{\perp j}\left( \Phi^{(1)}  + \Psi^{(1)} \right),\\
\gamma_{ij} &=& - \Perp_{ij} v^{(1)}_{\|\, o}-n_{(j}  v_{\perp i) \, o}^{(1)}   -  \int_0^{\bar \chi} \ud \tilde \chi \left[\left(\bar \chi-\tilde \chi\right) \frac{\tilde \chi}{ \bar \chi}  \tilde \p_{\perp (i}  \tilde \p_{\perp j)}\left( \Phi^{(1)}  + \Psi^{(1)} \right)\right]-\Perp_{ij} \kappa^{(1)},\label{gamp}\\
\vartheta_{ij}^{(1)}\vartheta^{ij(1)} &=&   +\frac{1}{2}v_{\perp i o}^{(1)}v_{\perp  o}^{i(1)} + \frac{1}{ \bar \chi}v_{\perp i \, o}^{(1)} \int_0^{\bar \chi} \ud \tilde \chi\bigg[ \left(\bar \chi-\tilde \chi\right)  \tilde \p^i_{\perp } \left(  \Phi^{(1)}  + \Psi^{(1)} \right)\bigg]\nonumber \\
  &&+\frac{1}{2\bar \chi^2}  \int_0^{\bar \chi} \ud \tilde \chi\bigg[ \left(\bar \chi-\tilde \chi\right)  \tilde \p_{\perp i}  \left(  \Phi^{(1)}  + \Psi^{(1)} \right)\bigg]  \int_0^{\bar \chi} \ud \tilde \chi\bigg[ \left(\bar \chi-\tilde \chi\right)  \tilde \p^i_{\perp}  \left(  \Phi^{(1)}  + \Psi^{(1)} \right)\bigg] \;. \label{varp}
 \end{eqnarray}

Assuming that galaxy velocities follow the matter velocity field, we find
\begin{eqnarray}
\label{Poiss-Deltag-4}
 \Delta_g^{(2)} &=&  \delta_g^{(2)}+v_{\|}^{(2)}-3  \Psi^{(2)}  +   b_e \, \Delta \ln a^{(2)} +  \p_{\parallel} \Delta x_{\parallel}^{(2)}  + \frac{2}{\bar \chi} \Delta x_{\parallel}^{(2)} - 2\kappa^{(2)} +\left(\Delta_g^{(1)} \right)^2  - \left(\delta_g^{(1)}\right)^2 - \left(\Phi^{(1)}\right)^2-\left(v^{(1)}_{\| }\right)^2 \nonumber \\
&&+2\Phi^{(1)}v^{(1)}_{\| } - 7\left(\Psi^{(1)}\right)^2-\frac{1}{\cH^2}\left( \p_\| v_\|^{(1)}\right)^2-\frac{1}{\cH^2}\left( \Psi^{(1)}{'} \right)^2-2\Phi^{(1)}\Psi^{(1)} -2\big|\gamma^{(1)}\big|^2-2\left(\kappa^{(1)}\right)^2+\vartheta_{ij}^{(1)}\vartheta^{ij(1)}     \nonumber \\
&&-\frac{2}{\cH}\Phi^{(1)} \Psi^{(1)}{'} -\frac{2}{\cH}v_\|^{(1)}\p_\| v_\|^{(1)}+\frac{2}{\cH}v_\|^{(1)} \Psi^{(1)}{'} +\frac{2}{\cH}\Psi^{(1)}\p_\| v_\|^{(1)}-\frac{2}{\cH}\Psi^{(1)} \Psi^{(1)}{'} +\frac{2}{\cH^2} \Psi^{(1)}{'} \p_\| v_\|^{(1)}+\frac{2}{\cH}\Phi^{(1)} \p_\| v_\|^{(1)}\nonumber \\
&&\nonumber \\
&&+   v^{(1)}_{\perp i} v^{i (1)}_{\perp}+2\p_{\parallel}\left( +3\Psi^{(1)} - v^{(1)}_{\| } -\delta_g^{(1)} \right)T^{(1)} -  \frac{4}{\bar \chi}  \kappa^{(1)}T^{(1)}- \frac{2}{\bar \chi^2} \left( T^{(1)} \right)^2 \nonumber \\
&&+ \frac{2}{\cH} \left(-\p_\| v_\|^{(1)}-\cH v_\|^{(1)} -\p_\| \Phi^{(1)}+3\frac{\ud \,}{\ud \bar \chi} \Psi^{(1)} -\frac{\ud \,}{\ud  \bar \chi} \delta_g^{(1)} - \frac{2}{\bar \chi^2}   T^{(1)} - \frac{2}{\bar \chi}    \kappa^{(1)}\right) \Delta \ln a^{(1)} \nonumber \\
&&  +2  \frac{\cH'}{\cH^2}\left(\Phi^{(1)}  - v^{(1)}_{\| }+\Psi^{(1)}- \frac{1}{\cH} \p_\| v_\|^{(1)} +\frac{1}{\cH} \Psi^{(1)}{'} \right) \Delta \ln a^{(1)}+ \left(- b_e +  \frac{\ud \ln b_e}{\ud  \ln \bar a} -  \left(\frac{\cH'}{\cH^2} \right)^2 - \frac{2}{\bar \chi^2 \cH^2}\right) \left( \Delta \ln a^{(1)}\right)^2  \nonumber \\
&&  -2 \p_{\perp i}\left(v^{(1)}_{\|}-3 \Psi^{(1)}  +\delta_g^{(1)} \right)\int_0^{\bar \chi} \ud \tilde \chi \left[ \left(\bar \chi-\tilde \chi\right)  \tilde \p^i_\perp \left( \Phi^{(1)}+ \Psi^{(1)} \right) \right]  + 2v^{i (1)}_{\perp } \p_{\perp i} T^{(1)}\nonumber \\
&&+ 4S_{\perp}^{i(1)}  \bigg\{ - \frac{1}{\bar \chi}\int_0^{\bar \chi} \ud \tilde \chi \left[ \left(\bar \chi-\tilde \chi\right) \tilde \p_{\perp i} \left( \Phi^{(1)}+ \Psi^{(1)} \right)\right] +\p_{\perp i}   \left(\frac{1}{\cH}  \Delta \ln a^{(1)}+   T^{(1)}  \right) \bigg\} +2v^{ (1)}_{\perp i \, o }  v^{i (1)}_{\perp \, o }  \nonumber \\
  & & -2v^{ (1)}_{\perp i \, o } \bigg\{ \bar \chi \p^i_{\perp}\left(v^{(1)}_{\|} -3  \Psi^{(1)} + \delta_g^{(1)} \right)+ 2S_{\perp}^{i(1)}+ \p^i_{\perp}   \left(\frac{1}{\cH}  \Delta \ln a^{(1)}+   T^{(1)}  \right)  - \frac{1}{\bar \chi} \int_0^{\bar \chi} \ud \tilde \chi \left[  \left(\bar \chi-\tilde \chi\right) \tilde \p^i_\perp \left( \Phi^{(1)} +\Psi^{(1)} \right)\right]\bigg\} \nonumber \\
 &&-2 \frac{\bar a^2}{\cH \bar \rho_m} \left( \mathcal{E}_m^{\| (1)}- \frac{\cH}{\bar a^2} \bar \rho_m b_m v_\|^{(1)} \right)\left[\Phi^{(1)}  - v^{(1)}_{\| }+\Psi^{(1)}- \frac{1}{\cH} \p_\| v_\|^{(1)} +\frac{1}{\cH} \Psi^{(1)}{'}- \left(1+\frac{\cH'}{\cH^2}\right)\Delta \ln a^{(1)}\right]\nonumber \\
&&-\left( \frac{\bar a^2}{\cH \bar \rho_m} \right)^2\left(\mathcal{E}_m^{\| (1)}- \frac{\cH}{\bar a^2} \bar \rho_m b_m v_\|^{(1)}\right)^2\;. 
 \end{eqnarray}

\end{document}